\documentclass[a4paper,11pt,english,longbibliography]{article}
\usepackage{jheppub}

\renewcommand{\Im}{{\rm Im \, }}
\renewcommand{\Re}{{\rm Re \, }}

\renewcommand{\vec}[1]{\mathbf{#1}}
\newcommand{\CO}{{\cal O}}

\newcommand{\CD}{{\cal D}}
\newcommand{\CL}{{\cal L}}

\newcommand{\CK}{{\cal K}}

\newcommand{\CV}{{\cal V}}

\newcommand{\CF}{{\cal F}}
\newcommand{\CP}{{\cal P}}

\newcommand{\mY}{{\mathbb Y}}

\newcommand{\so}{{\mathfrak{so}}}

\newcommand{\SO}{{\text{SO}}}

\newcommand{\CJ}{{\cal J}}
\newcommand{\CM}{{\cal M}}

\newcommand{\Real}{\mathbb{R}}

\newcommand{\bFF}[4]{\,\mathbf{F}\left(#1,#2,#3,#4\right)}

\usepackage{tikz}
\usetikzlibrary{calc,decorations.markings}
\usetikzlibrary{shapes.misc}
\tikzset{cross/.style={cross out, draw=black, minimum size=2*(#1-\pgflinewidth), inner sep=0pt, outer sep=0pt},
cross/.default={3pt}}
\newcommand{\limu}[1]{\mathrel{\mathop{\sim}\limits_{\scriptstyle{#1}}}}

\usepackage{amsthm}

\makeatletter
\newcommand*{\rom}[1]{\expandafter\@slowromancap\romannumeral #1@}
\makeatother
\newcommand{\ket}[1]{\bigl| #1 \bigr\rangle}
\newcommand{\bra}[1]{\bigl\langle #1 \bigr|}

\usepackage{verbatim}
\usepackage{prettyref}
\usepackage[numbers,sort&compress]{natbib}
\usepackage{amsmath}
\usepackage{amssymb}
\usepackage{framed}
\usepackage{tikz-cd}
\tikzset{commutative diagrams/row sep/huge=4cm}
\tikzset{commutative diagrams/column sep/huge=4cm}
\tikzcdset{scale cd/.style={every label/.append style={scale=#1},
    cells={nodes={scale=#1}}}}
\usepackage{lmodern}
\usepackage{tcolorbox}
\usepackage{cases}
\usepackage{bbold}
\usepackage{bm}
\usepackage{ytableau}
\usepackage{youngtab}
\usepackage{mathtools}
\usepackage{graphicx}
\usepackage[nottoc]{tocbibind}
\usepackage{mathrsfs}
\usepackage{caption}
\usepackage{subcaption}
\usepackage{cancel}
\usepackage{float}
\usepackage{tikz}
\usepackage{lipsum}
\usepackage{adjustbox}
\usepackage{mathtools}
\usepackage{braket}
\usepackage{pgfplots}
\pgfplotsset{width=10cm,compat=1.9}
\usetikzlibrary{snakes}

\newcommand{\Kallen}{K{\"a}ll{\'e}n-Lehmann }
\def\ba{\begin{equation}\begin{aligned}}
\def\ea{\end{aligned}\end{equation}}
\newcommand{\reef}[1]{(\ref{#1})}

\tikzset{
    Witten diagram/.style={
        execute at begin picture={
            \draw[black,line width=1.5pt] circle[radius=\pgfkeysvalueof{/tikz/Witten/radius}];
            \path node (X){\phantom{X}};
        },
        baseline={(X.base)}
    },
    vertex/.style={circle,fill,inner sep=3pt,node contents={}},
    Witten/.cd,
    radius/.initial=3cm
}
\newenvironment{wittendiagram}[1][]{\begin{tikzpicture}[Witten diagram,#1]}{\end{tikzpicture}}

 \usepackage{slashed}
\usepackage{esvect}
\usepackage{dsfont}

\usepackage{color}
\definecolor{shadecolor}{gray}{0.95}
\definecolor{darkgreen}{rgb}{0,0.5,0}
\definecolor{darkblue}{rgb}{0,0,0.6}
\definecolor{purple}{rgb}{0.4,.2,0.7}
\definecolor{light-gray}{gray}{0.97}

\numberwithin{equation}{section}
\numberwithin{figure}{section}
\numberwithin{table}{section}

\def\KL{{K\"all\'en-Lehmann}}
\def\CG{{\cal G}} 
 
\def\CH{{\cal H}}

\def\CM{{\cal M}}
\def\CN{{\cal N}}
\def\CD{{\cal D}}

\def\D{\Delta}
\def\be{\begin{equation}}
\def\ee{\end{equation}}
\def\half{\frac{1}{2}}
\def\hd{\frac{d}{2}}

\def\IR{{\mathbb R}}

\newcommand{\CA}{\mathcal{A}}

\DeclareFontShape{OT1}{cmr}{mx}{n}{<->cmr10}{}

\title{The Käll\'en–Lehmann representation in de Sitter spacetime}
\author[a]{Manuel Loparco,} 
\author[a]{João Penedones,}
\author[a,b]{Kamran Salehi Vaziri}
\author[c]{and Zimo Sun}

\affiliation[a]{Fields and Strings Laboratory, Institute of Physics\\
École Polytechnique Fédéral de Lausanne (EPFL)\\
Route de la Sorge, CH-1015 Lausanne, Switzerland}
\affiliation[b]{Institute of Physics, University of Amsterdam, Amsterdam, 1098 XH, The Netherlands}
\affiliation[c]{Princeton Gravity Initiative, Princeton University, Princeton, NJ 08544, USA}
\emailAdd{manuel.loparco@epfl.ch, joao.penedones@epfl.ch, k.salehivaziri@uva.nl, zs8479@princeton.edu}

\abstract{
We study two-point functions of  symmetric traceless local operators in the bulk of de Sitter spacetime. We derive the \KL \, spectral decomposition for any spin and show that unitarity implies its spectral densities are nonnegative.
In addition, we recover the \KL\ decomposition in Minkowski space by taking the flat space limit.
Using  harmonic analysis and the Wick rotation to Euclidean Anti de Sitter, we derive  an inversion formula to compute the spectral densities.
Using the inversion formula, we 
relate  
the analytic structure of the spectral densities to the late-time boundary operator content. 
We apply our technical tools to study two-point functions of composite operators in free and weakly coupled theories.
In the weakly coupled case, we show how the \KL\ decomposition is useful to find the anomalous dimensions of the late-time boundary operators. 
We also derive the \Kallen representation of two-point functions of spinning primary operators of a Conformal Field Theory on de Sitter.
}

\begin{document}
\maketitle
\flushbottom
\section{Introduction}
\label{sec:introduction}

de Sitter (dS) spacetime is the simplest model of an expanding universe.
Therefore, understanding Quantum Field Theory (QFT) in dS spacetime is the first step towards a description of quantum effects in Cosmology.

The main ingredients of QFT are states in the Hilbert space and local operators labeled by points in spacetime.
We have recently  given a systematic account of the Hilbert space of free QFTs and Conformal Field Theories (CFTs) in dS \cite{Penedones:2023uqc} and its decomposition in  Unitary Irreducible Representations (UIRs) of the isometry group $SO(d+1,1)$ of dS$_{d+1}$ (see section \ref{subsec:UIRs} for a brief review).
In general, this leads to the  decomposition of the identity as a sum/integral over projectors into UIRs,
\begin{align}
   \mathbb 1 =  \sum_{\ell=0}
   \int_{\mathbb R} d\lambda 
   \ \mathbb 1_{\mathcal P_{\Delta  ,\ell}} +\cdots
   \label{eq:resolution1}
\end{align}
where we show explicitly the contribution from principal series with dimension $\Delta=\frac{d}{2}+i\lambda$ and $SO(d)$ spin $\ell$ and the dots stand for other UIRs.

In this article, we continue the groundwork and derive the \Kallen representation of two-point functions of bulk local operators in the Bunch-Davies vacuum of dS. 
We systematize and extend the results of previous works \cite{Bros:2009bz,Bros_1998,BROS_1996,DiPietro:2021sjt,Hogervorst:2021uvp,Schaub:2023scu,BROS199122,Hollands_2011,Epstein:2012zz}. 
In particular, 
we employ the embedding space formalism to efficiently  treat the case of bosonic traceless symmetric  operators in arbitrary spacetime dimensions.

The \Kallen decomposition of a two-point function is simply obtained by inserting the resolution of the identity 
\eqref{eq:resolution1} in the middle of a Wightman two-point function. 
For example, for a two-point function of operators of spin $J$ in $d\geq 2$ we find
\small
\begin{align}
\langle \CO^{(J)}(Y_1,W_1) \CO^{(J)}(Y_2,W_2) \rangle 
&=\sum_{\ell=0}^J\int_{\mathbb R}\, d\lambda\,  
\langle \CO^{(J)}(Y_1,W_1) 
\mathbb{1}_{\mathcal P_{\Delta =\frac{d}{2}+i\lambda ,\ell}}
\CO^{(J)}(Y_2,W_2) \rangle 
 +\cdots \label{eq:KLIntro}
\\
 &=\sum_{\ell=0}^J\int_{\mathbb R}\, d\lambda\, \rho_{\CO^{(J)}}^{\mathcal P,\ell}( \lambda) \left[\left(W_1\cdot\nabla_{1}\right) \left(W_2\cdot\nabla_{2}\right)\right]^{J-\ell}G_{\lambda,\ell}(Y_1, Y_2; W_1, W_2) 
 +\cdots \nonumber
\end{align}
\normalsize
As explained in section \ref{subsec:embeddingspace}, $Y$  encodes the position in dS and $W$ encodes the indices of a tensor field. 
In section \ref{sec:kallanlehmann}, we show that $\langle \CO^{(J)}(Y_1,W_1) 
\mathbb{1}_{\mathcal P_{\Delta=\frac{d}{2}+i\lambda  ,\ell}}
\CO^{(J)}(Y_2,W_2) \rangle $ can be written in terms of the propagator
$G_{\lambda,\ell}(Y_1, Y_2; W_1, W_2)$  of a free field of spin $\ell$ and mass squared 
\begin{equation}
    m^2R^2=\lambda^2+\left(\frac{d}{2}+\ell-2+2\delta_{\ell,0}\right)^2\,,
\end{equation}
with $R$ the curvature radius of dS and $\delta_{\ell,0}$ is a Kronecker delta.
Therefore, all the dynamical information is encoded in the spectral densities $\rho_{\CO^{(J)}}^{\mathcal P,\ell}( \lambda)$ associated to intermediate states in  principal series UIRs.
The dots in (\ref{eq:KLIntro}) stand for the contribution of other UIRs. In particular, we also determine the contributions from the complementary series\footnote{A practical way to think about the contribution from complementary series states is as poles of $\rho_{\CO^{(J)}}^{\mathcal P,\ell}( \lambda)$ that crossed the integration contour. We give several examples of this phenomena in section \ref{sec:applications}.}
and the discrete series in the case of dS$_2$. This completes the picture in dS$_2$ and dS$_3$\footnote{In dS$_3$ there are no non-trivial UIRs beyond the principal and complementary series.} where we have derived all the contributions to the \KL\ representation.
In section \ref{sec:kallanlehmann}, we prove the positivity of the dS spectral densities and show how they morph into the standard flat space spectral densities in the limit $R\to \infty$.

In section \ref{sec:inversionformula}, we present explicit inversion formulae that give the spectral densities $\rho_{\CO^{(J)}}^{\mathcal P,\ell}( \lambda)$ as integrals over the associated two-point functions. 
To derive these formulae we analytically continue the two-point functions to Euclidean Anti-de Sitter (EAdS) space and then use harmonic analysis.
The inversion formulae imply a strip of analyticity in the $\lambda$ complex plane centered around the integration contour in (\ref{eq:KLIntro}).\footnote{The width of the strip is fixed by the asymptotic behavior of the two-point function or, equivalently, by the leading boundary operator in the late time expansion. See sections \ref{subsec:completeness} and 
  \ref{subsubsec:boundarytheory}.}
 In addition, we predict the presence of spurious (or kinematical) poles with fixed residues in the spectral densities. Assuming meromorphicity in $\lambda$, we derive the late time expansion of two-point functions and interpret it as a Boundary Operator Expansion (BOE). It would be interesting to understand the convergence properties of this BOE, and whether the same BOE can be used inside all correlation functions.
 
In section \ref{sec:applications}, we study   many examples (CFTs, free and weakly coupled QFTs) and always find spectral densities that are meromorphic functions of $\lambda\in \mathbb{C}$ and have the predicted spurious poles.  For  weakly coupled QFTs, we show how the \Kallen
decomposition can be used to compute  anomalous dimensions of late time boundary operators.

In section \ref{sec:conclusion}, we discuss possible future directions, and in the appendices we elaborate on many technical details.

Throughout the paper, we will guide the eye of the reader by highlighting important equations. Here we list some of them as a summary of our main results:
\begin{itemize}\setlength\itemsep{-0.1em}
    \item \KL\ decomposition of spin $J$ operators in
    \vspace{-0.3cm}
    \begin{itemize}\setlength\itemsep{-0.1em}
        \item dS$_2$ \reef{2DspinJfull},
        \item dS$_{d+1}$ with $d\geq 2$ \reef{eq:KL any J};
    \end{itemize}
    \item Flat space limit of the spectral densities (\ref{eq:rholimit});
    \item Inversion formula for the spectral densities of
     \begin{itemize}\setlength\itemsep{-0.1em}
     \vspace{-0.3cm}
        \item principal and discrete series in dS$_2$ (\ref{eq:2Dinver}),
        \item principal series in dS$_{d+1}$ with $d\geq 2$ (\ref{eq:inversionformulal});
    \end{itemize}
    \item Boundary operator expansion (\ref{eq:BOE}).
\end{itemize}

\noindent This work fits within the recent efforts to constrain QFT observables in dS by using general principles such as unitarity and symmetries ~\cite{Baumann:2022jpr,Arkani-Hamed:2015bza,Arkani-Hamed:2018kmz,Pajer:2020wxk,Goodhew:2020hob,Green:2020whw,Goodhew:2021oqg,Melville:2021lst, Hogervorst:2021uvp,DiPietro:2021sjt,Kravchuk:2021akc,Bissi:2023bhv}. 
\section{Preliminaries}
\label{sec:preliminaries}

In this section, we review two mathematical tools that will be very useful in the derivation of the \KL\, decomposition and the computation of spectral densities. The first topic concerns UIRs of the de Sitter isometry group $SO(d+1, 1)$, and the second topic is  the embedding space formalism.  

\subsection{Representation theory of de Sitter isometry group}\label{subsec:UIRs}
The $(d+1)$ dimensional de Sitter spacetime is a hypersurface in the embedding space $\mathbb{R}^{d+1, 1}$
\begin{align}\label{embdS}
-Y_0^2+Y_1^2+\cdots+ Y_{d+1}^2=R^2,
\end{align}
where $R$ is the de Sitter radius. The embedding (\ref{embdS}) manifests the  isometry group  $SO(d+1,1)$ of $\text{dS}_{d+1}$, which is generated by $L_{AB}=-L_{BA}, 0\le A,B\le d+1$ satisfying commutation relations 
\begin{align}\label{defiso}
[L_{AB}, L_{CD}]=\eta_{BC} L_{AD}-\eta_{AC} L_{BD}+\eta_{AD} L_{BC}-\eta_{BD} L_{AC},
\end{align} 
where $\eta_{AB}=\text{diag}(-1,1,\cdots, 1)$ is the metric on $\mathbb{R}^{d+1,1}$. In a unitary representation, $L_{AB}$ are realized as anti-hermitian operators on some Hilbert space.
The isomorphism between $\so(d+1,1)$ and the $d$-dimensional Euclidean conformal algebra is realized as
\begin{align}\label{defconf}
L_{ij}=M_{ij}~, \,\,\,\,\, L_{0, d+1}=D~, \,\,\,\,\, L_{d+1, i}=\frac{1}{2}(P_i+K_i)~, \,\,\,\,\, L_{0, i}=\frac{1}{2}(P_i-K_i)
\end{align}
where $D$ is the dilatation, $P_i$ ($i=1, 2,\cdots d$) are translations, $K_i$ are special conformal transformations and $M_{ij}=-M_{ji}$ are  rotations.
The commutation relations of the conformal algebra following from (\ref{defiso}) and (\ref{defconf}) are
\begin{align}\label{confalg}
&[D, P_i]=P_i~, \,\,\,\,\, [D, K_i]=-K_i~, \,\,\,\,\, [K_i, P_j]=2\delta_{ij}D-2M_{ij}~,\nonumber\\
&[M_{ij}, P_k]=\delta_{jk} P_i-\delta_{ik}P_j~, \,\,\,\,\,[M_{ij}, K_k]=\delta_{jk} K_i-\delta_{ik}K_j~,\nonumber\\
&[M_{ij}, M_{k\ell}]=\delta_{jk} M_{i\ell}-\delta_{ik} M_{j\ell}+\delta_{i\ell} M_{jk}-\delta_{j\ell} M_{ik}~.
\end{align}
The quadratic Casimir of $SO(d+1,1)$, which commutes with all $L_{AB}$, is chosen to be 
\begin{align}\label{generalcas}
C^{SO( d+1,1)}&=\frac{1}{2}L_{AB}L^{AB}=D(d-D)+P_i K_i+\frac{1}{2} M_{ij}^2~.
\end{align}
Here $\frac{1}{2}M_{ij}^2\equiv \frac{1}{2} M_{ij}M^{ij}$ is the quadratic Casimir of $SO(d)$ and it is negative-definite for a unitary representation since $M_{ij}$ are anti-hermitian. For example, for a spin-$s$ representation of $SO(d)$, it takes the value of $-s(s+d-2)$.

\subsubsection{Classification of UIRs}
An irreducible infinite dimensional representation of $SO(d+1,1)$ is  fixed by a complex parameter $\Delta$ \footnote{We will often call this parameter a scaling dimension. But it does not have the same group theoretical meaning as scaling dimensions in unitary CFT, since it is not associated to any operator bounded from below.} and a highest-weight vector $\vec S$ of  $SO(d)$. Throughout the paper, we will only consider $\vec S=(s, 0,\cdots,0)$, i.e. spin $s$ representation of $SO(d)$. Such representations corresponds to the single-particle Hilbert space of a free spin $s$ field in dS$_{d+1}$.
More general $\vec{S}$ describes fields of mixed symmetry, including form fields, spinors, tensor spinors, etc. See \cite{Basile:2016aen,A_Letsios_2021, Pethybridge_2022, https://doi.org/10.48550/arxiv.2206.09851} for recent discussions on these fields.
Fixing $\Delta$ and $s$, the quadratic Casimir is equal to $\Delta(d-\Delta)-s(d+s-2)$. For any $d\ge 3$, there are four types of UIRs apart from the trivial representation \cite{Dobrev:1977qv,Basile:2016aen,Sun:2021thf}:
\begin{itemize}
\item \textbf{Principal series} $\mathcal P_{\Delta, s}$: $\Delta\in\frac{d}{2}+i\mathbb R$ and $s\ge 0$.
\item \textbf{Complementary series} $\mathcal C_{\Delta, s}$: $0<\Delta<d$ when $s=0$ and $1<\Delta<d-1$ when $s\ge 1$. Both principal  and complementary series describe free massive particles in dS$_{d+1}$. 
\item \textbf{Type \rom{1} exceptional series } $\mathcal V_{p,0}$: $\Delta=d+p-1$ and $s=0$ for $p\in\mathbb Z_{>0}$. They correspond to shift symmetric scalars in dS$_{d+1}$ \cite{Bonifacio:2018zex}. 
\item \textbf{Type \rom{2} exceptional series } $\mathcal U_{s, t}$: $\Delta=d+t-1$ and $s\ge 1$ with $t=0,1,2\cdots, s-1$. 
The single-particle Hilbert space of a partially massless field of spin $s$ and depth $t$ in dS$_{d+1}$ furnishes the representation $\mathcal U_{s,t}$. 
\end{itemize}
When $d=2$, there are only principal series and complementary series up to isomorphism \cite{Dirac:1945cm,10.2307/97833,10.2307/1969129,GelNai47}, and  the complementary series representations always have $s=0$.  
When $d=1$, since the $SO(d)$ group becomes degenerate, the Casimir of $SO(2,1)$ can always be written as $\Delta(1-\Delta)$. The classification of UIRs is  as follows:
\begin{itemize}
\item \textbf{Principal series} $\mathcal P_{\Delta}$: $\Delta\in\frac{1}{2}+i\mathbb R$. Its restriction to $SO(2)$ yields \cite{Anous_2020,Marolf_2009}
\begin{align}\label{CPcomp2}
\left. \mathcal P_{\Delta}\right|_{SO(2)}=\bigoplus_{n\in\mathbb Z}(n)
\end{align}
where $(n)$ denotes the (one-dimensional) spin $n$ representation of $SO(2)$.
\item \textbf{Complementary series} $\mathcal C_{\Delta}$: $0<\Delta<1$. It has the same $SO(2)$ content as  $\mathcal P_{\Delta}$.
\item \textbf{Lowest-weight discrete series $\mathcal D^+_p$}: $C^{\SO(2,1)}=p(1-p),\, p\in\mathbb Z_+$. Its $SO(2)$ spectrum has a lower bound $p$.
\item \textbf{Highest-weight discrete series $\mathcal D^-_p$}: $C^{\SO(2,1)}=p(1-p),\, p\in\mathbb Z_+$.  Its $SO(2)$ spectrum has an upper bound $-p$.
\end{itemize}
There is an isomorphism between representations of scaling dimension $\Delta$ and $\bar\Delta=d-\Delta$ in the principal and complementary series, which is established by the shadow transformation. To remove such redundancy, one can further impose, for example, $\Im (\Delta)\ge 0$ in the principal series and $\Delta>\frac{d}{2}$ in the complementary series.

\subsubsection{Hilbert spaces of the UIRs} \label{sec:Hilbert}
As we will see in section \ref{sec:kallanlehmann}, the derivation of the \KL\, representation in de Sitter spacetime requires a detailed knowledge of the Hilbert space of each UIR listed above. The complementary series can be treated as a simple analytical continuation of the principal series in the derivation of the \KL\, representation (see Appendix \ref{sec:comp} for more details). The two exceptional series are absent in all examples considered in this paper. So we will only briefly review the Hilbert space of the principal series representation $\mathcal P_{\Delta, s}$ in any dimension (including $d=1$), and the discrete series representation $\mathcal D^\pm_p$ in dS$_2$,  by  following \cite{Sun:2021thf}. Given a principal series representation $\mathcal P_{\Delta, s}$, its Hilbert space is spanned by a continuous family of $\delta$ function normalized kets $|\Delta,\vec y\,\rangle_{i_1 \cdots i_s}$. Here $\vec y$ labels a point in $\mathbb R^d$, and the indices $\{i_1,i_2,\cdots i_s\}$, being symmetric and traceless, carry the spin $s$ representation of $SO(d)$. The action of the $\so(d+1,1)$ algebra on these states is realized by 
\ba\label{actonstate}
& P_i |\Delta, \vec y\,\rangle_{i_1\cdots i_s}=\partial_i |\Delta, \vec y\, \rangle_{i_1\cdots i_s}~,\\
& D|\Delta, \vec y\, \rangle_{i_1\cdots i_s}=(\vec y\cdot\partial_{\vec y} +\Delta)|\Delta, \vec y\,\rangle_{i_1\cdots i_s}~,\\
& M_{k\ell}|\Delta, \vec y\,\rangle_{i_1\cdots i_s}=\left(y_\ell \partial_k-y_k\partial_\ell+\CM^{(s)}_{k\ell }\right)|\Delta, \vec y\, \rangle_{i_1\cdots i_s}~,\\
& K_k |\Delta, \vec y\,\rangle_{i_1\cdots i_s}=\left(2y_k(\vec y\cdot \partial_{\vec y}+\Delta)-y^2\partial_k-2y^\ell \CM_{k\ell}^{(s)}\right)|\Delta, \vec y\,\rangle_{i_1\cdots i_s}~,
\ea
where $\CM^{(s)}_{k\ell}$ denotes the spin-$s$ representation of $\so(d)$
\begin{align}
\CM^{(s)}_{k\ell}|\Delta, \vec y\, \rangle_{i_1\cdots i_s}=\sum_{j=1}^s |\Delta, \vec y\,\rangle_{i_1\cdots i_{j-1} k \,i_{j+1}\cdots i_s}\delta_{\ell i_j}-|\Delta, \vec y\, \rangle_{i_1\cdots i_{j-1} \ell\, i_{j+1}\cdots i_s}\delta_{k i_j}~.
\end{align} 
By introducing an auxiliary null vector $z^i\in\mathbb C^d$, we define $|\vec y,\vec z\,\rangle_{\Delta,s} \equiv |\Delta, \vec y\,\rangle_{i_1\cdots i_s}\, z^{i_1}\cdots z^{i_s}~$,
which packages all the tensor components of $|\Delta, \vec y\,\rangle_{i_1\cdots i_s}$ into a generating function and also allows us to state the normalization condition of $|\Delta, \vec y\,\rangle _{i_1\cdots i_2}$ concisely 
\begin{align}
    _{\Delta,s}\langle\vec y_1, \vec z_1| \vec y_2, \vec z_2\rangle_{\Delta,s}=\delta^d(\vec y_1-\vec y_2)(\vec z_1\cdot \vec z_2)^s~.
\end{align}
Fixing this normalization, the resolution of the identity of $\mathcal P_{\Delta, s}$ is given by 
\begin{align}\label{eq:residen}
    \mathbb 1_{\mathcal P_{\Delta,s}}=\int d^d\vec y\, |\Delta, \vec y\,\rangle_{i_1\cdots i_s}\,^{i_1\cdots i_s}\langle \Delta, \vec y\,|=\frac{1}{\left(\frac{d}{2}-1\right)_s}\int d^d\vec y\, |\vec y, D_{\vec z}\rangle_{\Delta,s} \,_{\Delta,s}\langle \vec y, \vec z|~,
\end{align}
where $D_{\vec z}$ is the analogue of the ordinary derivative while preserving the nullness condition of $\vec z$
\begin{align}\label{intder}
D_{z^i}\equiv \left(\frac{d}{2}-1+\vec z\cdot\partial_{\vec z}\right)\partial_{z^i}-\frac{1}{2} z_i\, \partial_z^2~.
\end{align}
A generic normalizable state in the Hilbert space of $\mathcal P_{\Delta,s}$ can be expressed as a linear combination of $|\Delta,\vec y\,\rangle_{i_1\cdots i_s}$
\begin{align}
|\psi\rangle \equiv \int_{\mathbb R^d}\,d^d\vec y \, \psi_{i_1\cdots i_s}(\vec y)\,|\Delta, \vec y\,\rangle_{i_1\cdots i_s}~,
\end{align}
where $ \psi_{i_1\cdots i_s}(\vec y\,)$ is a smooth tensor valued wavefunction on $\mathbb R^d$, satisfying a certain fall-off condition at $\infty$ \cite{Sun:2021thf}.

\paragraph{In the $d=1$ case,} it is easier to describe the UIRs by using the following basis of $SO(2,1)$
\begin{align}\label{defineLpm}
    L_0=-\frac{i}{2} (P+K), \,\,\,\,\, L_\pm=-\frac{i}{2} (P-K)\mp D~,
\end{align}
where $L_0$ is the (hermitian) generator of the $SO(2)$ subgroup, and hence has integer eigenvalues in any single-valued representation of $SO(2,1)$. The new basis satisfy the commutation relations 
\begin{align}
[L_0,L_\pm]=\pm L_\pm, \,\,\,\,\, [L_-, L_+]=2 L_0~,
\end{align}
and reality conditions $L_0^\dagger=L_0, L_\pm^\dagger = L_\mp$. The principal series representation $\mathcal P_\Delta$ is spanned by eigenstates $\{|n\rangle_\Delta, n\in\mathbb Z\}$ of $L_0$, on which $L_\pm$ act as 
\begin{align}\label{Lpmact}
 L_\pm |n\rangle_\Delta=(n\pm\Delta)|n\pm 1\rangle_\Delta~.   
\end{align}
The inner product compatible with the reality conditions and the action (\ref{Lpmact}) is of the form $_\Delta\langle n|m\rangle_\Delta = c\,\delta_{nm}$, where  $c$ is a positive constant. We can simply choose $c=1$. With this choice fixed, the continuous $|y\rangle$ basis reviewed above is related to the discrete $|n\rangle_\Delta$ basis via the wavefunction $\psi_n (y)=\left(\frac{1-i y}{1+i y}\right)^n\frac{\pi^{-\frac{1}{2}}}{(1+y^2)^{\Delta}}$.

When $\Delta$ is a positive integer, say $\Delta=p$, 
the action of $L_\pm$ is  truncated at $n=\mp p$, leading to  two irreducible representations. These two representations are actually $\mathcal D^\pm_p$:
\begin{align}
    \mathcal D_p^+=\text{Span}\{|n\rangle_p,\, n\ge p\}, \,\,\,\,\,\  \mathcal D_p^-=\text{Span}\{|n\rangle_p,\, n\le -p\}~.
\end{align}
In this case, with the action (\ref{Lpmact}) being fixed, the simple normalization $_p\langle n|m\rangle_p=\delta_{nm}$ is not consistent with the reality condition $L_\pm^\dagger = L_\mp$. Instead, we need to use
\begin{align}
    _p\langle n|m\rangle_p=\frac{\Gamma(|n|+1-p)}{\Gamma(|n|+p)}\delta_{n,m}
\end{align}
So the resolution of the identity of $ \mathcal D_k^\pm$ becomes
\begin{align}\label{resodis}
\mathbb 1_{ \mathcal D_p^\pm } = \sum_{\pm n\ge p} \frac{\Gamma(|n|+p)}{\Gamma(|n|+1-p)} |n\rangle_p\,_p\langle n|
\end{align}

\subsection{Embedding space formalism}
\label{subsec:embeddingspace}
In this paper, we study symmetric traceless  tensor fields in dS. The embedding space formalism turns out to be very useful in the derivation of the \Kallen representation in section~\ref{sec:kallanlehmann} and the inversion formula for the spectral densities in section~\ref{sec:inversionformula} and~\ref{sec:applications}. In this section, we briefly describe the embedding space formalism for tensor fields in dS$_{d+1}$ \cite{Schaub:2023scu,Pethybridge_2022} following the similar construction in EAdS$_{d+1}$~\cite{Costa_2014}, and for the principal series representations of $SO(d+1, 1)$ adapting a  similar construction in CFT$_d$~\cite{Costa:2011mg}. We also notice that the construction in \cite{Costa_2014} is  degenerate when $d=1$. So we will give a separate and self-contained discussion about the embedding space formalism in this case.

\subsubsection{Coordinate systems}
As mentioned in eq.~\reef{embdS}, de Sitter spacetime can be seen as a hypersurface 
in embedding space $\mathbb{R}^{d+1, 1}$
. Among the different slicings and coordinate systems, we will use (conformal) global coordinates and planar  coordinates throughout this paper. Global coordinates are defined as 
\begin{equation}
Y^0 = R\, \sinh t~, \qquad Y^a = R\,\Omega^a \cosh t \,
\end{equation}
in which $t\in\IR$, $a=1,\ldots,d+1$ and $\Omega^a \in S^d \subset \mathbb{R}^{d+1}$ is a unit vector ($\Omega^a \Omega_a = 1$). The induced metric in global coordinates is given by
\begin{align}
  \label{eq:globalMetric}
ds^2 = R^2 \left(-dt^2 + \cosh^2 t\, d\Omega_d^2\right)
\end{align}
where $d\Omega^2_d$ denotes the standard metric of the unit $S^d$.
With a change of coordinate $\sinh t= \tan \tau$, the metric eq. (\ref{eq:globalMetric}) becomes 
\begin{align}\label{conformalglobal}
    ds^2=R^2\frac{-d\tau^2+d\Omega^2_d}{\cos^2\tau}  
\end{align}
    which is conformally equivalent to a finite cylinder $(-\frac{\pi}{2}, \frac{\pi}{2})\times S^d$. The coordinates $(\tau,\Omega^a)$ are called conformal global coordinates. 
    
\begin{figure}
\centering
\includegraphics[scale=1.2]{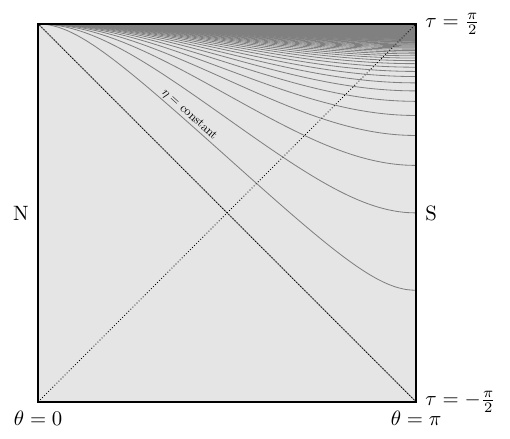}
\caption{Penrose diagram of de Sitter spacetime. We represent global conformal time $\tau$ on the vertical axis and the azimuthal angle $\theta$ on the horizontal axis. We indicate with S the south pole and N the north pole of the Cauchy slices of constant $\tau$, which are spheres. We represent Cauchy slices of constant planar time $\eta\in(-\infty,0)$ in dark gray. Planar coordinates only cover the top right half of global de Sitter.}
\label{fig:penrose}
\end{figure}

The planar coordinates $y^\mu=(\eta, \vec y)\in\mathbb R_{-}\times \mathbb R^d$, cover the causal future of
an observer at the south pole of the global $S^d$ (i.e. $Y^i=0$ for $i=1,2,\cdots,d$ with $Y^{d+1}<0$). They are given by
\begin{equation}
Y^0 =R\, \frac{\eta^2 - \vec{y}^2 -1}{2\eta}~,\qquad
Y^{i} = -R\,\frac{y^i}{\eta}~, \qquad
Y^{d+1} = R\,\frac{\eta^2 - \vec{y}^2 +1}{2\eta}
\label{Xetax}
\end{equation}
for which the induced metric is 
\begin{align}
ds^2=R^2\frac{-d\eta^2+d \vec y^2}{\eta^2}~.
\end{align}
At $\eta\to 0^-$, where the metric blows up, is the future boundary of dS$_{d+1}$. The region covered by $y^\mu$ corresponds to $Y^-\equiv Y^0-Y^{d+1}>0$, and hence it covers half of de Sitter spacetime, also called the Expanding Poincar\'e Patch (EPP). In figure \ref{fig:penrose}, we draw the Penrose diagram of de Sitter space with Cauchy slices of constant $\eta$.

From now on, we set the de Sitter radius to $R=1$, and will restore it when discussing the flat space limit in section \ref{subsec:flatspacelimit}. Let us emphasize that the results presented in this paper apply in any coordinate patch one chooses to work with. 

\subsubsection{Fields in embedding space}\label{fieldembedding}
Consider a spin-$J$ symmetric traceless  tensor\footnote{Let us emphasize that we use $\ell$ for the $SO(d)$ spin and $J$ for the $SO(d+1,1)$ spin. States carry $SO(d)$ spin while operators have $SO(d+1,1)$ indices.} $T_{A_1\cdots A_J}$(Y) in the embedding space $\IR^{d+1,1}$. Asking $Y^2=1$ and the tangential condition 
\be\label{tang}
Y^{A_1} T_{A_1\cdots A_J}(Y) = 0
\ee
defines a traceless symmetric tensor in de Sitter. The projection
\be\label{eq:pullback}
T_{\mu_1\cdots \mu_J}(y) = \frac{\partial Y^{A_1} }{\partial y^{\mu_1}} \cdots  \frac{\partial Y^{A_J} }{\partial y^{\mu_J}} T_{A_1\cdots A_J}(Y)
\ee
pulls back this tensor to the desired local coordinates $y^\mu$. Moreover,  we can represent a symmetric and traceless tensor in the index free formalism as a polynomial by contracting its indices with a null vector $W^A$
\be
T(Y,W) = W^{A_1} \ldots W^{A_J} T_{A_1\cdots A_J}(Y)~.
\ee
Due to the tangential condition (\ref{tang}), we can restrict to  $W^A$ such that  $Y\cdot W=0$.  
Altogether, a spin $J$ tensor $T_{\mu_1\cdots\mu_J}(y)$ is uniquely encoded in a degree $J$ homogeneous polynomial $T(Y, W)$, with $W^A$ satisfying $W^2=Y\cdot W=0$. The above discussion extends to differential operators. For example, the embedding space realization of the Levi-Civita connection $\nabla_\mu$ is given by 
\begin{align}\label{nablaA}
    \nabla_A = \partial_{Y^A}-Y_A \, Y\cdot \partial_Y-W_A\, Y\cdot\partial_W ~.
\end{align}

To recover the tensor $T_{A_1\cdots A_J}$ with indices in $d\geq 2$, one needs to act with the  differential operator 
\ba
K_A &= \left(\frac{d-1}{2}\right) \left[\frac{\partial}{\partial W^A} - Y_A\left(Y\cdot\frac{\partial}{\partial W}\right)\right]+ \left(W\cdot\frac{\partial}{\partial W}\right) \frac{\partial}{\partial W^A}  \\
& -Y_A \left(Y\cdot\frac{\partial}{\partial W}\right) \left(W\cdot\frac{\partial}{\partial W}\right) - \half \,W_A \left[\frac{\partial^2}{\partial W\cdot\partial W } - \left(Y\cdot\frac{\partial}{\partial W}\right) \left(Y\cdot\frac{\partial}{\partial W}\right) \right]
\label{eq:actionofKdS}
\ea
on the polynomial $T(Y,W)$, which is defined to be interior to the submanifold $Y^2-1=W\cdot Y=W^2=0$. Given this definition, it is straightforward to check that  $K_{[A}K_{B]}=K\cdot K=Y\cdot K=0$, and hence its action induces a symmetric and traceless tensor on dS$_{d+1}$. More precisely, it acts on any monomial of $W^A$ as 

\be
    K_{A_1}\cdots K_{A_J}W^{B_1}\cdots W^{B_J}=J!\left(\frac{d-1}{2}\right)_J\left[\frac{1}{J!}\sum_{\pi}G_{A_{\pi_1}}^{\ B_1}\cdots G^{\ B_J}_{A_{\pi_J}}-\text{traces}\right]\,~
    \label{eq:KWWW}
\ee
where $G_{AB}=\eta_{AB}-Y_AY_B$, and the sum is over all permutations $\pi$ of the indices $A_j$. As a simple application of $K_A$, the divergence of a tensor is implemented by $\nabla\cdot K$.

Treating $T(Y, W)$ as quantum fields, then the action of $\so(d+1,1)$ is defined as 
\begin{align}\label{bulkaction}
    [L_{AB}, T(Y, W)]=- \left(Y_A\partial_{Y_B}-Y_B\partial_{Y_A}+W_A\partial_{W_B}-W_B\partial_{W_A}\right)T(Y, W)
\end{align}
where the overall minus sign is introduced to ensure it to be a left action.

\paragraph{When $\mathbf{d=1}$,} the differential operator $K_A$ becomes purely second-order, and thus annihilates any vector fields in dS$_2$. The failure of $K_A$ to recover tensor indices in this case, is related to some subtleties of $SO(2)$ representations in contrast to higher dimensional $SO(d)$. We will discuss such subtleties and show how to modify the embedding space formalism of dS$_2$ accordingly. First, it is well-known that the spin $s$ representation of $SO(n)$ with $n\ge 3$, carried by a symmetric and traceless tensor $F_{i_1\cdots i_s}$, can be encoded in a degree $s$ polynomial $F(z)\equiv F_{i_1\cdots i_s}z^{i_1}\cdots z^{i_s}$, where $z^i$ is a null vector in $\mathbb C^n$. When $n=2$,  proceeding as in higher dimensions, the nullness condition
yields $(z^1+i z^2)(z^1-i z^2)=0$. So there are two different types of $z$, namely $z_\pm = (1, \pm i)$, which are related by $O(2)$ but not $SO(2)$.  $F(z_\pm)$ capture the only independent components of $F_{i_1\cdots i_s}$. Altogether, a symmetric and traceless tensor $T$ in 2D carries a two dimensional representation of $O(2)$, corresponding to the two chiral components of $F$. Each chirality furnishes an irreducible representation of $SO(2)$. In the index free formalism, the two chiralities are encoded in $F(z_\pm)$, where $z_\pm$ are two $SO(2)$-inequivalent null vectors.  Similarly in dS$_2$, a spin $J$ tensor $T_{\mu_1\cdots\mu_J}(y)$ also has two independent components, which should correspond to two $SO(2,1)$-inequivalent $W^A$ in embedding space. Indeed, we find that when $d=1$, the conditions $Y\cdot W=W^2=0$ are equivalent to $\epsilon_{ABC}Y^B W^C\pm W_A=0$, where $\epsilon_{ABC}$ is the totally antisymmetric tensor in $\mathbb R^{2,1}$ normalized as $\epsilon_{012}=1$.  Define $W_\pm^A$ such that  
\begin{align}\label{eYW}
    \epsilon^A_{\,\, BC}Y^B W_\pm^C\pm W^A_\pm=0~.
\end{align}
They are the analogue of $z_\pm$ defined above, in the sense that the two chiral componetns of $T_{\mu_1\cdots \mu_J}(y)$ are encoded in $T(Y, W_\pm)$ respectively. To prove this statement more precisely, let's consider the tensor $T_{\mu_1\cdots \mu_J}(y)$ in conformal global coordinates $Y^A=(\tan \tau, \frac{\cos\varphi}{\cos \tau},\frac{\sin\varphi}{\cos \tau})$.  Define lightcone coordinates $y^\pm = \tau\pm \varphi$, and then the two linearly independent components of $T_{\mu_1\cdots \mu_J}$ are  
\begin{align}
    T_{\pm\cdots\pm} = \partial_\pm Y^{A_1}\cdots \partial_\pm Y^{A_J} T_{A_1\cdots A_J}(Y)= T(Y, \partial_\pm Y)~.
\end{align}
It can be checked by direct computation that $W_\pm^A =\partial_\pm Y^A$ solves eq. (\ref{eYW}), which verifies our statement. Altogether, the tensor $T_{\mu_1\cdots \mu_J}$ in dS$_2$ is encoded in $T(Y, W_\pm)$, with $W_\pm^A$ satisfying (\ref{eYW}). The pull-back to the conformal global coordinates is easily implemented by the substitution $W_\pm^A\to \partial_\pm Y^A$.

Eq. (\ref{eYW}) can lead to some useful identities. For example, given  two distinct points $Y_1$ and $Y_2$ in $\mathbb R^{2,1}$, it implies 
\begin{align}\label{userela}
    &\left(Y_1\cdot W_2^\pm\right) \left(Y_2\cdot W_1^\pm\right)= \left(Y_1\cdot Y_2+1\right)\left(W_1^\pm\cdot W^\pm_2\right)~,\nonumber\\
    &\left(Y_1\cdot W_2^\mp\right) \left(Y_2\cdot W_1^\pm\right)= \left(Y_1\cdot Y_2-1\right)\left(W_1^\pm\cdot W^\mp_2\right)~.
\end{align}

\subsubsection{States in embedding space}\label{embstates}
Now let us proceed to describe the embedding space definition of the states $\ket{\Delta, \vec y\,}_{i_1\cdots i_s}$, which are defined in section~\ref{sec:Hilbert} as a basis of the principal series representation $\mathcal P_{\Delta,s}$. For this purpose, we'd like to make a detour  and review the physical realization of these abstractly defined states, focusing on the  $s=0$ and $s=1$ cases \cite{Sun:2021thf}. 

First consider  a free scalar $\phi$ of mass $m^2=\Delta(d-\Delta)$ in dS$_{d+1}$ in planar coordinates. Its leading late time behavior is given by 
\begin{align}
    \phi(\eta, \vec y\,)\stackrel{ \eta\to 0^-}{\sim} (-\eta)^\Delta\CO(\vec y\,)+(-\eta)^{\bar\Delta}\widetilde\CO(\vec y\,)~,
\end{align}
where $\CO(\vec y\,)$ and $\widetilde\CO(\vec y\,)$ are different linear combinations of the bulk creation and annihilation operators. More importantly, they are primary operators in the sense that 
\begin{align}
    [K_i,\CO(0)]=[K_i,\widetilde\CO(0)]=0, \,\,\,\,\, [D,\CO(0)]=\Delta\CO(0), \,\,\,\,\, [D,\widetilde\CO(0)]=\bar\Delta\widetilde\CO(0)~.
    \end{align}
These facts allow us to identify $|\Delta, \vec y\rangle$ as the (single-particle) state created by $\CO(\vec y\, )$ on the free Bunch-Davies vacuum $|0\rangle$, i.e. $|\Delta, \vec y\, \rangle = \CO(\vec y\, )|0\rangle$. Indeed, the action ~\reef{actonstate} is consistent with this identification. Similarly, for a free spin 1 field $V_\mu$ of mass $m^2=(\Delta-1)(\bar\Delta-1)$, the late time behavior of its spatial components $V_i$ is 
\begin{align}
    V_{i}(\eta, \vec y\, ) \stackrel{ \eta\to 0^-}{\sim} (-\eta)^{\Delta-1}\CA_i(\vec y\,)+(-\eta)^{\bar\Delta-1}\widetilde\CA_i(\vec y\,)~.
\end{align}
It can be checked that $\CA_i(\vec y\,)$ and $\widetilde\CA_i(\vec y\,)$ are spin 1 primary operators of scaling dimension $\Delta$ and $\bar\Delta$ respectively. In addition, $\CA_i(\vec y\,)|0\rangle$ can be identified as $|\Delta,\vec y\,\rangle_i$, whose transformation under the dS isometry group is given by ~\reef{actonstate}.

Altogether, the physical picture discussed  here implies that the embedding space formalism for  $|\Delta,\vec y\,\rangle_{i_1\cdots i_s}$ is essentially the same as that of primary operators in conformal field theory \cite{Costa:2011mg}. 
Based on this observation, we define $|\Delta, P\rangle_{A_1\cdots A_s}$, the embedding space realization of $|\Delta,\vec y\,\rangle_{i_1\cdots i_s}$ as follows:
\begin{itemize}
\item \textit{Nullness:} $P^A\in\mathbb R^{d+1,1}$ is a null vector, i.e. $P^A P_A=0$. We will focus on the $P^0>0$ part of the lightcone.
\item {\it Spin $s$ condition}: $|\Delta, P\rangle_{A_1\cdots A_s}$ is a symmetric and traceless tensor of $SO(d+1,1)$, on which  $SO(d+1,1)$ acts as Lie derivatives \footnote{More explicitly, $L_{AB}|\Delta, P\rangle_{A_1\cdots A_s}\equiv -\CL_{AB}|\Delta, P\rangle_{A_1\cdots A_s}$, where $\CL_{AB}$ denotes the derivative along the vector $P_A\partial_{P^B}-P_B\partial_{P^A}$.}.
\item \textit{Homogeneity:} $|\Delta, \lambda P\rangle_{A_1\cdots A_s}=\lambda^{-\Delta} |\Delta, P\rangle_{A_1\cdots A_s}$ with $\lambda>0~$.
\item \textit{Tangential condition:}  $P^{A_1} |\Delta, P\rangle_{A_1\cdots A_s}=0$.
\end{itemize}
Due to the homogeneity condition, $|\Delta,P\rangle_{A_1\cdots A_s}$ is completely fixed by its value on a section of the lightcone. We choose this section to be the future boundary of de Sitter (in planar coordinates)
\begin{align}
    P^0_{\vec y}=\frac{1}{2}\left(1+\vec y^2\right), \,\,\,\,\, P^i_{\vec y}= y^i, \,\,\,\,\, P^{d+1}_{\vec y}=\frac{1}{2}\left(\vec y^2-1\right)~,
\end{align}
since it realizes the state $ |\Delta, \vec y\,\rangle_{i_1\cdots i_s}$ as the pull-back of $|\Delta, P\rangle_{A_1\cdots A_s}$
\begin{align}
    |\Delta, \vec y\,\rangle_{i_1\cdots i_s}=\frac{\partial P^{A_1}_{\vec y}}{\partial y^{i_1}}\cdots \frac{\partial P^{A_s}_{\vec y}}{\partial y^{i_s}}|\Delta, P_{\vec y}\,\rangle_{A_1\cdots A_s}~.
\end{align}
In particular, the $SO(d+1,1)$ action on $|\Delta, P\rangle_{A_1\cdots A_s}$ induces eq. (\ref{actonstate}) via this pull-back.

Because of the nullness of $P^A$, there is a class of states that satisfy all the requirements of $|\Delta, P\rangle_{A_1\cdots A_2}$ but vanish when pulled back to the local coordinates $\vec y$. They are of the form $P_{(A_1} |\Delta+1, P\rangle_{A_2\cdots A_s)}$. We can kill these states, and implement the spin $s$ condition at the same time, by introducing an auxiliary vector $Z^A\in\mathbb R^{d+1,1}$ satisfying $Z^2=Z\cdot P=0$. Given such a vector $Z^A$, the state $|\Delta, P\rangle_{A_1\cdots A_s}$ is encoded in a degree $s$ polynomial in $Z$
\be
\ket{P,Z}_{\Delta,s} \equiv Z^{A_1} \cdots Z^{A_s} \ket{\Delta, P}_{A_1\cdots A_s}~.
\ee
In this index-free formalism, the tangential condition takes the form 
\begin{align}\label{btang}
    P\cdot\partial_Z \ket{P,Z}_{\Delta,s}=0
\end{align}
The resolution of the identity of $\mathcal P_{\Delta, s}$, c.f. ~\reef{eq:residen}, can be rewritten as a conformal integral defined in \cite{Simmons-Duffin:2012juh}
\be
 \mathbb 1_{\mathcal P_{\Delta,s}} = \frac{2}{s! \left(\hd-1\right)_s \text{Vol GL} (1,\IR)^+ } \int_{P^0>0} d^{d+2}P \, \delta(P^2)  \ket{P,D_{Z}}_{\Delta,s\, \Delta,s}\bra{P,Z}~.
\ee
where 
\be
D^A_Z = \left(\hd -1 + Z\cdot \frac{\partial}{\partial Z}\right) \frac{\partial}{\partial Z_A}-\half Z^A \frac{\partial^2}{\partial Z \cdot \partial Z}
\label{eq:DZoperator}
\ee 
is the interior derivative used to strip off $Z^A$. We will use the shorthand notation ``$\int_P$'' to denote the integral measure 
\begin{align}
 \int_P (\dots) \equiv   \frac{2}{\text{Vol GL} (1,\IR)^+ } \int_{P^0>0} d^{d+2}P \, \delta(P^2) (\dots)
\end{align}
in the remainder of this paper.

We also want to mention that although the states $|\Delta, \vec y\rangle$ or equivalently $|\Delta, P\rangle$ are introduced as boundary excitations in this section, they do not have to ``live" on the future boundary. Given a generic interacting theory, it is sometimes more convenient to think of them as  special states in the Hilbert space which transform in a particular way under the isometry group, with $\vec y$ being an abstract label of the states  that does not necessarily have the meaning of boundary coordinates.
\newpage
\section{The \Kallen decomposition in de Sitter}
\label{sec:kallanlehmann}
In this section, we give a  derivation of the \KL\, representation in de Sitter spacetime for two-point functions in the Bunch-Davies vacuum. Before starting, let us emphasize our starting assumptions.
\paragraph{The choice of state} In de Sitter, there is a one complex parameter family of states (the so-called $\alpha$-vacua) that are invariant under all the isometries \cite{Birrell:1982ix,Bunch:1978yq,Allen:1985ux,Banks:2002nv,Collins:2003zv,Einhorn:2002nu,Einhorn:2003xb,deBoer:2004nd}. In a free theory, among all these states only the Bunch-Davies vacuum leads to two-point functions which satisfy the Hadamard condition (commutators of fields inserted at space-like separation vanish). When interactions are turned on, it was shown \cite{Einhorn:2002nu,Einhorn:2003xb} that only the Bunch-Davies vacuum $|\Omega\rangle$ leads to sensible results in perturbation theory, while $\alpha$-vacua require the introduction of nonlocal counterterms. For these reasons we exclude $\alpha$-vacua from our discussions. 

In contrast to $\alpha$-vacua, there exists a family of potentially interesting states called $\alpha$-states $|\alpha\rangle$\footnote{The notation in the literature is unfortunate: sometimes $|\alpha\rangle$ indicates an $\alpha$-vacuum, and sometimes an $\alpha$-state. We use it to indicate the latter.} \cite{Einhorn:2003xb,deBoer:2004nd}. These states are squeezed excitations on top of the Bunch-Davies vacuum, and are thus not de Sitter invariant. They live in the same Hilbert space as $|\Omega\rangle$, and in particular they decompose into the same UIRs we reviewed in section \ref{sec:preliminaries}, and so they should not be independently included in the resolution of the identity. In contrast to $\alpha$-vacua, $|\alpha\rangle$ are well behaved in perturbation theory. Phenomenologically, $\alpha$-states are interesting because they leave an imprint on the late-time power spectrum. The universe might have indeed started in $|\alpha\rangle$ rather than in $|\Omega\rangle$, and only observations will allow us to prefer one over the other. Nevertheless, in this paper we focus on two-point functions on the Bunch-Davies vacuum 
$|\Omega\rangle$.

Let us summarize some properties of $|\Omega\rangle$:
\begin{itemize}
\item Starting from a Euclidean field theory on the sphere in which correlation functions are regular at non-coincident points, $SO(d+2)$ invariant and reflection positive, their continuation to de Sitter yields expectation values in the Bunch-Davies vacuum. This is the definition of the Bunch-Davies vacuum for a general interacting QFT in de Sitter.
\item $|\Omega\rangle$ is a strong late-time attractor, meaning that excitations on top of it get quickly washed out as the universe expands \cite{deBoer:2004nd}. 
\item Free propagators in the Bunch-Davies vacuum reduce to the canonically normalized flat space propagators (as we review in appendix \ref{subsec:flatspaceG}) when taking $R\to\infty$. 
\end{itemize}
\paragraph{Completeness of the Hilbert space} We assume that the full Hilbert space $\CH$ of a unitary quantum field theory in a fixed dS$_{d+1}$ background can be decomposed into a direct sum/integral of $SO(d+1,1)$ UIRs.
In other words, we assume that there is a  resolution of the identity in $\CH$, which takes the following form schematically \footnote{In principle, the direct integral over $\Delta$ should only be defined on the fundamental domain $\frac{d}{2}+i\mathbb R_{\ge 0}$, since there is an isomorphism of between $\mathcal P_{\Delta,\ell}$ and $\mathcal P_{\bar\Delta, \ell}$. Here, the equation (\ref{reso}) can be understood as a doubling trick. With that being said, $1_{\mathcal{P}_{\Delta, \ell}}$ and $ 1_{\mathcal{P}_{\bar\Delta, \ell}}$ are identified, and the overcounting is absorbed into the measure $[d\Delta]_\ell$, which is also invariant under the shadow symmetry by construction.\label{doubling}}
\begin{align}\label{reso}
\mathbb{1}_\CH=|\Omega\rangle \langle\Omega|+\sum_{\ell\ge 0}\int_{\frac{d}{2}+i\mathbb R}\left[d\Delta\right]_\ell\int_P\, |\Delta,P\rangle_{A_1\cdots A_\ell}\,^{A_1\cdots A_\ell}\langle\Delta,P|+\rm {other \,\, UIRs}~,
\end{align}
where $|\Omega\rangle$ is the interacting BD vacuum 
and  $\int_P\, |\Delta,P\rangle_{A_1\cdots A_\ell}\,^{A_1\cdots A_\ell}\langle\Delta,P|$ gives the identity operator $\mathbb 1_{\mathcal{P}_{\Delta, \ell}}$ in $\mathcal{P}_{\Delta, \ell}$ as discussed in section \ref{embstates}. The symbol $[d\Delta]_\ell $ denotes some unknown measure over the principal series, which roughly speaking, counts the ``multiplicity'' of $\mathcal P_{\Delta,\ell}$ in $\CH$ . Of course, this is still an oversimplification because there can be multiple copies of $\mathcal P_{\Delta, \ell}$, that are distinguished by other quantum numbers. Then, in principle, we  should integrate or sum over these quantum numbers. To avoid cluttering, we suppress labels of such quantum numbers, since it is easy to adapt our derivation to include them and the final expression of the \KL\, decomposition will not be changed.

With this assumption in mind, we start in $d\ge 2$, where we will mainly focus on the contribution of the principal series to the \Kallen representation. The complementary series part can be derived similarly, as is discussed in Appendix \ref{sec:comp}. In particular, for dS$_3$, principal and complementary series representations lead to a full \Kallen decomposition since they are the only UIRs of $SO(3,1)$. In higher dimensional de Sitter, there are many more UIRs, apart from the principal and complementary series, as reviewed in section \ref{subsec:UIRs}. For two-point functions of scalar operators, we conjecture that such representations do not contribute. For two-point functions of spinning operators, we do not have a general formula to incorporate all these representations, but we do not see them contributing to any example of two-point functions considered in this work. 

\subsection{Dimension $d\ge 2$}
Consider the Wightman two-point function $G_{\CO^{(J)}}$ of a generic spin $J$ operator $\CO^{(J)}$ in dS$_{d+1}$ ($d\ge 2$) in the embedding space formalism
\begin{align}\label{OO}
    G_{\CO^{(J)}}(Y_1,Y_2; W_1, W_2)=\langle\Omega|\CO^{(J)}(Y_1, W_1)\CO^{(J)}(Y_2, W_2)|\Omega\rangle~.
\end{align}
Inserting the resolution of the identity (\ref{reso}) into (\ref{OO}) yields
\small
\begin{align}\label{OO1}
    G_{\CO^{(J)}}(Y_1,Y_2; W_1, W_2)&=\langle\Omega|\CO^{(J)}(Y_1, W_1)|\Omega\rangle \langle\Omega|\CO^{(J)}(Y_2, W_2)|\Omega\rangle\nonumber\\
    &+\sum_{\ell\ge 0}\int_{\frac{d}{2}\!+\!i\mathbb R}\left[d\Delta\right]_\ell\int_P\, \langle\Omega|\CO^{(J)}(Y_1,\! W_1)|\Delta,P\rangle_{A_1\cdots A_\ell}\,^{A_1\cdots A_\ell}\langle \Delta, P|\CO^{(J)}(Y_2,\! W_2)|\Omega\rangle\nonumber\\
        &+\text{possible contributions from other UIRs}~.
\end{align}
\normalsize
Since $|\Omega\rangle$ is a dS invariant vacuum, the one-point function $\langle\Omega|\CO^{(J)}(Y, W)|\Omega\rangle$ has to be an  $SO(d+1,1)$ scalar. Using  $Y^2-1=W^2=Y\cdot W$, one can easily conclude that the one-point function vanishes when $J\ge 1$, and has to be a constant when $J=0$. In the latter case, we redefine the operator $\CO$ by a constant shift such that its vacuum expectation value vanishes. Altogether, we always consider the case $\langle\Omega|\CO^{(J)}(Y, W)|\Omega\rangle=0$. For the second line in eq.~(\ref{OO1}), the problem is reduced to computing the matrix elements $\langle\Omega|\CO^{(J)}(Y,W)|\Delta,P\rangle_{A_1\cdots A_\ell}$, or equivalently $\langle\Omega|\CO^{(J)}(Y,W)|P, Z\rangle_{\Delta, \ell}$ in the index-free formalism. We will show that such matrix elements are  fixed by symmetry up to a normalization constant, and then use them to derive the \KL\, decomposition. Let us start with the $J=0$ case.

\subsubsection{Scalar operators}
\label{subsec:scalaropsKLd}
First we  show that for a scalar operator $\CO(Y)$, the matrix element $\langle\Omega|\CO(Y)|P, Z\rangle_{\Delta,\ell}$ vanishes when $\ell\ge 1$. Because of its $SO(d+1,1)$ invariance, $\langle\Omega|\CO(Y)|P, Z\rangle_{\Delta,\ell}$ has to be a function of scalar bilinears of the three vectors $Y^A, P^A$ and $Z^A$.
On the other hand, since $P\cdot Z=P^2=Z^2=0$ and $Y^2=1$, it can only depend on $Y\cdot P$ and $Y\cdot Z$. The dependence is fixed by the homogeneity of $|P,Z\rangle_{\Delta, \ell}$ up to a constant
\begin{align}
\langle\Omega|\CO(Y)|P,Z\rangle_{\Delta,\ell}\propto\frac{(Y\cdot Z)^\ell}{(-2\, Y\cdot P)^\Delta}~.
\end{align}
We then impose the tangential condition (\ref{btang}) of the state $|P, Z\rangle_{\Delta, \ell}$. Noticing that $P\cdot\partial_Z (Y\cdot Z)^\ell\not=0$ for any $\ell\ge 1$, the proportional constant has to be zero for the tangential condition to be satisfied. Therefore, $\langle \Omega|\CO(Y)|P, Z\rangle_{\Delta,\ell}$ vanishes identically when $\ell\ge 1$.

\noindent When $\ell=0$, by a similar argument, we find that 
\begin{align}\label{O0}
    \langle\Omega|\CO(Y)|P\rangle=c_\CO(\Delta)\CK_\Delta(Y,P),\,\,\,\,\,\,
    \CK_\Delta(Y, P)=
    \frac{\Gamma(\Delta)}{2\pi^\frac{d+1}{2}}\frac{1}{(-2 Y\cdot P)^\Delta}~,
\end{align}
where $c_\CO(\Delta)$ is a $\Delta$-dependent constant. Plugging (\ref{O0}) into (\ref{OO1}) yields 
\begin{align}\label{GP}
    G_\CO(Y_1,Y_2)=\int_{\mathbb R}d\lambda\, \rho_\CO^{\mathcal P, 0}(\lambda)\,\int_P \CK_{\frac{d}{2}+i\lambda}(Y_1, P) \CK_{\frac{d}{2}-i\lambda}(Y_2, P)+\cdots
\end{align}
where $\rho_\CO^{\mathcal P,0}(\lambda)$ is a {\it nonnegative} function defined by absorbing $|c_\CO(\Delta)|^2$ into the measure $[d\Delta]_0$, i.e. $[d\Delta]_0|c_\CO(\Delta)|^2\equiv d\lambda\, \rho_\CO^{\mathcal P,0}(\lambda)$, with $\Delta=\frac{d}{2}+i\lambda$. It is also an even function of $\lambda$ by construction, because of the reason mentioned in footnote \ref{doubling}. The function $\CK_\Delta (Y, P)$ is the analogue of the bulk-to-boundary propagator in EAdS, but with a singularity at $Y\cdot P=0$. Therefore, we have to specify an $i\epsilon$ prescription to make sense of the $P$-integral in (\ref{GP}). The $i\epsilon$ prescription is chosen such that $G_\CO$ given by (\ref{GP}) reproduces the standard Wightman two-point function of a free field $\phi$ when $\CO=\phi$, which is reviewed in appendix \ref{scalarfield}:
\begin{align}\label{freeprop}
   G_{\lambda, 0}\left(Y_1, Y_2\right)= (\eta_1\eta_2)^{\frac{d}{2}}\int\frac{ d^d\vec k}{(2\pi)^d} e^{-i\vec k\cdot (\vec y_1-\vec y_2\,)}\bar h_{i\lambda}(|\vec k|\eta_1) h_{i\lambda}(|\vec k|\eta_2)
\end{align}
To match (\ref{freeprop}), we first write $\CK_\Delta$ in local coordinates
\begin{align}
    \CK_\Delta(\eta_1, \vec y_1; \vec y\,)=\frac{\Gamma(\Delta)}{2\pi^\frac{d+1}{2}} \frac{(-\eta_1)^\Delta}{\left((\vec y_1-\vec y)^2-\eta_1^2\right)^\Delta}~.
\end{align}
Then add a small imaginary part to the planar patch time $\eta_1$, i.e. $\eta_1\to e^{\pm i\epsilon}\eta_1$, and perform a Fourier transformation for the boundary coordinates $\vec y$. This Fourier transformation for both $\pm i\epsilon$ can be obtained by analytic continuation of the corresponding Wick rotated integral \footnote{Here we assume $z>0$. It can be thought as the radial component of the Poincar\'e coordinates of EAdS.}
\begin{align}\label{Eucint}
    \int\,d^d \vec y \frac{z^\Delta\, e^{-i\vec k\cdot \vec y}}{(z^2+\vec y^2)^\Delta}=\frac{2(\pi z)^{\frac{d}{2}}}{\Gamma(\Delta)}\left(\frac{2}{k}\right)^{-i\lambda} K_{i\lambda}(k z)~,
\end{align}
where $K$ is the Bessel K function. Put $z=-e^{i\theta}\eta_1$, and analytically continue from $\theta=0$ to $\theta=\epsilon-\frac{\pi}{2}$, i.e. $z=i e^{i\epsilon}\eta_1$. Using the relation $K_\nu (-i\xi)= \frac{i^{\nu+1}\pi}{2} H_\nu^{(1)}(\xi)$ between the Bessel K functions and Hankel functions, this gives
\begin{align}\label{CK1}
   \int d^d \vec y \,e^{-i\vec k\cdot \vec y}\,\CK_\Delta(e^{i\epsilon}\eta_1, \vec y_1; \vec y)=ie^{-\frac{\pi\lambda}{2}}(-\eta_1)^{\frac{d}{2}}\left(\frac{k}{2}\right)^{i\lambda} \, \bar h_{i\lambda}(|\vec k|\eta_1)e^{-i\vec k\cdot \vec y_1}~.
\end{align}
Similarly, by the Wick rotation $z=-i e^{-i\epsilon}\eta_1$, we obtain 
\begin{align}\label{CK2}
    \int d^d \vec y \,e^{-i\vec k\cdot \vec y}\,\CK_\Delta(e^{-i\epsilon}\eta_1, \vec y_1; \vec y)=-ie^{\frac{\pi\lambda}{2}}(-\eta_1)^{\frac{d}{2}}\left(\frac{k}{2}\right)^{i\lambda} \, h_{i\lambda}(|\vec k|\eta_1)e^{-i\vec k\cdot \vec y_1}~.
\end{align}
Thus, the only $i\epsilon$ prescription consistent with (\ref{freeprop}) is $\eta_1\to e^{i\epsilon}\eta_1$ and $\eta_2\to e^{-i\epsilon} \eta_2$, which in embedding space is equivalent to $Y_1\in\mathcal T_-$ and $Y_2\in\mathcal T_+$ for $\epsilon\in (0, \frac{\pi}{2}]$.
 This choice of $i\epsilon$ prescription should be understood in any Wightman function in this paper, although we will suppress $i\epsilon$ most of the time to avoid clutter of notations. As a byproduct of the discussion of the $i\epsilon$ prescription, eq. (\ref{CK1}) and eq. (\ref{CK2}) also lead to the dS split representation \cite{Sleight:2019mgd,Sleight_2020}, namely
 \begin{align}\label{scalarsplit}
    \int_P \CK_{\frac{d}{2}+i\lambda}(Y_1, P) \CK_{\frac{d}{2}-i\lambda}(Y_2, P)= G_{\lambda,0}(Y_1, Y_2)~, 
 \end{align}
 where we have used the fact that  $\int_P$ reduces to the flat measure $\int d^d\vec y$  on $\mathbb R^d$. Altogether, plugging eq. (\ref{scalarsplit}) into eq. (\ref{GP}), we obtain the \KL\, decomposition of the scalar operator $\CO(Y)$
\begin{align}
    G_\CO(Y_1, Y_2)=\int_{\mathbb R}d\lambda\, \rho_\CO^{\mathcal P}(\lambda)\, G_{\lambda, 0}(Y_1, Y_2)+\cdots~,
\end{align}
where $\rho_\CO^{\mathcal P, 0}(\lambda)$ is a nonnegative (even) function, and ``$\cdots$'' denotes possible contributions from other UIRs. For example, the contribution of the complementary series is computed explicitly in appendix \ref{sec:comp}. For the two exceptional series, we can argue that they do not contribute to scalar two-point functions. In the $\mathcal U_{s, t}$ case, it suffices to use  the fact that the $SO(d+1)$ content of $\mathcal U_{s, t}$  is \cite{Dobrev:1977qv, Sun:2021thf}
\begin{align}
    \mathcal U_{s, t}|_{SO(d+1)} =\bigoplus_{n\ge s} \bigoplus_{t+1\le m\le s}\mathbb Y_{n,m}~,
\end{align}
where $\mY_{n,m}$ denotes the two-row Young diagram with $n$ boxes in the first row and $m$ boxes in the second row. For example, $\mY_{2,1}=\Yvcentermath1\tiny\yng(2,1)$. On the other hand, it is clear that a scalar operator $\CO$ in dS$_{d+1}$ cannot generate any such two-row representation of $SO(d+1)$ when acting on the vacuum. It means that the matrix element of $\CO(Y)$ between $|\Omega\rangle$ and an arbitrary state in $\mathcal U_{s, t}$ vanishes. This excludes all $\mathcal U_{s, t}$. 

For $\mathcal V_p$, let's consider $\CG_p(Y_1, Y_2)=\langle\Omega|\CO(Y_1)|\mathbb 1_{\mathcal V_p}|\CO(Y_2)|\Omega\rangle$. Because $\mathbb 1_{\mathcal V_p}$ commutes with $SO(d+1,1)$ actions, $\CG_p(Y_1,Y_2)$ is a function of $\sigma\equiv Y_1\cdot Y_2$.  For the same reason, the $SO(d+1,1)$ Casimir operator which is equal to $(1-p)(d+p-1)$ acting on $\mathcal V_p$, yields a second order differential equation of $\CG_p(\sigma)$:
    \begin{align}\label{nnn}
        (1-\sigma^2)\partial_{\sigma}^2 \mathcal G_p(\sigma)-(d+1)\sigma\partial_\sigma \mathcal G_p(\sigma)=(1-p)(d+p-1)\mathcal G_p(\sigma)
    \end{align}
    The two linearly independent solutions of this equation are
\begin{align}
   & f_p(\sigma)=F\left(d+p-1,1-p,\frac{d+1}{2},\frac{1-\sigma}{2}\right)\nonumber\\
   &  g_p(\sigma)=\left(\frac{2}{1-\sigma}\right)^{d+p-1}F\left(d+p-1,p+\frac{d-1}{2},2p+d-1,\frac{2}{1-\sigma}\right)
\end{align}
Since $p\in\mathbb Z_{>0}$, the first solution $f_p$ is a polynomial of degree $p-1$ in $\sigma$, and hence it blows up as $\sigma\to -\infty$ (or remains a constant when $p=1$). 
Notice that  $\sigma<1$ corresponds to spacelike separated points, and the limit $\sigma\to -\infty$ means that the two points are very far away separated. However, any physical two-point function should decay in this limit. The other solution $g_p$ decays like $(-\sigma)^{-(d+p-1)}$ for large negative $\sigma$. But it has a  singularity at $\sigma=-1$, i.e. when $Y_1$ and $Y_2$ are antipodal points. Such antipodal singularities would violate our assumption of the Bunch-Davies vacuum since their continuation to the sphere would not be regular in the whole Euclidean domain with coincident points excluded $-1\leq\sigma<1$. Therefore,  eq. (\ref{nnn}) does not have a nontrivial solution that is free of singularity at both $\sigma=-\infty$ and $\sigma=-1$. At the same time, the sign of the antipodal singularity depends on $p$. There is thus a possibility that contributions associated to different values of $p$ conspire to cancel the singularity, resulting in a physically admissible two-point function. In the rest of this work, we adopt the conjecture phrased in appendix A of \cite{loparco2023radial} and assume no state in $\mathcal{V}_{p,0}$ can appear in the K\"allén-Lehmman decomposition of a scalar two-point function.

The full \KL\, decomposition of the scalar operator $\CO$ in $d\geq2$ is thus
\begin{shaded}
\begin{align}\label{KLscalar}
    G_\CO(Y_1, Y_2)=\int_{\mathbb R}d\lambda\, \rho_\CO^{\mathcal P, 0}(\lambda)\, G_{\lambda, 0}(Y_1,Y_2)+\int_{-\frac{d}{2}}^{\frac{d}{2}}d\lambda\, \rho_\CO^{\mathcal C, 0}(\lambda)\, G_{i\lambda, 0}(Y_1,Y_2)\,,
\end{align}
\end{shaded}
\noindent 
where $\rho_\CO^{\mathcal P, 0}(\lambda)$ and $\rho_\CO^{\mathcal C, 0}(\lambda)$ are the spectral densities corresponding to principal series and complementary series contributions respectively. They are nonnegative by construction.

In total generality, we thus expect the appearance of a continuum of states in the principal series and in the complementary series in the \KL\ decomposition of a scalar two-point function. What instead we observe in practice, in every example we have explored in section \ref{sec:applications}, is that the complementary series appears as a discrete sum of states corresponding to specific values of $\lambda$. Group theory arguments point to the fact that this is the case in free theories and CFTs \cite{Penedones:2023uqc}, but we do not have a proof to exclude a continuum of complementary series states in scalar two-point functions of generic interacting QFTs.

As a final comment, let us mention that special constructions of the two-point function of a free massless scalar ($p=1$ in (\ref{nnn})) with the zero mode removed are present in the literature (see for example \cite{Bros_2010,Folacci_1992,Epstein:2014jaa,Tarek}), but these are not true gauge invariant observables\footnote{Here the gauge symmetry is the shift symmetry of the free massless scalar.}. In other words, the operators constructed in these examples do not correspond to physical observables and thus we do not expect them to appear in the \KL\ decomposition of a physical scalar operator. This is in analogy with the case of free massless scalars in 2D flat space. Just like in that scenario, the two-point function of the derivatives of a free massless scalar in dS is instead a good observable. In fact, we expect it to contribute to the spinning version of the \KL\ decomposition in higher dimensions (\ref{eq:KL any J}), and we explicitly see it contributing to the spinning \KL\ decomposition in 2D (\ref{2DspinJfull}).

\subsubsection{Spinning operators}\label{sec:Spinning operators}
Given a spin $J$ bulk operator $\CO^{(J)}(Y, W)$, the main step towards its K\"all\'en-Lehmann representation is computing the matrix element $\CF_{J,\ell}(Y, P; W, Z)\equiv  \langle\Omega|\CO^{(J)}(Y, W)|P, Z\rangle_{\Delta,\ell}$,
for any $\ell\ge 0$. Due to the various constraints imposed on the four vectors $\{Y^A, W^A, P^A, Z^A\}$, the most general form  of $\CF_{J,\ell}$ is 
\begin{align}\label{CJsl0}
\CF_{J,\ell}(Y, P;W, Z)=\sum_{b=0}^{\min(J, \ell)} f_{\CO^{(J)}}^b(\Delta,\ell)\,\frac{(Y\cdot Z)^{\ell-b}(2P\cdot W)^{J-b}(W\cdot Z)^b}{(-2Y\cdot P)^{\Delta+J-b}}~,
\end{align}
 To find the coefficients $f_{\CO^{(J)}}^b$, we use the tangential condition of $P\cdot \partial_Z\CF_{J,\ell}=0$, which yields the following recurrence relation 
\begin{align}\label{frec}
(b+1)f_{\CO^{(J)}}^{b+1}=(\ell-b)f_{\CO^{(J)}}^b,   \,\,\,\,\,\, (\ell-\min(J,\ell))f_{\CO^{(J)}}^{\min(J,\ell)}=0~.
\end{align}
When $\ell>J$, the initial condition $(\ell-J)f_{\CO^{(J)}}^J=0$ gives $f_{\CO^{(J)}}^J=0$, which further implies that all the rest $f_{\CO^{(J)}}^b$ vanish because $\ell-b\ge \ell-J$ is always nonzero. So principal series of spin larger than $J$ cannot contribute to the two-point function of $\CO^{(J)}$.\footnote{The same argument also works for complementary series.} When $\ell\le J$, eq. (\ref{frec}) has a nontrivial solution instead
\begin{align}\label{fsolve}
f_{\CO^{(J)}}^b(\Delta,\ell)=\frac{(\Delta+\ell-1)\Gamma(\Delta)(\Delta+\ell)_{J-\ell}\,c_{\CO^{(J)}}(\Delta,\ell)}{2\pi^{\frac{d+1}{2}} (\Delta-1)}\binom{\ell}{ b} , \,\,\,\,\, b=0,1,\cdots, \ell~.
\end{align}
where the complicated normalization factor is inserted for later convenience.
Plugging this solution into eq. (\ref{CJsl0}) gives
\begin{align}
\CF_{J,\ell}(Y, P; W, Z)= \frac{(\Delta+\ell-1)\Gamma(\Delta)(\Delta+\ell)_{J-\ell}\,c_{\CO^{(J)}}(\Delta,\ell)}{2\pi^{\frac{d+1}{2}} (\Delta-1)} \frac{\Phi^{\ell}\,(2W\cdot P)^{J-\ell}}{(- 2 Y\cdot P)^{\Delta+J}}~,
\end{align}
where 
\begin{align}
    \Phi(Y,P;W,Z)\equiv 2(Y\cdot Z)(W\cdot P)-2(Y\cdot P)(W\cdot Z)~.
\end{align}
For $J=\ell$, $\CF_{J,J}$ reduces to $c_{\CO^{(J)}}(\Delta, J)\,\CK_{\Delta,J}$, with $\CK_{\Delta,J}(Y,P;W,Z)$ being the  bulk-to-boundary propagator of a spin $J$ field, given by 
\begin{align}
     \CK_{\Delta,J}(Y,P;W,Z)= \frac{(\Delta+J-1)\Gamma(\Delta)}{2\pi^{\frac{d+1}{2}} (\Delta-1)}\frac{\Phi^J}{(-2Y\cdot P)^{\Delta+J}}~.
\end{align}
For $\ell<J$, noticing that $\Phi$ is annihilated by $W\cdot\nabla_Y=W\cdot\partial_Y$, we can realize $\CF_{J,\ell}$ as derivatives of $\CK_{\Delta,\ell}$:
\begin{align}\label{eq:Split G}
\CF_{J,\ell}=c_{\CO^{(J)}}(\Delta, \ell)\,(W\cdot\nabla_Y)^{J-\ell}\CK_{\Delta,\ell}~.
\end{align}
Finally, using the de Sitter split representation of a spin $\ell$ Wightman function $G_{\lambda,\ell}$ \footnote{It is defined as the symmetric, traceless and transverse Green's function, satisfying 
\begin{align}
    \left(-\nabla^2+\frac{d^2}{4}+\lambda^2+
    \ell\right)G(Y_1,Y_2;W_1,W_2)=\delta(Y_1,Y_2)(W_1\cdot W_2)^\ell~.
\end{align}}
\cite{Sleight_2020}
\begin{align}\label{eq:InFormGen}
G_{\lambda,\ell}(Y_1,Y_2;W_1,W_2)=\frac{1}{\ell!\, (\frac{d}{2}-1)_\ell}\int_P\, \CK_{\Delta,\ell}(Y_1, P; W_1, D_Z)\CK_{\bar\Delta,\ell}(Y_2, P; W_2, Z)~,
\end{align}
we obtain the \KL\, decomposition of $\CO^{(J)}$:
\begin{shaded}
\begin{align} \label{eq:KL any J}
G_{\CO^{(J)}}(Y_1,\!Y_2; W_1,\!W_2)
=&\sum_{\ell=0}^J\int_{\mathbb R}\, d\lambda\, \rho_{\CO^{(J)}}^{\mathcal P,\ell}( \lambda) \left[\left(W_1\cdot\nabla_{1}\right) \left(W_2\cdot\nabla_{2}\right)\right]^{J-\ell}G_{\lambda,\ell}(Y_1, Y_2; W_1, W_2)\nonumber\\
&+\cdots
\end{align}
\end{shaded}
\noindent 
where $\rho_{\CO^{(J)}}^{\mathcal P,\ell}( \lambda) $ is a nonnegative function of $\lambda$ and $d\lambda\,\rho_{\CO^{(J)}}^{\mathcal P,\ell}( \lambda)$ is a product of the measure  $[d\Delta]_\ell$ and the factor $|c_{\CO^{(J)}}(\Delta, \ell)|^2$ and the dots stand for contributions coming from exceptional and complementary series.

\subsection{dS$_2$}
We have derived the \KL\, decomposition for generic operators in higher dimensional dS, focusing on the contribution of the principal series. The derivation is based on the resolution of the identity (\ref{reso}) in the full Hilbert space. In two dimensional dS, we need to modify (\ref{reso}) in several ways. First, because the principal series of $\SO(2,1)$ has only one label, namely the scaling dimension $\Delta$, the sum over $\ell$ in eq.~(\ref{reso}) cannot appear when $d=1$. The second modification is closely related to the discussion regarding the embedding space formalism of dS$_2$ in section \ref{subsec:embeddingspace}. A spin $J$ tensor operator $\CO^{(J)}_{\mu_1\cdots\mu_J}$ in dS$_2$ has two independent components, i.e. chiralities. The two chiralities  can be mapped to each other by parity, denoted by $\Theta$, which belongs to $O(2,1)$ instead of $SO(2,1)$. More precisely, $\Theta$ is defined to flip the sign of $Y^1$ in embedding space, or $\vec y$ in planar coordinates. 
We will focus on parity invariant QFTs.
From the representation side, it means that we should decompose the full Hilbert space into UIRs of $O(2,1)$. It is very easy to describe such UIRs. Given a fixed $\Delta$, there are two principal series (or complementary series depending on the value of $\Delta$) representations $\mathcal{P}_\Delta^\pm$, distinguished by the intrinsic parity under $\Theta$, i.e. 
\begin{align}
    \Theta |\Delta, y\rangle_\pm \equiv \pm |\Delta, -y\rangle_\pm
\end{align}
where $|\Delta, y\rangle_\pm$ is a basis of $\mathcal{P}_\Delta^\pm$. For the discrete series, $\mathcal D^-_p$ is the image of $\mathcal D^+_p$ under $\Theta$, because $\Theta$ flips the sign of $L_0$. So the direct sum $\mathcal D_p\equiv \mathcal D_p^+\oplus\mathcal D_p^-$ furnishes an $O(2,1)$ representation, while each summand does not. Altogether, the resolution of the identity in dS$_2$ can be formulated as
\begin{align}\label{reso2D}
    \mathbb 1_\CH=|\Omega\rangle\langle\Omega|+\sum_\pm \int_{\frac{1}{2}+i\mathbb R} [d\Delta]_\pm \int_P\, |\Delta, P\rangle_\pm \,_\pm\langle \Delta,P|+\sum_{p\ge 1} \mathbb 1_{\mathcal D_p}+\cdots
\end{align}
Before using it to derive the \KL\, decomposition in dS$_2$, let us make some remarks on this formula.
\begin{itemize}
    \item $\mathbb 1_{\mathcal D_p}$ is the identity operator in the representation $\mathcal D_p$:
\begin{align}
    \mathbb 1_{\mathcal D_p}=\sum_{|n|\ge p}\frac{\Gamma(|n|+1-p)}{\Gamma(|n|+p)} \, |n\rangle_p\,_p\langle n |~.
\end{align}
where the states $|n\rangle_p$ are introduced in section \ref{subsec:UIRs}.
\item ``$\sum_{p\ge 1}$'' is a formal sum of discrete series. It is possible that there are several copies of each $\CD_p$ distinguished by other quantum numbers. Sums over such quantum numbers are also implicitly included in $\sum_{p\ge 1}$.
\item The dots correspond to the contribution from the complementary series. It can be derived using the same approach as the principal series, see Appendix \ref{sec:comp}. 
\item We always shift the operator under consideration such that its vacuum one-point function vanishes. It means that the first term $|\Omega\rangle \langle\Omega|$ of (\ref{reso2D}) does not contribute.
\end{itemize}

\subsubsection{Scalar operators}
Let $\CO(Y)$ be a scalar operator in dS$_2$. The derivation of the principal and complementary series part of its \KL\, decomposition is exactly the same as in higher dimensions, except for an extra sum over two chiralities.  For discrete series states, we can show that they do not contribute and the argument is exactly the one we used for the exceptional series $\CV_p$ in higher dimensions.
So the full \KL\, decomposition of the scalar operator $\CO$ in dS$_2$ is 
    \begin{align}
        G_\CO(Y_1, Y_2)=\int_{\mathbb R}d\lambda\, \rho_\CO^{\mathcal P}(\lambda) G_{\lambda, 0}(Y_1, Y_2)+\int_{ -\frac{1}{2}}^{\frac{1}{2}}d\lambda\, \rho_\CO^{\mathcal C}(\lambda) G_{i\lambda, 0}(Y_1, Y_2)
    \end{align}
\noindent The functions $\rho_\CO^{\mathcal P}(\lambda)$ and $\rho_\CO^{\mathcal C}(\lambda)$ are nonnegative by construction.

\subsubsection{Spinning operators}
The distinction between $|\Delta, P\rangle_\pm$ becomes crucial when the bulk operator carries a nonzero spin. For example, let us consider a vector operator $V_A(Y)$. In higher dimensions, the matrix element $\langle\Omega|V_A(Y)|\Delta, P\rangle$ is a linear combination of $Y_A$ and $P_A$, and the former is killed in the index-free formalism. When $d=1$, there can be one more type of tensor structure in this matrix element, namely $\epsilon_{ABC} Y^B P^C$, where $\epsilon_{ABC}$ is the totally antisymmetric tensor in $\mathbb R^{2,1}$, normalized as $\epsilon_{012}=1$. It is a pseudo vector in contrast to $Y_A$ and $P_A$. So $\epsilon_{ABC}Y^BP^C$ can only appear in $\langle\Omega|V_A(Y)|\Delta, P\rangle_-$, while $Y_A$ and $P_A$ can only appear in $\langle\Omega|V_A(Y)|\Delta, P\rangle_+$ \footnote{Here we have assumed $V_A$ to be a vector instead of pseudo vector. In latter case, $\epsilon_{ABC}Y^BP^C$ is in $\langle\Omega|V_A(Y)|\Delta, P\rangle_+$, while $Y_A$ and $P_A$ are in $\langle\Omega|V_A(Y)|\Delta, P\rangle_-$. We will always consider tensors instead of pseudo tensors in the following discussion. It is easy to check that their \KL\, representations take the same form.}. Next, we will generalize this simple example to any spinning 
operators in dS$_2$. 

\,

\noindent{}\textbf{Principal series part}

Let $\CO^{(J)}$ be a spin $J$ operator in dS$_2$. Deriving its \KL\, decomposition amounts to computing the matrix elements $\CF_{J,\pm}(Y,P;W)\equiv \langle\Omega|\CO^{(J)}(Y, W)|\Delta, P\rangle_\pm$. $\CF_{J,+}$ is a scalar and hence its $W$ dependence can only be $(P\cdot W)^J$. In contrast, $\CF_-$ is a pseudo scalar, so its $W$ dependence should be $(P\cdot W)^{J-1} \epsilon(W,Y,P)$, where $ \epsilon(W,Y,P)\equiv \epsilon_{ABC}W^A Y^B P^C$. 
Altogether, the most general form of $\CF_\pm$ is 
\begin{align}
    &\CF_+(Y,P;W)=c^+_{\CO^{(J)}}(\Delta)\left(W\cdot\nabla_Y\right)^J
    \CK_\Delta(Y,P)\nonumber\\
    &
    \CF_-(Y,P;W)=\frac{c^-_{\CO^{(J)}}(\Delta)}{\Delta}\left(W\cdot\nabla_Y\right)^{J-1}\epsilon(W,Y,\nabla_Y)\CK_\Delta(Y,P)~,
\end{align}
where we have replaced any $P$ in the numerator by derivatives of $Y$.  Then, using the $d=1$ version of the split representation (\ref{scalarsplit}), we obtain
\small
\begin{align}
\int_P\langle\Omega|\CO^{(J)}(Y_1,W_1)|\Delta, P\rangle_+ \,_+\langle \Delta, P|\CO^{(J)}(Y_2, W_2)|\Omega\rangle=|c^+_{\CO^{(J)}}|^2
(W_1\cdot\nabla_1)^J(W_2\cdot\nabla_2)^J G_{\lambda,0}(Y_1,Y_2)~,
\end{align}
\normalsize
and
\small
\begin{align}
\int_P\langle\Omega|\CO^{(J)}(Y_1,W_1)&|\Delta, P\rangle_- \,_-\langle \Delta, P|\CO^{(J)}(Y_2, W_2)|\Omega\rangle\nonumber\\
&=|c^-_{\CO^{(J)}}|^2
(W_1\!\cdot\!\nabla_1)^{J-1}(W_2\!\cdot\!\nabla_2)^{J-1} G_{\lambda,1}(Y_1, Y_2; W_1, W_2)~,
\end{align}
\normalsize
where $G_{\lambda, 1}$ is the free two-point function of a Proca field of mass $m^2=\Delta\bar\Delta=\frac{1}{4}+\lambda^2$, and it is related to  the scalar two-point function  by eq.
(\ref{G1in2D}). Altogether, the principal series part of the \KL\, decomposition of $\CO^{(J)}$ is 
\begin{align}\label{2DspinJ}
    G_{\CO^{(J)}}(Y_1,Y_2; W_1, W_2)&=\int_{\mathbb R}d\lambda\, \rho^{\mathcal P, 0}_{\CO^{(J)}}(\lambda)(W_1\cdot\nabla_1)^J(W_2\cdot\nabla_2)^J G_{\lambda,0}(Y_1,Y_2)\nonumber\\
    &+\int_{\mathbb R}d\lambda\,\rho^{\mathcal P, 1}_{\CO^{(J)}}(\lambda)(W_1\cdot\nabla_1)^{J-1}(W_2\cdot\nabla_2)^{J-1} G_{\lambda,1}(Y_1, Y_2; W_1, W_2)+\cdots
\end{align}
In this equation, $\rho^{\mathcal P, 0}_{\CO^{(J)}}(\lambda)$ and $\rho^{\mathcal P, 1}_{\CO^{(J)}}(\lambda)$ are two nonnegative (even) functions of $\lambda$, defined by
\begin{align}
   |c^+_{\CO^{(J)}}(\Delta)|^2[d\Delta]_+ = \rho^{\mathcal P, 0}_{\CO^{(J)}}(\lambda)\,d\lambda, \,\,\,\,\,  |c^-_{\CO^{(J)}}(\Delta)|^2[d\Delta]_- = \rho^{\mathcal P, 1}_{\CO^{(J)}}(\lambda)\,d\lambda~.
\end{align}
The contribution of the complementary series takes the same form as eq. (\ref{2DspinJ}), except that the integral domain should be $-\frac{1}{2}<i\lambda<\frac{1}{2}$.

\,

\noindent{}\textbf{Discrete series part}

For scalar operators in dS$_2$, we have shown  that the discrete series cannot appear in the \KL\, decomposition. The argument is based on some second order differential equation of $\langle\Omega|\CO(Y_1)|\mathbb{1}_{\CD_p}|\CO(Y_2)|\Omega\rangle$, induced by the $SO(2,1)$ Casimir. 
Nontrivial solutions of such  differential equations always have unphysical singularities, and hence $\langle\Omega|\CO(Y_1)|\mathbb{1}_{\CD_p}|\CO(Y_2)|\Omega\rangle$ has to vanish. 
In the spin $J$ case, by leveraging this Casimir method, we are able to exclude all $\CD_p$ with $p>J$ in the \KL\, decomposition of $\CO^{(J)}$ in a similar way. We leave details of this argument to appendix \ref{p<J}. 
For $p\le J$, the Casimir equations have physical solutions, so $\CD_p$ does contribute to the two-point function of $\CO^{(J)}$. However, to prove the positivity of this contribution requires some extra input, for example, reflection positivity after a Wick rotation to the sphere \cite{DiPietro:2021sjt}. We will not give this type of arguments. Instead, we adopt the same strategy as in the principal series case, i.e. using the resolution of the identity operator $\mathbb{1}_{\CD^\pm_p}=\sum_\psi |\psi\rangle\langle \psi|$ in  $\mathcal D_p^\pm$ and computing the matrix elements of $\CO^{(J)}$ between the BD vacuum and $|\psi\rangle $.  This method  guarantees the positivity automatically, but meanwhile it also  leads to certain technical difficulties compared to the principal series case because discrete series representations do not admit ($\delta$-function) normalizable continuous basis such as $|y\rangle$ or $|P\rangle$. Instead, its resolution of the identity is formulated in terms of the discrete basis $|n\rangle_p$, c.f. eq. (\ref{resodis}). This basis diagonalizes $L_0$, so unlike $|P\rangle$, it is not $SO(2,1)$ covariant. Due to the loss of the manifest covariance, the embedding space formalism stops being an efficient computational tool, so it is much more difficult to calculate the matrix elements, e.g. $\langle\Omega|\CO^{(J)}(Y,W)|n\rangle_p$. With that being said, we choose to  directly work in the conformal global coordinates $(\tau, \varphi)$ (\ref{conformalglobal}), since they admit $L_0\sim \partial_\varphi$ as a Killing vector. As mentioned in section \ref{fieldembedding}, we also introduce lightcone coordinates $y^\pm =\tau\pm\varphi$. Then the two nonvanishing components of  $\CO^{(J)}$ are $\CO^{(J)}_{++\cdots +}$ and $\CO^{(J)}_{--\cdots -}$. The matrix elements of interest are $ \CF_{J,p}^{(n,\pm )}(y^+, y^-)\equiv \langle\Omega|\CO^{(J)}_{\pm\cdots \pm}(y^\pm )|n\rangle_p$.

Let's start with $\CF_{p,p}^{(p,\pm)}$ which corresponds to $J=n=p$. It should satisfy two first order differential equations induced by the conditions $L_0|p\rangle_p=p|p\rangle_p$ and $L_-|p\rangle_p=0$, since $|p\rangle_p$ is the lowest-weight state in the representation $\mathcal D_p^+$. To find such differential equations, we need to know how $\so(2,1)$ generators act on bulk operators. Recall that $\{L_0, L_\pm\}$ are defined by eq. (\ref{defineLpm}) and their associated Killing vectors are computed in \cite{Sun:2021thf}
\begin{align}\label{dS2killingvect}
   V_0=i\partial_\varphi, \,\,\,\,\, V_+=- i \left( e^{-i y^+}\partial_++e^{i y^-}\partial_-\right), \,\,\,\,\, V_-=- i \left( e^{i y^+}\partial_++e^{-i y^-}\partial_-\right)~,
\end{align}
where $\partial_\pm =\partial_{y^\pm}=\frac{1}{2}(\partial_\tau\pm\partial_\varphi)$.
Because of the convention (\ref{bulkaction}), the action of $-L_\alpha$ on $\CO^{(p)}$ is realized by the Lie derivative along $V_\alpha$, i.e. $ [L_\alpha, \CO^{(p)}_{\mu_1\cdots \mu_p}]= - \CL_{V_\alpha}\CO^{(p)}_{\mu_1\cdots \mu_p}$, where $\alpha=0, \pm$. For example, for $\alpha=0$, it implies 
\begin{align}
    i\partial_\varphi \CF_{p,p}^{(p,\pm)}=-\langle\Omega|[L_0,\CO^{(p)}_{+\cdots +}(y^\pm )]|p\rangle_p= p\CF_{p,p}^{(p,\pm)}~.
\end{align}
So the $\varphi$ dependence in $\CF_{p,p}^{(p,\pm)}$ is simply $e^{-i p\varphi}$. Similarly,  for $\alpha=-$, we obtain $\partial_\mp\CF_{p,p}^{(p,\pm)}=0$. Therefore, $\CF_{J,p}^{(p,\pm)}$ are determined up to normalization constants
$\CF_{J,p}^{(p,\pm)}(y^\pm) = c_{p,p}^{\pm} e^{\mp i p y^\pm}$.
With this lowest-weight mode known, any $\CF_{p,p}^{(n,\pm)} $ with $n\ge p$ can be obtained by acting $n-p$ times with $\CL_{V_+}$ since $L_+^{n-p}|p\rangle_p=(2p)_{n-p}|n\rangle_p$ (c.f. eq. (\ref{Lpmact})):
\begin{align}\label{dismatrix}
    \CF_{p,p}^{(n,\pm)}(y^\pm)= \frac{1}{(2p)_{n-p}} \CL^{n-p}_{V_+}  \CF_{p,p}^{(p,\pm)}(y^\pm)= (\mp)^{n-p}\, c_{p,p}^{\pm} \, e^{\mp i n y^\pm}~,
\end{align}
which allows us  to compute the contribution of $\mathcal D_{p}^+$ in the two-point function of $\CO^{(p)}$. For example, for the $(+,+)$ component, we have 
\begin{align}\label{ioio}
\langle \Omega | \CO^{(p)}_{+\cdots +}(y_1)|\mathbb 1_{\CD^+_p}| \CO^{(p)}_{+\cdots +}(y_2) |\Omega\rangle = |c_{p,p}^+|^2 \,\sum_{n\ge p} \frac{\Gamma(n+p)}{\Gamma(n-p+1)} e^{-i n y^+_{12}}~,
\end{align}
where $y_{12}^+= y_1^+- y_2^+$. The infinite sum over $n$ in (\ref{ioio}) looks  divergent because we have suppressed the explicit $i\epsilon$ prescription. Using $e^{i y_j^+}=\frac{i-(y_j+\eta_j)}{i+(y_j+\eta_j)}$, which relates the planar coordinates $(\eta, y)$ and conformal global coordinates $(\tau, \varphi)$ or equivalently $(y^+, y^-)$, and restoring the  $i\epsilon$ prescription $(\eta_1\to e^{i\epsilon}\eta_1, \eta_2\to e^{-i\epsilon}\eta_2)$, then $e^{-i y_{12}^+}$ in eq. (\ref{ioio}) should be replaced by 
\begin{align}\label{jjj}
r(\epsilon)\equiv \frac{i-(y_2+e^{-i\epsilon}\eta_2)}{i+(y_2+e^{-i\epsilon}\eta_2)}\frac{i+(y_1+e^{i\epsilon}\eta_1)}{i-(y_1+e^{i\epsilon}\eta_1)}~.
\end{align}
It is straightforward to check that $|r(\epsilon)|<1$ for small $\epsilon$, and hence the sum in (\ref{ioio}) is convergent given the $i\epsilon$ prescription $(\eta_1\to e^{i\epsilon}\eta_1, \eta_2\to e^{-i\epsilon}\eta_2)$. Evaluating the sum yields 
\begin{align}
\langle \Omega | \CO^{(p)}_{+\cdots +}(y_1)|\mathbb 1_{\mathcal D^+_p}| \CO^{(p)}_{+\cdots +}(y_2) |\Omega\rangle = \frac{\Gamma(2p)|c_{p,p}^+|^2 }{(-4)^p}\left(\sin\frac{y_{12}^+}{2}\right)^{-2p}~.
\end{align}
Similarly, for other components, we have  
\begin{align}
&\langle \Omega | \CO^{(p)}_{+\cdots +}(y_1)|\mathbb 1_{\mathcal D^+_p}| \CO^{(p)}_{-\cdots -}(y_2) |\Omega\rangle = \frac{\Gamma(2p)c_{p,p}^+ (c_{p,p}^-)^*}{4^p}\left(\cos\frac{y_{1}^++y^-_2}{2}\right)^{-2p}\nonumber\\
&\langle \Omega | \CO^{(p)}_{-\cdots -}(y_1)|\mathbb 1_{\mathcal D^+_p}| \CO^{(p)}_{+\cdots +}(y_2) |\Omega\rangle = \frac{\Gamma(2p)c_{p,p}^- (c_{p,p}^+)^* }{4^p}\left(\sin\frac{y_1^-+y_{2}^+}{2}\right)^{-2p}\nonumber\\
&\langle \Omega | \CO^{(p)}_{-\cdots -}(y_1)|\mathbb 1_{\mathcal D^+_p}| \CO^{(p)}_{-\cdots -}(y_2) |\Omega\rangle = \frac{\Gamma(2p)|c_{p,p}^-|^2 }{(-4)^p}\left(\sin\frac{y_{12}^-}{2}\right)^{-2p}~.
\end{align}
The contribution of $\mathcal D_{p}^-$ does not require any extra computation since it is the image of $\mathcal D_{p}^+$ under the parity $\Theta$. Noticing that $\Theta$ also flips chiralities, it is easy to obtain relations like 
\begin{align}
\langle \Omega | \CO^{(p)}_{+\cdots +}(y^+_1, y_1^-)|\mathbb 1_{\mathcal D^-_p}| \CO^{(p)}_{+\cdots +}(y^+_2, y_2^-) |\Omega\rangle = \langle \Omega | \CO^{(p)}_{-\cdots -}(y^-_1, y_1^+)|\mathbb 1_{\mathcal D^+_p}| \CO^{(p)}_{-\cdots -}(y^-_2, y_2^+) |\Omega\rangle~. 
\end{align}
Altogether, the contribution of $\mathcal D_p=\mathcal D_p^+\oplus \mathcal D_p^-$ to the $\CO^{(p)}$ two-point function can be summarized as 
\begin{align}\label{kkk}
   & \langle \Omega | \CO^{(p)}_{\pm\cdots \pm}(y_1)|\mathbb 1_{\mathcal D_p}| \CO^{(p)}_{\pm\cdots \pm}(y_2) |\Omega\rangle = \frac{\Gamma(2p)\left(|c_{p,p}^+|^2 + |c_{p,p}^-|^2\right)}{(-4)^p}\left(\sin\frac{y_{12}^\pm }{2}\right)^{-2p}\nonumber\\
    & \langle \Omega | \CO^{(p)}_{\pm\cdots \pm}(y_1)|\mathbb 1_{\mathcal D_p}| \CO^{(p)}_{\mp\cdots \mp}(y_2) |\Omega\rangle = \frac{\Gamma(2p)\left(c^\pm_{p,p}\left(c^\mp_{p,p}\right)^*+\left(c^\pm_{p,p}\right)^*c^\mp_{p,p}\right)}{4^p}\left(\cos\frac{y_1^\pm+y_2^\mp }{2}\right)^{-2p}~.
\end{align}
The $(\pm, \mp)$ component blows up when $\cos\frac{y_1^\pm+y_2^\mp}{2}$ vanishes. On the other hand, we have $1+Y_1\cdot Y_2\propto \prod_{\pm} \cos\frac{y_1^\pm+y_2^\mp}{2}$, which implies that the (+, -) and (-, +) components in (\ref{kkk}) have an antipodal singularity. So these components have to vanish, and this requirement imposes a nontrivial constraint on the coefficients $c_{p,p}^\pm$, namely $c_{p,p}^+ (c_{p,p}^-)^*+c_{p,p}^- (c_{p,p}^+)^*=0$. Comparing  eq. (\ref{kkk}) with (\ref{ppcom}) and (\ref{other}), we can make the following identification
\begin{align}
     \langle \Omega | \CO^{(p)}_{\alpha\cdots\alpha}(y_1)|\mathbb 1_{\mathcal D_p}| \CO^{(p)}_{\beta\cdots\beta}(y_2) |\Omega\rangle= 4\pi\left(|c_{p,p}^+|^2 + |c_{p,p}^-|^2\right)\left(\nabla^{(1)}_\alpha\right)^p \left(\nabla^{(2)}_\beta\right)^p G_{-i(p-\frac{1}{2})}(y_1, y_2)
\end{align}
where $\alpha,\beta\in\{+, -\}$, and in embedding space it means
\small
\begin{align}\label{ppfinal}
     \langle \Omega | \CO^{(p)}(Y_1, W_1)|\mathbb 1_{\mathcal D_p}| \CO^{(p)}(Y_2, W_2) |\Omega\rangle= 4\pi\left(|c_{p,p}^+|^2 + |c_{p,p}^-|^2\right)\left(W_1\cdot\nabla_1\right)^p \left(W_2\cdot\nabla_2\right)^p G_{-i(p-\frac{1}{2})}(Y_1, Y_2)
\end{align}
\normalsize

The remaining task is to generalize the computation above to the case of  $J>p$. As before, we start with building the lowest-weight modes $\CF^{(p,\pm)}_{J,p}$, which is fixed up to normalization  by the defining properties of $|p\rangle_p$.
With a short computation, we find 
\begin{align}\label{Lk}
\CF^{(p,+)}_{J,p}=(2p)_{J-p} \,c^+_{J,p}\, \frac{e^{-i J t -i p\varphi}}{(2i\cos t)^{J-p}}, \,\,\,\,\, \CF^{(p,-)}_{J,p}=(2p)_{J-p} \,c^-_{J,p}\, \frac{e^{i J t -i p\varphi}}{(-2i\cos t)^{J-p}}~,
\end{align}
where $c_{J, p}^\pm$ are unknown normalization factors.
Unlike in the $J=p$ case, $\CF^{(p,\pm)}_{J,p}$ are not chiral or anti-chiral functions. This fact makes it hard to compute the repeated action of $\CL_{V_+}$ on these modes. How we deal with this technical difficulty is based on several important observations. First we notice that  $\CF^{(p,\pm)}_{J,p}$  can be realized as covariant derivatives of $\CF^{(p,\pm)}_{p,p}$.
\small
\begin{align}
&\CF^{(p,+)}_{J, p}= c_{J,p}^+\left(\partial_+-(J-1)\tan t\right)\cdots \left(\partial_+-p\tan t\right)e^{-i p y^+}=c_{J,p}^+\nabla_+^{J-p}e^{-i p y^+}\nonumber\\
&\CF^{(p,-)}_{J, p}= c_{J,p}^-\left(\partial_--(J-1)\tan t\right)\cdots \left(\partial_--p\tan t\right)e^{i p y^-}=c_{J,p}^-\nabla_-^{J-p}e^{i p y^-}~,
\end{align}
\normalsize
where  the Christoffel symbols $\Gamma^+_{++}=\Gamma^-_{--}=\tan t$ have been used. Here we want to emphasize that $e^{-i p y^+}$ and $e^{i p y^-}$ are not normal functions. They should be treated as the two lightcone components of a symmetric and traceless spin $p$ tensor. The next observation is that $[\CL_{V_+}, \nabla_\pm] \xi_{\pm\cdots \pm}=0$, for any symmetric and traceless $\xi$. It allows us to commute the Lie derivatives and covariant derivatives when computing $\CF_{J,p}^{n,\pm}$. In the end, the Lie derivatives effectively act on $e^{\mp i y^\pm}$, and this action has already been figured out in the $J=p$ case:
\begin{align}
    &\CF^{(n,\pm)}_{J, p}=  c_{J,p}^\pm\nabla_\pm^{J-p}(\mp)^{n-p}\, e^{\mp i n y^\pm}~.
\end{align}
Compared to the $J=p$ case, the only difference is the extra covariant derivative $\nabla_\pm^{J-p}$. So the previous analysis can be applied here in the exactly same way, which yields
\begin{align}\label{Jpfinal}
     \langle 0 | \CO^{(J)}(Y_1, W_1)|&\mathbb 1_{\mathcal D_p}| \CO^{(J)}(Y_2, W_2) |\Omega\rangle
     \nonumber\\
     &= 4\pi\left(|c_{J,p}^+|^2 + |c_{J,p}^-|^2\right)\left(W_1\cdot\nabla_1\right)^J
     \left(W_2\cdot\nabla_2\right)^J G_{-i(p-\frac{1}{2})}(Y_1, Y_2)~.
\end{align}
Altogether, combining (\ref{reso2D}), (\ref{2DspinJ}) and (\ref{Jpfinal}), we obtain the full \KL\, decomposition of $\CO^{(J)}$ in dS$_2$:
\begin{shaded}
\begin{align}\label{2DspinJfull}
    G_{\CO^{(J)}}(Y_1,Y_2; W_1, W_2)&=\int_{\mathbb R}d\lambda\, \rho^{\mathcal P, 0}_{\CO^{(J)}}(\lambda)(W_1\cdot\nabla_1)^J(W_2\cdot\nabla_2)^J G_{\lambda,0}(Y_1,Y_2)\nonumber\\
    &+\int_{\mathbb R}d\lambda\,\rho^{\mathcal P, 1}_{\CO^{(J)}}(\lambda)(W_1\cdot\nabla_1)^{J-1}(W_2\cdot\nabla_2)^{J-1} G_{\lambda,1}(Y_1, Y_2; W_1, W_2)\nonumber\\
   & +\int_{-\frac{1}{2}}^{\frac{1}{2}}d\lambda\, \rho^{\mathcal C, 0}_{\CO^{(J)}}(\lambda)(W_1\cdot\nabla_1)^J(W_2\cdot\nabla_2)^J G_{i\lambda,0}(Y_1,Y_2)\nonumber\\
    &+\int_{-\frac{1}{2}}^{\frac{1}{2}}d\lambda\,\rho^{\mathcal C, 1}_{\CO^{(J)}}(\lambda)(W_1\cdot\nabla_1)^{J-1}(W_2\cdot\nabla_2)^{J-1} G_{i\lambda,1}(Y_1, Y_2; W_1, W_2)\nonumber\\
    &+\sum_{p=1}^J \rho^{\mathcal D_p}_{\CO^{(J)}}\left(W_1\cdot\nabla_1\right)^J
     \left(W_2\cdot\nabla_2\right)^J G_{-i(p-\frac{1}{2})}(Y_1, Y_2)~, 
\end{align}
\end{shaded}
\noindent where the nonnegative function $\rho^{\mathcal D_p}_{\CO^{(J)}}$ is obtained by absorbing $ 4\pi\left(|c_{J,p}^+|^2 + |c_{J,p}^-|^2\right)$ into the formal sum over the discrete series in (\ref{reso2D}). It is worth mentioning that the term with $p=J$ in the sum in the last line of (\ref{2DspinJfull}) is actually  
proportional to the CFT two-point function of a spin $p$ conserved current in a dS$_2$ background.

\subsection{Flat space limit}
\label{subsec:flatspacelimit}
We have derived the \KL\ decomposition for de Sitter spinning two-point functions in $d\geq2$ and $d=1$. Now, let us consider how (\ref{eq:KL any J}) and (\ref{2DspinJfull}) reduce to the \KL\ decomposition in Minkowski space when taking the radius of de Sitter to infinity. What we expect to happen is that, given that free scalar fields with $\Delta$ in the principal series correspond to the range of masses $m^2>\frac{d^2}{4R^2}$, in the flat space limit the principal series range will be extended to account for all massive representations. The complementary series, accounting for $0<m^2<\frac{d^2}{4R^2}\,,$ is reduced to only massless representations. The same is true for the discrete series, because keeping $m^2R^2$ fixed to some discrete value in the flat space limit necessarily implies $m\to0$. Apart from these distinctions between the various dS unitary irreps, taking the flat space limit of the \KL\ decomposition in dS is analogous to how it is done in AdS \cite{meineri2023renormalization}. 

In $d+1$ dimensional Minkowski spacetime, the \KL\ decomposition of Wightman two-point functions of traceless symmetric spin $J$ operators organizes itself in blocks that are labeled by the eigenvalue of $P^\mu P_\mu\equiv-m^2$ and the spin of the little group $SO(d)$, denoted by $\ell$ \cite{Karateev:2020axc}
\begin{equation}
    \langle\Omega| \CO^{(J)}(x_1,w_1)\CO^{(J)}(x_2,w_2)|\Omega\rangle=\sum_{\ell=0}^J\int_0^\infty dm^2\ \rho^{\mathbb{M},\ell}_{\mathcal{O}^{(J)}}(m^2)\Delta^{(J)}_{m^2,\ell}(x_1,x_2;w_1,w_2),
\end{equation}
where $w_i,x_i\in\mathbb{R}^{d,1}$, $w_i$ are some null vectors to contract all indices, and $\rho^{\mathbb{M},\ell}_{\CO^{(J)}}(m^2)$ are the positive flat space spectral densities. $\Delta^{(J)}_{m^2,\ell}(x_1,x_2;w_1,w_2)$ are the free Wightman propagators with Lorentz spin $J$, little group spin $\ell$ and mass squared $m^2$
\begin{equation}
    \Delta^{(J)}_{m^2,\ell}(x_1,x_2;w_1,w_2)=(-)^{J-\ell}m^{2J}\int\frac{d^{d+1}p}{(2\pi)^d}e^{ip\cdot x}\theta(p^0)\delta(p^2+m^2)\Pi_{\ell}^{(J)}(p,w_1,w_2)\,,
    \label{eq:flatspacewightman}
\end{equation}
where $\Pi_{\ell}^{(J)}(p,w_1,w_2)$ are the projectors on the little group irrep of spin $\ell$. The prefactor $m^{2J}$ is inserted  following \cite{Karateev:2020axc}, such that $\Delta^{(J)}_{m^2,\ell}$ does not diverge in the massless limit when $d\ge 2$. This way, massless representations are smoothly connected to massive ones, and they appear with spectral densities that are proportional to $\delta(m^2)$. In contrast to \cite{Karateev:2020axc}, we also include a spin-dependent sign $(-)^{J-\ell}$ in $\Delta^{(J)}_{m^2,\ell}$. This choice is consistent with the positivity of $\rho^{\mathbb{M},\ell}_{\CO^{(J)}}(m^2)$.

In the large $R$ limit, the conformal dimension $\Delta=\frac{d}{2}+i\lambda$ is connected to the mass $m$ (which is kept fixed) as $\lambda^2\approx m^2 R^2$. 
In Appendix \ref{subsec:flatspaceG} we argue that, when taking $R\to\infty$ while keeping $Y_1\cdot Y_2-R^2$ fixed, the free propagators become 
\begin{equation}
    [(W_1\cdot\nabla_1)(W_2\cdot\nabla_2)]^{J-\ell}G_{R m,\ell}(Y_1,Y_2;W_1,W_2)\approx\,\beta_{J,\ell}\, m^{-2\ell}\Delta_{m^2,\ell}^{(J)}(x_1,x_2;w_1,w_2)~,
    \label{eq:propagatorasympt}
\end{equation}
where $\beta_{J,\ell}$ are normalization factors, known for $0\le \ell\le J\le 2$. For example, it is equal to $1$ when $0\le J\le 1$, and  is given by eq. (\ref{flat3}) when $J=2$. 
Now consider the principal series contribution to the \KL\ decomposition after performing the change of variables $\lambda=Rm$\footnote{Notice that, assuming the two-point function we are decomposing has mass dimensions $2\mathbf{\Delta}$, and given that the mass dimensions of the free propagators are
\begin{equation}
    [[(W_1\cdot\nabla_1)(W_2\cdot\nabla_2)]^{J-\ell}G_{Rm,\ell}(Y_1,Y_2;W_1,W_2)]=d-1+2(J-\ell)\,,
\end{equation}
the correct way to restore dimensions to the spectral densities is to reintroduce an extra factor of $R^{d-1+2(J-\ell)-2\mathbf{\Delta}}$ in $\rho_{\mathcal{O}^{(J)}}^{\mathcal{P,\ell}}(Rm)$. }
\begin{equation}
    \sum_{\ell=0}^J\int_0^\infty dm^2\frac{R\, }{m}\rho_{\mathcal{O}^{(J)}}^{\mathcal{P},\ell}(Rm)[(W_1\cdot\nabla_1)(W_2\cdot\nabla_2)]^{J-\ell}G_{Rm,\ell}(Y_1,Y_2;W_1,W_2)~.
    \label{eq:KLwithR}
\end{equation}
If the two-point function we are decomposing does not diverge as $R\to\infty$, (\ref{eq:KLwithR}) becomes under this limit
\begin{equation}
    (\ref{eq:KLwithR})\approx\sum_{\ell=0}^J\int_0^\infty dm^2\ \rho^{\mathbb{M},\ell}_{\mathcal{O}^{(J)}}(m^2)\Delta_{m^2,\ell}^{(J)}(x_1,x_2;w_1,w_2)\,,\label{eq:flatspace}
\end{equation}
where we read off the connection between de Sitter principal series and flat space spectral densities 
\begin{shaded}
\begin{equation}
    \rho^{\mathbb{M},\ell}_{\mathcal{O}^{(J)}}(m^2)=\lim_{R\to\infty}\frac{\beta_{J,\ell}\, R}{m^{2\ell+1}}\rho^{\mathcal{P},\ell}_{\mathcal{O}^{(J)}}(Rm)+\cdots \label{eq:rholimit}
\end{equation}
\end{shaded}
\noindent where the dots stand for contributions coming from other UIRs. In practice this means that, in the large $R$ limit, de Sitter spectral densities grow with a power of $R$ which is fixed by dimensional analysis, and its coefficient is the associated flat space spectral density. We check that (\ref{eq:rholimit}) is true for our CFT examples in section \ref{subsec:cfts} by comparing with the flat space CFT spectral densities computed in \cite{Karateev:2020axc}, and find perfect agreement. Let us also discuss the discrete series contributions in $d=1$:
\begin{equation}
    \sum_{p=0}^J\rho_{\mathcal{O}^{(J)}}^{\mathcal{D}_p}(W_1\cdot\nabla_1)^J(W_2\cdot\nabla_2)^JG_{-i(p-\frac{1}{2})}(Y_1,Y_2)\,.
\end{equation}
To restore dimensions correctly, we need to redefine $\rho_{\mathcal{O}^{(J)}}^{\mathcal{D}_p}$ with a factor of $R^{-2\mathbf{\Delta}+2J}$ where $2\mathbf{\Delta}$ is the mass dimension of the two-point function we are decomposing. Then, these contributions can be incorporated in (\ref{eq:rholimit}) as 
\begin{shaded}
\begin{equation}
    \rho^{\mathbb{M},0}_{\mathcal{O}^{(J)}}(m^2)=
    \delta(m^2)\sum_{p=0}^J\lim_{R\to\infty}R^{-2(\mathbf{\Delta}-J)}\rho^{\mathcal{D}_p}_{\mathcal{O}^{(J)}}\,, \qquad \text{if }d=1\,.
\end{equation}
\end{shaded}
\noindent We find agreement between these statements and the massless representations that appear in $d=1$ in the CFT spectral densities in \cite{Karateev:2020axc} when the CFT primary being decomposed is a conserved current. We show this explicitly in the examples in section \ref{subsec:cfts}.
\section{Inversion formulae}
\label{sec:inversionformula}

In this section, we find inversion formulae that extract the spectral densities from the \Kallen decompositions~\reef{eq:KL any J} and~\reef{2DspinJfull}.
For the scalar two-point function, using the analytic continuation from the sphere, an inversion formula was found in~\cite{Hogervorst:2021uvp}. There, it was shown that the spectral density can be computed by carrying out an integral over the discontinuity  of the two-point function in the region where the two points are timelike separated. In this section, we propose an alternative and more convenient procedure to derive the spectral densities for spinning de Sitter two-point functions using harmonic analysis in Euclidean Anti de Sitter (EAdS).  We will first derive an inversion formula for the principal series spectral densities in $d\geq 2$ and then one for the principal series and discrete series contributions in $d=1$. The main idea is to continue the \KL\ decomposition (\ref{eq:KL any J}) from dS to EAdS and to exploit the orthogonality of harmonic functions under integrals over EAdS. We emphasize that this method is a mathematical trick, and that all spectral densities we derive in this way lead to \KL\ integral representations which we numerically test directly in de Sitter. In section \ref{subsec:completeness}, we argue that, for two-point functions satisfying certain criteria, there are no more contributions to the \KL\ decomposition other than principal series contributions. 

Our derivation of the inversion formula relies on a specific assumption on the analytic structure of spinning two-point functions.
\paragraph{Analyticity of spinning two-point functions} In \cite{loparco2023radial}, it was shown that scalar two-point functions with a convergent K\"allén-Lehmann decomposition on the sphere are analytic within the ``maximal analyticity" domain $\sigma\equiv Y_1\cdot Y_2\in\mathbb{C}/[1,\infty)$. In \cite{Bros:1995js}, it was shown that this domain also follows from assuming a smaller domain of analyticity, within the ``forward tube" domain, defined as 
\begin{equation}\label{eq:forwardtube}
\begin{split}
\eta_{AB}Y^A_1Y_1^B=\eta_{AB}Y^A_2Y_2^B=1, \quad \mathrm{Im}(Y_{21})\in V_{+},
\end{split}
\end{equation}
where we are considering complexified $Y^A\in\mathbb{C}^{d+2}$, $Y_{ij}\equiv Y_i-Y_j$ for convenience, and $V_+$ represents the forward lightcone, defined as:
\begin{equation}
\begin{split}
V_+:=\left\{Y\in \mathbb{R}^{1,d+1}\ \Big{|}\ Y^0>\sqrt{\sum\limits_{a=1}^{d+1}\left(Y^a\right)^2}\right\}.
\end{split}
\end{equation}
We are not aware of any generalization of either of these two statements to spinning two-point functions. In general, we can say that spinning two-point functions only depend on a few dot products
\begin{equation}
    G_{\mathcal{O}^{(J)}}(Y_1,Y_2;W_1,W_2)=G_{\mathcal{O}^{(J)}}(\sigma,(Y_1\cdot W_2)(Y_2\cdot W_1),W_1\cdot W_2)\,,
\end{equation}
where the dependance on $((Y_1\cdot W_2)(Y_2\cdot W_1),W_1\cdot W_2)$ is by construction purely polynomial.

\begin{figure}
\centering
\includegraphics[scale=1.2]{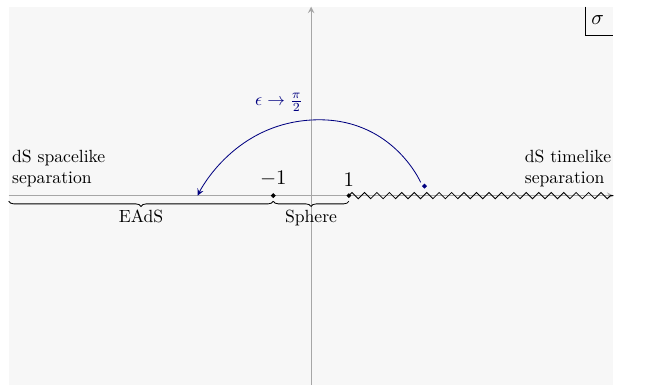}
\caption{The maximal analyticity domain in the complex $\sigma\equiv Y_1\cdot Y_2$ plane. We indicate the range of values taken by $\sigma$ in EAdS and on the sphere. In de Sitter, $\sigma$ can take all real values, with $\sigma>1$ for timelike separation, $\sigma<1$ for spacelike separation and $\sigma=1$ for null separation. 
On the sphere, $\sigma=1$ for coincident points and $\sigma=-1$ for antipodal points. In EAdS, $\sigma=-1$ for coincident points. The Wick rotation from dS to EAdS which we implement in this paper (see section \ref{subsec:wick}), maps points in $\sigma=\mathbb{R}\pm i\epsilon$ to points in $\sigma\in(-\infty,-1]$ through rotations which avoid the cut at $\sigma\in[1,\infty)$.} 
\label{fig:chordal}
\end{figure}

We are then going to phrase the following conjecture
\begin{shaded}
    \noindent\textbf{Conjecture:} \emph{Let $G_{\mathcal{O}^{(J)}}$ be the two-point function of a spin $J$ traceless symmetric operator with a positive and convergent K\"allén-Lehmann decomposition of the form (\ref{eq:KL any J}) with principal and complementary series contributions only. Then, $G_{\mathcal{O}^{(J)}}$ is analytic in $\sigma\in\mathbb{C}/[1,\infty)$. 
    }
\end{shaded}
\noindent
We will call this domain of analyticity ``maximal analyticity" like in the scalar case. Notice that this domain includes two-point configurations on the Euclidean sphere $S^{d+1}$ and the 
Euclidean Anti de Sitter space $\text{EAdS}_{d+1}$. The range that $\sigma$ takes in these Euclidean spaces is reported in Figure \ref{fig:chordal}. Importantly, the Wick rotation which we will make use of in this paper, discussed in section \ref{subsec:wick}, moves points in the complex $\sigma$ plane from de Sitter to EAdS without crossing the cut.

For in depth discussions on the analyticity of two-point functions in complexified de Sitter, we refer the reader to \cite{Bros:1995js,Bros_1998,Schlingemann:1999mk,Sleight:2019mgd,Sleight_2020,Sleight_2021Exch,Sleight:2021plv,loparco2023radial}. 
\subsection{Wick rotation to EAdS}
\label{subsec:wick}
As mentioned above, the first step to invert the \KL\ decomposition is continuing (\ref{eq:KL any J}) to Euclidean Anti de Sitter space, of which we review various coordinate systems and for which we set up notation in Appendix \ref{subsec:coordinates}. Here we describe the precise way in which we realize this continuation, inspired by what was done in \cite{DiPietro:2021sjt,Sleight:2021plv,Anninos_2015,Sleight_2021Exch,Sleight_2020,Sleight:2019mgd}.By $SO(d+1,1)$ invariance, Wightman functions only depend on the following dot products
\begin{equation}
    G_{\mathcal{O}^{(J)}}(Y_1,Y_2;W_1,W_2)=G_{\mathcal{O}^{(J)}}(Y_1\cdot Y_2,(Y_1\cdot W_2)(Y_2\cdot W_1),W_1\cdot W_2)
\end{equation}
As discussed in \ref{subsec:scalaropsKLd}, Wightman functions in de Sitter are defined with an $i\epsilon$ prescription which is realized in planar coordinates (\ref{Xetax}) as 
\begin{equation}
    Y_1=Y_1(\eta_1 e^{i\epsilon},\vec y_1)\,,\ Y_2=Y_2(\eta_2 e^{-i\epsilon},\vec{y}_2)\,.
\end{equation}
We Wick rotate to EAdS by simply taking $\epsilon\to\frac{\pi}{2}$ and identifying the EAdS Poincar\'e coordinates $z=|\eta|$ and $\vec x=\vec y$, so that the dot products transform as
\begin{equation}
    \begin{aligned}
        Y_1\cdot Y_2\underset{\epsilon\to\frac{\pi}{2}}{\longrightarrow}X_1\cdot X_2\,, \qquad (Y_1\cdot W_2)(Y_2\cdot W_1)\underset{\epsilon\to\frac{\pi}{2}}{\longrightarrow}(X_1\cdot W_2)(X_2\cdot W_1)\,,
    \end{aligned}
\end{equation}
where $X\in\text{EAdS}_{d+1}$ and $X\cdot W=0$ as is reviewed in \ref{subsec:coordinates}. It can be checked that, under this particular Wick rotation, $Y_1\cdot Y_2$ will move through the domain of analyticity discussed in section \ref{sec:kallanlehmann} and will not cross the cut at $Y_1\cdot Y_2\in[1,\infty)$. In Figure \ref{fig:chordal} we show an example of how $Y_1\cdot Y_2$ moves in the complex plane under this rotation. Moreover, this Wick rotation maps Wightman functions in dS for free traceless symmetric tensor fields to harmonic functions in EAdS (eq. (2.70) in \cite{Sleight_2020}):
\begin{equation}\label{eq:G to Omega}
    G_{\lambda,\ell}(Y_1,Y_2;W_1,W_2)\underset{\epsilon\to\frac{\pi}{2}}{\longrightarrow}\Gamma(\pm i\lambda)\,\Omega_{\lambda,\ell}(X_1,X_2;W_1,W_2)\,,
\end{equation} 
where throughout this paper we use the shorthand convention that, inside gamma functions and Pochhammer symbols, $\Gamma(a\pm b)\equiv\Gamma(a+b)\Gamma(a-b).$

In Appendix 
\ref{sec:harmonicanalysis}, we review some of the useful properties of harmonic functions. Among them, the orthogonality relation~\reef{eq:Omega ortho}
will play a crucial role in the derivation of our inversion formula.

\subsection{Inversion formula for $d\geq2$}
We start with the spinning \KL\ decomposition in $d\geq2$ that we proved in section \ref{sec:kallanlehmann}. After the Wick rotation to EAdS, it reads
\small
\begin{align}
    G_{\CO^{(J)}}(X_1,X_2; W_1, W_2)=&\sum_{\ell=0}^J\int_{\mathbb{R}}d\lambda\ \rho^{\mathcal{P},\ell}_{\mathcal{O}^{(J)}}(\lambda)\,[(W_1\cdot\nabla_1)(W_2\cdot\nabla_2)]^{J-\ell}\Gamma(\pm i\lambda)\Omega_{\lambda,\ell}(X_1,W_1;X_2,W_2)\nonumber\\
    &+\text{possible contributions from other UIRs}
    \label{eq:KLharmonic}
\end{align}
\normalsize
As we will discuss in section~\ref{subsec:completeness}, the harmonic functions form a complete and orthogonal basis of square-integrable two-point functions \cite{Costa_2014}. In other words, if a two-point function is square-integrable, its \KL\ decomposition only includes principal series contributions. 
In $d\geq 2$, we find that the two-point functions we considered in our examples in section \ref{sec:applications} can always be studied in a regime where the principal series contributions reproduce the full two-point function, and then by analytic continuation away from that regime we could recover any complementary series part as poles that cross the contour of integration in (\ref{eq:KLharmonic}). We have not encountered exceptional series contributions in any of our examples. 
 Given these facts, let us for now focus on inverting the decomposition over the principal series. To exploit the orthogonality of the harmonic functions, we act on both sides of \reef{eq:KLharmonic} with the integro-differential operator
\begin{equation}
    \int_{X_1}\Omega_{\lambda',m}(X_3,X_1,W_3,K_1)[(K_1\cdot \nabla_1)(K_2\cdot\nabla_2)]^{J-m}\,.
\end{equation}
where we use the shorthand notation for integrating $X_1$ over EAdS defined in~(\ref{eq:intX}).
The right hand side of (\ref{eq:KLharmonic}) becomes
 \begin{equation}
    \sum_{\ell=0}^J\int d\lambda\ \Gamma(\pm i\lambda)\rho^{\mathcal{P},\ell}_{\mathcal{O}^{(J)}}(\lambda)\int_{X_1}\Omega_{\lambda',m}[(K_1\cdot \nabla_1)(K_2\cdot\nabla_2)]^{J-m}[(W_1\cdot\nabla_1)(W_2\cdot\nabla_2)]^{J-\ell}\Omega_{\lambda,\ell}\,,
     \label{eq:step0inversion}
 \end{equation}
where we are omitting the arguments of the harmonic functions to avoid clutter. Let us focus on the quantity
\begin{equation}
     (K_1\cdot \nabla_1)^{J-m}(W_1\cdot\nabla_1)^{J-\ell}\Omega_{\lambda,\ell}(X_1,X_2;W_1,W_2)\,.
     \label{eq:step1inversion}
\end{equation}
The fact that divergences of harmonic functions vanish, implies that we can express (\ref{eq:step1inversion}) in terms of commutators
\begin{equation}
    [(K_1\cdot \nabla_1)^{J-m},(W_1\cdot\nabla_1)^{J-\ell}]\Omega_{\lambda,\ell}(X_1,X_2;W_1,W_2)\,.
\end{equation}
Using basic properties of commutators together with the divergenceless condition, we can write this as
\begin{equation}
 (K_1\cdot \nabla_1)^{J-m-1}[(K_1\cdot \nabla_1),(W_1\cdot\nabla_1)^{J-\ell}]\Omega_{\lambda,\ell}(X_1,X_2;W_1,W_2)\,.
\label{eq:step2inversion}
\end{equation}
Evaluating this commutator (\ref{eq:commuteKW}), we get that (\ref{eq:step0inversion}) can be written as 
\small
\begin{equation}
 \sum_{\ell=0}^J\kappa^2_{J-\ell,\ell}\int d\lambda\ \Gamma(\pm i\lambda)\rho^{\mathcal{P},\ell}_{\mathcal{O}^{(J)}}(\lambda)\int_{X_1}\Omega_{\lambda',m}[(K_1\cdot \nabla_1)(K_2\cdot\nabla_2)]^{J-m-1}[(W_1\cdot\nabla_1)(W_2\cdot\nabla_2)]^{J-\ell-1}\Omega_{\lambda,\ell}\,,
\end{equation}
\normalsize
with
\begin{equation}
    \kappa_{n,m}\equiv-\frac{n}{2}(d+n+2m-2)\left(\left(\frac{d}{2}+m+n-1\right)^2+\lambda^2\right)\,.
\end{equation}
By iteration, we find three possible scenarios that can happen to the integral over $X_1$:
\begin{itemize}
    \item If $\ell>m$, the spatial integral would eventually be proportional to
         \begin{equation}
             \int_{X_1}\Omega_{\lambda',m}(X_3,X_1;W_3,K_1)(K_1\cdot\nabla_1)(K_2\cdot\nabla_2)\Omega_{\lambda,\ell}(X_1,X_2;W_1,W_2)=0
         \end{equation}
     \item If $\ell<m$, instead, one eventually obtains
     \begin{equation}
             \int_{X_1}\Omega_{\lambda',m}(X_3,X_1;W_3,K_1)(W_1\cdot\nabla_1)(W_2\cdot\nabla_2)\Omega_{\lambda,\ell}(X_1,X_2;W_1,W_2)=0\,,
         \end{equation}
         which vanishes because integrating by parts the derivative $\nabla_1$, the integrand becomes a divergence on $\Omega_{\lambda',m}\,.$
     \item The case $\ell=m$ is the only one that does not vanish. Instead, it gives for the spatial integral
     \begin{equation}
     \begin{aligned}
         \int_{X_1}\Omega_{\lambda',m}[(K_1\cdot \nabla_1)&(K_2\cdot\nabla_2)]^{J-m}[(W_1\cdot\nabla_1)(W_2\cdot\nabla_2)]^{J-\ell}\Omega_{\lambda,\ell}\\
         &=\delta_{\ell,m}\delta(\lambda-\lambda')\left(\prod_{n=1}^{J-\ell}\kappa^2_{n,\ell}\right)\Omega_{\lambda,\ell}(X_2,X_3;W_2,W_3)\,.
     \end{aligned}
     \end{equation}
 \end{itemize}
This procedure allows us to isolate the $\ell$-th contribution in the sum in (\ref{eq:KLharmonic})
\begin{equation}
\begin{aligned}
    \widetilde{\mathcal{N}}_{J,\ell}\ &\Omega_{\lambda,\ell}(X_2,X_3,W_2,W_3)\rho^{\mathcal{P},\ell}_{\mathcal{O}^{(J)}}(\lambda)\\
    &=\int_{X_1}\Omega_{\lambda,\ell}(X_3,X_1;W_3,K_1)[(K_1\cdot \nabla_1)(K_2\cdot\nabla_2)]^{J-\ell}G_{\mathcal{O}^{(J)}}(X_1,X_2;W_1,W_2)\,,
    \label{eq:KL43}
\end{aligned}
\end{equation}
where 
\small
\begin{equation}
    \widetilde{\mathcal{N}}_{J,\ell}\equiv \Gamma(\pm i\lambda)\prod_{n=1}^{J-\ell}\kappa_{n,\ell}^2=\frac{\ell!\Gamma(\pm i\lambda)}{2^{2(J-\ell)}}\left(\frac{d-1}{2}\right)_\ell\left((J-\ell)!(d+2\ell-1)_{J-\ell}\left(\frac{d}{2}\pm i\lambda+\ell\right)_{J-\ell}\right)^2~.
\end{equation}
\normalsize
Equation (\ref{eq:KL43}) is valid for all $X_2$ and $X_3$ in EAdS as well as any null $W_2$ and $W_3$ that satisfy the tangential condition. We therefore pick the convenient choice of $X_2=X_3$. Note that, unlike bulk-to-bulk propagators, harmonic functions are regular at coincident points in EAdS. In addition, we take a trace over the free indices by performing the substitution $W_3\rightarrow K_2~$. After all these operations, we find an \textbf{inversion formula} for the principal series spectral densities appearing in the \KL\ decomposition of spinning two-point functions in dS
\begin{shaded}
\begin{equation}
\label{eq:inversionformulal}
    \rho^{\mathcal{P},\ell}_{\mathcal{O}^{(J)}}(\lambda)=\frac{1}{\mathcal{N}_{J,\ell}}\int_{X_1}\Omega_{\lambda,\ell}(X_2,X_1;K_2,K_1)[(K_1\cdot \nabla_1)(K_2\cdot\nabla_2)]^{J-\ell}G_{\CO^{(J)}}(X_1,X_2; W_1, W_2)\,,
\end{equation}
\end{shaded}
\noindent with
\begin{equation}
    \mathcal{N}_{J,\ell}\equiv\widetilde{\mathcal{N}}_{J,\ell}\ \Omega_{\lambda,\ell}(X,X;K,W)\,
    \label{eq:calN}.
\end{equation}
The trace and coincident point limit of $\Omega_{\lambda,\ell}$ was computed in \cite{Costa_2014}. Here, we report the result
\begin{equation}
    \Omega_{\lambda,J}(X,X;K,W)=\frac{J!\left(\frac{d-1}{2}\right)_J g(J)}{(4\pi)^{\frac{d+1}{2}}\Gamma(\frac{d+1}{2})\Gamma(\pm i\lambda)}  \left(\frac{d}{2}+J-1 \pm i \lambda\right)\Gamma\left(\frac{d}{2}-1\pm i\lambda\right)
\end{equation}
with
\ba
    g(J)&=\frac{(2J+d-2)(J+d-3)!}{(d-2)!J!}\,, \qquad &&d\geq 3\,,\\
    g(0)&=1\,, \qquad g(J\neq 0)=2\,, \qquad &&d=2~.
\ea 
Altogether, the overall normalization factor $\CN_{J,\ell}$ is equal to 
\begin{align}\label{CNJl}
    \CN_{J,\ell}= \frac{g(\ell)\left[ \ell!\,(J-\ell)!\,(\frac{d-1}{2})_\ell\,(d+2\ell-1)_{J-\ell}\,\Gamma(\frac{d}{2}+J\pm i\lambda)\right]^2}{4^{J-\ell}(4\pi)^{\frac{d+1}{2}}\Gamma(\frac{d+1}{2})\Gamma(\frac{d}{2}+\ell\pm i\lambda)\prod_{t=0}^{\ell-1}\left(\frac{d}{2}\pm i\lambda +t-1\right)}~.
\end{align}

In practice, one might conveniently evaluate \reef{eq:inversionformulal} by placing $X_2$ at the origin of EAdS. This choice makes the angular part of the integral trivial after carrying out all derivatives and index contractions. Therefore, we will be left with a one-dimensional integral over the EAdS chordal distance. We spell out the  explicit formulae of these one dimensional integrals  for $J=0,1$ in Appendix \ref{sec:explicitinversion}.

\subsubsection{Spurious poles}
\label{subsubsec:spurious}
The inversion formula (\ref{eq:inversionformulal}) implies that the spectral density $\rho_{\mathcal{O}^{(J)}}^{\mathcal{P},\ell}(\lambda)$ may contain $\CO^{(J)}$-independent poles in the complex $\lambda$ plane coming from the normalization factor $\CN^{-1}_{J,\ell}$ and the harmonic function $\Omega_{\lambda,\ell}$, which  we will refer to as \emph{spurious poles}. First, we claim that the poles of $\Omega_{\lambda,\ell}$ actually do not lead to poles in $\rho_{\mathcal{O}^{(J)}}^{\mathcal{P},\ell}$. 
More precisely, focusing on the $\lambda$-dependent part of $\mathcal{N}_{J,\ell}$, c.f. eq. (\ref{CNJl}) 
\begin{equation}
    \mathcal{N}_{J,\ell}^{-1}\sim\frac{\Gamma(\frac{d}{2}\pm i\lambda+\ell)}{\Gamma(\frac{d}{2}\pm i\lambda+J)^2}\prod_{t=0}^{\ell-1}\left(\left(\frac{d}{2}-1+t\right)^2+\lambda^2\right)\,,
\end{equation}
one can show that the factors in the product cancel out all the poles of $\Omega_{\lambda, \ell}$. To illustrate this cancellation, we write down the poles and residues of $\Omega_{\lambda, \ell}$ for $\ell$ up to 2, using the explicit expressions of the harmonic functions given in appendix \ref{subsec:hfunc}
\begin{align}
        &id\underset{\lambda=-i\frac{d-2}{2}}{\text{Res}}\left[\Omega_{\lambda,1}(X_1,X_2;W_1,W_2)\right]=(W_1\cdot\nabla_1)(W_2\cdot\nabla_2)\Omega_{-i\frac{d}{2},0}(X_1,X_2)\,,\label{eq:propagatorpoles}\\
         &i\frac{d+2}{2}\underset{\lambda=-i\frac{d}{2}}{\text{Res}}\left[\Omega_{\lambda,2}(X_1,X_2;W_1,W_2)\right]=(W_1\cdot\nabla_1)(W_2\cdot\nabla_2)\Omega_{-i\frac{d+2}{2},1}(X_1,X_2;W_1,W_2)\,,\nonumber\\
        &-id(d+2)\underset{\lambda=-i\frac{d-2}{2}}{\text{Res}}\left[\Omega_{\lambda,2}(X_1,X_2;W_1,W_2)\right]=(W_1\cdot\nabla_1)^2(W_2\cdot\nabla_2)^2\Omega_{-i\frac{d+2}{2},0}(X_1,X_2)\,.\nonumber
    \end{align}
These relations have very clear physical meanings. For example, in the first line of (\ref{eq:propagatorpoles}), evaluating the residue of $\Omega_{\lambda,1}$ at $\lambda=-i\frac{d-2}{2}$ amounts to 
approaching the massless limit of the free two-point function of a Proca field, recalling that the Proca mass in dS$_{d+1}$ is given by $\sqrt{(\frac{d}{2}-1)^2+\lambda^2}$ \footnote{Although we state the relations (\ref{eq:propagatorpoles}) in terms of EAdS harmonic functions, it apparently also holds (up to the proportional constant) for de Sitter free two-point function $G_{\lambda, \ell}$, as a direct result of the Wick rotation (\ref{eq:G to Omega}). We will use the dS version of (\ref{eq:propagatorpoles}) when explaining the underlying physical picture.}. The same as in flat space, the longitudinal part of a Proca two-point function diverges in the massless limit and can be removed by a gauge transformation, with the ghost field being a massless scalar. This explains the appearance of $(W_1\cdot\nabla_1)(W_2\cdot\nabla_2)\Omega_{-i\frac{d}{2},0}$. Similarly, the second and third lines of (\ref{eq:propagatorpoles}) correspond to taking the massless and partially massless limit \cite{Deser:1983mm, Deser:2001pe, Deser:2001us,Deser:2001wx, Zinoviev:2001dt} of a free spin 2 field respectively. In the latter case, the ghost field is a tachyonic scalar of mass square $m^2=-(d+1)$ and that is why we have  $(W_1\cdot\nabla_1)^2(W_2\cdot\nabla_2)^2\Omega_{-i\frac{d+2}{2},0}$.
More generally, $\Omega_{\lambda, \ell}$ has a simple pole at the partially massless point of depth $t\in\{0,1,\cdots, \ell-1\}$, i.e. $\lambda= - i (\frac{d}{2}+t-1)$, and at this point the residue is proportional to  $(W_1\cdot\nabla_1)^{\ell-t}(W_2\cdot\nabla_2)^{\ell-t}\Omega_{-i(\frac{d}{2}+\ell-1),t}$:
\begin{align}\label{Omegares}
    \underset{\lambda=-i(\frac{d}{2}+t-1)}{\text{Res}}\Omega_{\lambda,\ell} = \frac{1}{\alpha_{\ell, t}}(W_1\cdot\nabla_1)^{\ell-t}(W_2\cdot\nabla_2)^{\ell-t}\Omega_{-i(\frac{d}{2}+\ell-1),t}~,
\end{align}
where $\alpha_{\ell, t}$ is a constant, e.g. $\alpha_{1, 0}= id, \alpha_{2,1}= i \frac{d+2}{2}, \alpha_{2, 0}=-id(d+2)$. Apparently, such poles are precisely cancelled by the corresponding zeros in $\CN_{J, \ell}^{-1}$. Before proceeding to discuss the poles of $\CN_{J, \ell}^{-1}$, we'd like to make a conjecture about the explicit form of  $\alpha_{\ell,t}$ for generic $\ell$ and $t$:
\begin{align}\label{aconj}
   \alpha_{\ell,t}= -i (-2)^{\ell-t} \binom{\ell}{t}^{-1} \,\frac{\Gamma \left(\frac{d}{2}+\ell\right) \Gamma (\ell-t)}{\Gamma \left(\frac{d}{2}+t\right)}~.
\end{align}
It matches the known results of $\alpha_{\ell, t}$ for $\ell\le 2$.

The remaining ratio of gamma functions in $\CN^{-1}_{J,\ell}$ has poles at  
\begin{equation}
\lambda=\pm i\left(\frac{d}{2}+\ell+q-1\right)\,,\qquad q\in\{1,2,\cdots,J-\ell\}\,.
\label{eq:positionspurious}
\end{equation}
Combing the conjecture (\ref{aconj}) and the inversion formula (\ref{eq:inversionformulal}), we can derive a relation between the residue of  $\rho^{\CP,\ell}_{\CO^{(J)}}$ at these spurious poles and the value of $\rho^{\CP,\ell+q}_{\CO^{(J)}}$ at $\lambda=-i\left(\frac{d}{2}+\ell-1\right)$:
\begin{shaded}
\begin{align}\label{rhorelation}
  \rho^{\CP,\ell+q}_{\CO^{(J)}}\left(-i\left(\frac{d}{2}+\ell-1\right)\right)=i \, 2^q \, \Gamma(q)\binom{\ell+q}{\ell}^{-1} \left(\frac{d}{2}+\ell-1\right)_q\, \underset{\lambda=-i(\frac{d}{2}+\ell+q-1)}{\text{Res}}\rho^{\CP,\ell}_{\CO^{(J)}}(\lambda)
\end{align}
\end{shaded}
\noindent
We note that these identities are very similar to relations between conformal blocks and partial amplitudes of different spins and conformal dimension found in the AdS and CFT literature \cite{Costa_2014,Costa:2012cb,Cornalba:2007fs}. 
In all the examples we have tested in section \ref{sec:applications}, the relations (\ref{rhorelation}) are verified to hold.

In section \ref{subsubsec:boundarytheory}, we will argue that closing the contour of integration over the principal series in (\ref{eq:KL any J}) and taking the late time limit, turns the \KL\ decomposition into a sum over boundary operators. The identities (\ref{Omegares}) and (\ref{rhorelation}) ensure
\begin{align}
    \underset{\lambda=-i(\frac{d}{2}+\ell+q-1)}{\text{Res}}\Big[\rho^{\CP,\ell}_{\CO^{(J)}}(\lambda) &(W_1\cdot\nabla_1\, W_2\cdot \nabla_2)^{J-\ell} G_{\lambda, \ell}\Big]\nonumber\\
    &= - 
      \underset{\lambda=-i(\frac{d}{2}+\ell-1)}{\text{Res}}\Big[\rho^{\CP,\ell+q}_{\CO^{(J)}}(\lambda) (W_1\cdot\nabla_1\, W_2\cdot \nabla_2)^{J-\ell-q} G_{\lambda, \ell+q}\Big]~,
\end{align}
where the residue on the L.H.S comes from $\rho^{\CP,\ell}_{\CO^{(J)}}(\lambda)$, and the residue on the R.H.S comes from $G_{\lambda, \ell+q}$. This relation ensures that the spurious poles picked up when closing the contour of integration do not contribute to the two-point function of $\CO^{(J)}$, implying the absence of boundary operators with the spurious conformal dimensions $\Delta=d+\ell+q-1$ in the Boundary Operator Expansion of $\CO^{(J)}$.

\subsubsection{Relation to the inversion formula from the sphere}
In this section, we compare the explicit form of (\ref{eq:inversionformulal}) in $J=0$ case with the inversion formula obtained from analytical continuation from the sphere in \cite{Hogervorst:2021uvp}. In Appendix \ref{sec:explicitinversion}, we show that the inversion formula~\reef{eq:inversionformulal} for some scalar operator $\mathcal{O}$ simplifies to
\be\label{eq:rhoAdS}
\rho^{\mathcal{P},0}_{\mathcal{O}}(\lambda) = \frac{2\pi^\frac{d+1}{2}}{\Gamma(\pm i\lambda)} \int_{-\infty}^{-1} d\sigma \, (\sigma^2-1)^\frac{d-1}{2} \, \mathbf{F}\left(\hd+i\lambda,\hd-i\lambda,\frac{d+1}{2},\frac{1+\sigma}{2}\right) \, G_{\mathcal{O}}(\sigma)~.
\ee
where $G_{\mathcal{O}}(\sigma)$ is the two-point function of $\mathcal{O}$ and by symmetry it can only depend on the $SO(d+1,1)$ invariant $\sigma\equiv Y_1\cdot Y_2$. The spectral density can thus be derived from an integral over a range of $\sigma$ that corresponds to a part of the spacelike separated region in de Sitter. 
This means one would be able to reconstruct the whole two-point function just having access to its value in the region $\sigma \in (-\infty,-1)$.\footnote{Here we assume the two-point function is well-defined and single-valued and satisfies the appropriate conditions for the completeness of the principal series discussed in section~\ref{subsec:completeness}}

Equation~\reef{eq:rhoAdS} is another version of the inversion formula~\cite{Hogervorst:2021uvp} 
\small
\be\label{eq:RhoFormulaNew}
\rho^{\mathcal{P},0}_{\mathcal{O}}(\lambda)=\frac{ (4\pi)^{\frac{d-1}{2}}\Gamma(1-\hd\pm i\lambda)}{i \Gamma(\pm i\lambda)} \int_1^\infty d\sigma \;  \mathbf{F}\left(1-\hd + i\lambda,1-\hd - i\lambda, \frac{3-d}{2},\frac{1-\sigma}{2}\right) \text{Disc}\left[G_{\mathcal{O}}(\sigma)\right]
\ee
\normalsize
which was found by analytical continuation from the sphere. Here the discontinuity is defined as $ \text{Disc}\left[G_{\mathcal{O}}(\sigma)\right] = \lim_{\epsilon \to 0} G_{\mathcal{O}}(\sigma+i\epsilon)-G_{\mathcal{O}}(\sigma-i\epsilon)$. The integral in~\reef{eq:RhoFormulaNew} is over the timelike separated region ($\sigma\in[1,\infty)$) where the two-point function has a branch cut and the integration is over its discontinuity. 

\begin{figure}
\centering
\includegraphics[scale=1.2]{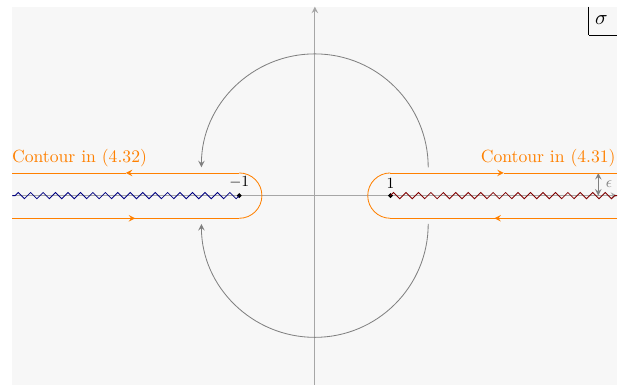}
\caption{The contour deformation that illustrates the equivalence of the two inversion formulae -- eq.~\reef{eq:RhoFormulaNew} from the sphere and eq.~\reef{eq:rhoAdS} from EAdS. In orange, we represent the contour which in the inversion formula derived from the sphere is around the cut at $\sigma\in[1,\infty)$, represented by the red zigzag line. We deform the contour as indicated by the gray arrows to the $\sigma\in(-\infty,-1]$ interval where there is the cut of the hypergeometric in eq.~\reef{eq:RhoFormulaNew} represented by the blue zigzag line. We assume the two-point function satisfies the analyticity properties discussed in section \ref{sec:kallanlehmann}.}
\label{fig:contour}
\end{figure}

We now argue that these two formulae are simply equivalent assuming that the Wightman two-point function $G_{\mathcal{O}}$ satisfies the analyticity properties discussed at the beginning of Section \ref{sec:inversionformula}. 
Consider the integral~\reef{eq:RhoFormulaNew}. It can be written as a contour integral that goes around the branch cut $\sigma\in[1,\infty)$. One can deform this contour until it surrounds the region $\sigma\in(-\infty,-1]$, as is illustrated in Figure \ref{fig:contour}.

In this contour deforming process we assumed $G_{\mathcal{O}}(\sigma)$ is analytic everywhere except for the mentioned branch cut and it decays sufficiently fast so that the contribution from the arc at infinity vanishes\footnote{The hypergeometric in \reef{eq:RhoFormulaNew} falls like $\sigma^{-1+\hd}$, so $G_{\mathcal{O}}(\sigma)$ has to fall faster than $\sigma^{-\hd}$ for this contribution to vanish. This is the same condition for the completeness of the principal series discussed in section \ref{subsec:completeness}.}. This new contour surrounds the branch cut of the regularized hypergeometric function in~\reef{eq:RhoFormulaNew}, which is precisely in the region $\sigma \in (-\infty,-1)$:
\small
\begin{equation}
    \rho^{\mathcal{P},0}_{\mathcal{O}}(\lambda)=\frac{ (4\pi)^{\frac{d-1}{2}}\Gamma(1-\hd\pm i\lambda)}{i\Gamma(\pm i\lambda)} \int_{-\infty}^{-1} d\sigma \;  \text{Disc}\Big[\!\bFF{1-\hd + i\lambda}{1-\hd - i\lambda}{\frac{3-d}{2}}{\frac{1-\sigma}{2}}\!\Big]G_{\mathcal{O}}(\sigma)\,.
\end{equation}
\normalsize
The discontinuity of the hypergeometric function around its branch cut is given by
\begin{multline}
\text{Disc}\left[ \bFF{1-\hd+i\lambda}{1-\hd-i\lambda}{\frac{3-d}{2}}{\frac{1-\sigma}{2}}\right] =\\
\frac{ 2^{2-d} \pi i  }{\Gamma(1-\hd\pm i\lambda)} (\sigma^2-1)^{\frac{d-1}{2}} \bFF{\hd+i\lambda}{\hd-i\lambda}{\frac{d+1}{2}}{ \frac{1+\sigma}{2}}~.$$
\end{multline}
Using this, one finds that~\reef{eq:rhoAdS} and~\reef{eq:RhoFormulaNew} are equivalent.

\subsection{Completeness of principal series and analyticity of the spectral densities}
\label{subsec:completeness}
In this section,
we will spell out the conditions under which the \KL\ decomposition of a spinning two-point function in EAdS$_{d+1}$ with $d\geq 2$ only contains principal series representations. Moreover, by analytical continuation of the inversion formula derived in~\reef{eq:inversionformulal}, we study analytic properties of the spectral densities. 

Let us start from the fact that harmonic functions $\Omega_{\lambda,\ell} (X_1,X_2;W_1,W_2)$ with $\lambda \in \Real$ 
are a complete basis for square-integrable two-point functions in EAdS$_{d+1}$ with $d\geq 2$~\cite{Camporesi:1994ga,Costa_2014}.
In other words any square-integrable spin-$J$ two-point function in EAdS can be written as 
\be
G_{\mathcal{O}^{(J)}}(X_1,X_2;W_1,W_2) = \sum_{\ell=0}^J \int_{\mathbb{R}} d\lambda \, c_{\ell,J}(\lambda)  \left((W_1\cdot \nabla_1)(W_2\cdot \nabla_2)\right)^{J-\ell} \Omega_{\lambda , \ell}(X_1,X_2;W_1,W_2)
\ee
for some coefficients  $c_{\ell,J}(\lambda)$   that do not depend on $X_1$ and $X_2$. The right hand side has exactly the form of the principal series contributions in the \KL\ decomposition in de Sitter (\ref{eq:KL any J}). 
Therefore, if a de Sitter two-point function after the Wick rotation to EAdS is square-integrable, we expect that only contributions from representations in the principal series appear in its \Kallen decomposition. 

Let us see how square-integrability in EAdS   translates into specific conditions on two-point functions in de Sitter. A generic spin-$J$ two-point function in the index-free formalism can be organized as a polynomial in $W_1$ and $W_2$ as follows
\be\label{eq:gen GG}
G_{\CO^{(J)}}(X_1,X_2;W_1,W_2)=\sum_{n=0}^J \left(W_1\cdot W_2\right)^{J-n} \left((W_1\cdot X_2) (W_2\cdot X_1)\right)^{n} \mathcal{G}_{\mathcal{O}^{(J)}}^{(n)}(\sigma)~.
\ee
Its square-integrability can be phrased in terms of the convergence of the following integral\footnote{For instance, in the case of a scalar two-point function ($J=0$)  this condition simplifies to 
\be\label{eq: expansion of G in curly G}
\int_{X_1}  |G_{\mathcal{O}}(X_1,X_2)|^2 = \int_0^\infty dr \, \sinh^d r \,|\mathcal{G}^{(0)}_{\mathcal{O}}(-\cosh r)|^2 = \int_{-\infty}^{-1} d\sigma\, (\sigma^2-1)^\frac{d-1}{2} |\mathcal{G}^{(0)}_{\mathcal{O}}(\sigma)|^2 < \infty ~.
\ee} 
\begin{equation}\label{eq:Integrablity condition}
    \int_{X_1} G_{\CO^{(J)}}(X_1,X_2;K_1,K_2) G_{\CO^{(J)}}(X_1,X_2;W_1,W_2) < \infty\,.
\end{equation}
Substituting~\reef{eq:gen GG} into this condition, assuming that $\mathcal{G}_{\mathcal{O}^{(J)}}^{(n)}(\sigma)$ are regular on the interval $\sigma\in(-\infty,-1)$ (which corresponds to spacelike separation in de Sitter), we can keep the leading terms in the large $\sigma$ limit and obtain the following inequality\footnote{This comes from (\ref{eq:KWWWAdS}) and counting powers of $X_1$ and $X_2$ in 
\begin{equation}
    (K_1\cdot K_2)^{J-n}((K_1\cdot X_2)(K_2\cdot X_1))^n(W_1\cdot W_2)^{J-m}((W_1\cdot X_2)(W_2\cdot X_1))^m\,.
\end{equation}}
\begin{equation}
    \int_{-\infty}^{-1}d\sigma\ |\sigma|^{d-1+2J+m+n}\mathcal{G}_{\mathcal{O}^{(J)}}^{(n)}(\sigma)\mathcal{G}_{\mathcal{O}^{(J)}}^{(m)}(\sigma)<\infty\,, \qquad \forall m,n=0,\ldots,J.
\end{equation}

Now let us assume that, in the large distance limit, these functions decay as power-law\footnote{As discussed in section~\ref{subsubsec:boundarytheory}, this statement follows from the existence of the bulk-to-boundary operator expansion.}:
\begin{equation}\mathcal{G}_{\mathcal{O}^{(J)}}^{(n)}(\sigma)\limu{\sigma\to -\infty} |\sigma|^{-\omega_{J,n}-n}\,,
\end{equation} 
Then, the square-integrability of a spinning two-point function and therefore the completeness of the principal series in its \KL\ decomposition is ensured if 
\begin{shaded}
\begin{equation}
   \min_n[\text{Re}(\omega_{J,n})]>\hd+J~,\qquad \text{completeness of principal series} 
     \label{eq:completenesscondition}
\end{equation}
\end{shaded}
\noindent 
where by $\displaystyle\min_n[x_n]$ we mean the minimum value of the set $\{x_n\}$.
When the fall-offs of a two-point function violate this condition, other representations than the principal series might appear in its \KL\ decomposition. In the
examples in section \ref{sec:applications}, we observe that in the limit cases in which this inequality is saturated, the principal series is still enough to reconstruct the full two-point function.

Now let us consider a two-point function which satisfies the condition \reef{eq:completenesscondition}, so that only the principal series contributes to its \KL\ decomposition. Given the inversion formula~\reef{eq:inversionformulal}, we can analytically continue in $\lambda$ and study the analyticity properties of the principal series spectral densities by studying the convergence of the inversion integrals. For instance, consider the scalar case, in which the only spectral density is given by the inversion formula~\reef{eq:rhoAdS}. If we analytically continue this equation in the complex $\Delta=\frac{d}{2}+i\lambda$ plane, we would see that the integral in~\reef{eq:rhoAdS} is convergent if 
\begin{equation}
    d-\text{Re}(\omega_{0,0})<\text{Re}(\Delta)<\text{Re}(\omega_{0,0})\,,
    \label{eq:strip}
\end{equation}
where we used the fact that the hypergeometric in~\reef{eq:rhoAdS} has large distance fall-offs with powers $\Delta$ and $d-\Delta$. We thus expect the spectral density $\rho^{\mathcal{P}}_{\mathcal{O}^{(0)}}(\lambda)$ to be fully analytic in the strip defined in (\ref{eq:strip}). 

\begin{figure}[h!]
\centering
\begin{tikzpicture}[scale=1.25]
\coordinate (origin) at (0,0);
\fill[light-gray] (-1.5,-2) -- (-1.5,3) -- (5.5,3) -- (5.5,-2);
\draw[line width=0.3mm,black,-stealth] (origin) -- ++(5.5,0);
\draw[line width=0.3mm,black,-stealth] (origin) -- ++(0,3);
\draw[line width=0.3mm,black,-] (origin) -- ++(-1.5,0) node (zminus) {};
\draw[line width=0.3mm,black,-] (origin) -- ++(0,-2) node (xminus) {};
\filldraw[blue!70,opacity=0.3] (0.5,-2) rectangle (2.5,3);
\draw[line width=0.3mm,red!70,-,dashed] (1.5,-2) node [red, below] {$\frac{d}{2}$} --(1.5,3);
\draw[line width=0.3mm,blue!70,-,dashed] (0.5,-1);
\draw[line width=0.3mm,blue!70,-,dashed] (2.5,-1);
\draw (2.5,1) node[cross,black!70] {};
\draw (2.5,-1) node[cross,black!70] {};
\draw (4,1.5) node[cross,black!70] {};
\draw (4,-1.5) node[cross,black!70] {};
\draw (3.5,0) node[cross,black!70] {};
\draw (3-2.5,1) node[cross,black!70] {};
\draw (3-2.5,-1) node[cross,black!70] {};
\draw (3-4,1.5) node[cross,black!70] {};
\draw (3-4,-1.5) node[cross,black!70] {};
\draw (3-3.5,0) node[cross,black!70] {};
\draw[black!50, thick,stealth-] (1.5,3.1)--(2,3.1) node[black!50, above] {$\text{Re}(\omega)-\frac{d}{2}$};
\draw[black!50, thick,-stealth] (2,3.1)--(2.5,3.1); 
\draw[line width=0.2mm,black,-] (5,2.5)--(5.5,2.5);
\draw[line width=0.2mm,black,-] (5,3)--(5,2.5)node [black, above right] {$\Delta$};
\end{tikzpicture}
\caption{The analytic structure of the spectral density of a scalar two-point function with a power-law large distance behavior $G_{\CO^{(0)}}(\sigma)\limu{\sigma \to -\infty}  |\sigma|^{-\omega}$. There is a strip of analyticity (the blue shaded region) if $\Re(\omega)>\hd$. Because of the shadow symmetry of the spectral density, the position of the possible poles (grey crosses) are also shadow symmetric. Moreover if the operator in the two-point function is Hermitian, the poles come in complex conjugate pairs i.e. reflection symmetric with respect to the x-axis.}
\label{fig:rho}
\end{figure}
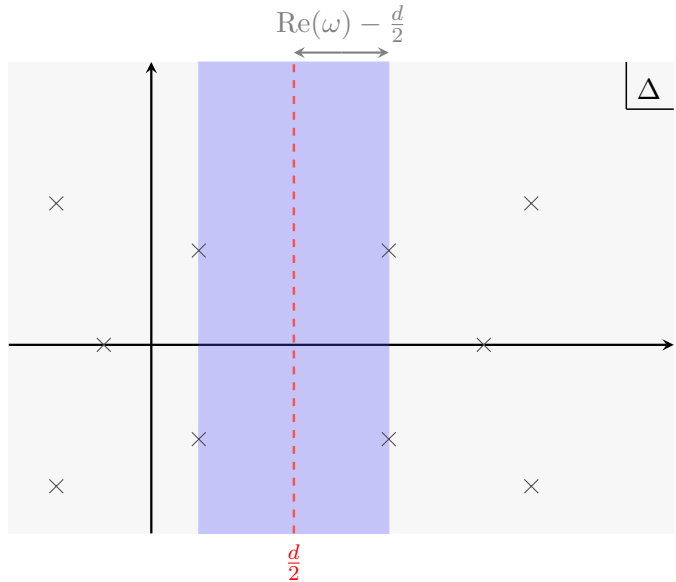

In the spin $1$ case, the explicit inversion formulae for $\rho^{\CP,1}_{\CO^{(1)}}$ and $\rho^{\CP,0}_{\CO^{(1)}}$ are given by~\reef{eq:inversion spin 1 explicit}. In the large $\sigma$ limit the inversion integrals converge if 
\begin{equation}
    \begin{aligned}
      \ell=1:\qquad& & d-\underset{n}{\min}\ \text{Re}(\omega_{1,n})&<\ \text{Re}(\Delta)<\underset{n}{\min}\ \text{Re}(\omega_{1,n})\\
     \ell=0:\qquad& &d+1-\underset{n}{\min}\ \text{Re}(\omega_{1,n})&<\ \text{Re}(\Delta)<\underset{n}{\min}\ \text{Re}(\omega_{1,n})-1
    \end{aligned}
\end{equation}
For arbitrary spin, we conjecture that $\rho^{\CP,\ell}_{\CO^{(J)}}(\lambda)$ is analytic in 
\begin{equation}
d-(\underset{n}{\min}\ \text{Re}(\omega_{J,n})+\ell-J)<\Re(\D)<\underset{n}{\min}\ \text{Re}(\omega_{J,n})+\ell-J\,.
\label{eq:stripofanalyl}
\end{equation}
We have explicitly checked this conjecture for $J=2$.

Let us now discuss the appearance of other UIRs than the principal series. If one has control over the fall-offs $\omega_{J,n}$ of the two-point function $G_{\mathcal{O}^{(J)}}$ by tuning some parameters of the theory, then one can reach a regime where (\ref{eq:completenesscondition}) is violated. In the process of this analytic continuation, poles or branch points of the spectral densities cross the principal series integrals in the \KL\, decomposition, resulting in additional sums and integrals over other UIRs. 
Group theory results in~\cite{Penedones:2023uqc} as well as the examples in section \ref{sec:applications} suggest that additional representations contributing solely as isolated points rather than as a continuum of states, but at the moment we cannot rule out their presence as a continuum in a generic interacting QFT. 
 
In some examples in section \ref{sec:applications}, we tune $\omega_{J,n}$ by tuning the masses in the theories we are considering and we see how, when (\ref{eq:completenesscondition}) is violated, poles in the spectral densities cross the contour of integration over the principal series at $\text{Im}(\Delta)=0$, so that they lead to the appearance of complementary series states. Before the continuation, these poles appear in the complex $\Delta$ plane in symmetric pairs with respect to the $\D$-real axis. So either they are on the real line or they come in pairs when they are off of the real-line c.f. figure~\ref{fig:rho}. In the latter case, considering that the complementary series corresponds to real $\D$, when we are performing the analytic continuation in $\omega_{J,n}$ by first decreasing its imaginary part, in the examples in section \ref{sec:applications}, the complex conjugate pairs of poles merge on the real line where they meet a simple zero. 
Then, one of them moves towards the contour and ultimately crosses it, introducing a complementary series contribution in the \KL\ decomposition, while the other typically moves in the opposite direction. 

Let us finally remark that the boundary of the strip of analyticity mentioned above is not necessarily saturated by poles. In other words, (\ref{eq:stripofanalyl}) is just the \emph{minimum} region of analyticity of $\rho_{\mathcal{O}^{(J)}}^{\mathcal{P},\ell}(\lambda)$. Moreover,  for a fixed $J$, the thinnest strip is for $\ell=0$. In this case the strip of analyticity disappears when $\underset{n}{\min}\ \Re(\omega_{J,n}) =\hd +J$, which is exactly in agreement with the completeness condition.

\subsection{Boundary operator expansion} 
\label{subsubsec:boundarytheory}
In this section we assume the following  about  the spectral densities $\rho^{\mathcal{P},\ell}_{\mathcal{O}^{(J)}}(\lambda)$ of a two-point function:
\begin{enumerate}
    \item Meromorphicity in $\lambda$.
    \item Growth that is at most exponential in the limit $\text{Im}(\lambda)\to -\infty$.
    \item Presence of zeroes at $\lambda=-in$ for $n\in\mathbb{N}$.
\end{enumerate}
Then, we can show that the spinning operator $\mathcal{O}^{(J)}$ appearing in the two-point function can be expanded around the late time surface in terms of boundary operators. These boundary operators transform as primaries and descendants under the $d$-dimensional Euclidean conformal group. They will in general have complex scaling dimensions, and as such, the putative Euclidean CFT on the boundary that they define is non-unitary. The discussion in this section is analogous to what was argued in \cite{Hogervorst:2021uvp} for the scalar \KL\ decomposition, we just generalize it to higher spins. We do not claim these are necessary conditions, but they are sufficient. Some of these conditions might be relaxed while maintaining the existence of the Boundary Operator Expansion, but all of them are satisfied by the spectral densities in the examples we studied in section \ref{sec:applications}.
Let us start from the following identity, which should be understood with the $i\epsilon$ prescription
\begin{equation}
    G_{\lambda,\ell}(Y_1,Y_2;W_1,W_2)=\Gamma(\pm i\lambda)\Omega_{\lambda,\ell}(Y_1,Y_2;W_1,W_2)\,.
\end{equation}
At the same time, harmonic functions can be expressed in terms of EAdS bulk-to-bulk propagators \cite{Costa_2014}
\begin{equation}
    \Omega_{\lambda,\ell}(Y_1,Y_2;W_1,W_2)=\frac{i\lambda}{2\pi}\left(\Pi_{\lambda,\ell}(Y_1,Y_2;W_1,W_2)-\Pi_{-\lambda,\ell}(Y_1,Y_2;W_1,W_2)\right)\,.
\end{equation}
We stress that these are just functional relations and that we are still in de Sitter space.
Using these relations we can write the principal series contributions to the \KL\ decomposition (\ref{eq:KL any J}) as
\begin{equation}
    G_{\mathcal{O}^{(J)}}=\sum_{\ell=0}^J\int_{\mathbb{R}}d\lambda\ \rho_{\mathcal{O}^{(J)}}^{\mathcal{P},\ell}(\lambda)\frac{i\lambda}{\pi}\Gamma(\pm i\lambda)((W_1\cdot\nabla_1)(W_2\cdot\nabla_2))^{J-\ell}\Pi_{\lambda,\ell}(Y_1,Y_2;W_1,W_2)\,,
    \label{eq:KLPis}
\end{equation}
where we are omitting the arguments of $G_{\mathcal{O}^{(J)}}$ to avoid clutter. This representation is convenient for our purposes because the $\Pi_{\lambda,\ell}$ become bulk-to-boundary propagators when we send one of their coordinates to the boundary \cite{Costa_2014}
\begin{equation}
   \Pi_{\lambda,\ell}(Y,-P/\eta;W,-Z/\eta)\underset{\eta\to0^-}{\approx}(-\eta)^{\Delta-\ell}\Pi_{\lambda,\ell}(Y,P;W,Z)+O(\eta^{\Delta-\ell+1})
   \label{eq:latetimeexp}
\end{equation}
where the explicit expression of the bulk-to-boundary propagator is (\ref{eq:defbulktoboundaryEAdS}), $\Delta\equiv\frac{d}{2}+i\lambda$ as usual and $P$ and $Z$ are the embedding space realization of boundary vectors; we introduced them in section \ref{embstates}.
Moreover, by using the recursion relations in \cite{Costa_2014}, it is possible to show that the bulk-to-bulk propagators $\Pi_{\lambda,\ell}$ have the following large $\text{Re}(\Delta)$ behavior
\begin{equation}
\begin{aligned}
    \Pi_{\lambda,\ell}(Y_1,Y_2;W_1,W_2)\underset{\text{Re}(\Delta)\to\infty}{\sim}\sum_{n=0}^\ell &c_n(\sigma)\frac{2^{\Delta}\Delta^{\frac{d}{2}-1}}{(1-\sigma)^\D} \left[1+ \sqrt{\frac{\sigma+1}{\sigma-1}}\right]^{-2\D}\\
    &\times(W_1\cdot W_2)^{\ell-n}((Y_1\cdot W_2)(Y_2\cdot W_1))^{n}\,,
\end{aligned}
\end{equation}
for some coefficients $c_n(\sigma)$ which are independent of $\Delta$.

Now consider the fact that taking one of the time coordinates to late times $\eta\to 0^-$ corresponds to $|\sigma|\to\infty$ (c.f. eq. (\ref{eq:sigmadsAdS})). 
Assuming the spectral densities satisfy the properties which we have listed at the beginning of this section, we can consider the two sides of (\ref{eq:KLPis}) at some fixed $0<|\sigma|^{-1}\ll1$ and close the contour of integration in the lower side of the complex $\lambda$ plane and the contribution from the arc at infinity will vanish\footnote{Even if the spectral densities grow exponentially with $\lambda$, we will always be able to pick a $\sigma$ that is large enough such that the contribution of the arc at infinity vanishes.}. Spurious poles will give contributions that cancel with each other as discussed in section \ref{subsubsec:spurious}. The poles at $\lambda=-in$ in the gamma function appearing in (\ref{eq:KLPis}) are canceled by the zeroes in the spectral density. We are thus left with the contributions of the non-spurious (let us call them physical) poles of the spectral densities
\begin{equation}
    G_{\mathcal{O}^{(J)}}=2\sum_{\ell=0}^J\sum_{\lambda_*}\underset{\lambda=\lambda_*}{\text{Res}}\left[\rho^{\mathcal{P},\ell}_{\mathcal{O}^{(J)}}(\lambda)\right]\lambda_*\Gamma(\pm i\lambda_*)((W_1\cdot\nabla_1)(W_2\cdot \nabla_{2}))^{J-\ell}\Pi_{\lambda_*,\ell}(Y_1,Y_2;W_1,W_2)\,.
    \label{eq:KL as sum}
\end{equation}
Now, we take $Y_2$ to a point $P$ on the late time boundary and $W$ to a null vector $Z$ such that $P\cdot Z=0$, as in (\ref{eq:latetimeexp}).
On the right hand side of (\ref{eq:KL as sum}), we obtain a sum of bulk-to-boundary propagators and their derivatives. By comparison with the left hand side, this suggests that a spin $J$ operator $\mathcal{O}^{(J)}(Y,W)$ in de Sitter satisfies the following late time expansion in terms of boundary operators
\begin{shaded}
\begin{equation}
    \mathcal{O}^{(J)}(-P/\eta,-Z/\eta)\underset{\eta\to0^-}{\approx}\sum_{\ell=0}^J\sum_{\Delta_*}c_{\mathcal{O}^{(J)}O^{(\ell)}_{\Delta_*}}(-\eta)^{\Delta_*-\ell}(Z\cdot\partial_{P})^{J-\ell}O^{(\ell)}_{\Delta_*}(P,Z)+\cdots\,,
    \label{eq:BOE}
\end{equation}
\end{shaded}
\noindent where $O^{(\ell)}_{\Delta_*}(P,Z)$ are boundary CFT primaries of spin $\ell$ and we call $c_{\mathcal{O}^{(J)}O^{(\ell)}_{\Delta_*}}$ the Boundary Operator Expansion (BOE) coefficients. The dots stand for descendants, and $\Delta_*\equiv\frac{d}{2}+i\lambda_*$ with $\lambda_*$ being the position of the physical poles of the spectral density $\rho_{\mathcal{O}^{(J)}}^{\mathcal{P},\ell}(\lambda)$. By comparing with (\ref{eq:KL as sum}), we relate the BOE coefficients to the residues of the spectral density
\begin{equation}
    c_{\mathcal{O}^{(J)}O^{(\ell)}_{\Delta_*}}=2\underset{\lambda=\lambda^*}{\text{Res}}\left[\rho^{\mathcal{P},\ell}_{\mathcal{O}^{(J)}}(\lambda)\right]\lambda^*\Gamma(\pm i\lambda^*)\,.
\end{equation}
Let us stress that the existence of this BOE is dependant on the assumptions we stated at the beginning of this section. It would be interesting to understand what are its convergence properties and whether any of our assumptions can be relaxed while maintaining its validity. We leave these for future work. 
In the examples in section \ref{sec:applications}, where these assumptions are verified, we will draw precise connections between the poles of the spectral densities we will be studying and the associated boundary operators. When extra representations other than the principal series appear in our examples, their contributions are canceled when closing the contour of integration and landing on the sum in (\ref{eq:KL as sum}). In practice this means that the BOE, once derived by closing the contour of integration over the principal series, can be trusted even if we continue the two-point function beyond the regime in which it decomposes in principal series representations only. If more representations than the principal series appeared in the \KL\ decomposition, then we expect to find boundary operators with $\text{Re}(\Delta)<\frac{d}{2}$.

Let us also note that, if this BOE exists, then the bulk two-point function of $\mathcal{O}^{(J)}$ has to have a power law decay at late times, justifying the discussion in section \ref{subsec:completeness}. Moreover, given (\ref{eq:BOE}), the power of this decay corresponds to the conformal dimension of the lowest lying primary in the BOE of $\mathcal{O}^{(J)}$.

Finally, an important open question is whether the same BOE (\ref{eq:BOE}) of a bulk local operator can be used inside different correlation functions.

\subsection{Inversion formula in dS$_2$}\label{2Dinv}
In this section, we will derive an inversion formula to extract the spectral densities in the dS$_2$ \KL\ decomposition (\ref{2DspinJfull}). For   simplicity, we first assume that the spectral density associated with the complementary series is vanishing, and we will later discuss under what conditions such an assumption is valid.

In general dimensions, the tensor structure of $G_{\CO^{(J)}}(Y_1, Y_2; W_1, W_2)$ has two building blocks, namely $W_1\cdot W_2$ and $(Y_1\cdot W_2)(Y_2\cdot W_1)$. In dS$_2$, because of the relations in eq.(\ref{userela}), $G_{\CO^{(J)}}$ is actually a scalar function of $\sigma=Y_1\cdot Y_2$, multiplied by $(W_1\cdot W_2)^J$, and the scalar function depends on whether $W_1$ and $W_2$ have the same chirality. Without loss of generality, fixing $W_1= W_1^+$,  $G_{\CO^{(J)}}$ is encoded in two scalar functions $G^{\pm}_{\CO^{(J)}}(\sigma)$, defined by 
\begin{align}\label{GOcomponents}
    G_{\CO^{(J)}}(Y_1, Y_2; W_1^+, W_2^\pm )=(W_1^+\cdot W_2^\pm )^J\,G^{\pm }_{\CO^{(J)}}(\sigma)~.
\end{align}
Plugging it into (\ref{2DspinJfull}), we should have 
\begin{align}\label{2Dchi}
    (W_1^+\cdot W_2^\pm )^J\,G^{\pm }_{\CO^{(J)}}(\sigma) &=\int_{\mathbb R}d\lambda\, \rho^{\mathcal P, 0}_{\CO^{(J)}}(\lambda)(W^+_1\cdot\nabla_1)^J(W^\pm_2\cdot\nabla_2)^J G_{\lambda,0}(Y_1,Y_2)\nonumber\\
    &+\int_{\mathbb R}d\lambda\,\rho^{\mathcal P, 1}_{\CO^{(J)}}(\lambda)(W^+_1\cdot\nabla_1)^{J-1}(W^\pm_2\cdot\nabla_2)^{J-1} G_{\lambda,1}(Y_1, Y_2; W^+_1, W^\pm_2)\nonumber\\
    &+\sum_{p=0}^J \rho^{\mathcal{D}_p}_{\CO^{(J)}}\left(W^+_1\cdot\nabla_1\right)^J
     \left(W^\pm_2\cdot\nabla_2\right)^J G_{-i(p-\frac{1}{2})}(Y_1, Y_2)~.
\end{align}
The next task is to reduce the tensor structure on the R.H.S. For the first line, it is actually solved in appendix \ref{scalarfield}, c.f. eq. (\ref{j1}) and eq. (\ref{j2})
\begin{align}
    (W^+_1\cdot\nabla_1)^J(W^\pm_2\cdot\nabla_2)^J G_{\lambda,0}(Y_1,Y_2)=(W^+_1\cdot W_2^\pm)^J\, \phi^\pm_{\lambda, J}(\sigma)   ~,
\end{align}
where 
\begin{align}
    \phi^\pm_{\lambda, J}(\sigma)\equiv \partial^J_\sigma((\sigma\pm 1)^J \partial_\sigma^J) G_{\lambda, 0}(\sigma), \,\,\,\,\, G_{\lambda, 0}(\sigma)=\frac{\Gamma(\frac{1}{2}\pm i\lambda)}{4\pi} F\left(\frac{1}{2}+i\lambda, \frac{1}{2}-i\lambda, 1,\frac{1+\sigma}{2}\right)~.
\end{align}
For the second line, using the definition of $G_{\lambda, 1}$ given by eq. (\ref{G1in2D}), we have
\begin{align}
   \left(\frac{1}{4}+\lambda^2\right) G_{\lambda,1}(Y_1, Y_2; W^+_1, W^\pm_2)= \pm (W^+_1\cdot\nabla_1)(W^\pm_2\cdot\nabla_2)G_{\lambda, 0}(\sigma)~.
\end{align}
So it is equivalent to the first line. The reduction of the third line is given by eq. (\ref{disder}) and eq. (\ref{psipq}). Altogether, the spin $J$ \KL\, decomposition (\ref{2DspinJfull}) is equivalent to the following two scalar equations:
\begin{align}\label{Gp}
    G^{+}_{\CO^{(J)}}(\sigma)=\int_{\mathbb R}\,d\lambda\, \rho^{\mathcal P, +}_{\CO^{(J)}}(\lambda)  \phi^+_{\lambda, J}(\sigma)+\sum_{p=0}^J \rho^{\mathcal{D}_p}_{\CO^{(J)}}\psi_{p,J}(\sigma), \,\,\,\,\, \rho^{\mathcal P, +}_{\CO^{(J)}}=\rho^{\mathcal P,0}_{\CO^{(J)}}+\frac{1}{\frac{1}{4}+\lambda^2}\rho^{\mathcal P, 1}_{\CO^{(J)}}~,
    \end{align}
    and 
    \begin{align}\label{G-}
     G^{-}_{\CO^{(J)}}(\sigma)=\int_{\mathbb R}\,d\lambda\, \rho^{\mathcal P, -}_{\CO^{(J)}}(\lambda)  \phi^-_{\lambda, J}(\sigma), \,\,\,\,\, \, \,\,\,\,\, \rho^{\mathcal P, -}_{\CO^{(J)}}=\rho^{\mathcal P,0}_{\CO^{(J)}}-\frac{1}{\frac{1}{4}+\lambda^2}\rho^{\mathcal P, 1}_{\CO^{(J)}}~,
\end{align}
where $\psi_{p,J}(\sigma)$ is defined in eq. (\ref{psipq}). To invert these two equations, we introduce $J$-dependent inner products for real functions defined on $(-\infty, -1)$:
\begin{align}\label{2Dinner}
(f, g)_J^\pm=\int_{-\infty}^{-1}d\sigma (\sigma\mp 1)^{2J} f(\sigma) g (\sigma)~.
\end{align}
In appendix \ref{tech}, we show that $\{\phi_{\lambda, J}^+\}\cup\{\psi_{p,J}\}$ is an orthogonal basis with respect to $(\,,\,)_J^+$, and $\{\phi^-_{\lambda, J}\}$ is an orthogonal basis with respect to $(\,,\,)_J^-$. Using the orthogonality relations, c.f. eq. (\ref{phipnorm}), (\ref{psinorm}) and (\ref{phimnorm}),  we obtain the following \textbf{inversion formulae} for dS$_2$
\begin{shaded}
\begin{align}\label{eq:2Dinver}
    &\rho^{\mathcal P, \pm}_{\CO^{(J)}}(\lambda)
=\frac{4\lambda\sinh(2\pi\lambda)}{(\frac{1}{2}+i\lambda)_J^2(\frac{1}{2}-i\lambda)_J^2}\int_{-\infty}^{-1}d\sigma(\sigma\mp 1)^{2J}  G^{\pm}_{\CO^{(J)}}(\sigma)\phi_{\lambda, J}^\pm(\sigma)\nonumber\\
&\rho^{\mathcal{D}_p}_{\CO^{(J)}}=\frac{8\pi^2 \,(2p-1)}{\Gamma(J+p)^2\Gamma(1+J-p)^2}\int_{-\infty}^{-1}d\sigma(\sigma-1)^{2J}  G^{+}_{\CO^{(J)}}(\sigma)\psi_{p, J}(\sigma)
\end{align}
\end{shaded}
\noindent and $\rho^{\mathcal P, 0}_{\CO^{(J)}}(\lambda), \rho^{\mathcal P, 1}_{\CO^{(J)}}(\lambda)$ can be recovered by taking linear combinations of $\rho^{\mathcal P, \pm}_{\CO^{(J)}}(\lambda)$. 

The expansions (\ref{Gp}) and (\ref{G-}) are valid and unique when $G^\pm_{\CO^{(J)}}$ is integrable with respect to $(\,,\,)_J^\pm$. Alternatively, it means that complementary series does not contribute to the two-point function of $\CO^{(J)}$ if $G^\pm_{\CO^{(J)}}(\sigma)$ decays faster than $(-\sigma)^{-J-\frac{1}{2}}$ at large $-\sigma$.

\section{Applications} 
\label{sec:applications}
In this section, we apply the inversion formulae (\ref{eq:inversionformulal}) and (\ref{eq:2Dinver}) to compute the \KL\, decomposition of a variety of two-point functions. We study two-point functions of composite operators in free theories, primary operators in de Sitter Conformal Field Theories and composite operators in a weakly coupled Quantum Field Theory. In the free theory case, studying which terms appear in the \KL\ decomposition informs us on the decomposition of tensor products of  UIRs of the de Sitter group. In the CFT case, we study how $SO(d+1,2)$ UIRs decompose into $SO(d+1,1)$ UIRs. In the weakly coupled case, we use the \KL\ representation to compute anomalous dimensions of boundary operators.  All throughout, we compare the decomposition in $d>1$ with the one in $d=1$, where the discrete series states contribute up to $\Delta=J$ for spin $J$ two-point functions. The two exceptional series never appear in our examples. 

\subsection{Free QFTs} 
\label{subsec:freeqfts}
One of the uses of the  Käll\'en–Lehmann decomposition is to study the decomposition of multi-particle states into single particle UIRs. By studying the contents of the \KL\ decomposition of a two-point function of a composite operator made of products of elementary fields, we infer the complete set of UIRs that is generated by the action of that operator on the Bunch-Davies vacuum. As shown in \cite{Penedones:2023uqc}, the UIRs that appear in such a decomposition are almost exclusively belonging to the principal series, except for a few isolated complementary series states that we recover by analytic continuation.In this section, to avoid clutter, we will write
\begin{equation}
    \langle\mathcal{O}(Y_1)\mathcal{O}(Y_2)\rangle\equiv\langle\Omega|\mathcal{O}(Y_1)\mathcal{O}(Y_2)|\Omega\rangle\,.
\end{equation}

\subsubsection{Spin 0 Examples}
\label{subsubsec:freeqft0}
Let us start with the simplest possible case: the two-point function of a free elementary 
 massive scalar field $\phi$ with $\Delta_\phi=\frac{d}{2}+i\lambda_\phi$ in the principal series 
 \begin{equation}
     \langle\phi(Y_1)\phi(Y_2)\rangle=G_{\lambda_\phi,0}(Y_1,Y_2)\,.
 \end{equation}
The \KL\ decomposition of this two-point function should read
\begin{equation}
    \langle\phi(Y_1)\phi(Y_2)\rangle=\int_{\mathbb{R}}d\lambda\ \rho^{\mathcal{P},0}_{\phi}(\lambda)G_{\lambda,0}(Y_1,Y_2)\,.
\end{equation}
It is then immediate to see that, necessarily, 
\begin{equation}
\begin{aligned}
    \rho_\phi^{\mathcal{P},0}(\lambda)&=\frac{1}{2}(\delta(\lambda+\lambda_\phi)+\delta(\lambda-\lambda_\phi))\\
    &=\lim_{\epsilon\to0}\frac{\epsilon}{2\pi(\epsilon^2+(\lambda^2-\lambda_\phi^2)^2)}\,,
\end{aligned}
\end{equation}
which is a manifestly real and positive quantity. It has two poles in the lower half of the complex $\lambda$ plane, signaling the presence of two primary boundary operators in the BOE of $\phi$
\begin{equation}
    \phi(-P/\eta)\underset{\eta\to0^-}{\approx}(-\eta)^{\Delta_\phi}\mathcal{O}(P)+(-\eta)^{\bar\Delta_\phi}\widetilde{\mathcal{O}}(P)+\ldots
\end{equation}
where $\mathcal{O}(P)$ and $\widetilde{\mathcal{O}}(P)$ are CFT$_d$ primaries with scaling dimensions $\Delta_\phi$ and $d-\Delta_\phi$ respectively, and the dots stand for descendants.
The fact that the spectral density is a delta function in this case makes sense, since a free field already falls into a single particle UIR.
To see more interesting features, like a decomposition into a continuum of states, one has to instead consider two-point functions of composite operators, such as $\phi_1\phi_2(Y)$ in a free theory of two massive scalars
\begin{equation}
    \langle \phi_1\phi_2(Y_1)\phi_1\phi_2(Y_2)\rangle=\langle \phi_1(Y_1)\phi_1(Y_2)\rangle\langle \phi_2(Y_1)\phi_2(Y_2)\rangle = G_{\lambda_1}(Y_1, Y_2) G_{\lambda_2}(Y_1, Y_2)
    \label{eq:Gphi1phi2}
\end{equation}
    where we take the two fields to have scaling dimensions $\Delta_1=\frac{d}{2}+i\lambda_1$ and $\Delta_2=\frac{d}{2}+i\lambda_2$ in the principal series, so $\lambda_1,\lambda_2\in\mathbb{R}$. This two-point function is free of antipodal singularities and decays at large distances as
    \begin{equation}
        |G_{\phi_1\phi_2}(\sigma)|\underset{\sigma\to-\infty}{\sim}|\sigma|^{-d}\,, \qquad \sigma\equiv Y_1\cdot Y_2
    \end{equation}
    Given the discussion in section \ref{subsec:completeness}, this means the \KL\ decomposition of this two-point function will only include states in the principal series, as long as $d>2$. We observe through numerical checks of (\ref{eq:densityphi1phi2}) that even in the limit case $d=2$ this two-point function decomposes into principal series representations only. To apply the inversion formula (\ref{eq:inversionformulal}) to this two-point function, we analytically continue it to EAdS as discussed in section \ref{subsec:wick}. Under this continuation, (\ref{eq:Gphi1phi2}) becomes a product of two harmonic functions
    \begin{equation}
        \langle\phi_1\phi_2(X_1)\phi_1\phi_2(X_2)\rangle=\Gamma(\pm i\lambda_1)\Gamma(\pm i\lambda_2)\Omega_{\lambda_1,0}(X_1,X_2)\Omega_{\lambda_2,0}(X_1,X_2)\,.
    \end{equation}
    Then, the inversion formula reads 
\begin{equation}
    \rho^{\mathcal{P},0}_{\phi_1\phi_2}(\lambda)=\frac{\Gamma(\pm i\lambda_1)\Gamma(\pm i\lambda_2)}{\mathcal{N}_{0,0}}\int_{X_1}\Omega_{\lambda,0}(X_1,X_2)\Omega_{\lambda_1,0}(X_1,X_2)\Omega_{\lambda_2,0}(X_1,X_2)\,,
\label{eq:scalarrhoint}
\end{equation}
where $\mathcal{N}_{J,\ell}$ is defined in (\ref{eq:calN}).
To make progress, we use the split representation (\ref{eq:defsplitrep}) on the three harmonic functions, following what was first done in \cite{Bros:2009bz}. Defining $\Delta_3\equiv \frac{d}{2}+i\lambda$, we have
\begin{equation}
    \rho^{\mathcal{P},0}_{\phi_1\phi_2}(\lambda)=\frac{\lambda^2\lambda_1^2\lambda_2^2\Gamma(\pm i\lambda_1)\Gamma(\pm i\lambda_2)}{\pi^3\mathcal{N}_{0,0}}\int_{X_1}\prod_{k=1}^3\int_{P_i}\Pi_{\Delta_k,0}(X_1,P_k)\Pi_{\bar\Delta_k,0}(X_2,P_k)\,,
\end{equation}
where $\Pi_{\Delta,0}(X,P)$ is a EAdS scalar bulk-to-boundary propagator, of which we report the definition in (\ref{eq:defbulktoboundaryEAdS}). The integral over $X_1$ leads to a CFT three point function 
\begin{equation}
    \rho^{\mathcal{P},0}_{\phi_1\phi_2}(\lambda)=\frac{\lambda^2\lambda_1^2\lambda_2^2\Gamma(\pm i\lambda_1)\Gamma(\pm i\lambda_2)b(\Delta_1,\Delta_2,\Delta,0)}{\pi^3\mathcal{N}_{0,0}}\int_{P_1,P_2,P_3}\frac{\prod_{k=1}^3\Pi_{\bar\Delta_k,0}(X_2,P_k)}{{(P_{12})^{\Delta_{123}}(P_{13})^{\Delta_{132}}(P_{23})^{\Delta_{231}}}}\,, \label{eq:phi1phi2step}
\end{equation}
where the notation and all the coefficients are explicit in the Appendix \ref{subsec:masterfreeqft}. There, we also show how to solve the remaining integrals over the three boundary points $P_1$, $P_2$ and $P_3$. Importantly, the spectral density of every free QFT two-point function of composite operators made of two fundamental fields with spin can be reduced to linear combinations of this specific integral, so that in the spinning examples we will make extensive use of it. In Appendix \ref{subsec:masterfreeqft} we show how to eventually obtain 
\begin{equation}
    \rho^{\mathcal{P},0}_{\phi_1\phi_2}(\lambda)=\frac{\lambda\sinh(\pi\lambda)}{32\pi^{\frac{d}{2}+3}\Gamma(\frac{d}{2})\Gamma(\frac{d}{2}\pm i\lambda)}\prod_{\pm,\pm,\pm}\Gamma\left(\frac{\frac{d}{2}\pm i\lambda\pm i\lambda_1\pm i\lambda_2}{2}\right)\,.
    \label{eq:densityphi1phi2}
\end{equation}
\begin{figure}
\centering
\begin{tikzpicture}[scale=1]
\fill[light-gray] (-2.3,-3) -- (-2.3,3.5) -- (6,3.5) -- (6,-3);
\coordinate (origin) at (0,0);
\draw[line width=0.3mm,black,-stealth] (origin) -- ++(6,0);
\draw[line width=0.3mm,black,-stealth] (origin) -- ++(0,3.5);
\draw[line width=0.3mm,black,-] (origin) -- ++(-2.3,0) node (zminus) {};
\draw[line width=0.3mm,black,-] (origin) -- ++(0,-3) node (xminus) {};
\draw[line width=0.3mm,red!70,-,dashed] (1.5,-3) node [red, below] {$\frac{d}{2}$} --(1.5,3.5);
\draw[line width=0.3mm,blue!70,-,dashed] (3,-3) node [blue!70, below] {$d$} --(3,3.5);
\draw[line width=0.3mm,blue!70,-,dashed] (0.5,-1);
\draw[line width=0.3mm,blue!70,-,dashed] (2.5,-1);
\draw (3,2.5) node[cross,black] {};
\draw[orange, thick,->] (3,2.2)--(3,1.6); 
\draw[orange, thick,->] (3,1)--(3,0.4); 
\draw (3,-2.5) node[cross,black] {};
\draw[gray, thick,->] (3,-2.2)--(3,-1.6); 
\draw[gray, thick,->] (3,-1)--(3,-0.4); 
\draw (0,2.5) node[cross,black] {};
\draw[gray, thick,->] (0,2.2)--(0,1.6); 
\draw[gray, thick,->] (0,1)--(0,0.4); 
\draw (0,-2.5) node[cross,black] {};
\draw[orange, thick,->] (0,-2.2)--(0,-1.6); 
\draw[orange, thick,->] (0,-1)--(0,-0.4); 
\draw[orange, thick,->] (2.5,0)--(1.9,0); 
\draw[orange, thick,->] (0.5,0)--(1.1,0); 
\draw[gray, thick,->] (3.5,0)--(4.1,0); 
\draw[gray, thick,->] (-0.5,0)--(-1.1,0); 
\draw (-2,2.5) node[cross,black] {};
\draw (-2,-2.5) node[cross,black] {};
\draw (-1,2.5) node[cross,black] {};
\draw (-1,-2.5) node[cross,black] {};
\draw (4,2.5) node[cross,black] {};
\draw (5,2.5) node[cross,black] {};
\draw (4,-2.5) node[cross,black] {};
\draw (5,-2.5) node[cross,black] {};
\draw[line width=0.2mm,black,-] (5.5,3)--(6,3);
\draw[line width=0.2mm,black,-] (5.5,3.5)--(5.5,3)node [black, above right] {$\Delta$};
\end{tikzpicture}
\caption{The analytic structure of $\rho^{\mathcal{P},0}_{\phi_1\phi_2}(\lambda)$, the spectral density for $\langle \phi_1\phi_2 (Y_1)\phi_1\phi_2 (Y_2)\rangle$. Here we represent the poles at $\lambda=\pm(\lambda_1+\lambda_2)\pm i(\frac{d}{2}+2n)$ for each combination of signs. As we continue $\lambda_1$ and $\lambda_2$ to the complementary series, we indicate with colored arrows how the leading poles corresponding to $n=0$ move in the complex $\Delta=\frac{d}{2}+i\lambda$ plane. Eventually, some of these poles can cross the integration contour at $\text{Re}(\Delta)=\frac{d}{2}$ (we highlight their path in orange). Their residues need to be summed, leading to the discrete sum of complementary series states in (\ref{eq:KLphi1phi2}).}
\label{fig:rhophiqphiw}
\end{figure}
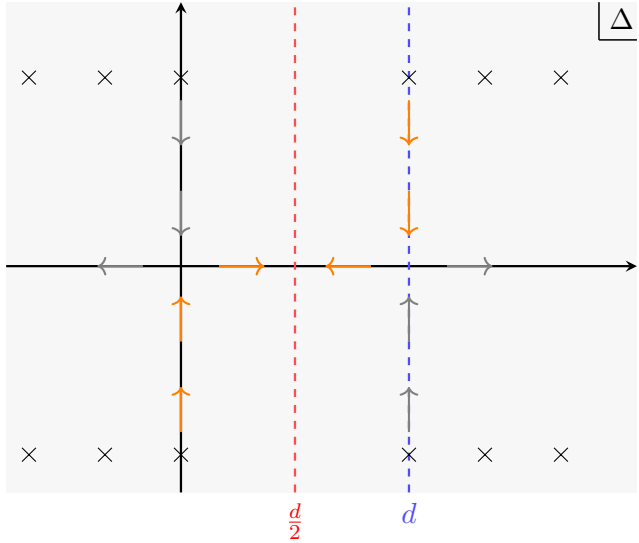
It can be checked numerically that if $\lambda_1,\lambda_2\in\mathbb{R}$, the integral of (\ref{eq:densityphi1phi2}) fully reproduces (\ref{eq:Gphi1phi2})
\begin{equation}
    \langle \phi_1\phi_2(Y_1)\phi_1\phi_2(Y_2)\rangle=\int_{\mathbb{R}}\mathrm{d}\lambda\ \rho^{\mathcal{P},0}_{\phi_1\phi_2}(\lambda)G_{\lambda,0}(Y_1,Y_2)\,, \qquad \text{if }\lambda_1,\lambda_2\in\mathbb{R}\,.
\end{equation}
Analytically continuing $\lambda_1$ and $\lambda_2$ to the complementary series such that $i\lambda_1\in(0,\frac{d}{2})$ and $i\lambda_2\in(0,\frac{d}{2})$
, poles of $\rho_{\phi_1\phi_2}^{\mathcal{P},0}(\lambda)$ can cross the contour of integration over the principal series, so that their residues need to be added by hand, introducing some complementary series contributions to the \KL\ decomposition of this two-point function. This is in agreement with what is discussed in section \ref{subsec:completeness}. By studying the gamma functions in (\ref{eq:densityphi1phi2}) we see that poles cross the contour if there exists some $n\in\mathbb N$ such that 
\begin{equation}
    \frac{d}{2}+2n<i\lambda_1+i\lambda_2<d\,,
    \label{eq:existancecomp}
\end{equation}
where the second inequality comes from the fact that $\lambda_1$ and $\lambda_2$ are constrained to be on the complementary series. Let us assume more specifically that 
\begin{equation}
    \frac{d}{2}+2N<i\lambda_1+i\lambda_2<\frac{d}{2}+2(N+1)\,,
\end{equation}
for some $N<\frac{d}{4}$. Then, the full decomposition reads 
\begin{equation}
\begin{aligned}
\label{eq:KLphi1phi2}
    \langle \phi_1\phi_2(Y_1)\phi_1\phi_2(Y_2)\rangle=&\int_{\mathbb{R}}\mathrm{d}\lambda\ \rho^{\mathcal{P},0}_{\phi_1\phi_2}(\lambda)G_{\lambda,0}(Y_1,Y_2)+\sum_{n=0}^{N}\rho^{\mathcal {C},0}_{\phi_1\phi_2}(n)G_{\lambda_1+\lambda_2+i\left(\frac{d}{2}+2n\right),0}(Y_1,Y_2)
\end{aligned}
\end{equation}
where 
\begin{equation}
\begin{aligned}
    \rho^{\mathcal {C},0}_{\phi_1\phi_2}(n)=&\frac{(-1)^n(\frac{d}{2})_n\Gamma(-n+i\lambda_{12})\Gamma(\frac{d}{2}+n-i\lambda_{12})\prod_{j=1,2}\Gamma(-n+i\lambda_j)\Gamma(\frac{d}{2}+n-i\lambda_j)}{4\pi^{1+\frac{d}{2}}n!\Gamma(-2n+i\lambda_{12})\Gamma(-\frac{d}{2}-2n+i\lambda_{12})\Gamma(d+2n-i\lambda_{12})\Gamma(\frac{d}{2}+2n-i\lambda_{12})}
\end{aligned}
\end{equation}
with $\lambda_{12}\equiv\lambda_1+\lambda_2$. The complementary series densities are simply the residues on the poles of $\rho_{\phi_1\phi_2}^{\mathcal{P},0}(\lambda)$
\begin{equation}
   \rho^{\mathcal {C},0}_{\phi_1\phi_2}(n)=4\pi i\underset{\lambda=\lambda_1+\lambda_2+i\left(\frac{d}{2}+2n\right)}{\text{Res}}\rho^{\mathcal{P},0}_{\phi_1\phi_2}(\lambda)\,.
\end{equation}
As expected from the proof of the \KL\ decomposition, by studying the sign of $\rho_{\phi_1\phi_2}^{\mathcal {C},0}(n)$, it can be verified that these functions are positive as long as $\lambda_1$ and $\lambda_2$ are in the complementary series and lie in the interval (\ref{eq:existancecomp}). The appearance of this discrete sum of complementary series UIRs is in agreement with \cite{Penedones:2023uqc}, see Table 1.3 there.
Moreover, this sum was derived before in \cite{Akhmedov:2017ooy} and we checked that our $\rho_{\phi_1\phi_2}^{\mathcal {C},0}(n)$ matches their equation (48)\footnote{To verify the matching one needs to substitute $\lambda_1\rightarrow -i\alpha$, $\lambda_2\rightarrow -i\beta$ and $d\rightarrow D-1$.}. The sum over $n$ runs up to $N<\lfloor\frac{d}{4}\rfloor$ because only in this range  the interval $(\frac{d}{2}+2n,d)$ is non-vanishing.
\paragraph{Boundary Operator Expansion}
The spectral density (\ref{eq:densityphi1phi2}) satisfies all the assumptions of section \ref{subsubsec:boundarytheory}. From its poles in the lower half of the complex $\lambda$ plane we can thus read off the primary operators which appear in the 
BOE of $\phi_1\phi_2$, namely 
\begin{equation}
\begin{aligned}
    \phi_1\phi_2(-P/\eta)\underset{\eta\to0^-}{\approx}\sum_{n=0}^\infty \Big[&c_{\Delta_1\Delta_2}(-\eta)^{\Delta_1+\Delta_2+2n}[\mathcal{O}_1\mathcal{O}_2]_n(P)\\
    &+c_{\Delta_1\bar\Delta_2}(-\eta)^{\Delta_1+\bar\Delta_2+2n}[\mathcal{O}_1\widetilde{\mathcal{O}}_2]_n(P)+\cdots\Big]+\cdots
    \end{aligned}
\end{equation}
where notation like $[\mathcal{O}_1\mathcal{O}_2]_n$ should be understood to stand for all the boundary scalar primaries that can be constructed with $\mathcal{O}_1, \mathcal{O}_2$ and $2n$ contracted derivatives while being symmetric under $1\leftrightarrow 2$. The dots in the square brackets stand for contributions from primaries like $[\widetilde{\mathcal{O}}_1\mathcal{O}_2]_n$ and $[\widetilde{\mathcal{O}}_1\widetilde{\mathcal{O}}_2]_n$, while the dots outside of the brackets stand for descendants. The operators $\mathcal{O}_i(P)$ are defined as the leading late time behavior of the free fields $\phi_1(Y)$ and $\phi_2(Y)$
\begin{equation}
\begin{aligned}
    \phi_{1,2}(-P/\eta)\underset{\eta\to0^-}{\sim}(-\eta)^{\Delta_{1,2}}\mathcal{O}_{1,2}(P)+(-\eta)^{\bar\Delta_{1,2}}\widetilde{\mathcal{O}}_{1,2}(P)\,.
\end{aligned}
\end{equation}
so that $\mathcal{O}_1(P)$ and $\widetilde{\mathcal{O}}_1(P)$ transform as CFT scalar primaries with scaling dimensions $\Delta_1=\frac{d}{2}+i\lambda_1$ and $\bar\Delta_1=\frac{d}{2}-i\lambda_1$ respectively (and analogously $\mathcal{O}_2(P)$ and $\widetilde{\mathcal{O}}_2(P)$). An extra comment: when $\lambda_1$ and $\lambda_2$ satisfy (\ref{eq:existancecomp}), the poles of (\ref{eq:densityphi1phi2}) at $\lambda=\lambda_1+\lambda_2+i(\frac{d}{2}+2n)$ can be picked up when closing the contour of integration to find the boundary operators in the late time limit. One could thus expect there to be operators on the boundary with $\Delta=d-\Delta_1-\Delta_2-2n$. But the residues on these poles are precisely canceled by the complementary series sum in (\ref{eq:KLphi1phi2}), and thus such operators are actually not appearing in the bulk-boundary OPE of $\phi_1\phi_2(Y)$.

\subsubsection{Spin 1 Examples}
\label{subsubsec:freeqft1}
Consider the correlator
\begin{equation}
    \langle V\phi(Y_1;W_1)V\phi(Y_2;W_2)\rangle=\langle V(Y_1;W_1)V(Y_2;W_2)\rangle\langle \phi(Y_1)\phi(Y_2)\rangle
\end{equation}
in a free theory of a massive vector with $\Delta_V=\frac{d}{2}+i\lambda_V$ and a massive scalar $\Delta_\phi=\frac{d}{2}+i\lambda_\phi\,,$ both on the principal series. This two-point function has two scalar components
\begin{equation}
    G_{V\phi}(Y_1,Y_2;W_1,W_2)=\mathcal{G}_{V\phi}^{(0)}(\sigma)(W_1\cdot W_2)+\mathcal{G}_{V\phi}^{(1)}(\sigma)(W_1\cdot Y_2)(W_2\cdot Y_1)\,,
\end{equation}
which decay at large distances as
\begin{equation}
    |\mathcal{G}_{V\phi}^{(0)}(\sigma)|\underset{\sigma\to-\infty}{\sim}|\sigma|^{-d}\,, \qquad |\mathcal{G}_{V\phi}^{(1)}(\sigma)|\underset{\sigma\to-\infty}{\sim}|\sigma|^{-d-1}\,.
\end{equation}
Following the discussion in section \ref{subsec:completeness}, we can thus state that the \KL\ decomposition of this two-point function will only include principal series contributions, as long as $d>2$. We verified that this is true also in the limit case $d=2$. 
These will be organized in two terms, related to transverse and longitudinal degrees of freedom. 
 In Appendix \ref{subsec:masterfreeqft} we show in detail how to apply the inversion formula to this case and how to express the two spectral densities as linear combinations of (\ref{eq:scalarrhoint}). Here we report the result
    \begin{align}
    \rho^{\mathcal{P},0}_{V\phi}(\lambda)&=\frac{2^{-1}\pi^{-3-\frac{d}{2}}\lambda\sinh(\pi\lambda)}{(\Delta_V\!-\!1)(\bar\Delta_V\!-\!1)(d^2\!+\!4\lambda^2)\Gamma(\frac{d}{2})\Gamma(\frac{d}{2}\pm i\lambda+1)}\prod_{\pm,\pm,\pm}\Gamma\left(\frac{\frac{d}{2}+1\pm i\lambda\pm i\lambda_V\pm i\lambda_\phi}{2}\right)\nonumber\\
     \rho^{\mathcal{P},1}_{V\phi}(\lambda)&=\frac{2^{-12}\pi^{-3-\frac{d}{2}}\lambda\sinh(\pi\lambda)f_{\lambda,\lambda_V,\lambda_\phi}}{\Gamma(\frac{d+2}{2})(\Delta_V-1)(\bar\Delta_V-1)\Gamma(\frac{d}{2}\pm i\lambda+1)}\prod_{\pm,\pm,\pm}\Gamma\left(\frac{\frac{d}{2}\pm i\lambda\pm i\lambda_\phi\pm i\lambda_V}{2}\right)\,,
     \label{eq:rhoVphi}
    \end{align}
with 
\begin{align}
f_{\lambda,\lambda_V,\lambda_\phi}=&16 \left(\lambda_\phi^2-(\lambda^2+\lambda_V^2)\right)^2+64 (d-1) \lambda ^2 \lambda_V^2\nonumber\\
&+8 d (3 d-4) \lambda_\phi ^2+8 d \left(2 d^2-5 d+4\right) \left(\lambda ^2+\lambda_V^2\right)+d^3 \left(4 d^2-11 d+8\right)\,,
\end{align}
where we see the appearance of the spurious pole predicted in section \ref{subsubsec:spurious}.
When $\phi$ and $V$ have scaling dimensions in the principal series, the integrals over the principal series reproduce the full two-point function. We elaborate on how exactly to carry out numerical checks in section \ref{subsec:hfunc}.
Continuing their scaling dimensions to the complementary series, $i\lambda_V\in\left(0,\frac{d}{2}-1\right)$ and $i\lambda_\phi\in(0,\frac{d}{2})$, instead leads to different poles of $\rho_{V\phi}^{\mathcal{P},0}(\lambda)$ and $\rho_{V\phi}^{\mathcal{P},1}(\lambda)$ crossing the contour of integration. This happens when the following conditions are satisfied for some integers $N_0$ and $N_1$:
\begin{equation}
\begin{aligned}
    &\rho_{V\phi}^{\mathcal{P},0}:\qquad \frac{d}{2}+2N_0+1<i\lambda_\phi+i\lambda_V<\frac{d}{2}+2(N_0+1)+1\,,\\
    &\rho_{V\phi}^{\mathcal{P},1}:\qquad \frac{d}{2}+2N_1<i\lambda_\phi+i\lambda_V<\frac{d}{2}+2(N_1+1)\,,\label{eq:compexistance2}
\end{aligned}
\end{equation}
where the unitarity bounds for the complementary series impose $N_0<\frac{d-4}{4}$ and $N_1<\frac{d-2}{4}\,.$ The complementary series contributions to this two-point function when (\ref{eq:compexistance2}) are satisfied, then, are given by the sum over the residues of $\rho_{V\phi}^{\mathcal{P},0}(\lambda)$ and $\rho_{V\phi}^{\mathcal{P},1}(\lambda)$ on those poles. Moreover, in $d=1$ the discrete series contributes as well. We can explicitly derive this extra contribution by analytically continuing in the dimension until $d=1$, and keeping track of any poles that cross the contour of integration over the principal series. Specifically, what happens is that the poles at $\lambda=\pm i\frac{d-2}{2}$ in the spin-1 free propagator (see section \ref{subsubsec:spurious} for a discussion about these poles) cross the contour of integration. 
At the precise value $d=2$ these poles at $\lambda=\pm i\frac{d-2}{2}$ are canceled by zeroes of the form $\lambda\sinh(\pi\lambda)$ which are present in the propagator, but when continuing all the way to $d=1$, the poles need to be taken into account. The complete decomposition thus reads
\begin{align}
    \langle V\phi(Y_1;W_1)V\phi(Y_2;W_2)\rangle=&\sum_{\ell=0}^1\int_{\mathbb{R}}\mathrm{d}\lambda\ \rho^{\mathcal{P},\ell}_{V\phi}(\lambda)[(W_1\cdot\nabla_1)(W_2\cdot\nabla_2)]^{1-\ell}G_{\lambda,\ell}(Y_1,Y_2;W_1,W_2)\nonumber\\
    &+\sum_{n=0}^{N_0}\rho^{\mathcal {C},0}_{V\phi,n}(W_1\cdot\nabla_1)(W_2\cdot\nabla_2)G_{\lambda_\phi+\lambda_V+i(\frac{d}{2}+2n+1),0}(Y_1,Y_2)\nonumber\\
    &+\sum_{n=0}^{N_1}\rho^{\mathcal {C},1}_{V\phi,n}G_{\lambda_\phi+\lambda_V+i(\frac{d}{2}+2n),1}(Y_1,Y_2;W_1,W_2)\label{eq:KLVphi}\\
    &+\delta_{d,1}\rho^{\mathcal{D}_1}_{V\phi}\left(W_1\cdot\nabla_1\right)
     \left(W_2\cdot\nabla_2\right) G_{-\frac{i}{2},0}(Y_1,Y_2)\,,\nonumber
\end{align}
where $\delta_{d,1}$ is a Kronecker delta, because the discrete series term only contributes in $d=1$. We stress that the complementary series contributions appear only if (\ref{eq:compexistance2}) are satisfied for some $N_0$ and $N_1$, and are absent otherwise. The spectral densities of the complementary series contributions are, specifically,
\begin{equation}
    \begin{aligned}
        \rho^{\mathcal {C},0}_{V\phi,n}&=4\pi i\underset{\lambda=\lambda_\phi+\lambda_V+i(\frac{d}{2}+2n+1)}{\text{Res}}\rho^{\mathcal{P}}_{V\phi,0}(\lambda)\,,\\
        \rho^{\mathcal {C},1}_{V\phi,n}&=4\pi i\underset{\lambda=\lambda_\phi+\lambda_V+i(\frac{d}{2}+2n)}{\text{Res}}\rho^{\mathcal{P}}_{V\phi,1}(\lambda)\,,
    \end{aligned}
\end{equation}
and we verify that they are positive functions for $\lambda_V$ and $\lambda_\phi$ in (\ref{eq:compexistance2}). The discrete series density is instead given by 
\begin{equation}
    \rho^{\mathcal{D}_1}_{V\phi}=\frac{\pi(\lambda_V^2-\lambda_\phi^2)}{(1+4\lambda_V^2)\sinh(\pi(\lambda_V-\lambda_\phi))\sinh(\pi(\lambda_V+\lambda_\phi))}\,.
\end{equation}
Now let us discuss the spectrum of \textbf{boundary operators} that we can infer from this two-point function. As reviewed in \cite{Sun:2021thf} and discussed in section \ref{embstates}, bulk free vector fields have the following asymptotic behavior
\begin{equation}
    V_i(\eta,\vec{y})\underset{\eta\rightarrow 0^-}{\sim}(-\eta)^{\Delta_V-1}\mathcal{A}_i(\vec{y})+(-\eta)^{\bar\Delta_V-1}\widetilde{\mathcal{A}}_i(\vec{y})\,,
\end{equation}
with $\mathcal{A}_i(\vec{y})$ and $\widetilde{\mathcal{A}}_i(\vec{y})$ transforming as CFT primaries with scaling dimensions $\Delta_V$ and $\bar\Delta_V$. Using the fact that $\nabla_\mu V^{\mu}=0$ we can fix the asymptotic behavior of $V_\eta(\eta,\vec y)$
\begin{equation}
    V_\eta(\eta,\vec{y})\underset{\eta\rightarrow 0^-}{\sim}\frac{1}{\bar\Delta_V-1}(-\eta)^{\Delta_V}\partial\cdot\mathcal{A}(\vec{y})+\frac{1}{\Delta_V-1}(-\eta)^{\bar\Delta_V}\partial\cdot\widetilde{\mathcal{A}}(\vec{y})\,.
    \label{eq:boundaryVeta}
\end{equation}
We recognize the appearance of these boundary operators in the poles of the two spectral densities. Specifically, we can write the following BOE (c.f. eq. (\ref{eq:BOE}))
\begin{align}
    V\phi(-P/\eta,-Z/\eta)\underset{\eta\to0^-}{\approx}\sum_{n=0}^\infty\Big[&c^{(0)}_{\Delta_\phi\Delta_V}(-\eta)^{\Delta_V+\Delta_\phi+1+2n}(Z\cdot\partial_P)[(D_Z\cdot\partial_P)\mathcal{A}\mathcal{O}]_n(P,Z)\nonumber\\
    &c^{(1)}_{\Delta_\phi\Delta_{V}}(-\eta)^{\Delta_V+\Delta_\phi-1+2n}[\mathcal{A}\mathcal{O}]_n(P,Z)
    +\cdots\Big]+\cdots\,,
\end{align}
where we see both spin $0$ and spin $1$ boundary operators appearing and the dots stand for double trace primaries like $[\widetilde{\mathcal{A}}\mathcal{O}]_n(P,Z)$ and descendants. The boundary operator $\mathcal{O}$ is defined through the late time limit of the free field
\begin{equation}
    \phi(-P/\eta)\underset{\eta\to0^-}{\approx}(-\eta)^{\Delta_\phi}\mathcal{O}(P)+(-\eta)^{\bar\Delta_\phi}\widetilde{\mathcal{O}}(P)\,.
\end{equation}
Finally, we verify that the contributions from the spurious poles at $\lambda=-i\frac{d}{2}$ and $\lambda=-i\frac{d-2}{2}$ in (\ref{eq:KLVphi}) exactly cancel when closing the contour of integration due to the identities in \ref{subsubsec:spurious}. That means they are not associated to any boundary operator.
We also computed the decomposition of the correlator 
\begin{equation}
\begin{aligned}
    \langle \phi_1\nabla\phi_2(Y_1;W_1)\phi_1\nabla\phi_2(Y_2;W_2)\rangle=\langle \phi_1(Y_1)\phi_1(Y_2)\rangle(W_1\cdot\nabla_1)(W_2\cdot\nabla_2)\langle \phi_2(Y_1)\phi_2(Y_2)\rangle\,.
\end{aligned}
\end{equation}
When $\lambda_1,\lambda_2\in\mathbb{R},$ the following principal series spectral densities account for the full \KL\ decomposition of this two-point function (see appendix \ref{subsec:masterfreeqft} for more details) 
    \begin{align}
    \rho^{\mathcal{P},0}_{\phi_1\nabla\phi_2}(\lambda)&=\frac{(d^2+4(\lambda^2-\lambda_1^2+\lambda_2^2))^2\Gamma(\frac{d+1}{2})\lambda\sinh(\pi\lambda)}{2^{10-d}\pi^{\frac{d+7}{2}}(d^2+4\lambda^2)\Gamma(d)\Gamma(\frac{d}{2}+1\pm i\lambda)}\prod_{\pm,\pm,\pm}\Gamma\left(\frac{\frac{d}{2}\pm i\lambda\pm i\lambda_1\pm i\lambda_2}{2}\right)\nonumber\\
    \rho^{\mathcal{P},1}_{\phi_1\nabla\phi_2}(\lambda)&=\frac{\lambda\sinh(\pi\lambda)}{2^4\pi^{3+\frac{d}{2}}\Gamma(\frac{d+2}{2})\Gamma(\frac{d}{2}+1\pm i\lambda)}\prod_{\pm,\pm,\pm}\Gamma\left(\frac{\frac{d}{2}+1\pm i\lambda\pm i\lambda_1\pm i\lambda_2}{2}\right)
    \label{eq:rhophi1delphi2}
    \end{align}
When analytically continuing the conformal weights of $\phi_1$ and $\phi_2$ to the complementary series $i\lambda_1\in(0,\frac{d}{2})$ and $i\lambda_2\in(0,\frac{d}{2})$, poles of $\rho_{\phi_1\nabla\phi_2}^{\mathcal{P},0}(\lambda)$ and $\rho_{\phi_1\nabla\phi_2}^{\mathcal{P},1}(\lambda)$ cross the integration contour when the following conditions are satisfied for some $N_0<\frac{d-2}{4}$ and $N_1<\frac{d-4}{4}$
\begin{equation}
\begin{aligned}
    &\rho_{\phi_1\nabla\phi_2}^{\mathcal{P},0}:\qquad \frac{d}{2}+2N_0<i\lambda_1+i\lambda_2<\frac{d}{2}+2(N_0+1)\,,\\
    &\rho_{\phi_1\nabla\phi_2}^{\mathcal{P},1}:\qquad \frac{d}{2}+2N_1+1<i\lambda_1+i\lambda_2<\frac{d}{2}+2(N_1+1)+1\,.\label{eq:compexistance3}
\end{aligned}
\end{equation}
Notice that these are slightly different poles than (\ref{eq:compexistance2}). 
Moreover, in $d=1$ there is a discrete series state appearing corresponding to a massless scalar, with $\Delta=1$. The full decomposition reads
\small
\begin{align}
    \langle \phi_1\nabla\phi_2(Y_1;W_1)\phi_1\nabla\phi_2(Y_2;W_2)\rangle=&\sum_{\ell=0}^1\int_{\mathbb{R}}\mathrm{d}\lambda\ \rho^{\mathcal{P},\ell}_{\phi_1\nabla\phi_2}(\lambda)[(W_1\cdot\nabla_1)(W_2\cdot\nabla_2)]^{1-\ell}G_{\lambda,\ell}(Y_1,Y_2;W_1,W_2)\nonumber\\
    &+\sum_{n=0}^{N_0}\rho^{\mathcal {C},0}_{\phi_1\nabla\phi_2,n}(W_1\cdot\nabla_1)(W_2\cdot\nabla_2)G_{\lambda_\phi+\lambda_V+i(\frac{d}{2}+2n),0}(Y_1,Y_2)\nonumber\\
    &+\sum_{n=0}^{N_1}\rho^{\mathcal {C},1}_{\phi_1\nabla\phi_2,n}G_{\lambda_\phi+\lambda_V+i(\frac{d}{2}+2n+1),1}(Y_1,Y_2;W_1,W_2)\nonumber\\    
    &+\delta_{d,1}\rho^{\mathcal{D}_1}_{\phi_1\nabla\phi_2}\left(W_1\cdot\nabla_1\right)
     \left(W_2\cdot\nabla_2\right) G_{-\frac{i}{2},0}(Y_1,Y_2)\label{eq:phi1phi2KL}\,.
\end{align}
\normalsize
The complementary series densities are once again positive functions, given by the residues of the principal series densities on the poles that cross the contour
\begin{equation}
    \begin{aligned}
        \rho^{\mathcal {C},0}_{\phi_1\nabla\phi_2,n}&=4\pi i\underset{\lambda=\lambda_1+\lambda_2+i(\frac{d}{2}+2n)}{\text{Res}}\rho^{\mathcal{P},0}_{\phi_1\nabla\phi_2}(\lambda)\,,\\
        \rho^{\mathcal {C},1}_{\phi_1\nabla\phi_2,n}&=4\pi i\underset{\lambda=\lambda_1+\lambda_2+i(\frac{d}{2}+2n+1)}{\text{Res}}\rho^{\mathcal{P},1}_{\phi_1\nabla\phi_2}(\lambda)\,,
    \end{aligned}
\end{equation}
and their contribution is instead absent when $\lambda_\phi$ and $\lambda_V$ are real or imaginary but outside of the intervals (\ref{eq:compexistance3}).
The discrete series density is again obtainable by analytically continuing in the spacetime dimension and adding the residue on the pole that crosses the contour of integration  
\begin{equation}
    \rho^{\mathcal{D}_1}_{\phi_1\nabla\phi_2}=\frac{\pi(\lambda_1^2-\lambda_2^2)}{4\sinh(\pi(\lambda_1-\lambda_2))\sinh(\pi(\lambda_1+\lambda_2))}\,.
\end{equation}
The difference in the pole structure of (\ref{eq:rhophi1delphi2}) compared to (\ref{eq:rhoVphi}) is explained when we consider the boundary operators appearing in the BOE of $\phi_1\nabla\phi_2(Y)$
\begin{equation}
\begin{aligned}
    \phi_1(Z\cdot\partial_P)\phi_2(-P/\eta)\underset{\eta\to0^-}{\approx}\sum_{n=0}^\infty\Big[&c_{\Delta_1\Delta_2}^{(0)}(-\eta)^{\Delta_1+\Delta_2+2n}(Z\cdot\partial_P)[\mathcal{O}_1\mathcal{O}_2]_n(P)\\
    &c_{\Delta_1\Delta_2}^{(1)}(-\eta)^{\Delta_1+\Delta_2+2n}[\mathcal{O}_1(Z\cdot\partial_P)\mathcal{O}_2]_n(P)\Big]
\end{aligned}
\end{equation}
To form a boundary scalar, in fact, $V\phi$ needs the action of a derivative, such that the scalar boundary operators with the lowest scaling dimension have $\Delta=\Delta_V+\Delta_\phi+1$. The operator $\phi_1\nabla\phi_2$, instead, can form a boundary scalar operator without the use of derivatives and with scaling dimension $\Delta=\Delta_1+\Delta_2$. Vice versa for the boundary vector operators. Finally, we verify that the contributions of the spurious poles exactly cancel also in this case, because of the identities in \ref{subsubsec:spurious}.
\label{subsubsec:freeqft2}
\subsection{Conformal Field Theories}
\label{subsec:cfts}
We have shown some examples of \KL\ decompositions of two-point functions of composite operators in free QFTs. In this section, we use the Käll\'en–Lehmann decomposition to study how states generated by the action of bulk CFT primaries on the Euclidean vacuum decompose into UIRs of the de Sitter group. That corresponds to decomposing irreps of $SO(d+1,2)$ into irreps of $SO(d+1,1)\,.$ We test examples up to spin 2 and recover the fact that for general $d>1$, CFT states decompose into principal series states and complementary series states, while for $d=1$ there is the appearance of discrete series states up to $\Delta=J,$ as in the free theory case. We verify the validity of our results by comparing their flat space limit as described in section \ref{subsec:flatspacelimit} to the results presented in \cite{Karateev:2020axc}.
\paragraph{Spinning CFT two-point functions in de Sitter}
To start, let us review the form of CFT two-point functions of traceless symmetric primary operators with spin in the bulk of de Sitter. The relevant group of symmetries of these correlators is $SO(d+1,2).$ We thus embed the $d+1$ dimensional de Sitter CFT in $\mathbb{R}^{d+1,2}$ with metric $\eta=\text{diag}(-1,1,\ldots,1,-1)$. We denote points in this embedding space $\mathcal{Y}\in\mathbb{R}^{d+1,2}$, and the invariance under $SO(d+1,2)$ is enforced by $\mathcal{Y}^2=0$. Explicitly, 
\begin{equation}
    \mathcal{Y}^2=Y^2-(\mathcal{Y}^{d+2})^2=0\,,
\end{equation}
where $Y\in\mathbb{R}^{d+1,1}$. The de Sitter hyperboloid constraint is enforced by $(\mathcal{Y}^{d+2})^2=1$. That is the section of the lightcone in $\mathbb{R}^{d+1,2}$ on which we will focus. We also embed the polarization tensors as $\mathcal{Z}=(W,0),$ so that $\mathcal{Y}_1\cdot \mathcal{Z}_2=Y_1\cdot W_2\,.$ In \cite{Costa:2011mg} it is shown that CFT two-point functions of a spin $J$ primary operator of conformal dimension $\mathbf{\Delta}$ is, in embedding space,
\begin{equation}
    \langle \CO^{(J)}(\mathcal{Y}_1,\mathcal{Z}_1)\CO^{(J)}(\mathcal{Y}_2,\mathcal{Z}_2)\rangle=c_{\CO^{(J)}}\frac{(-2[(\mathcal{Z}_1\cdot \mathcal{Z}_2)(\mathcal{Y}_1\cdot \mathcal{Y}_2)-(\mathcal{Y}_1\cdot \mathcal{Z}_2)(\mathcal{Y}_2\cdot \mathcal{Z}_1)])^J}{(-2\mathcal{Y}_1\cdot \mathcal{Y}_2)^{\mathbf{\Delta}+J}}\,.
\end{equation}
Projecting to de Sitter by using the lightcone constraint 
\begin{equation}
    \mathcal{Y}_1\cdot \mathcal{Y}_2=Y_1\cdot Y_2-1\,,
\end{equation}
we obtain the general form of a spin $J$ CFT two-point function in de Sitter expressed in the embedding space formalism
\begin{equation}\label{CFTL}
    \langle \mathcal{O}^{(J)}(Y_1,W_1)\mathcal{O}^{(J)}(Y_2,W_2)\rangle=c_{\mathcal{O}^{(J)}}\frac{[(W_1\cdot W_2)(1-Y_1\cdot Y_2)+(Y_1\cdot W_2)(Y_2\cdot W_1)]^J}{2^{\mathbf{\Delta}}(1-Y_1\cdot Y_2)^{\mathbf{\Delta}+J}}\,,
\end{equation}
\subsubsection{Spin 0 Example}
\label{subsubsec:cft0}
Let us start by reviewing the Käll\'en–Lehmann decomposition of the CFT two-point function of a scalar primary operator $\CO$ of conformal dimension $\mathbf{\Delta}$ in de Sitter, which was computed before in \cite{Hogervorst:2021uvp}. The two-point function has the form
\begin{equation}
    \langle \CO(Y_1)\CO(Y_2)\rangle=\frac{c_{\CO}}{2^{\mathbf{\Delta}}(1-Y_1\cdot Y_2)^{\mathbf{\Delta}}}\,.
\end{equation}
It has been argued in section \ref{sec:kallanlehmann} that only scalar principal series and complementary series can contribute to such a two-point function. In addition, using the criterion found in section \ref{subsec:completeness}, we know that complementary series is also absent when $\mathbf{\Delta}>\frac{d}{2}$.  In this case, 
the inversion formula for the principal series contribution reads
\begin{equation}
    \rho^{\mathcal{P},0}_\CO(\lambda)=\frac{c_{\CO}}{2^{\mathbf{\Delta}}\mathcal{N}_{0,0}}\int_{X_1}\Omega_{\lambda,0}(X_1,X_2)(1-X_1\cdot X_2)^{-\mathbf{\Delta}}\,.
\end{equation}
This integral can be solved by using the Mellin-Barnes representation of the hypergeometric function and Barnes' first lemma, as explained in Appendix \ref{subsec:mastercft}. We obtain
\begin{equation}
    \rho^{\mathcal{P},0}_\CO(\lambda)=c_{\CO}\frac{2^{1+d-2\mathbf{\Delta}}\pi^{\frac{d-1}{2}}\Gamma(-\frac{d}{2}+\mathbf{\Delta}\pm i\lambda)}{\Gamma(\mathbf{\Delta})\Gamma(\frac{1-d}{2}+\mathbf{\Delta})}\lambda\sinh(\pi\lambda)\,.
    \label{eq:rhoCFT0}
\end{equation}
We numerically verified that the Käll\'en–Lehmann integral over the principal series of this spectral density fully reproduces the two-point function, if $\mathbf{\Delta}>\frac{d}{2}$. When instead $\mathbf{\Delta}<\frac{d}{2}$, the poles at
\begin{equation}
    \lambda=\pm i\left(\mathbf{\Delta}-\frac{d}{2}\right)
\end{equation}
cross the contour of integration and need to be added. This corresponds to a complementary series state since $\lambda$ is imaginary. The contribution of this state when $\mathbf{\Delta}<\frac{d}{2}$ is in agreement with the decomposition of $SO(d+1,2)$ irreps into $SO(d+1,1)$ as shown in table 1.4 of \cite{Penedones:2023uqc}. Unitarity fixes $\rho_\CO^{\mathcal{P}}(\lambda)$ to be positive, and from this fact we can infer the usual unitarity bound on $\mathbf{\Delta}$
\begin{equation}
    \mathbf{\Delta}>\frac{d-1}{2}\,.
\end{equation}
In general, we can write the full decomposition of a CFT scalar two-point function as
\begin{equation}
    \langle \CO(Y_1)\CO(Y_2)\rangle=\int_{\mathbb{R}}\mathrm{d}\lambda\ \rho_\CO^{\mathcal{P},0}(\lambda)G_{\lambda,0}(Y_1,Y_2)+\theta\left(\frac{d}{2}-\mathbf{\Delta}\right)\rho^{\mathcal {C},0}_\CO G_{i\left(\mathbf{\Delta}-\frac{d}{2}\right),0}(Y_1,Y_2)\,,
\end{equation}
where $\theta(x)$ is a Heaviside theta, and 
\begin{equation}
    \rho_\CO^{\mathcal {C},0}=c_\CO\frac{4\pi^{\frac{d}{2}}\Gamma(1-\frac{d}{2}+\mathbf{\Delta})\sin\left(\frac{\pi}{2}(d-2\mathbf{\Delta})\right)}{\Gamma(\mathbf{\Delta})}=4\pi i\underset{\lambda=i(\mathbf{\Delta}-\frac{d}{2})}{\text{Res}}\rho_\CO^{\mathcal{P}}(\lambda)\,.
\end{equation}
Notice that the complementary series contribution is the two-point function of a free field with $\Delta=d-\mathbf{\Delta}$. Furthermore, if $\mathbf{\Delta}=\frac{d-1}{2}$, we have that $\rho_\CO^{\mathcal{P},0}(\lambda)=0$ and 
\begin{equation}
    \langle \CO(Y_1)\CO(Y_2)\rangle=\rho_\CO^{\mathcal{C},0}G_{i\left(\mathbf{\Delta}-\frac{d}{2}\right),0}(Y_1,Y_2)\Big|_{\mathbf{\Delta}=\frac{d-1}{2}}\,.
\end{equation}
This is the two-point function of a conformally coupled scalar in dS$_{d+1}$. Finally, we comment on \textbf{boundary operators}. The poles in $\rho_\mathcal{O}^{\mathcal{P},0}(\lambda)$ are at $\Delta=\mathbf{\Delta}+\mathbb{N}$, signaling the appearance of boundary operators with these weights in the Boundary Operator Expansion of the bulk CFT primary $\mathcal{O}$. 
Let us explain their origin by considering a $d+1$ dimensional Lorentzian CFT on the Minkowski cylinder. The discussion is entirely analogous in de Sitter because they are conformally equivalent spacetimes. Consider a scalar primary $\mathcal{O}(t,\vec x)$ in this CFT with conformal dimension $\mathbf{\Delta}$ and let us focus on the timeslice at $t=0$, where the primaries of the smaller $SO(d+1,1)$ group at that timeslice are those operators $\mathcal{O}_n(\vec x)$ which satisfy 
\begin{equation}
[D,\mathcal{O}_n(0)]=\Delta_n\mathcal{O}_n(0)\,, \qquad [K_i,\mathcal{O}_n(0)]=0 \qquad i=1,\ldots,d\,,
\label{eq:boundaryprims}
\end{equation}
It can be checked that particular linear combinations of operators like $(\boldsymbol{\partial}^2)^m\partial_t^n\mathcal{O}(0,\vec x)$ for different $m$ and $n$ such that $\frac{m}{2}+n$ is constant, satisfy this condition. For example, the operator $\mathcal{O}_0(\vec x)\equiv\mathcal{O}(0,\vec x)$ trivially satisfies (\ref{eq:boundaryprims}) with $\Delta_0=\mathbf{\Delta}$. But also
\begin{equation}
    [K_i,[P_0,\mathcal{O}(0,0)]]=-2[M_{i0},\mathcal{O}(0,0)]=0
\end{equation}
so that $\mathcal{O}_1(\vec x)\equiv\partial_t\mathcal{O}(0,\vec x)$ at the timeslice $t=0$ is a boundary primary with $\Delta_1=\mathbf{\Delta}+1$. If we go further, we find
\begin{equation}
\begin{aligned}
    [K_i,[P_0^2,\mathcal{O}(0,0)]]=2[P_i,\mathcal{O}(0,0)]\,, 
\end{aligned}
\end{equation}
but also
\begin{equation}
    [K_i,[P_jP^j,\mathcal{O}(0,0)]]=2(2(\mathbf{\Delta}+1)-d)[P_i,\mathcal{O}(0,0)]
\end{equation}
so that a good boundary primary is 
\begin{equation}
    \mathcal{O}_2(\vec x)\equiv \partial_t^2\mathcal{O}(0,\vec x)-\frac{1}{2(\mathbf{\Delta}+1)-d}\boldsymbol{\partial}^2\mathcal{O}(0,\vec x)\,,
\end{equation}
with $\Delta_2=\mathbf{\Delta}+2$. This can be iterated to find higher and higher primaries with $\Delta=\mathbf{\Delta}+\mathbb{N}$. By Weyl equivalence, the discussion in de Sitter space is entirely analogous and we can thus explain the presence of the poles at $\Delta=\mathbf{\Delta}+\mathbb{N}$ in the spectral density (\ref{eq:rhoCFT0}).

\paragraph{Flat space limit}
To compute the flat space limit of the \KL\ decomposition of this two-point function, we follow the discussion in section \ref{subsec:flatspacelimit} by 
 first restoring the dimensions of the spectral densities by adding the appropriate factors of the de Sitter radius $R$: $\rho_{\CO}^{\mathcal{P},0}(\lambda)\to R^{-2\mathbf{\Delta}+d-1}\rho_{\CO}^{\mathcal{P},0}(\lambda)$. Then, after changing variables to $\lambda=R m$, we take the limit
\begin{equation}
   \lim_{R\to\infty}\frac{R}{m}\rho_{\CO}^{\mathcal{P},0}(Rm)=c_\CO\frac{2^{d+1-2\mathbf{\Delta}}\pi^{\frac{d+1}{2}}}{\Gamma(\mathbf{\Delta})\Gamma(\frac{1-d}{2}+\mathbf{\Delta})}m^{2\mathbf{\Delta}-d-1}\,,
\end{equation}
which precisely matches the flat space CFT scalar spectral density (eq. (4.3) in \cite{Karateev:2020axc}).\footnote{To match conventions $d_{\text{here}}=(d-1)_{\text{there}}$ and $\mathbf{\Delta}_{\text{here}}=\Delta_{\mathcal{\CO}\text{there}}$.} Once adjusted for dimensions, the complementary series spectral density instead reads
\begin{equation}
    \rho^{\mathcal{C},0}_\CO=c_\CO\frac{4\pi^{\frac{d}{2}}\Gamma(1-\frac{d}{2}+\mathbf{\Delta})\sin\left(\frac{\pi}{2}(d-2\mathbf{\Delta})\right)}{R^{2\mathbf{\Delta}-d+1}\Gamma(\mathbf{\Delta})}\,.
\end{equation}
In the large $R$ limit it survives only if $\mathbf{\Delta}=\frac{d-1}{2}$. That is in agreement with the fact that in flat space, a free massless scalar is a CFT primary operator. We can thus write that in flat space we have
\begin{equation}
    \rho^{\mathbb{M},0}_\CO(m^2)=c_\CO\frac{2^{d+1-2\mathbf{\Delta}}\pi^{\frac{d+1}{2}}}{\Gamma(\mathbf{\Delta})\Gamma(\frac{1-d}{2}+\mathbf{\Delta})}m^{2\mathbf{\Delta}-d-1}\,,
\end{equation}
and instead, if $\mathbf{\Delta}=\frac{d-1}{2}$,
\begin{equation}
    \rho^{\mathbb{M},0}_\CO(m^2)=c_\CO\delta(m^2)\frac{4\pi^{\frac{d+1}{2}}}{\Gamma(\frac{d-1}{2})}\,.
\end{equation}
\subsubsection{Spin 1 Example}
\label{subsubsec:cft1}
For a spin-1 primary operator $J$ of conformal weight $\mathbf{\Delta}$, the two-point function is 
\begin{equation}
    \langle J(Y_1;W_1)J(Y_2;W_2)\rangle=\frac{c_J}{2^{\mathbf{\Delta}}}\left[\frac{W_1\cdot W_2}{(1-Y_1\cdot Y_2)^{\mathbf{\Delta}}}+\frac{(Y_1\cdot W_2)(Y_2\cdot W_1)}{(1-Y_1\cdot Y_2)^{\mathbf{\Delta}+1}}\right]\,.
\end{equation}
In Appendix \ref{subsec:mastercft} we show explicitly how to invert this two-point function and find the principal series spectral densities: 
\begin{equation}
    \begin{aligned}
        \rho_{J}^{\mathcal{P},0}(\lambda)&=c_J\frac{2^{3+d-2\mathbf{\Delta}}\pi^{\frac{d-1}{2}}(\mathbf{\Delta}-d)\Gamma\left(-\frac{d}{2}+\mathbf{\Delta}\pm i\lambda\right)}{(d^2+4\lambda^2)\Gamma(\mathbf{\Delta}+1)\Gamma(\frac{1-d}{2}+\mathbf{\Delta})}\lambda\sinh(\pi\lambda)\,,\\
        \rho_{J}^{\mathcal{P},1}(\lambda)&=c_J\frac{2^{1+d-2\mathbf{\Delta}}\pi^{\frac{d-1}{2}}(\mathbf{\Delta}-1)\Gamma(-\frac{d}{2}+\mathbf{\Delta}\pm i\lambda)}{\Gamma(\mathbf{\Delta}+1)\Gamma(\frac{1-d}{2}+\mathbf{\Delta})}\lambda\sinh(\pi\lambda)\,.
        \label{eq:rhoCFT1}
    \end{aligned}
\end{equation}
For $d\geq2$, this is the complete Käll\'en–Lehmann decomposition of this two-point function, as can be verified numerically by performing the integral over the principal series. In $d=1$, instead, we see the appearance of discrete series states. We can either compute those directly from the inversion formula in $d=1$, or just derive them by analytically continuing the higher dimensional result all the way to $d=1$ while keeping track of any poles that cross the principal series integration contour. The results one obtains in these two ways agree. The full \KL\ decomposition for the two-point function of a spin 1 primary CFT operator is thus
\begin{align}
    \langle J(Y_1;W_1)J(Y_2;W_2)\rangle=&\sum_{\ell=0}^1\int_{\mathbb{R}}\mathrm{d}\lambda\ \rho_{J}^{\mathcal{P},\ell}(\lambda)[(W_1\cdot\nabla_1)(W_2\cdot\nabla_2)]^{1-\ell}G_{\lambda,\ell}(Y_1,Y_2;W_1,W_2)\nonumber\\
    &+\delta_{d,1}\rho_{J}^{\mathcal{D}_1}(W_1\cdot\nabla_1)(W_2\cdot\nabla_2)G_{-i/2,0}(Y_1,Y_2)
    \label{eq:KLspin1J}
\end{align}
where
\begin{equation}
    \rho_{J}^{\mathcal{D}_1}=c_J\frac{2^{3-2 \mathbf{\Delta} }\pi}{\mathbf{\Delta} }\,.
\end{equation}
When $d=1$ and $\mathbf{\Delta}=1$, we have that $\rho^{\mathcal{P},0}_{J}=\rho^{\mathcal{P},1}_{J}=0$ and the left hand side matches exactly the extra contribution from the discrete series. The discrete series term is in fact the two-point function of the operator $W\cdot\nabla\varphi$ with $\varphi$ being the massless scalar, corresponding to the $\Delta=1$ operator in the discrete series
\begin{equation}
(W_1\cdot\nabla_1)(W_2\cdot\nabla_2)\langle\varphi(Y_1)\varphi(Y_2)\rangle\,.
\end{equation}
In two-dimensional Minkowski space, the operator $\partial_{\mu}\varphi$ with $\varphi$ being a free massless scalar is a CFT primary. It should not surprise us then that the same is true in de Sitter, by Weyl equivalence. One last observation is that from (\ref{eq:rhoCFT1}) we can recover unitarity bounds for spin $1$ CFT primaries
\begin{equation}
    \mathbf{\Delta}\geq d\,.
\end{equation}
Moreover, we observe the expected feature that for a conserved current $\mathbf{\Delta}=d$, only transverse states propagate, since $\rho^{\mathcal{P},0}_{J}(\lambda)$ vanishes. Finally, we study the boundary operators appearing in the Boundary Operator Expansion of the bulk CFT primary $J(Y;W).$ The pole structure of the two spectral densities (\ref{eq:rhoCFT1}), after verifying that the contributions of the spurious poles cancel as expected, signals the appearance of scalar and vector boundary operators with conformal dimensions $\mathbf{\Delta}+\mathbb{N}.$ Analogously to the scalar case, these come from the late time expansion of the bulk operator and of linear combinations of $\partial_t^n(\boldsymbol{\partial}^2)^{m}J^\mu(0,\vec x)$. Scalar boundary primary operators then come from particular linear combinations of $\partial_t^n(\boldsymbol{\partial}^2)^{m}J^0(0,\vec x)$ and $\partial_t^n(\boldsymbol{\partial}^2)^{m}\boldsymbol{\partial}\cdot\vec J(0,\vec x)$. Vector operators come from $\partial_t^n(\boldsymbol{\partial}^2)^{m}J^i(0,\vec x)$.

\paragraph{Flat space limit}
We start by restoring the dimensions $\rho_J^{\mathcal{P},0}(\lambda)\to R^{2\mathbf{\Delta}-d-1}\rho_J^{\mathcal{P},0}(\lambda)$ and $\rho_J^{\mathcal{P},1}(\lambda)\to R^{2\mathbf{\Delta}-d+1}\rho_J^{\mathcal{P},1}(\lambda)$. Then, we have
\begin{equation}
\begin{aligned}
    \lim_{R\to\infty}\frac{R}{m}\rho^{\mathcal{P},0}_J(Rm)&=c_J\frac{2^{d+1-2\mathbf{\Delta}}\pi^{\frac{d+1}{2}}(\mathbf{\Delta}-d)}{\Gamma(1+\mathbf{\Delta})\Gamma(\frac{1-d}{2}+\mathbf{\Delta})}m^{2\mathbf{\Delta}-d-3}\,,\\
    \lim_{R\to\infty}\frac{R}{m^3}\rho^{\mathcal{P},1}_J(Rm)&=c_J\frac{2^{d+1-2\mathbf{\Delta}}\pi^{\frac{d+1}{2}}(\mathbf{\Delta}-1)}{\Gamma(1+\mathbf{\Delta})\Gamma(\frac{1-d}{2}+\mathbf{\Delta})}m^{2\mathbf{\Delta}-d-3}\,.
\end{aligned}
\end{equation}
which match equations (4.10) in \cite{Karateev:2020axc}. The discrete series contribution with dimensions restored reads
\begin{equation}
    \rho^{\mathcal{D}_1}_J=c_J\frac{2^{3-2\mathbf{\Delta}}\pi}{R^{2\mathbf{\Delta}-2}\mathbf{\Delta}}\,.
\end{equation}
In the flat space limit, this survives only if $\mathbf{\Delta}=1$, corresponding to the case in which we are decomposing a two-point function of a conserved current in $d=1$. We can thus write, for $\mathbf{\Delta}>1$
\begin{equation}
    \begin{aligned}
        \rho^{\mathbb{M},0}_J(m^2)&=c_J\frac{2^{d+1-2\mathbf{\Delta}}\pi^{\frac{d+1}{2}}(\mathbf{\Delta}-d)}{\Gamma(1+\mathbf{\Delta})\Gamma(\frac{1-d}{2}+\mathbf{\Delta})}m^{2\mathbf{\Delta}-d-3}\,,\\
        \rho^{\mathbb{M},1}_J(m^2)&=c_J\frac{2^{d+1-2\mathbf{\Delta}}\pi^{\frac{d+1}{2}}(\mathbf{\Delta}-1)}{\Gamma(1+\mathbf{\Delta})\Gamma(\frac{1-d}{2}+\mathbf{\Delta})}m^{2\mathbf{\Delta}-d-3}\,,
    \end{aligned}
\end{equation}
which precisely match eq. (4.10) in \cite{Karateev:2020axc}. In the $d=1$ and $\mathbf{\Delta}=1$ case, instead, the \KL\ decomposition is given by a massless state
\begin{equation}
    \begin{aligned}
        \rho^{\mathbb{M},0}_J(m^2)=2\pi c_J\delta(m^2)\,, \qquad \rho^{\mathbb{M},1}_J(m^2)=0\,.
    \end{aligned}
\end{equation}
\subsubsection{Spin 2 Example}
\label{subsubsec:cft2}
The two-point function of a spin-2 traceless and symmetric CFT primary with conformal weight $\mathbf{\Delta}$ is 
\small
\begin{equation}
    \langle T(Y_1;W_1)T(Y_2;W_2)\rangle=\frac{c_T}{2^{\mathbf{\Delta}}}\Big[\frac{(W_1\cdot W_2)^2}{(1-Y_1\cdot Y_2)^{\mathbf{\Delta}}}+2\frac{(W_1\cdot W_2)(Y_1\cdot W_2)(Y_2\cdot W_1)}{(1-Y_1\cdot Y_2)^{\mathbf{\Delta}+1}}
    \!+\!\frac{[(Y_1\cdot W_2)(Y_2\cdot W_1)]^2}{(1-Y_1\cdot Y_2)^{\mathbf{\Delta}+2}}\Big]
\end{equation}
\normalsize
In Appendix \ref{subsec:mastercft} we show how to apply the inversion formula to this case. The resulting principal series contributions to the \KL\ decomposition are
\begin{equation}
    \begin{aligned}
        \rho_{T}^{\mathcal{P},0}(\lambda)&=c_T\frac{2^{5+d-2\mathbf{\Delta}}(d+1)\pi^{\frac{d-1}{2}}(d-\mathbf{\Delta})(d+1-\mathbf{\Delta})\Gamma(-\frac{d}{2}+\mathbf{\Delta}\pm i\lambda)}{d(d^2+4\lambda^2)((d+2)^2+4\lambda^2)\Gamma(\mathbf{\Delta}+2)\Gamma(\frac{1-d}{2}+\mathbf{\Delta})}\lambda\sinh(\pi\lambda)\,,\\
        \rho_{T}^{\mathcal{P},1}(\lambda)&=c_T\frac{2^{4+d-2\mathbf{\Delta}}\pi^{\frac{d-1}{2}}(1-\mathbf{\Delta})(d+1-\mathbf{\Delta})\Gamma(-\frac{d}{2}+\mathbf{\Delta}\pm i\lambda)}{((d+2)^2+4\lambda^2)\Gamma(\mathbf{\Delta}+2)\Gamma(\frac{1-d}{2}+\mathbf{\Delta})}\lambda\sinh(\pi\lambda)\,,\\
        \rho_{T}^{\mathcal{P},2}(\lambda)&=c_T\frac{2^{1+d-2\mathbf{\Delta}}\pi^{\frac{d-1}{2}}(\mathbf{\Delta}-1)\mathbf{\Delta}\Gamma(-\frac{d}{2}+\mathbf{\Delta}\pm i\lambda)}{\Gamma(\mathbf{\Delta}+2)\Gamma(\frac{1-d}{2}+\mathbf{\Delta})}\lambda\sinh(\pi\lambda)\,.
        \label{eq:rhoCFT2}
    \end{aligned}
\end{equation}
Once again, for $d>1$ the Käll\'en–Lehmann integral over the principal series fully reproduces the two-point function, and the structure of spurious poles matches the predictions made in section \ref{subsubsec:spurious}. In $d=1$, through the inversion formula for the discrete series we find contributions from $\Delta=1$ and $\Delta=2$ states.
\begin{align}
    \langle T(Y_1;W_1)T(Y_2;W_2)\rangle=&\sum_{\ell=0}^2\int_{\mathbb{R}}\mathrm{d}\lambda\ \rho_{T}^{\mathcal{P},\ell}(\lambda)[(W_1\cdot\nabla_1)(W_2\cdot\nabla_2)]^{2-\ell}G_{\lambda,\ell}(Y_1,Y_2;W_1,W_2)\nonumber\\
    &+\delta_{d,1}\sum_{p=1}^2\rho_{T}^{\mathcal{D}_p}(W_1\cdot\nabla_1)^2(W_2\cdot\nabla_2)^2G_{i(\frac{1}{2}-p),0}(Y_1,Y_2)
    \label{eq:KLTT}
\end{align}
with 
\begin{equation}
    \rho_{T}^{\mathcal{D}_2}=c_T\frac{2^{3-2 \mathbf{\Delta} }\pi}{\mathbf{\Delta} +1}\,, \qquad \rho_{T}^{\mathcal{D}_1}=c_T\frac{2^{3-2\mathbf{\Delta}}\pi(\mathbf{\Delta}-2)}{\mathbf{\Delta}(\mathbf{\Delta}+1)}\,.
\end{equation}
The positivity of (\ref{eq:rhoCFT2}) implies the expected unitarity bound for a spin 2 CFT primary 
\begin{equation}
    \mathbf{\Delta}\geq d+1\,,
\end{equation}
and when $T$ is a conserved stress tensor, $\mathbf{\Delta}=d+1$, both $\rho^{\mathcal{P},0}_{T}$ and $\rho^{\mathcal{P},1}_{T}$ vanish. When $d=1$, $G_{\lambda,2}$ vanishes and the decomposition matches (\ref{2DspinJfull}) with vanishing complementary series contributions. The two discrete series contributions correspond to the two-point functions 
\begin{equation}
    (W_1\cdot\nabla_1)^2(W_2\cdot\nabla_2)^2\langle\varphi(Y_1)\varphi(Y_2)\rangle\,, \qquad (W_1\cdot\nabla_1)^2(W_2\cdot\nabla_2)^2\langle\varphi'(Y_1)\varphi'(Y_2)\rangle\,,
\end{equation}
with $\varphi$ being a massless scalar, corresponding to $\Delta=1$, and $\varphi'$ being the first tachyon in the discrete series, the $\Delta=2$ representation. When $d=1$ and $\mathbf{\Delta}=2$, all spectral densities vanish except for $\rho_{T}^{\mathcal{D}_2}$, and then 
\begin{equation}
    \langle T(Y_1;W_1)T(Y_2;W_2)\rangle\Big|_{d=1,\mathbf{\Delta}=2}=\rho_{T}^{\mathcal{D}_2}\Big|_{\mathbf{\Delta}=2}(W_1\cdot\nabla_1)^2(W_2\cdot\nabla_2)^2G_{-\frac{3i}{2},0}(Y_1,Y_2)\,,
\end{equation}    
meaning that in $d=1$ the operator $(W\cdot\nabla)^2\varphi'$ with $\varphi'$ being a tachyonic free field with $\Delta_{\varphi'}=2$
is a CFT primary. Now let us comment on the boundary operators. First of all, we verify that the contributions from the spurious poles cancel. Then, we observe that the physical poles imply the appearance of boundary operators with spins $\ell=0,1,2$ and conformal weight $\mathbf{\Delta}+\mathbb{N}$ in the BOE of $T(Y,W)$. These again come from the late time limit of linear combinations of $\partial_t^n(\boldsymbol{\partial}^2)^mT^{\mu\nu}(0,\vec x)$. The spin 2 boundary operators are of the form $\partial_t^n(\boldsymbol{\partial}^2)^mT^{ij}(0,\vec x)$, the spin 1 operators are combinations of derivatives and $\mu,\nu=0,i$ and the scalars are combinations of the trace and of divergences.

\paragraph{Flat space limit}
As done for the previous examples, we restore the factors of $R$ in the spectral densities and then we take the large $R$ limit 
\begin{equation}
\begin{aligned}
    \lim_{R\to\infty}\frac{d}{d+1}\frac{R}{m}\rho^{\mathcal{P},0}_T(Rm)&=c_T\frac{2^{d+1-2\mathbf{\Delta}}\pi^{\frac{d+1}{2}}(\mathbf{\Delta}-d)(\mathbf{\Delta}-d-1)}{\Gamma(2+\mathbf{\Delta})\Gamma(\frac{1-d}{2}+\mathbf{\Delta})}m^{2\mathbf{\Delta}-d-5}\,,\\
    \lim_{R\to\infty}\frac{1}{2}\frac{R}{m^3}\rho^{\mathcal{P},1}_T(Rm)&=c_T\frac{2^{d+1-2\mathbf{\Delta}}\pi^{\frac{d+1}{2}}(\mathbf{\Delta}-1)(\mathbf{\Delta}-d-1)}{\Gamma(2+\mathbf{\Delta})\Gamma(\frac{1-d}{2}+\mathbf{\Delta})}m^{2\mathbf{\Delta}-d-5}\,,\\
    \lim_{R\to\infty}\frac{R}{m^5}\rho^{\mathcal{P},2}_T(Rm)&=c_T\frac{2^{d+1-2\mathbf{\Delta}}\pi^{\frac{d+1}{2}}(\mathbf{\Delta}-1)\mathbf{\Delta}}{\Gamma(2+\mathbf{\Delta})\Gamma(\frac{1-d}{2}+\mathbf{\Delta})}m^{2\mathbf{\Delta}-d-5}\,,
\end{aligned}
\end{equation}
where we have used $\beta_{2,0}=\frac{d}{d+1}, \beta_{2,1}=\frac{1}{2}$ and $\beta_{2,2}=1$. These results match equations (4.13) in \cite{Karateev:2020axc}. 
Restoring the dimensions in the discrete series densities, instead, gives
\begin{equation}
    \rho^{\mathcal{D}_2}_T=c_T\frac{2^{3-2\mathbf{\Delta}}\pi}{R^{2\mathbf{\Delta}-4}(\mathbf{\Delta}+1)}\,, \qquad \rho^{\mathcal{D}_1}_T=c_T\frac{2^{3-2\mathbf{\Delta}}\pi(\mathbf{\Delta}-2)}{R^{2\mathbf{\Delta}-4}\mathbf{\Delta}(\mathbf{\Delta}+1)}\,.
\end{equation}
Under the flat space limit, only when $\mathbf{\Delta}=2$ they have a chance of surviving. The $p=1$ density vanishes even in this case, because of the factor of $\mathbf{\Delta}-2$ in the numerator, and so we are left with the following flat space densities for $\mathbf{\Delta}>2$
\begin{equation}
    \begin{aligned}
        \rho^{\mathbb{M},0}_T(m^2)&=c_T\frac{2^{d+1-2\mathbf{\Delta}}\pi^{\frac{d+1}{2}}(\mathbf{\Delta}-d)(\mathbf{\Delta}-d-1)}{\Gamma(2+\mathbf{\Delta})\Gamma(\frac{1-d}{2}+\mathbf{\Delta})}\,m^{2\mathbf{\Delta}-d-5}\,,\\
        \rho^{\mathbb{M},1}_T(m^2)&=c_T\frac{2^{d+1-2\mathbf{\Delta}}\pi^{\frac{d+1}{2}}(\mathbf{\Delta}-1)(\mathbf{\Delta}-d-1)}{\Gamma(1+\mathbf{\Delta})\Gamma(\frac{1-d}{2}+\mathbf{\Delta})}\,m^{2\mathbf{\Delta}-d-5}\,,\\
        \rho^{\mathbb{M},2}_T(m^2)&=c_T\frac{2^{d+1-2\mathbf{\Delta}}\pi^{\frac{d+1}{2}}(\mathbf{\Delta}-1)\mathbf{\Delta}}{\Gamma(2+\mathbf{\Delta})\Gamma(\frac{1-d}{2}+\mathbf{\Delta})}\,m^{2\mathbf{\Delta}-d-5}\,,
        \label{eq:CFT2}
    \end{aligned}
\end{equation}
and the special case $\mathbf{\Delta}=2$ and $d=1$ instead being
\begin{equation}
    \begin{aligned}
        \rho^{\mathbb{M},0}_T(m^2)=\frac{\pi}{12}c_T\delta(m^2)\,, \qquad \rho^{\mathbb{M},1}_T(m^2)=0\,.
        \label{eq:specialCFT2}
    \end{aligned}
\end{equation}
Equations (\ref{eq:CFT2}) and (\ref{eq:specialCFT2}) match  (4.13) in \cite{Karateev:2020axc}.\footnote{The placement of the massless state in $\rho^{\mathbb{M},0}_T$ or $\rho^{\mathbb{M},1}_T$ is arbitrary since the Wightman propagators $\Delta_{0,1}^{(2)}$ and $\Delta_{0,0}^{(2)}$ are the same in $d=1$ in the massless limit.}
\subsubsection{Higher spin examples in dS$_2$}
So far, we have derived the spectral densities for spin $J\in\{0,1,2\}$ CFT operators $\CO^{(J)}$ in dS$_2$ by an analytical continuation from higher $d$. This can also be done systematically using the dS$_2$ inversion formula developed in section \ref{2Dinv}. In general, the two-point function of a spin $J$ primary is given by eq. (\ref{CFTL}), and its corresponding chiral components as defined in eq. (\ref{GOcomponents}) are
\begin{align}\label{GpmCFT}
 G_{\CO^{(J)}}^+(\sigma)=\frac{2^{J-\mathbf{\Delta}}c_{\mathcal{O}^{(J)}}}{(1-\sigma)^{\mathbf{\Delta}+J}}, \,\,\,\,\, \,\,  G_{\CO^{(J)}}^-(\sigma)=0~.
\end{align}
Assuming the unitarity bound $\mathbf{\Delta}\ge J$, the asymptotic behavior $G_{\CO^{(J)}}^\pm (\sigma)\sim (-\sigma)^{-(\mathbf{\Delta}+J)}$ ensures the absence of complementary series in the \KL\, decomposition of $\CO^{(J)}$. In addition, as a direct result of eq. (\ref{G-}), the vanishing of $G_{\CO^{(J)}}^-(\sigma)$ implies 
\begin{align}
   \rho^{\mathcal P, 1}_{\CO^{(J)}} (\lambda)= \left(\frac{1}{4}+\lambda^2\right)\rho^{\mathcal P, 0}_{\CO^{(J)}}(\lambda)~.
\end{align}
Then we apply the inversion formula (\ref{eq:2Dinver}) to $G^{(+)}_{\CO^{(J)}}$ given by eq. (\ref{GpmCFT})
\begin{align}
    &\rho^{\mathcal P, 0}_{\CO^{(J)}}(\lambda)
=\frac{2^{J-\mathbf{\Delta}+1}c_{\mathcal{O}^{(J)}}\lambda\sinh(2\pi\lambda)}{(\frac{1}{2}+i\lambda)_J^2(\frac{1}{2}-i\lambda)_J^2}\int_{-\infty}^{-1}d\sigma\,\frac{\phi_{\lambda, J}^+(\sigma)}{(1-\sigma)^{\mathbf{\Delta}-J}}\nonumber\\
&\rho^{\mathcal D_p}_{\CO^{(J)}}=\frac{2^{J-\mathbf{\Delta}+3}c_{\mathcal{O}^{(J)}}\pi^2 \,(2p-1)}{\Gamma(J+p)^2\Gamma(1+J-p)^2}\int_{-\infty}^{-1}d\sigma\,\frac{\psi_{p, J}(\sigma)}{(1-\sigma)^{\mathbf{\Delta}-J}}~.
\end{align}
The evaluation of this type of integral is extensively  discussed in appendix \ref{subsec:mastercft}. Here we just report the final results
\begin{align}
    &\rho^{\mathcal P, 0}_{\CO^{(J)}} (\lambda)=\left(\frac{1}{4}+\lambda^2\right)^{-1}\rho^{\mathcal P, 1}_{\CO^{(J)}} (\lambda)= \frac{c_{\mathcal{O}^{(J)}}\Gamma(\pm i\lambda+\mathbf{\Delta}-\frac{1}{2})}{2^{2\mathbf{\Delta}-J-1}(\frac{1}{2}\pm i\lambda)_J \Gamma(\mathbf{\Delta}+J)\Gamma(\mathbf{\Delta}-J)}\lambda\sinh(\pi\lambda) \nonumber\\
    &\rho^{\mathcal D_p}_{\CO^{(J)}} =\frac{c_{\mathcal{O}^{(J)}}\pi  (2 p-1)  \Gamma (\mathbf{\Delta} -p) \Gamma (p+\mathbf{\Delta} -1)}{2^{2 \mathbf{\Delta}-J-2}\Gamma
   (\mathbf{\Delta} -J) \Gamma (\mathbf{\Delta}+J ) \Gamma (J-p+1) \Gamma (J+p)}~.\label{genJ}
\end{align}
When $\mathbf{\Delta}=J$, i.e. saturation of unitarity bound, only discrete series with $p=J$ contributes.
\paragraph{Flat space limit}
The spectral densities we have just derived allow us to make a prediction about flat space two dimensional CFTs. After restoring the correct dimensions, under the flat space limit we get
\begin{equation}
\begin{aligned}
    \lim_{R\to\infty}\frac{R\,\beta_{J,0}}{m}\rho_{\mathcal{O}^{(J)}}^{\mathcal{P},0}(Rm)&=c_{\mathcal{O}^{(J)}}\frac{2^{2-2\mathbf{\Delta}}\pi}{\Gamma(\mathbf{\Delta}-J)\Gamma(\mathbf{\Delta}+J)}m^{2(\mathbf{\Delta}-J-1)}\,,\\
    \lim_{R\to\infty}\frac{R\,\beta_{J,1}}{m^3}\rho_{\mathcal{O}^{(J)}}^{\mathcal{P},1}(Rm)&=c_{\mathcal{O}^{(J)}}\frac{2^{2-2\mathbf{\Delta}}\pi}{\Gamma(\mathbf{\Delta}-J)\Gamma(\mathbf{\Delta}+J)}m^{2(\mathbf{\Delta}-J-1)}\,,
\end{aligned}
\end{equation}
where we have used $\beta_{J,0}=\beta_{J, 1}= 2^{1-J}$, which can be easily read off from eq. (\ref{flat4}). For the discrete series contribution, since $R^{2(J-\mathbf{\Delta})}$ has to be inserted in $\rho^{\mathcal D_p}_{\CO^{(J)}}$ before taking $R\to \infty$, it can survive in the flat space limit only if $\mathbf{\Delta}=J$, which itself forces $p=J$. 

In total we thus have that in 2 spacetime dimensions, the flat space spectral densities of a CFT primary of spin $J$ and conformal dimension $\mathbf{\Delta}$ are, for $\mathbf{\Delta}>J$,
\begin{equation}
\begin{aligned}
\rho^{\mathbb{M},0}_{\mathcal{O}^{(J)}}(m^2)=\rho^{\mathbb{M},1}_{\mathcal{O}^{(J)}}(m^2)=c_{\mathcal{O}^{(J)}}\frac{2^{2-2\mathbf{\Delta}}\pi}{\Gamma(\mathbf{\Delta}-J)\Gamma(\mathbf{\Delta}+J)}m^{2(\mathbf{\Delta}-J-1)}\,.
\end{aligned}
\end{equation}
and for the special case $\mathbf{\Delta}=J$,
\begin{equation}
    \rho^{\mathbb{M},0}_{\mathcal{O}^{(J)}}(m^2)=\delta(m^2)c_{\mathcal{O}^{(J)}}\frac{2^{3-2 J}\pi}{\Gamma(2J)}\,, \qquad \rho^{\mathbb{M},1}_{\mathcal{O}^{(J)}}(m^2)=0\,.
\end{equation}
\subsection{Weakly coupled QFT}
\label{subsec:qftint}
The Käll\'en–Lehmann decomposition is a non-perturbative representation of two-point functions in de Sitter. At the same time, it can be used to decompose two-point functions order by order in a perturbative expansion, when the QFT is weakly coupled. Since poles in the spectral densities can be related to the conformal dimensions of boundary operators \cite{Hogervorst:2021uvp} (see discussion in section \ref{subsubsec:boundarytheory}), we will observe their position shifting as we turn on interactions in the bulk. From this shift, we can read off anomalous dimensions for the boundary operators. At the same time, the fact that we will stop to a definite order in the coupling expansion, means we will lose the positivity of the spectral densities as a side effect. This is just an artifact of perturbation theory and does not mean that the theory is not unitary. We will now review how to derive anomalous dimensions of boundary operators from the spectral densities and then show a practical example in a scalar weakly interacting theory. 

Consider the Käll\'en–Lehmann decomposition of a two-point function of a scalar operator $\Phi$ (which for simplicity we take to include only principal series contributions) in an interacting theory governed by a coupling $g$. It will present simple poles on the nonperturbative value of the conformal dimension of each boundary operator that appears in the bulk-boundary OPE of $\Phi$ \cite{Hogervorst:2021uvp}. Near a pole at $\Delta_*(g)$, it will behave as
\begin{equation}
    \rho(\Delta,g)\approx\frac{\underset{\Delta=\Delta_*(g)}{\text{Res}}\left[\rho(\Delta,g)\right]}{\Delta-\Delta_*(g)}\,.
\end{equation}
For a weakly coupled theory, we can expand this formula in $g$ and obtain to first order
\begin{align}\label{c2def}
    \rho(\Delta,g)&\approx\frac{\underset{\Delta=\Delta_*(0)}{\text{Res}}\left[\rho(\Delta,0)\right]}{\Delta-\Delta_*(0)}+g\left(\frac{\partial_{g}\underset{\Delta=\Delta_*(0)}{\text{Res}}\left[\rho(\Delta,0)\right]}{\Delta-\Delta_*(0)}+\frac{\underset{\Delta=\Delta_*(0)}{\text{Res}}\left[\rho(\Delta,0)\right]\partial_{g}\Delta_*(0)}{(\Delta-\Delta_*(0))^2}\right)+O(g^2)\nonumber\\
    &\equiv\frac{c_0}{\Delta-\Delta_*(0)}+g\left(\frac{c_1}{\Delta-\Delta_*(0)}+\frac{c_2}{(\Delta-\Delta_*(0))^2}\right)+O(g^2)\,,
\end{align}
where in the second line we simply set up notation.
If we consider the series expansion of $\Delta_*(g)$,
\begin{equation}
    \Delta_*(g)=\Delta_*(0)+g\partial_{g}\Delta_*(0)+O(g^2)\,,
\end{equation}
we recognize that the anomalous dimension of the boundary operator with $\Delta=\Delta_*(g)$ at order $g$ is given by
\begin{equation}
    \gamma_*=g\frac{c_2}{c_0}\,.
    \label{eq:anomalousdimsgen}
\end{equation}
Simply put, the anomalous shift in the dimension $\Delta_*(0)$ at first order in $g$ is given by the ratio between the coefficient of the double pole in $\rho(\Delta,g)$ at the position $\Delta=\Delta_*(0)$ appearing at order $g$ and the coefficient of the simple pole at the same position but in the free theory. Let us now consider a concrete example.
\subsubsection{Anomalous dimensions from quartic interactions}
\label{subsubsec:anomalous}
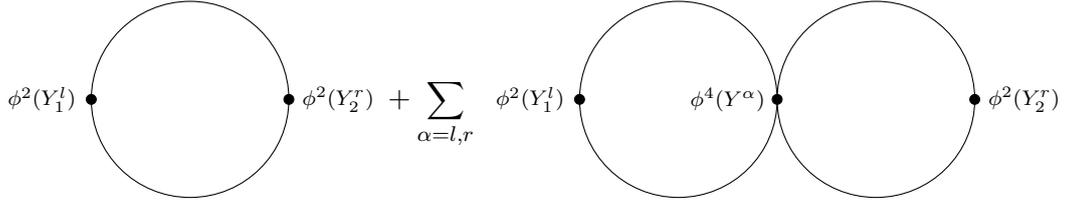
\begin{figure}
\centering
\begin{equation*}
\raisebox{-1.2cm}{
\begin{tikzpicture}[x=1.3cm,y=1.3cm]
\draw (0,0) circle (1);
\filldraw[color=black] (-1,0) circle (0.05);
\filldraw[color=black] (1,0) circle (0.05);
\begin{scriptsize}
\draw[color=black] (-1.5,0) node {$\phi^2(Y_1^l)$};
\draw[color=black] (1.5,0) node {$\phi^2(Y_2^r)$};
\end{scriptsize}
\end{tikzpicture}}+\sum_{\alpha=l,r}
\raisebox{-1.2cm}{
\begin{tikzpicture}[x=1.3cm,y=1.3cm]
\draw (0,0) circle (1);
\draw (2,0) circle (1);
\filldraw[color=black] (-1,0) circle (0.05);
\filldraw[color=black] (1,0) circle (0.05);
\filldraw[color=black] (3,0) circle (0.05);
\begin{scriptsize}
\draw[color=black] (-1.5,0) node {$\phi^2(Y_1^l)$};
\draw[color=black] (3.5,0) node {$\phi^2(Y_2^r)$};
\draw[color=black] (0.5,0) node {$\phi^4(Y^\alpha)$};
\end{scriptsize}
\end{tikzpicture}}
\end{equation*}
\caption{The free theory and order $g$ contributions to the Wightman two-point function $\langle\phi^2\phi^2\rangle$.}
\label{fig:Diagram}
\end{figure}

\begin{figure}
    \centering
    \includegraphics[width=0.8\textwidth]{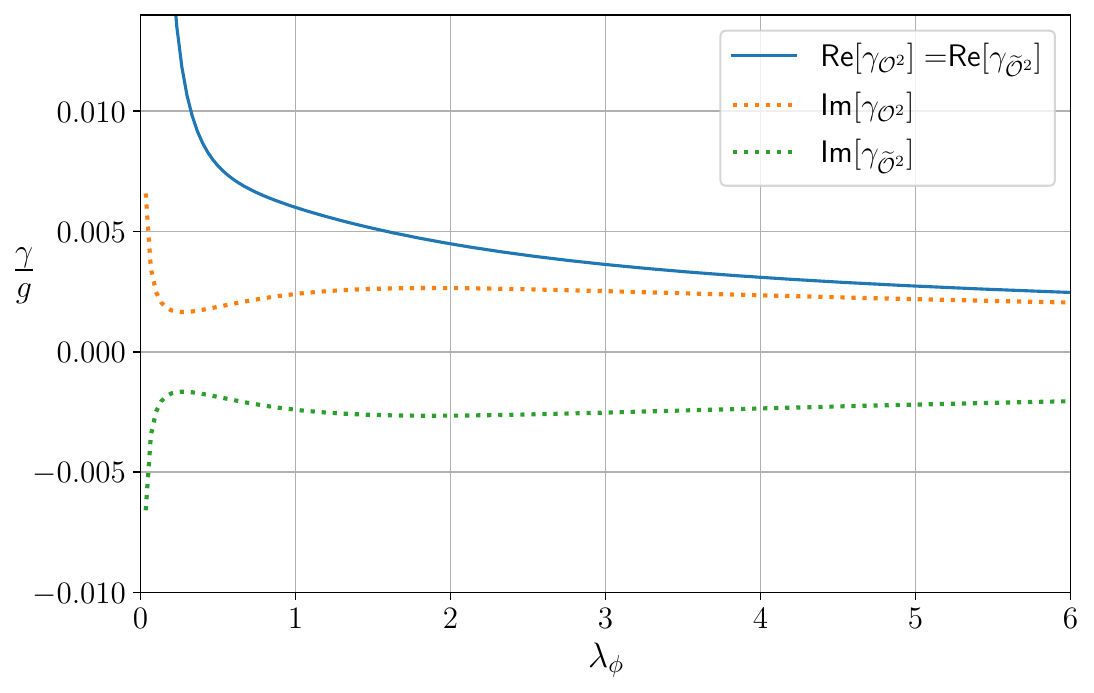}
    \caption{Anomalous dimensions of the boundary operators $\widetilde{\mathcal{O}}^2$ and $\mathcal{O}^2$ in $\frac{g}{4!}\phi^4$ theory in dS$_4$ for a field $\phi$ which, in the absence of interactions, has  $\Delta_\phi=\frac{d}{2}+i\lambda_\phi$ in the principal series. $\mathcal{O}$ and $\widetilde{\mathcal{O}}$ are the leading boundary operators in the BOE of $\phi$, and in the absence of interactions they transform as CFT scalar primaries with conformal dimensions $\Delta_\phi$ and $d-\Delta_\phi$ respectively. Notice that $(\gamma_{\mathcal{O}^2})^*=\gamma_{\widetilde{\mathcal{O}}^2}$. The divergence at $\lambda_\phi=0$ is due to the mixing of $\mathcal{O}^2$ and $\widetilde{\mathcal{O}}^2$ in that degenerate case. }
    \label{fig:principalanomalous}
\end{figure}

\begin{figure}
    \centering
    \includegraphics[width=0.8\textwidth]{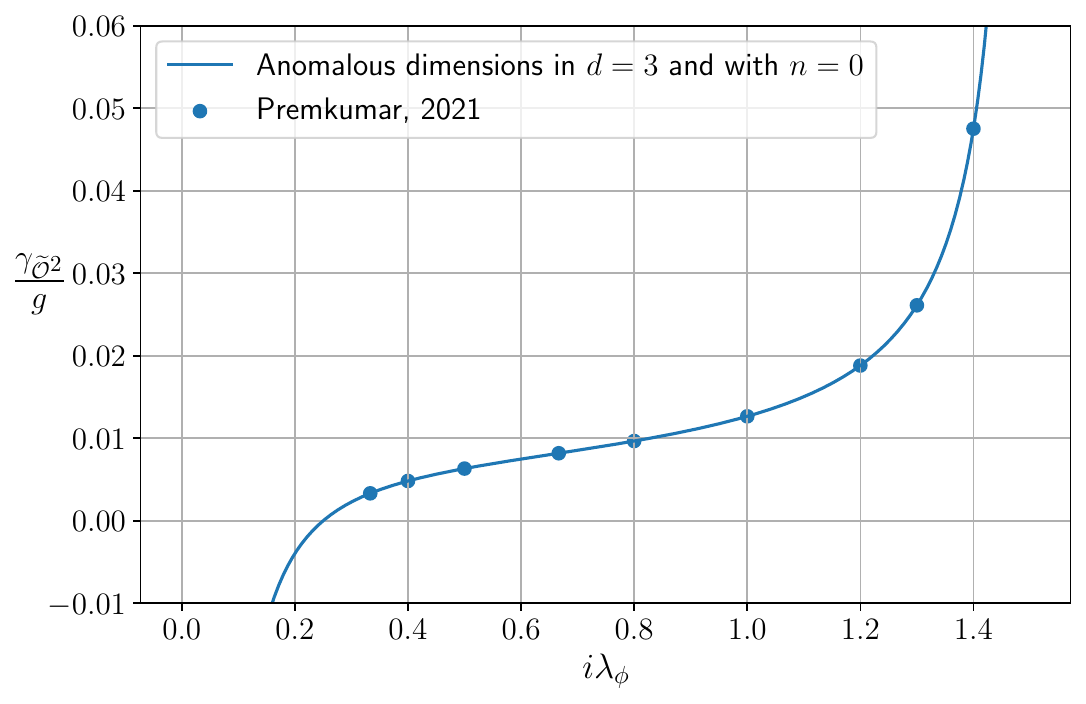}
    \caption{Anomalous dimensions of $\widetilde{\mathcal{O}}^2$ in $\frac{g}{4!}\phi^4$ theory in dS$_4$, where $\widetilde{\mathcal{O}}$ is the slowest decaying boundary operator associated to $\phi$ and $\phi$ is a field that, in the absence of interactions, has $\Delta_\phi=\frac{3}{2}+i\lambda_\phi$ in the complementary series. The dots are from Figure 8 of \cite{Premkumar:2021mlz}.}
    \label{fig:anomalousplotnew}
\end{figure}

Consider the following weakly coupled theory for a massive real scalar in de Sitter
\begin{equation}
    \mathcal{L}=-\frac{1}{2}g^{\mu\nu}\partial_\mu\phi\partial_\nu\phi-\frac{1}{2}m^2\phi^2-\frac{g}{4!}\phi^4\,,
\end{equation}
with $\Delta=\frac{d}{2}+i\lambda_\phi$ and $\lambda_\phi\in\mathbb{R}.$ We are going to be interested in this theory when the interaction is relevant or marginal, so $d=1,2,3$. We compute the correction to the free two-point function of the composite operator $\phi^2$ by using the in-in formalism, which we review in Appendix \ref{sec:ininformalism}. We choose to consider the Wightman function
\begin{equation}
    \langle\Omega|\phi^2(Y_1)\phi^2(Y_2)|\Omega\rangle\,,
\end{equation}
with $Y_1$ and $Y_2$ in the Expanding Poincar\'e Patch, $|\Omega\rangle$ is the interacting Bunch-Davies vacuum and, as we discussed in \ref{subsec:wick}, we are avoiding the branch cut by taking $\eta_1\rightarrow e^{i\epsilon}\eta_1$ and $\eta_2\rightarrow e^{-i\epsilon}\eta_2\,.$ In the notation from \cite{DiPietro:2021sjt}, which we are going to adopt for the rest of this subsection, it means we are selecting $Y_2\in r$ and $Y_1\in l$ (see Appendix \ref{sec:ininformalism} for more details).
When the coupling $g$ is turned off, the two-point function is given by the free theory contribution, which has the Käll\'en–Lehmann representation shown in \ref{subsubsec:freeqft0}, with $\lambda_1=\lambda_2=\lambda_\phi$ and a factor of 2 accounting for symmetry
\begin{equation}\label{rhofree}
    \rho^{\mathcal{P},0}_{\phi^2,\text{free}}(\lambda)=\frac{\lambda\sinh(\pi\lambda)}{16\pi^{3+\frac{d}{2}}\Gamma(\frac{d}{2})\Gamma(\frac{d}{2}\pm i\lambda)}\Gamma\left(\frac{\frac{d}{2}\pm i\lambda}{2}\right)^2\prod_{\pm,\pm}\Gamma\left(\frac{\frac{d}{2}\pm i\lambda\pm 2i\lambda_\phi}{2}\right)\,.
\end{equation}
Importantly, this spectral density has simple poles at 
\begin{equation}
\begin{aligned}
    \lambda&=2\lambda_\phi-i\left(\frac{d}{2}+2n\right)\longrightarrow\Delta=2\Delta_\phi+2n\,, \\
    \lambda&=-2\lambda_\phi-i\left(\frac{d}{2}+2n\right)\longrightarrow\Delta=2\bar\Delta_\phi+2n\,,
    \label{eq:phi2poles}
\end{aligned}
\end{equation}
due to the fact that boundary operators of the form $[\mathcal{O}\mathcal{O}]_n$ and $[\widetilde{\mathcal{O}}\widetilde{\mathcal{O}}]_n$\footnote{We remind the reader that $[\mathcal{O}_1\mathcal{O}_2]_n$ is a schematic notation to indicate all the scalar double trace operators one can form with $\mathcal{O}_1$, $\mathcal{O}_2$ and $2n$ derivatives, and $\phi(\eta,\vec 
x)\underset{\eta\to 0^-}{\sim}(-\eta)^{\Delta_\phi}\mathcal{O}(\vec x)+(-\eta)^{\bar\Delta_\phi}\widetilde{\mathcal{O}}(\vec x)$.} appear in the bulk-boundary OPE of $\phi^2$. We expect these operators to inherit anomalous dimensions once we turn on interactions. At leading order in the coupling, the two-point function is corrected by the diagram shown in Figure \ref{fig:Diagram}, which following the in-in formalism, corresponds to the following integrals
\begin{equation}
\label{eq:ininintegrals}
\langle\phi^2(Y_1)\phi^2(Y_2)\rangle^{lr}_{(g)}=ig\left[\int_{Y^l}(G^{ll}_{\lambda_\phi}(Y_1, Y)\, G^{lr}_{\lambda_\phi}(Y,Y_2))^2-\int_{Y^r}(G^{lr}_{\lambda_\phi}(Y_1,Y)\, G^{rr}_{\lambda_\phi}(Y,Y_2))^2\right]
\end{equation}
In Appendix \ref{sec:ininformalism} we analytically continue these integrals to EAdS and solve them. We obtain that the order $g$ contribution to the two-point function has the following spectral density
\begin{equation}
\begin{aligned}\label{eq:rhointeracting}
    \rho^{\mathcal{P},0}_{\phi^2,g}(\lambda)=&g\frac{\rho^{\mathcal{P},0}_{\phi^2,\rm free}(\lambda)}{4\sinh^2 (\pi\lambda_\phi)}\Bigg[\sin\left(\pi\left(\frac{d}{2}+2i\lambda_\phi\right)\right)B_{\Delta_\phi,\Delta_\phi}(\lambda)
    \\
    &+\sin\left(\pi\left(\frac{d}{2}-2i\lambda_\phi\right)\right)B_{\bar\Delta_\phi,\bar\Delta_\phi}(\lambda)-2\sin\left(\frac{d\pi}{2}\right)B_{\Delta_\phi,\bar\Delta_\phi}(\lambda)\Bigg]\,,
\end{aligned}
\end{equation}
with $B_{\Delta_1\Delta_2}(\lambda)$ 
defined as an infinite series 
in (\ref{eq:sumB}). 
This function is well-defined when $d<3$ and suffers from a UV divergence when $d\ge 3$. 
In $d=3$, as discussed in appendix \ref{subsec:anomalousappendix}, we can make sense of $B_{\Delta_1\Delta_2}(\lambda)$ by dimensional regularization, i.e. $d=3-\epsilon$, and absorbing the divergence into the wavefunction renormalization of $\phi^2$.
In the same appendix, we also show how to  extract the anomalous dimensions of $[\mathcal{O}\mathcal{O}]_n$ and $[\widetilde{\mathcal{O}}\widetilde{\mathcal{O}}]_n$ from eq. (\ref{eq:rhointeracting}), following the prescription outlined above. More precisely, we did that for $\lambda_\phi\in\mathbb R$ with the final expressions  given by eq. (\ref{gammaina}):
\begin{equation}
\begin{aligned}
\gamma_{[\mathcal{O}\mathcal{O}]_n}&=-g\frac{(\frac{d}{2})_n\Gamma(\frac{1}{2}+n+i\lambda_\phi)\Gamma(\frac{d}{2}+n+i\lambda_\phi)\Gamma(\frac{d}{2}+n+2i\lambda_\phi)\sin(\frac{\pi}{2}(d+4i\lambda_\phi))}{2^{d+3}\pi^{\frac{d}{2}}n!\sinh^2(\pi\lambda_\phi)\Gamma(1+n+i\lambda_\phi)\Gamma(\frac{d+1}{2}+n+i\lambda_\phi)\Gamma(1+n+2i\lambda_\phi)}\,,\\
\gamma_{[\widetilde{\mathcal{O}}\widetilde{\mathcal{O}}]_n}&=-g\frac{(\frac{d}{2})_n\Gamma(\frac{1}{2}+n-i\lambda_\phi)\Gamma(\frac{d}{2}+n-i\lambda_\phi)\Gamma(\frac{d}{2}+n-2i\lambda_\phi)\sin(\frac{\pi}{2}(d-4i\lambda_\phi))}{2^{d+3}\pi^{\frac{d}{2}}n!\sinh^2(\pi\lambda_\phi)\Gamma(1+n-i\lambda_\phi)\Gamma(\frac{d+1}{2}+n-i\lambda_\phi)\Gamma(1+n-2i\lambda_\phi)}\,.
\label{eq:anomalousdims}
\end{aligned}
\end{equation}
For an elementary field that in the absence of interactions is in the principal series ($\lambda_\phi\in\mathbb{R}$), these anomalous dimensions are complex, satisfying $(\gamma_{[\mathcal{O}\mathcal{O}]_n})^*=\gamma_{[\widetilde{\mathcal{O}}\widetilde{\mathcal{O}}]_n}$ and with a positive real part. The late time boundary operators associated to $\phi^2$ will thus decay faster once interactions are turned on (assuming that $g>0$, or in other words that the Hamiltonian is bounded from below). In \cite{Arkani-Hamed:2015bza,Marolf:2010zp,Krotov:2010ma,Bros:2009bz}, a similar phenomenon was observed for the boundary operators $\mathcal{O}$ and $\widetilde{\mathcal{O}}$ themselves. In figure \ref{fig:principalanomalous} we plot $\gamma_{[\mathcal{O}\mathcal{O}]_n}$ and $\gamma_{[\widetilde{\mathcal{O}}\widetilde{\mathcal{O}}]_n}$ for $\lambda_\phi\in(0,10)$ in $d=3$ and with $n=0$. 
If we naively continue (\ref{eq:anomalousdims}) to imaginary values of $\lambda_\phi$ to study the case in which $\phi$ is in the complementary series, we can match with known results in the literature \cite{Premkumar:2021mlz} on the anomalous dimension of $[\widetilde{\CO}\widetilde{\CO}]_0$ in dS$_4$, as shown in Figure \ref{fig:anomalousplotnew}. We believe this analytic continuation should be done with care, since many of the steps in Appendix \ref{subsec:anomalousappendix} do not trivially generalize to the complementary series, but simply making $\lambda_\phi$ imaginary seems to work, at least in this case. 
One can further compare (\ref{eq:anomalousdims}) with the anomalous dimensions of the corresponding boundary operators in AdS in a quartic theory \cite{Heemskerk:2009pn,Fitzpatrick:2010zm,gravina,Levine:2023ywq}:
\begin{equation}
    \gamma^{\text{dS}}_{[\mathcal{O}\mathcal{O}]_n}=-\frac{1}{2}\text{csch}^2(\pi \lambda_\phi)\sin\left(\pi\left(\frac{d}{2}+2i\lambda_\phi\right)\right)\gamma^{\text{AdS}}_{[\mathcal{O}\mathcal{O}]_n}\,.
\end{equation}
The trigonometric factors appearing in front have two separate origins. The hyperbolic cosecant factor comes from the different normalizations for bulk-to-boundary propagators in dS and AdS, while the sine factor originates from the interference of the two branches of the in-in contour. Interestingly, its role is to cancel unphysical singularities that are otherwise present in the AdS result when analytically continued to the complementary series (which is partly outside of the unitarity bounds in AdS). These facts were already pointed out in \cite{Sleight:2021plv,Sleight_2021Exch}. 

The anomalous dimensions of $\widetilde{\mathcal{O}}^2$ in dS diverge as one approaches the two endpoints of the complementary series. The divergence as $i\lambda_\phi\rightarrow \frac{d}{2}$ is a symptom of the breaking down of perturbation theory due to the IR divergences associated to massless fields. The divergence as $i\lambda_\phi\rightarrow 0$ instead corresponds to the degeneracy between the boundary operators $\mathcal{O}^2$ and $\widetilde{\mathcal{O}}^2$ when $\Delta_\phi=\frac{d}{2}\,.$

\section{Outlook}
\label{sec:conclusion}

In this paper we have derived the \KL\ decomposition for spinning traceless symmetric bulk operators in dS (see section \ref{sec:kallanlehmann}), and applied it to many examples in section \ref{sec:applications}. Here we outline potential future applications that we can imagine for this technology.
\begin{itemize}
\item
In the context of \textbf{bootstrapping} QFT in de Sitter \cite{baumann2022snowmass}, it may be useful to study a mixed system involving boundary and bulk correlation functions analogous to what was recently done in flat space \cite{Karateev:2019ymz, Correia:2022dyp} and AdS \cite{meineri2023renormalization}.
In particular, it should be possible to generalize \cite{Cardy:1988tj,Zamolodchikov:1986gt} and derive a sum rule for the spectral density of the trace of the stress tensor that gives the central charge of the UV CFT in a two dimensional de Sitter background.

\item In this work, we mostly focused on the contributions of principal and complementary series UIRs (except in dS$_2$ where we also considered the discrete series systematically). It would be interesting to study the effect of the \textbf{type \rom{2} exceptional series} UIRs, which include photons and gravitons in $d>2$. For example, in \cite{Loparco:2025azm} we show that photons appear generically in the spectral decomposition of antisymmetric operators with two indices in four dimensions. 

Understanding the Ward identities of the boundary operators associated to photons and gravitons is the natural next step towards understanding quantum gravity in de Sitter. 

\item It would be interesting to establish the convergence properties of the Boundary Operator Expansion (BOE) non-perturbatively for QFT in dS. It is likely that the connection between boundary operators and quasi-normal modes of the \textbf{static patch} of de Sitter \cite{Ng:2012xp,Jafferis:2013qia,Sun:2020sgn,Anninos_2010,Anninos_2012,Anninos_2022} 
will be useful in this context. 
\end{itemize}

\section*{Acknowledgements}

We are grateful to Tarek Anous, Miguel Correia, Frederik Denef, Victor Gorbenko, Aditya Hebbar, Matthijs Hogervorst, Austin Joyce, Shota Komatsu, Fedor Popov, Akhil Premkumar and Jiaxin Qiao for useful discussions.

JP and ML are supported by the Simons Foundation grant 488649 (Simons Collaboration on the Nonperturbative Bootstrap) and the Swiss National Science Foundation through the project
200020\_197160 and through the National Centre of Competence in Research SwissMAP. ZS is supported by the US
National Science Foundation under Grant No. PHY-2209997 and the Gravity Initiative at
Princeton University.
\newpage
\appendix

\section{Various properties of Green's functions in de Sitter}\label{scalarfield}
\subsection{Canonical quantization of a free scalar}
Let $\Phi$ be a free massive scalar of mass $m\ge \frac{d}{2}$ in dS$_{d+1}$. We parametrize the mass by $m^2=\frac{d^2}{4}+\lambda^2$, with $\lambda\in\mathbb R$ (principal series) or $-\frac{d}{2}<i \lambda<\frac{d}{2}$ (complementary series). The  standard bulk mode expansion of $\phi$ in planar coordinates is 
\begin{align}
    \Phi(\eta, \vec y\,)= \int \frac{d^d \vec y}{(2\pi)^{\frac{d}{2}}}\left (a_{\vec k}\, \phi_{\vec k}(\eta) e^{i\vec k\cdot \vec y}+a^\dagger_{\vec k}\, \phi_{\vec k}(\eta)^* e^{-i\vec k\cdot \vec y}\right)~,
    \end{align}
where $\phi_{\vec k}(\eta)$ satisfies the equations of motion
\begin{align}\label{eoms}
  \left(  (-\eta)^{d+1}\partial_\eta (-\eta)^{1-d}\partial_\eta+|\vec k|^2\eta^2+m^2\right)\phi_{\vec k}(\eta)=0~.
\end{align}
and the Klein-Gordon normalization condition (needed to ensure the canonical commutation relation between $\Phi$ and $\Pi_\Phi=(-\eta)^{1-d}\partial_\eta\Phi$)
\begin{align}\label{KG}
    i (-\eta)^{1-d}\left(\phi^*_{\vec k}\partial_\eta\phi_{\vec k}-\partial_\eta\phi^*_{\vec k}\phi_{\vec k}\right) = 1
\end{align}
This is solved by 
\begin{align}\label{modefun}
    \phi_{\vec k}(\eta)=(-\eta)^{\frac{d}{2}} \bar h_{i\lambda}(|\vec k| \eta), \,\,\,\,\,  \phi_{\vec k}(\eta)^*=(-\eta)^{\frac{d}{2}}  h_{i\lambda}(|\vec k| \eta)~,
\end{align}
where 
\begin{align}\label{defineh}
    h_{i\lambda}(\xi)=\frac{\sqrt{\pi}}{2}e^{\frac{\pi \lambda}{2}}H^{(2)}_{i\lambda}(-\xi), \,\,\,\,\,  \bar h_{i\lambda}(\xi)=\frac{\sqrt{\pi}}{2}e^{-\frac{\pi \lambda}{2}}H^{(1)}_{i\lambda}(-\xi)~.
\end{align}
The functions $h_{i\lambda}$ and  $\bar h_{i\lambda}$ are invariant under $\lambda\leftrightarrow -\lambda$.  This is consistent with the fact that $m$ is independent of the sign of $\lambda$. For light scalars, i.e. $0<m<\frac{d}{2}$, $\lambda$ is purely imaginary, and $e^{\pm\frac{\pi\lambda}{2}}$ becomes a phase.

The solution of eq. (\ref{eoms}) and eq. (\ref{KG}) is not unique. Different linear combinations of the Hankel functions also satisfy the equation of motion and the Klein-Gordon normalization condition. This is related to the usual ambiguity of choosing the vacuum in a curved spacetime. What singles out the above choice is the early time $\eta\to-\infty$ asymptotic behavior
\begin{align}
    \phi_{\vec k}(\eta)\approx e^{-i\frac{\pi}{4}} (-\eta)^{\frac{d-1}{2}}\frac{e^{-i|\vec k|\eta}}{\sqrt{2 |\vec k|}}~.
\end{align}
It means that at early time, the corresponding choice of vacuum looks like the canonical Minkowski vacuum. The vacuum selected in this way is called the Bunch-Davies  vacuum. 

Given the mode functions (\ref{modefun}), the Wightman two-point function of $\Phi$ in the Bunch-Davies vacuum can be expressed as 
\begin{align}\label{phiWigh}
   G_{\lambda,0}(\eta_1, y_1; \eta_2,\vec y_2\,)&\equiv \langle\Omega|\Phi(\eta_1, \vec y_1\,)\Phi(\eta_2, \vec y_2\,)|\Omega\rangle\nonumber \\
   &= (\eta_1\eta_2)^{\frac{d}{2}}\int\frac{ d^d\vec k}{(2\pi)^d} e^{-i\vec k\cdot (\vec y_1-\vec y_2\,)}\bar h_{i\lambda}(|\vec k|\eta_1) h_{i\lambda}(|\vec k|\eta_2)~,
\end{align}
Evaluating the Fourier transformation in eq. (\ref{phiWigh}) yields the hypergeometric representation of $G_{\lambda, 0}$:
\begin{equation}
\label{Glambda}
    G_{\lambda, 0}(\eta_1, \vec y_1; \eta_2,\vec y_2\,)=\frac{\Gamma(\Delta)\Gamma(\bar\Delta)}{(4\pi)^{\frac{d+1}{2}}}\mathbf{F}\left(\Delta,\bar\Delta, \frac{d+1}{2},\frac{1+\sigma}{2}\right)~.
\end{equation}
where $\Delta=\frac{1}{2}+i\lambda$,  $\sigma$ is the chordal distance between $(\eta_1, \vec y_1)$ and $(\eta_2, \vec y_2)$
\begin{equation}
    \sigma=Y_1\cdot Y_2=\frac{\eta_1^2+\eta_2^2-\vec y_{12}^{\,2}}{2\eta_1\eta_2},\,\,\,\,\, \vec y_{12}=\vec y_1-\vec y_2~,
\end{equation}
and $\mathbf{F}(a,b,c,z)\equiv \frac{1}{\Gamma(c)} F(a, b,c, z)$ is the regularized hypergeometric function.

\subsection{Proca fields in dS$_2$}
In dS$_2$, the mode expansion of a Proca field $A_\mu$ is closely related to that of $\Phi$. More precisely, consider the Proca Lagrangian
\begin{align}
\mathcal L=-\frac{1}{4}F_{\mu\nu} F^{\mu\nu}-\frac{1}{2}m^2 A_\mu A^\mu~,
\end{align}
where $m^2=\frac{1}{4}+\lambda^2$. In planar  coordinates, the equation of motion of $A_\mu$ is satisfied by 
\begin{align}
A_\eta= \alpha \partial_y \Phi, \,\,\,\,\, A_y=\alpha \partial_\eta \Phi~,
\end{align}
where $\alpha $ is a normalization constant and $\Phi$ is a canonically normalized scalar field of the same mass $m$. We fix $\alpha$ by requiring $A_y$ and its conjugate momentum $\Pi_y$ satisfy the standard commutation relation. The canonical momentum $\Pi_y$ is given by 
\begin{align}
\Pi_y=\eta^2 F_{\eta y}=\alpha\, \eta^2(\partial^2_\eta-\partial_y^2)\Phi=-\alpha m^2 \Phi~,
\end{align}
where in the last step we have used the equation of motion of $\Phi$. Using the fact that $\partial_\eta\Phi=\Pi_\Phi$ is the canonical momentum of $\Phi$, we get 
\begin{align}    
[\Pi_y, A_y]=-\alpha^2 m^2[\Phi, \Pi_\Phi]=-i\alpha^2 m^2~,
\end{align}
which implies $\alpha^2=\frac{1}{m^2}$ for canonical quantization. Thus the Green's function of $A_\mu$ can be summarized as 
\begin{align}
d=1: \,\,\, \langle\Omega|A_\mu(\eta_1, y_1) A_\nu (\eta_2,y_2)|\Omega\rangle=\frac{1}{m^2}\, \epsilon_{\mu}^{\,\,\,\,\alpha} \partial_{y_1^\alpha} \epsilon_{\nu}^{\,\,\,\,\beta} \partial_{y_2^\beta} G_{\lambda, 0}(\sigma)~,
   \end{align}
where $\epsilon_{\mu\alpha}$ and $\epsilon_{\nu\beta}$ are totally antisymmetric tensors at $(\eta_1, y_1)$ and $(\eta_2, y_2)$ respectively. To write the two-point function $\langle\Omega|A_\mu(\eta_1, y_1) A_\nu (\eta_2,y_2)|\Omega\rangle$  in terms of embedding space coordinates, we need the following relation
\begin{align}\label{epsilonrelation}
    \epsilon_{ABC}\frac{\partial Y^A}{\partial y^\mu}\frac{\partial Y^B}{\partial y^\nu} Y^C=-\epsilon_{\mu\nu}~,
\end{align}
which can be directly checked in any local coordinates $y^\mu$. It implies that the embedding space counterpart of $\epsilon_{\mu\nu}$ is $-\epsilon_{ABC}Y^C$. Then the uplift of $\epsilon_\mu^{\,\,\,\,\alpha}\partial_{y^\alpha}$ to embedding space should be 
\begin{align}
    \epsilon_\mu^{\,\,\,\,\alpha}\partial_{y^\alpha}\Longrightarrow -\epsilon_{ABC}Y^C (\partial_{Y^B}-Y_B\,Y\cdot\partial_Y)=\epsilon_{ABC}Y^B\partial_{Y^C}~.
\end{align}
Altogether, the two-point function $G_{\lambda, 1}(Y_1,Y_2;W_1, W_2)$ of $A(Y, W)$ is 
\begin{align}\label{G1in2D}
    G_{\lambda, 1}(Y_1,Y_2;W_1, W_2)&\equiv \langle\Omega|A(Y_1, W_1)A(Y_2, W_2)|\Omega\rangle\nonumber\\
    &=\left(\frac{1}{4}+\lambda^2\right)^{-1}\epsilon\left(W_1, Y_1,\partial_{Y_1}\right)\epsilon\left(W_2, Y_2,\partial_{Y_2}\right)G_{\lambda, 0}(Y_1, Y_2)~,
\end{align}
where $\epsilon(U_1,U_2,U_3)\equiv \epsilon_{ABC}U_1^AU_2^BU_3^C$.

\subsection{Analytical continuation of $G_{\lambda, 0}$ in dS$_2$}
In dS$_2$, the Green's function $G_{\lambda, 0}$ in eq. (\ref{Glambda}) becomes divergent when $\Delta\equiv \frac{1}{2}+i\lambda = p\in\mathbb Z_+$. Such divergences can be removed by acting with derivatives. More precisely, let's first apply the series expansion of hypergeometric functions to $G_{\lambda, 0}$
\begin{align}
    G_{\lambda, 0}(\sigma)=\sum_{n\ge 0}\frac{\Gamma(\Delta+n)\Gamma(\bar\Delta+n)}{(n!)^2}\left(\frac{1+\sigma}{2}\right)^n~.
\end{align}
Then we take the limit $\Delta\to p$. The problematic terms in this limit correspond to $n\le p-1$, which means that the divergent part is a polynomial in $\sigma=Y_1\cdot Y_2$ of degree $p-1$. This polynomial is obviously annihilated by the differential operator $(W_1\cdot\nabla_1)^p (W_2\cdot \nabla_2)^p$.
Therefore, the following function has a well-defined $\Delta\to p$ limit 
\begin{align}
   \Psi_{p,\Delta}(Y_1,Y_2;W_1,W_2)\equiv   (W_1\cdot\nabla_1)^p (W_2\cdot \nabla_2)^p\, G_{-i(\Delta-\frac{1}{2}), 0}(\sigma)~.
\end{align}
In the remaining part of the section, we are going to compute $\Psi_{p,q}\equiv \lim_{\Delta\to q}\Psi_{p,\Delta}$ for any $1\le q\le p$. As discussed in section \ref{fieldembedding}, it amounts to computing the matrix 
\begin{align}\label{chiralmatrix}
    \begin{pmatrix}
\Psi_{p,q}(Y_1,Y_2;W^+_1,W^+_2) & \Psi_{p,q}(Y_1,Y_2;W^+_1,W^-_2) \\
\Psi_{p,q}(Y_1,Y_2;W^-_1,W^+_2) &\Psi_{p,q}(Y_1,Y_2;W^-_1,W^-_2) 
    \end{pmatrix}~,
\end{align}
where $W^\pm$ are defined by eq. (\ref{eYW}) to encode the two chiral components of a symmetric and traceless tensor in dS$_2$.
Let's  start with the diagonal entries $\Psi_{p,\Delta}(Y_1,Y_2;W^\pm_1,W^\pm_2)$
\begin{align}
\Psi_{p,\Delta}(Y_1,Y_2;W^\pm_1,W^\pm_2)
&=\sum_{n=0}^p \binom{p}{n}  (W^\pm_1\cdot\nabla_1)^{p-n} (W^\pm_2\cdot Y_1)^p  (W^\pm_1\cdot\nabla_1)^{n} \partial_\sigma^p G_{-i(\Delta-\frac{1}{2}), 0}(\sigma)\nonumber\\
&=\sum_{n=0}^p \binom{p}{n}\frac{p!}{n!} (W_1^\pm \!\cdot\! W_2^\pm)^{p-n}(W_1^\pm \!\cdot\! Y_2)^n (W_2^\pm \!\cdot\! Y_1)^n  \partial_\sigma^{p+n}G_{-i(\Delta-\frac{1}{2}), 0}(\sigma)~.
\end{align} 
Using the relation $(W_1^\pm \cdot Y_2)(W_2^\pm \cdot Y_1)=(\sigma+1)W_1^\pm \cdot W_2^\pm$ established in  eq. (\ref{userela}), we get
\begin{align}\label{j1}
\Psi_{p,\Delta}(Y_1,Y_2;W^\pm_1,W^\pm_2)
&=(W_1^\pm \cdot W_2^\pm)^{p}\sum_{n=0}^p \binom{p}{n}\frac{p!}{n!} (\sigma+1)^n \partial_\sigma^{J+n}G_{-i(\Delta-\frac{1}{2}), 0}(\sigma)\nonumber\\
&=\frac{\Gamma(\Delta)\Gamma(\bar\Delta)}{4\pi}(W_1^\pm \cdot W_2^\pm)^{p}\partial_\sigma^p((\sigma+1)^p\partial_\sigma^p)F\left(\Delta,\bar\Delta,1,\frac{1+\sigma}{2}\right)~,
\end{align}
where the  $2p$-th order differential operator $\partial_\sigma^p((\sigma+1)^p\partial_\sigma^p)$ acting on $F\left(\Delta,\bar\Delta,1,\frac{1+\sigma}{2}\right)$ yields another hypergeometric function:
\begin{align}\label{ppdd}
\Psi_{p,\Delta}(Y_1,Y_2;W^\pm_1,W^\pm_2)=\frac{\Gamma(\Delta+p)\Gamma(\bar\Delta+p)}{2^{p+2}\pi}(W_1^\pm \cdot W_2^\pm)^{p}F\left(\Delta+p,\bar\Delta+p,1,\frac{1+\sigma}{2}\right)~.
\end{align}
It clearly has a finite limit when $\Delta$ hits any positive integer $q$ that is not larger than $p$:
\begin{align}\label{psipq}
\Psi_{p,q}(Y_1,Y_2;W^\pm_1,W^\pm_2)=(W_1^\pm \cdot W_2^\pm)^{p}\,\psi_{p,q}(Y_1,Y_2)~,
\end{align}
where
\begin{align}\label{psipq1}
\psi_{p,q}(Y_1,Y_2)=\frac{\Gamma(p+q)\Gamma(p+1-q)}{2^{p+2}\pi}F\left(p+q,p+1-q,1,\frac{1+Y_1\cdot Y_2}{2}\right) ~.
\end{align}
The off-diagonal matrix elements in (\ref{chiralmatrix}) can be computed similarly
\begin{align}\label{j2}
\Psi_{p,\Delta}(Y_1,Y_2;W^\pm_1,W^\mp_2)&=\frac{\Gamma(\Delta)\Gamma(\bar\Delta)}{4\pi}(W_1^\pm \cdot W_2^\mp)^{p}\partial_\sigma^p((\sigma-1)^p\partial_\sigma^p)F\left(\Delta,\bar\Delta,1,\frac{1+\sigma}{2}\right)\nonumber\\
&=\frac{\Gamma(\Delta\!+\!p)^2\Gamma(\bar\Delta\!+\!p)^2 (W_1^\pm \!\cdot\! W_2^\mp)^{p}}{(-2)^{p+2}(2p)!\pi\Gamma(\Delta)\Gamma(\bar\Delta)}F\left(\Delta\!+\!p,\bar\Delta\!+\!p,2p\!+\!1,\frac{1\!+\!\sigma}{2}\right)~.
\end{align} 
They vanish when $\Delta$ approaches at positive integer $q\in\{1,2,\cdots, p\}$, because $\lim_{\Delta\to q}(\bar\Delta)_p=0$. Altogether, we have 
\begin{align}\label{disder}
 &(W^\pm_1\cdot\nabla_1)^p (W^\pm_2\cdot \nabla_2)^p\, G_{-i(q-\frac{1}{2}), 0}(Y_1, Y_2)=(W_1^\pm \cdot W_2^\pm)^{p}\,\psi_{p,q}(Y_1,Y_2)~,\nonumber\\
  &   (W^\pm_1\cdot\nabla_1)^p (W^\mp_2\cdot \nabla_2)^p\, G_{-i(q-\frac{1}{2}), 0}(Y_1, Y_2)=0~.
\end{align}
When $q=p$, $\psi_{p,p}(Y_1, Y_2)$ becomes particularly simple
\begin{align}
    \psi_{p,p}(Y_1, Y_2)=\frac{2^{p-2}\Gamma(2p)}{\pi (1-Y_1\cdot Y_2)^{2p}}~.
\end{align}
To end this section, we show the pull-back  of $\Psi_{p,p}(Y_1, Y_2; W_1, W_2)$ to conformal global coordinates. According to the discussion in section \ref{fieldembedding}, it amounts to replacing $W^A_\pm$ by $\partial_\pm Y^A$ (where $\partial_\pm$ denotes the ordinary derivative with respect to the local lightcone coordinates $y^\pm = \tau\pm \varphi$),  for example, 
\begin{align}\label{ppcom}
    \left(\nabla_+^{(1)}\right)^p  \left(\nabla_+^{(2)}\right)^p G_{-i(p-\frac{1}{2})}(Y_1, Y_2)=  \Psi_{p,p}(Y_1, Y_2; \partial_+ Y_1, \partial_+ Y_2)~,\nonumber\\
    =\frac{2^{p-2}\Gamma(2p)}{\pi}\left(\frac{\partial_{y_1^+}\partial_{y_2^+}\sigma}{(1-\sigma)^2}\right)^p=\frac{\Gamma(2p)}{4\pi\left(-4\sin^2\frac{y^+_{12}}{2}\right)^p}~,
\end{align}
where $\nabla_\pm$ denotes covariant derivative along $y^\pm$. Similarly, the other rest components are 
\begin{align}\label{other}
    &\left(\nabla_-^{(1)}\right)^p  \left(\nabla_-^{(2)}\right)^p G_{-i(p-\frac{1}{2})}(Y_1, Y_2)= \frac{\Gamma(2p)}{4\pi \left(-4\sin^2\frac{y^-_{12}}{2}\right)^p}~,\nonumber\\
    &\left(\nabla_+^{(1)}\right)^p  \left(\nabla_-^{(2)}\right)^p G_{-i(p-\frac{1}{2})}(Y_1, Y_2)=\left(\nabla_-^{(1)}\right)^p  \left(\nabla_+^{(2)}\right)^p G_{-i(p+\frac{1}{2})}(Y_1, Y_2)=0~.
\end{align}

\subsection{Flat space limit of $G_{\lambda,\ell}$}
\label{subsec:flatspaceG}
In this subsection, we elaborate on the discussion in section \ref{subsec:flatspacelimit} and show that the canonically normalized de Sitter Green's functions $[(W_1\cdot\nabla_1)(W_2\cdot\nabla_2)]^{J-\ell}G_{\lambda,\ell}(Y_1,Y_2;W_1,W_2)$ reduce to the 
flat space Wightman functions  
$m^{-2\ell}\Delta_{m^2,\ell}^{(J)}(x_1,x_2;w_1,w_2)$ (which are introduced in \cite{Karateev:2020axc} for $0\le \ell\le J\le 2$), up to numerical normalization factors.  For example, in the $J=0$ case, we expect the flat space limit of the scalar Green's function $G_{\lambda, 0}$ in dS to reproduce $\Delta_{m^2,0}^{(0)}$, which is given by  \cite{Karateev:2020axc}
\begin{equation}
    \Delta_{m^2,0}^{(0)}(x_1,x_2)=\frac{1}{(2\pi)^{\frac{d+1}{2}}}m^{d-1}\frac{K_{\frac{d-1}{2}}\left(m\sqrt{ x_{12}^2}\right)}{\left(m\sqrt{x_{12}^2}\right)^{\frac{d-1}{2}}}\,,
    \label{eq:flatspacescalar}
\end{equation}
where $x_1^\mu, x_2^\mu$ are flat space coordinates of $\mathbb R^{ d,1}$, and 
$i\epsilon$ is suppressed in $\sqrt{x_{12}^2}$. 

Let's start with discussing the flat space limit of the coordinates and the metric of dS. The dS metric in planar coordinate is $ds^2=R^2\frac{-d\eta^2+d\vec y^{\,2}}{\eta^2}$, where the dS radius $R$ is restored. We consider the large $R$ limit, with $R(\eta+1)\equiv x^0$ 
and $R\,  y^i\equiv x^i$ being fixed. Then the dS metric reduces to the flat space metric $ds^2= \eta_{\mu\nu} dx^\mu dx^\nu$. It is also useful to show how $\sigma=Y_1\cdot Y_2$ is related to the flat space distance $x_{12}^2$ in this limit. For this purpose, we define a new variable 
\begin{align}
    \rho\equiv\frac{1}{2}\left(1-\frac{\sigma}{R^2}\right)=\frac{-\eta_{12}^2+\vec y_{12}^{\, 2}}{4\eta_1\eta_2}=\frac{-(x^0_{12})^2+\vec x_{12}^{\, 2}}{4\eta_1\eta_2 R^2}~.
\end{align}
In the flat space limit, we can simply replace $4\eta_1\eta_2$ by $4$, and hence obtain $\rho\approx \frac{x_{12}^2}{4R^2}$. In other words, $\rho\to 0$ in the flat space limit, but $4\rho R^2$ is finite and equal to the  flat space distance $x_{12}^2$.  Using the large $R$ relation $\lambda=mR$, we can also write $4\rho\lambda^2 \approx m^2 x_{12}^2$.

Next, we consider the de Sitter scalar propagator
\begin{equation}
    G_{\lambda,0}(\sigma)=\frac{\Gamma(\frac{d}{2}\pm i\lambda)}{2^{d+1}\pi^{\frac{d+1}{2}}R^{d-1}}\mathbf{F}\left(\frac{d}{2}-i\lambda,\frac{d}{2}+i\lambda,\frac{d+1}{2},1-\rho\right)\,,
\end{equation}
which is expressed in terms of the new variable $\rho$.
To retrieve the flat space propagator we will thus have to take $\lambda^2\to\infty$ and $\rho\to0$ while keeping the product fixed. 
This limit can be easily implemented if we rewrite $G_{\lambda,0}$ by using the following Mellin representation of the hypergeometric function
\begin{align}\label{Mellin2}
 \mathbf{F}(a,b,c, z)=\frac{\int_{\mathbb R+i\epsilon}\, dt\,\Gamma(a+i t)\Gamma(b+i t)\Gamma(c-a-b-i t)\Gamma(-i t)(1-z)^{it}}{2\pi \Gamma(a)\Gamma(b)\Gamma(c-a)\Gamma(c-b)}~.
\end{align}
which  yields 
\begin{align}\label{GM}
     G_{\lambda,0}(\sigma)=\frac{\int_{\mathbb R+i\epsilon}\, dt\,\Gamma(\frac{d}{2}+i t\pm i\lambda)\Gamma(\frac{1-d}{2}-i t)\Gamma(-i t)\rho^{it}}{2^{d+2}\pi^{\frac{d+3}{2}} \Gamma(\frac{1}{2}\pm i\lambda)R^{d-1}}~.
\end{align}
 Taking the large $\lambda$ limit in eq. (\ref{GM}) :
\begin{align}
\frac{1}{\Gamma(\frac{1}{2}\pm i\lambda)} \approx \frac{1}{2\pi}\, e^{\pi\lambda}, \,\,\,\,\, \Gamma\left(\frac{d}{2}+i t\pm i\lambda\right)\approx2\pi e^{-\pi\lambda}\lambda^{d-1+2i t}~,
\end{align}
it is then easy to see that $G_{\lambda,0}(\sigma)$ becomes 
\begin{align}\label{GM1}
     G_{\lambda,0}(\sigma)\approx\frac{m^{d-1}}{2^{d+2}\pi^{\frac{d+3}{2}}}\int_{\mathbb R+i\epsilon}\, dt\,\Gamma\left(\frac{1-d}{2}-i t\right)\Gamma\left(-i t\right)(\lambda^2\rho)^{it}~,
\end{align}
where we have used $\lambda/R =m$.
The remaining  integral can be evaluated by using the integral representation of $\Gamma$ functions 
\begin{align}\label{GM2}
     G_{\lambda,0}(\sigma)\approx\frac{m^{d-1}}{2^{d+2}\pi^{\frac{d+3}{2}}}
     \int_0^\infty \frac{du}{u} \,u^{\frac{1-d}{2}}\, e^{-u}
     \int_{0}^\infty \frac{dv}{v}\, e^{-v}\int_{\mathbb R} dt\, \left(\frac{\lambda^2\rho}{u v}\right)^{it}~.
\end{align}
Performing  the $t$ integral gives a delta function supported on $ uv =\lambda^2 \rho$, so (\ref{GM2}) reduces to a familiar form
\begin{align}
     G_{\lambda,0}(\sigma)\approx \frac{m^{d-1}}{2^{d+1}\pi^{\frac{d+1}{2}}} \int_0^\infty \frac{du}{u} \,u^{\frac{1-d}{2}}\, e^{-u-\frac{\lambda^2 \rho}{u}} =\frac{m^{d-1}}{(2\pi)^{\frac{d+1}{2}}}\frac{ K_{\frac{d-1}{2}}\left(2\lambda\sqrt{\rho}\right)}{(2\lambda\sqrt{\rho})^{\frac{d-1}{2}}}~.
\end{align}
This is exactly (\ref{eq:flatspacescalar}) after using the relation $4\rho\lambda^2\approx m^2x^2_{12}$. We want to mention that a EAdS version of this story was discussed in \cite{Carmi:2018qzm}.

The second example is $J=\ell=1$. On the flat space side, we have  \cite{Karateev:2020axc}
\begin{equation}\label{Delta11}
    m^{-2}\Delta_{m^2,1}^{(1)}(x_1,x_2;w_1,w_2)=(w_1\cdot w_2-m^{-2}(w_1\cdot\partial_1)(w_2\cdot\partial_2))\Delta_{m^2,0}^{(0)}(x_1,x_2)
\end{equation}
where $w_1^\mu, w_2^\mu$ are auxiliary null vectors.
On the dS side,  the spin 1 Green's function $G_{\lambda, 1}$ can be obtained by evaluating the split representation (\ref{eq:InFormGen}) for $\ell=1$ \footnote{One can check that when $d=1$, eq. (\ref{G1anyD}) is actually equivalent to eq.(\ref{G1in2D}) by using the equation of motion $(1-\sigma^2)\partial_\sigma^2 G_{\lambda, 0}-2\sigma\partial_\sigma G_{\lambda, 0}=(\frac{1}{4}+\lambda^2)G_{\lambda, 0}$.}:
\small
\begin{align}\label{G1anyD}
    G_{\lambda, 1}(Y_1, Y_2; W_1, W_2)=\frac{\left[\left(\frac{d^2}{4}\!+\!\lambda^2\right)(W_1\cdot W_2)\!+\!\sigma (W_1\cdot\partial_{Y_1}) (W_2\cdot\partial_{Y_2})\!+\!d(W_1\cdot Y_2)(W_2\cdot\partial_{Y_2})\right]G_{\lambda, 0}(\sigma)}{\frac{(d-2)^2}{4}+\lambda^2}
\end{align}
\normalsize
where the $R$ dependence is implicitly restored in $G_{\lambda, 0}(\sigma)$.
To take the flat space limit of $G_{\lambda, 1}$, we should pull $W^A$ back to local coordinates, which is realized by the relation $W^A= \frac{w^\mu}{R}\frac{\partial Y^A}{\partial y^\mu}$ \footnote{In general, the null vector $w^\mu$ here are different from the one used in the flat space two-point function (\ref{Delta11}), because $w^\mu$ is null respect to the dS metric $g_{\mu\nu}$. In this case, as we choose the planar coordinates, $g_{\mu\nu}$ is conformally equivalent to $\eta_{\mu\nu}$ and hence $w^\mu$ is also null with repsect to the flat metric.}. Here $y^\mu = (\eta, \vec y)$ denotes the planar coordinates. Applying this pull-back rule  to $W_1\cdot W_2$ yields $
    W_1\cdot W_2 = \frac{w_1^\mu w_2^\nu}{R^2} \partial_{y^\mu_1} \partial_{y^\mu_2} \sigma$.
Because of the identification $x^\mu = (R(\eta+1), R y^i)$, we can replace $R^{-1}\partial_{y^\mu}$ by $\partial_{x^\mu}$ and then in the flat space limit we get 
\begin{align}
 W_1\cdot W_2 = (w_1\cdot \partial_1)(w_2\cdot \partial_2)(-2\rho R^2) \approx w_1\cdot w_2 
\end{align}
where  the substitution $\sigma= R^2-2\rho R^2$ is made in the first step, and the relation $\rho R^2=\frac{1}{4}x_{12}^2$ is used in the second step.
Similarly, the term $\sigma (W_1\cdot\partial_{Y_1}) (W_2\cdot\partial_{Y_2})$ reduces to $\sigma (w_1\cdot\partial_1)(w_2\cdot\partial_2)$. 
Although $\sigma$ itself diverges in the flat space limit, the combination $\sigma/\lambda^2$ is actually finite and equal to $m^{-2}$ because of the large $R$ relations $\frac{\sigma}{R^2}=1$ and $m R= \lambda$. It is also easy to check that  the remaining term in $G_{\lambda, 1}(Y_1, Y_2; W_1, W_2)$ does not survive in the flat space limit. Altogether, the flat space limit of $G_{\lambda, 1}$ should be 
\begin{align}
    G_{\lambda, 1}\approx \left(w_1\cdot w_2-m^{-2}(w_1\cdot\partial_1)(w_2\cdot\partial_2)\right)\Delta^{(0)}_{m^2, 0}(x_1, x_2)=m^{-2}\Delta_{m^2,1}^{(1)}(x_1,x_2;w_1,w_2).
\end{align}
Let's also  consider the case $J=1$, $\ell=0$. In flat space, $ \Delta_{m^2,0}^{(1)}$  is given by 
\begin{equation}
    \Delta_{m^2,0}^{(1)}(x_1,x_2;w_1,w_2)=(w_1\cdot\partial_1)(w_2\cdot\partial_2)\Delta_{m^2,0}^{(0)}(x_1,x_2)~,
\end{equation}
and its de Sitter counterpart is $(W_1\cdot\nabla_1)(W_2\cdot\nabla_2)G_{\lambda,0}(\sigma)$. Since $W\cdot\nabla=W\cdot \partial_Y$ reduces to $w\cdot\partial_x$ in the flat space limit, 
we have
\begin{equation}
    (W_1\cdot\nabla_1)(W_2\cdot\nabla_2)G_{\lambda,0}(\sigma)\approx \Delta_{m^2,0}^{(1)}(x_1,x_2;w_1,w_2)~.
\end{equation}

We report the flat space limit in the $J=2$ case without showing details. On the flat space side, all $\Delta^{(2)}_{m^2, \ell}$ with $0\le \ell\le 2$ are given by \cite{Karateev:2020axc}
\small
\begin{align}\label{flat2}
   & \Delta^{(2)}_{m^2, 0}(x_1, x_2; w_1,w_2)\!=\! \frac{d+1}{d}(w_1\cdot \partial_1)^2(w_2\cdot\partial_2)^2\Delta^{(0)}_{m^2,0}(x_1, x_2)\,\nonumber\\
   & \Delta^{(2)}_{m^2, 1}(x_1, x_2; w_1,w_2)\!=\!
    2((w_1\cdot\partial_1)^2(w_2\cdot\partial_2)^2+m^2(w_1\cdot w_2)(w_1\cdot\partial_1)(w_2\cdot\partial_2))\Delta^{(0)}_{m^2,0}(x_1,x_2)\\
     &\Delta^{(2)}_{m^2, 2}(x_1,x_2; w_1,w_2)\!=\!\left((m^2\, w_1\cdot w_2+ w_1\cdot\partial_1\,w_2\cdot\partial_2)^2-\frac{1}{d}(w_1\!\cdot\!\partial_1)^2(w_2\!\cdot\!\partial_2)^2\right)\Delta^{(0)}_{m^2,0}(x_1, x_2)~.
    \nonumber
\end{align}
\normalsize
and on the de Sitter side we find 
\begin{align}\label{flat3}
   & (W_1\cdot\nabla_1)^2 (W_2\cdot\nabla_2)^2 G_{\lambda, 0}(Y_1, Y_2)\approx \frac{d}{d+1}\Delta^{(2)}_{m^2, 0}(x_1, x_2; w_1,w_2)\nonumber\\
   & (W_1\cdot\nabla_1) (W_2\cdot\nabla_2) G_{\lambda, 1}(Y_1, Y_2; W_1, W_2) \approx \frac{1}{2\, m^2}\Delta^{(2)}_{m^2, 1}(x_1, x_2; w_1,w_2)\nonumber\\
    &G_{\lambda, 2}(Y_1, Y_2; W_1, W_2) \approx \frac{1}{m^4} \Delta^{(2)}_{m^2, 2}(x_1,x_2; w_1,w_2)~.
\end{align}

For generic $J$, the construction of $\Delta^{(J)}_{m^2, \ell}$ is much more involved since it requires $J+1$ projections operators $\Pi_{J\ell}$ that effectively implement the branching rule from the spin $J$ representation of $SO(d,1)$ to spin $\ell$ representations of $SO(d)$. The explicit expressions of $\Pi_{J\ell}$ have been solved in   \cite{Dobrev:1977qv} but they are very complicated. We will not give these expressions. Instead, we will discuss $\Delta^{(J)}_{m^2, \ell}$ when $d=1$. This is a very degenerate case, because all $\Pi_{J\ell}$ with $\ell\ge 2$ vanish. To find $\Pi_{J0}$ and $\Pi_{J1}$, let's pick a vector $p^\mu$ in $\mathbb R^{1, 1}$. $\Pi_{J0}$ should only have longitudinal components along $p^\mu$, which in the index free formalism means $\Pi_{J0}\propto p^{-2 J}(p\cdot w_1)^J(p\cdot w_2)^J$. The proportional constant can be fixed by using the fact that $\Pi_{J0}$ is a projector. Indeed, we find 
\begin{align}\label{PiJ0}
    \Pi_{J0}(p,w_1,w_2)=2^{J-1}\frac{(p\cdot w_1)^J(p\cdot w_2)^J}{p^{2J}}~.
\end{align}
By the completeness of $\Pi_{J0}$ and $\Pi_{J1}$, we know immediately that $\Pi_{J1}=(w_1\cdot w_2)^J-\Pi_{J0}$, with $\Pi_{J0}$ given by eq. (\ref{PiJ0}). In $\mathbb R^{1,1}$, there exists an interesting relation $(w_1\cdot w_2)(p^2\, w_1\cdot w_2- 2 p\cdot w_1\, p\cdot w_2)=0$, which allows us to rewrite $\Pi_{J1}$ as 
\begin{align}\label{PiJ1}
     \Pi_{J1}(p,w_1,w_2)=2^{J-1}\frac{(p\cdot w_1)^{J-1}(p\cdot w_2)^{J-1}}{p^{2J}}\left( p^2\, w_1\cdot w_2-p\cdot w_1\, p\cdot w_2\right)~.
\end{align}
With these projectors known, we can define $\Delta^{(J)}_{m^2, \ell}$ following  the general recipe given in \cite{Karateev:2020axc}:
\begin{align}
    \Delta^{(J)}_{m^2,0}&=(-)^J m^{2J} \Pi_{J0}\left(p^2\to-m^2, p_\alpha\to i\partial_{x_1^\alpha} \right)  \Delta^{(0)}_{m^2,0}(x_1 ,x_2)\nonumber\\
    &=2^{J-1}(w_1\cdot\partial_1)^J (w_2\cdot\partial_2)^J  \Delta^{(0)}_{m^2,0}(x_1 ,x_2)~,
\end{align}
and 
\begin{align}
    \Delta^{(J)}_{m^2,1}&=(-)^{J-1} m^{2J} \Pi_{J1}\left(p^2\to-m^2, p_\alpha\to i\partial_{x_1^\alpha} \right)  \Delta^{(0)}_{m^2,0}(x_1 ,x_2)\nonumber\\
    &=2^{J-1}(w_1\cdot\partial_1)^{J-1} (w_2\cdot\partial_2)^{J-1}\left( m^2\, w_1\cdot w_2+w_1\cdot\partial_1\, w_2\cdot \partial_2\right)  \Delta^{(0)}_{m^2,0}(x_1 ,x_2)~,
\end{align}
where the extra signs $(-)^J$ and $(-)^{J-1}$ have been explained in section \ref{subsec:flatspacelimit}. On the de Sitter side, we thus find 
 \begin{align}\label{flat4}
   & (W_1\cdot\nabla_1)^J (W_2\cdot\nabla_2)^J G_{\lambda, 0}(Y_1, Y_2)\approx 2^{1-J}\Delta^{(J)}_{m^2, 0}(x_1, x_2; w_1,w_2)~,\nonumber\\
   & (W_1\cdot\nabla_1)^{J-1} (W_2\cdot\nabla_2)^{J-1} G_{\lambda, 1}(Y_1, Y_2; W_1, W_2) \approx \frac{2^{1-J}}{ m^2}\Delta^{(J)}_{m^2, 1}(x_1, x_2; w_1,w_2)~.
\end{align}
For $d=1$ and $J=2$, eq. (\ref{flat3}) and eq. (\ref{flat4}) are consistent.

\section{Complementary series in the \KL\, decompositions}\label{sec:comp}
In this appendix, we show how  complementary series contributes to the two-point function of a scalar operator $\CO(Y)$. We will only focus on $\mathcal C_{\Delta, 0}$ since $\mathcal C_{\Delta, \ell}$ with $\ell>0$ cannot contribute to $\langle\Omega|\CO(Y_1)\CO(Y_2)|\Omega\rangle$. 

Compared to the principal series, the main difference is the resolution of the identity. For a principal series representation $\mathcal P_{\Delta, 0}$, the identity operator in its Hilbert space can be expressed as 
\begin{align}
   \mathbb 1_{\mathcal P_{\Delta, 0}}=\int d^d\vec y\, |\Delta, \vec y\,\rangle\langle \Delta, \vec y\, | = \int_P |\Delta, P\rangle \langle \Delta, P|~.
\end{align}
which follows from the inner product $\langle \Delta, \vec y_1|\Delta, \vec y_2\,\rangle =\delta^d(\vec y_1-\vec y_2\,)$. However, for complementary series, we are not allowed to choose such an inner product since it does not respect the reality condition of $\so(d+1,1)$ generators \cite{Sun:2021thf}. Instead, $\langle \Delta, \vec y_1|\Delta, \vec y_2\,\rangle$ has to be proportional to the CFT two-point function of a scalar primary of scaling dimension $\Delta$, i.e. $\langle \Delta, \vec y_1|\Delta, \vec y_2\,\rangle\propto |\vec y_1-\vec y_2\,|^{-2\Delta}$. From the embedding space point of view, it is obvious that $\langle \Delta, P_1|\Delta, P_2\rangle\propto (-2 P_1\cdot P_2)^{-\Delta}$ is the only choice that is compatible with the $SO(d+1,1)$ invariance and  the scaling property imposed on $|\Delta, P\rangle$, when $\Delta $ is real. Also because of $-2P_1\cdot P_2=|\vec y_1-\vec y_2|^2$, it is consistent with the expression in local coordinates.

We fix the normalization of inner products in $\mathcal C_{\Delta, 0}$ by choosing
\begin{align}\label{innernormal}
    \langle \Delta, P_1|\Delta, P_2\rangle=\frac{N_\Delta}{(-2 P_1\cdot P_2)^\Delta}, \,\,\,\,\, N_\Delta=\frac{\Gamma(\Delta)}{\pi^{\frac{d}{2}}\,\Gamma(\frac{d}{2}-\Delta)}~,
\end{align}
It fully determines the resolution of identity 
\begin{align}\label{resocom}
    \mathbb 1_{\mathcal C_{\Delta, 0}}=\int_{P_1, P_2}\, |\Delta, P_1\rangle \, \frac{N_{\bar\Delta}}{(-2 P_1\cdot  P_2)^{\bar\Delta}}\, \langle \Delta, P_2|~.
\end{align}
For example, it is straightforward to check  $\langle \Delta, P_1|\mathbb 1_{\mathcal C_{\Delta, 0}}|\Delta, P_2\rangle =\langle \Delta, P_1|\Delta, P_2\rangle$ by using 
\begin{align}
    \int_{P_1}\frac{1}{(-2 P_0\cdot P_1)^\Delta} \frac{1}{(-2 P_1\cdot P_2)^{\bar\Delta}}=\int\,d^d \vec y_1 \frac{1}{|\vec y_0-\vec y_1|^{2\Delta}} \frac{1}{|\vec y_1-\vec y_2|^{2\bar\Delta}} = \frac{\delta^d(\vec y_0, \vec y_2\,)}{N_\Delta N_{\bar\Delta}}~.
\end{align}
We insert $\mathbb 1_{\mathcal C_{\Delta, 0}}$ into the two-point function of $\CO(Y)$
\begin{align}
    \langle\Omega|\CO(Y_1)|\mathbb 1_{\mathcal C_{\Delta, 0}}|\CO(Y_2)|\Omega\rangle=\int_{P_3, P_4}\, \langle\Omega|\CO(Y_1)|\Delta, P_3\rangle \, \frac{N_{\bar\Delta}}{(-2 P_3\cdot  P_4)^{\bar\Delta}}\, \langle \Delta, P_4|\CO(Y_2)|\Omega\rangle~,
\end{align}
where $\langle 0 |\CO(Y)|\Delta, P\rangle = c_\CO(\Delta)\, \CK_\Delta (Y, P)$. Next, we write the remaining double integral in local coordinates and use the Fourier transformation
\begin{align}
    \frac{N_{\bar\Delta}}{x^{2\bar\Delta}} =  \int\frac{d^d \vec k}{(2\pi)^d}\left(\frac{k}{2}\right)^{d-2\Delta} e^{i\vec k\cdot \vec x}~,
\end{align}
which leads to 
\begin{align}
    \langle\Omega|\CO(\eta_1, \vec y_1)|\mathbb 1_{\mathcal C_{\Delta, 0}}|\CO(\eta_2, \vec y_2)|\Omega\rangle &=|c_\CO(\Delta)|^2\int\frac{d^d \vec k}{(2\pi)^d}\left(\frac{k}{2}\right)^{d-2\Delta} \nonumber\\
  &\times \int\,d^d\vec y_3\, \CK_\Delta(\eta_1,\vec y_1; \vec y_3)  e^{i\vec k\cdot \vec y_3} \int \, d^d\vec y_4\, \CK_\Delta(\eta_2, \vec y_2; \vec y_4) e^{-i\vec k\cdot \vec y_4}~.
\end{align}
For the integral over $\vec y_3$ and $\vec y_4$, we use eq. (\ref{CK1}) and eq. (\ref{CK2})  respectively. In the end, we obtain the mode expansion of the free Green's function $G_{-i(\Delta-\frac{1}{2})}$, c.f. eq. (\ref{phiWigh}):
\begin{align}
    \langle\Omega|\CO(\eta_1, \vec y_1)|\mathbb 1_{\mathcal C_{\Delta, 0}}|\CO(\eta_2, \vec y_2)|\Omega\rangle &=|c_\CO(\Delta)|^2G_{-i(\Delta-\frac{1}{2})}(\eta_1, \vec y_1; \eta_2, \vec y_2)~.
\end{align}
Therefore the complementary series part of the $\CO$ \KL\, decomposition takes the form 
\begin{align}
    \int_{-\frac{d}{2}}^{\frac{d}{2}}\, d\lambda\, \rho_\CO^{\mathcal C, 0}(\lambda)\, G_{i\lambda}(Y_1, Y_2),
\end{align}
where $\rho_\CO^{\mathcal C, 0}(\lambda)$ is a nonnegative function by construction. For spinning operators, there is a similar result.

\section{Discrete series in free scalar theory in dS$_2$}\label{nodis}
Let $\Phi(Y)$ be a free scalar field of scaling dimension $\Delta=\frac{1}{2}+i\nu$. We have argued that discrete series representations  cannot contribute to  the two-point function of $\Phi^2(Y)$  because it would always lead to antipodal singularity while $\langle\Omega|\Phi^2(Y_1)\Phi^2(Y_2)|\Omega\rangle$ is clearly free of such a singularity given that $|\Omega\rangle$ is the BD vacuum. On the other hand, according to the group theoretical analysis in \cite{Penedones:2023uqc}, the two-particle Hilbert space $\CH_2$ of $\Phi$ should contain all $\CD^\pm_k$ for $k=2,4,6,\cdots$. To reconcile this contradiction, we will explicitly compute the matrix elements of $\Phi^2$ between the vacuum $|\Omega\rangle$ and 
discrete series states in $\CH_2$. More precisely, we will focus on the lowest-weight state $|k\rangle_k$ in each $\CD_k^+$, because all $\langle\Omega|\Phi^2(Y)|\ell\rangle_k$ should vanish once $\langle\Omega|\Phi^2(Y)|k\rangle_k$ vanishes as a simple result of the $SO(2,1)$ symmetry. For $\CD_k^-$, the argument is exactly the same.

Let's first describe the single-particle states of $\Phi$ in conformally  global coordinates, using the following mode expansion in BD vacuum \cite{Sun:2021thf}: 
\begin{align}
    \Phi=\sum_{n\in\mathbb Z} \phi_n a_n+\phi_n^* a_n^\dagger, \,\,\,\,\, \phi_n=g_n(\tau)\frac{e^{-in \varphi}}{\sqrt{2\pi}}~,
\end{align}
where\footnote{We remind the reader that $\mathbf{F}$ is our notation for the regularized hypergeometric function.}
\begin{align}
    g_n(\tau)=\frac{\Gamma(n+\bar\Delta)}{\sqrt{2}}e^{-i n \tau}\mathbf {F}\left(\Delta, \bar\Delta, n+1, \frac{1}{1+e^{2i \tau}}\right) ~.
\end{align}
The canonically normalized single-particle states are $|n\rangle_\Delta\equiv a_n^\dagger|\Omega\rangle$, and the action of $\so(2,1)$ on these states is computed in \cite{Sun:2021thf}
\begin{align}
L_\pm |n\rangle_\Delta =(n\pm \Delta)|n\pm 1\rangle_\Delta,\,\,\,\,\, L_0|n\rangle_\Delta =n|n\rangle_\Delta ~.
\end{align}
The two-particle Hilbert space $\CH_2$ is spanned by $|n, m\rangle_\Delta\equiv a_n^\dagger a_m^\dagger |\Omega\rangle$, 
and the  lowest-weight state $|k\rangle_k$ of $\CD^+_k$ in $\CH_2$ can be written as \cite{Penedones:2023uqc}
\begin{align}
    |k\rangle_k= c \sum_{\ell\in\mathbb Z}\frac{\Gamma(\Delta+\ell-k)}{\Gamma(\bar\Delta+\ell)}|\ell, k-\ell\rangle_\Delta~.
\end{align}
where $c$ is some unimportant normalization constant. 
With all these ingredients, we can now compute the matrix element of $\Phi^2$ between $\langle\Omega|$ and $|k\rangle_k$:
\small
\begin{align}\label{cd1}
    \langle 0 | \Phi^2(\tau,\varphi)|k\rangle_k&=2c\sum_{\ell} \frac{\Gamma(\Delta+\ell-k)}{\Gamma(\bar\Delta+\ell)} \phi_\ell(\tau,
    \varphi) \phi_{k-\ell} (\tau,
    \varphi)\nonumber\\
    &=\frac{c\,e^{-ik(\tau+\varphi)}}{2\cosh(\pi\nu)}\sum_{\ell}(-)^\ell\mathbf {F}\left(\Delta, \bar\Delta, \ell+1, \frac{1}{1+e^{2i \tau}}\right)\mathbf {F}\left(\Delta, \bar\Delta, k-\ell+1, \frac{1}{1+e^{2i \tau}}\right)~.
\end{align}
\normalsize
To evaluate the sum over $\ell$ in eq. (\ref{cd1}), we use the series expansion of the regularized hypergeometric function
\begin{align}
     \langle 0 | \Phi^2(\tau,\varphi)|k\rangle_k
    &=\frac{c\,e^{-ik(\tau+\varphi)}}{2\cosh(\pi\nu)}\sum_{n, m\ge 0}\frac{(\frac{1}{2}\pm i\nu)_n(\frac{1}{2}\pm i\nu)_m}{n!\, m!}\, \mathcal T(n\!+\!1, m\!+\!k\!+\!1) \left(1\!+\!e^{2i\tau}\right)^{-n-m}~,
\end{align}
where 
\begin{align}
    \mathcal T(a, b)\equiv \sum_{\ell\in\mathbb Z}\frac{(-)^\ell}{\Gamma(a+\ell)\Gamma(b-\ell)}, \,\,\,\,\, a, b\in\mathbb Z~.
\end{align}
Since $a, b$ are integers, $\mathcal T(a, b)$ is actually a finite sum supported on $1-a\le \ell\le b-1$, which implies that $\mathcal T$ vanishes when $a+b<2$. When $a+b\ge 2$, the function $f(z)\equiv \frac{1}{\Gamma(a+z)\Gamma(b-z)}$  decays fast enough at large $|z|$ such that we can use the Sommerfeld-Watson transformation to claim that $-\mathcal T(a, b)$ is equivalent to summing over residues of  $\frac{\pi f(z)}{\sin (\pi z)}$ at the poles of $f(z)$. On the other hand, it is clear that $f(z)$ is an entire function, and hence $\mathcal T(a, b)$ should vanish when $a+b\ge 2$. Altogether, $\mathcal T(a, b)$ vanishes identically for any integer $a$ and $b$, implying that $\langle 0 | \Phi^2(\tau,\varphi)|k\rangle_k=0$. This simple computation shows explicitly why discrete  series states do not appear in the two-point function of $\Phi^2$. In other words, although the two-particle Hilbert space of $\Phi$ contains irreducible components that furnish discrete series representations, it is impossible to excite states with such symmetry by acting with $\Phi^2$ on the BD vacuum $|\Omega\rangle$. 

Instead, if we consider two $\Phi$ fields that are separated in spacetime, it can be  checked similarly that $\langle\Omega|\Phi(\tau_1, \varphi_1)\Phi(\tau_2, \varphi_2)|k\rangle_k$ does not vanish, which means that  discrete series can have nonzero contribution to the  four-point function of $\Phi$. This is consistent with the analysis of the four-point function of late time operators in \cite{Hogervorst:2021uvp}.

\section{Properties of $\phi^\pm_{\lambda, J}$ and $\psi_{p, J}$}\label{tech}
In this appendix, we give various details of the functions $\phi^\pm_{\lambda, J}$ and $\psi_{p, J}$, focusing on their inner products with respect to $(\,,\,)_J^\pm$. We list the definitions of these functions 
\begin{align}
   & \phi^+_{\lambda, J}(\sigma)=\partial_\sigma^J((\sigma+1)^J\partial_\sigma^J)G_{\lambda, 0}(\sigma)=\frac{\Gamma(\frac{1}{2}\pm i\lambda+J)}{2^{J+2}\pi}F\left(\frac{1}{2}+i\lambda+J,\frac{1}{2}-i\lambda+J,1,\frac{1+\sigma}{2}\right) ~,\nonumber\\
   & \phi^-_{\lambda, J}(\sigma)=\partial_\sigma^J((\sigma-1)^J\partial_\sigma^J)G_{\lambda, 0}(\sigma)=\frac{\Gamma(\frac{1}{2}\pm i\lambda+J)^2 F\left(\frac{1}{2}\!+\!i\lambda\!+\!J,\frac{1}{2}\!-\!i\lambda\!+\!J,2J+1,\frac{1+\sigma}{2}\right)}{(-2)^{J+2} (2J)!\pi \Gamma(\frac{1}{2}\pm i\lambda)}~, \nonumber\\
   & \psi_{p,J}(\sigma)=\frac{\Gamma(J+p)\Gamma(J+1-p)}{2^{J+2}\pi}F\left(J+p,J+1-p,1,\frac{1+\sigma}{2}\right) ~,
    \end{align}
and  the two inner products   $(\,,\,)_J^\pm$ ( for real functions defined on $(-\infty, -1)$ )
\begin{align}
(f, g)_J^\pm = \int_{-\infty}^{-1}d\sigma\, \left(1\mp \sigma\right)^{2J}\,f(\sigma) g(\sigma)\,.
\end{align}
$\phi^\pm_{\lambda, J}$ share the same large $-\sigma$ behavior at the leading order: 
\begin{align}\label{phiasym}
&\phi^\pm_{\lambda, J}(\sigma)\approx \frac{1}{(-2)^{J+2}\pi}\left(\frac{\Gamma(-2 i\lambda)\Gamma(\frac {1}{2}+i\lambda)(\frac{1}{2}+i\lambda)_J^2}{\Gamma(\frac {1}{2}-i\lambda)}\left(-\frac{1+\sigma}{2}\right)^{-(\frac{1}{2}+i\lambda+J)}+c.c\right).
\end{align}
The asymptotic behavior of $ \psi_{p,J}(\sigma)$ can be easily obtained by using  the relation 
\begin{align}\label{f21t}
 \psi_{p,J}(\sigma)=\left(\frac{1-\sigma}{2}\right)^{-2 J} F\left(1-J-p, p-J, 1, \frac{1+\sigma}{2}\right)
\end{align}
Noticing that $F\left(1-J-p, p-J, 1, \frac{1+\sigma}{2}\right)$ is a polynomial of degree $J-p$ in $\frac{1+\sigma}{2}$, we find that $ \psi_{p,J}(\sigma) $ decays as 
$(-\sigma)^{-J-p}$ for large $-\sigma$.

 For each fixed $J$, define a second order differential operator  $\mathfrak D^{(J)}_+\equiv(1-\sigma^2)\partial_\sigma^2+2(1-(J+1)(\sigma+1))\partial_\sigma$, which is hermitian with respect to the inner product $(\,,\,)^+_J$. It admits $\phi^+_{\lambda, J}$ and $\psi_{p, J}$ as eigenfunctions, i.e. 
 \begin{align}
 \mathfrak D^{(J)}_+ \phi^+_{\lambda, J}=\left[\left(\frac{1}{2}+J\right)^2+\lambda^2\right]\phi^+_{\lambda, J}, \,\,\,\,\, \mathfrak D^{(J)}_+ \psi_{p, J}=(J+p)(J+1-p)\psi_{p, J}~.
 \end{align}
which follows from the hypergeometric nature of these functions. $\{\phi^+_{\lambda, J}\}$  consist of the continuous spectrum of $\mathfrak D^{(J)}_+$. They are $\delta$ function normalizable and their inner product can be easily extracted from the asymptotic behavior (\ref{phiasym})
\begin{align}\label{phipnorm}
(\phi_{\lambda, J}^+, \phi_{\lambda', J}^+)_J^+= \frac{\left(\frac{1}{2}\pm i\lambda\right)_J^2}{8\lambda \sinh (2\pi\lambda)}\left(\delta(\lambda-\lambda')+\delta(\lambda+\lambda')\right)~.
\end{align}
Similarly, $\{\psi_{p, J}\}$ is an orthogonal basis of the discrete spectrum of $\mathfrak D^{(J)}_+$. We can compute their  norm $(\psi_{p,J}, \psi_{p,J})^+_J$ by using eq. (\ref{f21t}) for one of the two $\psi_{p, J}$:
\begin{align}\label{ppJ}
    (\psi_{p,J},\psi_{p,J})^+_J&=\mathfrak{N}\int_{-\infty}^{-1}d\sigma(1-\sigma)^{2J}F\left(1+J-p,J+p,1,\frac{1+\sigma}{2}\right)^2\nonumber\\
    &=4^{J}\mathfrak{N}\,\int_{-\infty}^{-1}d\sigma F\left(1+J-p,J+p,1,\frac{1+\sigma}{2}\right) F\left(1-J-p, p-J, 1, \frac{1+\sigma}{2}\right)\nonumber\\
    &=2^{2J+1}\mathfrak{N}\,\int^{\infty}_{0}d s F\left(1+J-p,J+p,1,-s\right) F\left(1-J-p, p-J, 1, -s\right)~,
\end{align}
where $s=-\frac{1+\sigma}{2}$, and 
\begin{equation}
    \mathfrak{N}=\frac{1}{4^{2+J}\pi^2}\Gamma(J+p)^2\Gamma(1+J-p)^2~.
\end{equation}
As we have mentioned above, $F\left(1-J-p, p-J, 1, -s\right)$ is a polynomial of degree $J-p$ in $s$. One crucial point is that only the leading term of this polynomial contributes to the integral (\ref{ppJ}), as a result of the Mellin transformation of hypergeometric functions
\begin{align}\label{MF}
\int_0^\infty dx\, x^{t-1} F(a,b,c,-x)=\frac{\Gamma(c)\Gamma(a-t)\Gamma(b-t)}{\Gamma(a)\Gamma(b)\Gamma(c-t)}\Gamma(t)~,
\end{align}
which holds when $\min(\Re(a), \Re(b))>\Re s> 0$.
More precisely, performing the $s$ integral against the monomial $s^m$ in $F\left(1-J-p,p-J,1,-s\right)$ would lead to $(-m)_{J-p}$, and hence the integral vanishes when $m<J-p$. The leading term of $F\left(1-J-p,p-J,1,-s\right)$ can be easily extracted by the series expansion of hypergeometric function
\begin{align}
F\left(1-J-p,p-J,1,-s\right) = \frac{\Gamma(J+p)}{\Gamma(2p)\Gamma(J-p+1)}(-s)^{J-p}+\cdots
\end{align}
Therefore, the integral (\ref{ppJ}) reduces to 
\begin{align}\label{PPJ2}
(\psi_{p,J},\psi_{p,J})^+_J=\frac{(-)^{J-p}2^{2J+1}\Gamma(J+p)}{\Gamma(2p)\Gamma(J-p+1)}\mathfrak{N}\,\int_0^\infty\, ds F\left(1+J-p,J+p,1,-s\right)s^{J-p}~, 
\end{align}
which can be evaluated as the analytical continuation of (\ref{MF})
\begin{align}
\int_0^\infty\, ds F\left(1+J-p,J+p,1,-s\right)s^{J-p}&=\lim_{t\to J-p+1} \frac{\Gamma(J+p-t)\Gamma(J-p+1-t)}{\Gamma(J+k)\Gamma(J-p+1)\Gamma(1-t)}\Gamma(t)\nonumber\\
&=(-)^{J-p}\frac{\Gamma(2p-1)\Gamma(J-p+1)}{\Gamma(J+p)}~.
\end{align} 
Altogether, we obtain the norm of $\psi_{p, J}$ with respect to $(\,,\,)^+_J$
\begin{align}\label{psinorm}
(\psi_{p,J},\psi_{p,J})^+_J=\frac{2^{2J+1}}{2p-1}\mathfrak N=\frac{\Gamma(J+p)^2\Gamma(1+J-p)^2}{8\pi^2(2p-1)}~.
\end{align}

For $\phi^-_{\lambda, J}$, it is easy to see that they are eigenfunctions of the differential operator $\mathfrak D^{(J)}_-\equiv(1-\sigma^2)\partial_\sigma^2+2((2J+1)-(J+1)(\sigma+1))\partial_\sigma$, which is hermitian with respect to $(\, ,\, )_J^-$:
\begin{align}
\mathfrak D^{(J)}_- \phi^-_{\lambda, J}=\left[\left(\frac{1}{2}+J\right)^2+\lambda^2\right]\phi^-_{\lambda, J}~.
\end{align}
They are $\delta$ function normalizable. Because of eq. (\ref{phiasym}), their normalization should be the same as $\phi^+_{\lambda, J}$
\begin{align}\label{phimnorm}
(\phi_{\lambda, J}^-, \phi_{\lambda', J}^-)_J^-= \frac{\left(\frac{1}{2}\pm i\lambda\right)_J^2}{8\lambda \sinh (2\pi\lambda)}\left(\delta(\lambda-\lambda')+\delta(\lambda+\lambda')\right)~.
\end{align}
Unlike $\mathfrak D^{(J)}_+$, $\mathfrak D^{(J)}_-$ does not have a discrete spectrum.

\section{Discrete series $\CD_p$ in the two-point function of $\CO^{(J)}$ when $p<J$ }\label{p<J}
In this appendix, we prove that including $\CD_p$  in the two-point function of $\CO^{(J)}$ in dS$_2$ will lead to antipodal singularities when  $p\ge J$. More precisely, we are considering 
\begin{align}
    \CG^{(J)}_p(Y_1,Y_2; W_1, W_2) \equiv \langle\Omega|\CO^{(J)}(Y_1,W_1)|\mathbb 1_{\CD_p}|\CO^{(J)}(Y_2,W_2)|\Omega\rangle
\end{align}
which has two independent chiral components
\begin{align}
    \CG^{(J)}_p(Y_1,Y_2; W^+_1, W^\pm_2)= (W^+_1\cdot W_2^\pm)^J\,\CG^{(J, \pm)}_p(\sigma), \,\,\,\,\, \sigma =Y_1\cdot Y_2~.
\end{align}
After a short computation, one can show that \footnote{A similar result has been obtained in AdS \cite{sleight2017lectures}.}
\begin{align}
    \nabla_1^2  \CG^{(J)}_p(Y_1,Y_2; W_1, W_2) = \left(C_2^{SO(2,1 )} + J^2\right) \, \CG^{(J)}_p(Y_1,Y_2; W_1, W_2)~,
\end{align}
where $\nabla_1^2=\nabla_{1 A}\nabla_1^A$, with $\nabla_{1A}=\partial_{Y_1^A}-Y_{1A}\,Y_1\cdot \partial_{Y_1}-W_{1A}( Y_1\cdot\partial_{W_1})$ given by eq. (\ref{nablaA}). Due to the $SO(2,1)$ invariance of $\langle\Omega|$, the Casimir $C_2^{SO(2,1 )}$ effectively acts on the projection operator $\mathbb 1_{\CD_p}$ and yields $p(1-p)$. Therefore, $ \CG^{(J)}_p(Y_1,Y_2; W_1, W_2)$ is an eigenmode of $\nabla_1^2$ with eigenvalue $p(1-p)+J^2$. Using the explicit expression of $\nabla^2_1$, we can further obtain a second order differential equation for $\CG^{(J, \pm)}_p(\sigma)$
\begin{align}\label{p>JODE}
  (1-\sigma^2)\partial_\sigma^2 \CG^{(J, \pm)}_p- 2\left((1+J)\sigma\pm J \right)\partial_\sigma \CG^{(J, \pm)}_p(\sigma) = (1+J-p)(J+p) \CG^{(J, \pm)}_p(\sigma) 
\end{align}
For each fixed sign, i.e. chiral component, the ODE (\ref{p>JODE}) has two linearly independent solutions, one of which has growing behavior for large $-\sigma$. Such solutions are clearly unphysical. The other decaying solution is given by 
\begin{align}
    \CG^{(J, \pm )}_p(\sigma) = \left(\frac{2}{1-\sigma}\right)^{J+p} F\left(p\mp J, p+J, 2 p, \frac{2}{1-\sigma}\right)
\end{align}
They have power law decay $(-\sigma)^{-J-p}$ for two points with large spacelike separation.  However, they are not regular when the two points are antipodal. Their leading singular behaviors around $\sigma=-1$ are
\begin{align}
   & \CG^{(J, + )}_p(\sigma) \stackrel{\sigma\to -1}{\approx} -\frac{\Gamma(2p)}{\Gamma(p-J)\Gamma(p+J)}\log \left(-\frac{\sigma+1}{2}\right)~,\nonumber\\
   &  \CG^{(J, -)}_p(\sigma) \stackrel{\sigma\to -1}{\approx} -\frac{\Gamma(2p)\Gamma(2J)}{\Gamma(p-J)\Gamma(p+J)}\left(\frac{2}{1+\sigma}\right)^{2J}~.
\end{align}
Therefore, the discrete series $\CD_p$ should not contribute to $\langle\Omega|\CO^{(J)}(Y_1, W_1)\CO^{(J)}(Y_2, W_2)|\Omega\rangle$ when $p>J$.

\section{Harmonic Analysis in EAdS}
\label{sec:harmonicanalysis}
In this appendix, we review some facts about Euclidean Anti de Sitter. We mostly follow the notations in \cite{Costa_2014}.
\subsection{Coordinates in Euclidean Anti de Sitter}
\label{subsec:coordinates}
$(d+1)$-dimensional Euclidean Anti de Sitter spacetime can be defined as a set of points embedded in Minkowski space $\mathbb{R}^{d+1,1}$:
\be\label{eq:def EAdS}
\text{EAdS}_{d+1}: -({X^0})^2  \,+\,  ({X^1})^2  \,+\,  \ldots \,+\, ({X^{d+1}})^2= -R^2~,
\ee
which defines two disconnected hypersurfaces. In our convention, we pick EAdS to be the one with $X^0>0$. This definition makes it manifest that EAdS is invariant under $SO(d+1,1)$ rotation and boosts and hence its $(d+1)(d+2)/2$ generators satisfy the commutation relations of $SO(d+1,1)$ algebra in~\reef{confalg}. Let us present two coordinate systems that are useful for us in this paper. One is the \textit{global coordinate} system that is given by
\be
X^0 = R \cosh r \cosh \zeta~, \quad X^i= R \sinh r\, \Omega^i~, \quad X^{d+1} = R \cosh r  \sinh\zeta~, 
\label{eq:globalAdScoords}
\ee
where $i=1,\ldots,d$, Euclidean time $\zeta\in\mathbb{R}$, $r\in\mathbb{R}^+$ \footnote{This is true for $ d\geq 2$. In $d=1$ the range is instead $r\in \Real$.} and $\Omega^i \in S^{d-1} \subset \mathbb{R}^{d}$ is a unit vector ($\Omega^i \Omega_i = 1$).  
This coordinate system leads to the induced metric of
\be
ds^2 = R^2(\cosh^2r \, \mathrm{d}\zeta^2 + \mathrm{d}r^2 + \sinh^2r\, \mathrm{d}\Omega^2_{d-1})~
\ee
Other useful coordinates are \textit{Poincar\'e coordinates}, defined as
\be
X^0 = R \frac{z^2+\vec x^2+1}{2z}~, \quad X^{i}= R \frac{x^i}{z}~, \quad X^{d+1} = R \frac{z^2+\vec x^2-1}{2z}~,
\ee
with metric
\be
ds^2= R^2 \frac{dz^2+d\vec{x}^2}{z^2}
\ee
where $\vec x \in \mathbb{R}^d$ with $i=1,\cdots,d$ make a flat $d$-dimensional Euclidean spatial slice and $z>0$ to satisfy the $X^0>0$ condition. 

Let us also define the two-point invariants $\sigma$ in both dS and EAdS as
\be
\sigma^\text{dS}= \frac{Y_1\cdot Y_2}{R^2}~, \qquad \sigma^{\text{EAdS}}= \frac{X_1\cdot X_2}{R^2}~
\label{eq:sigmadsAdS}
\ee
which for instance in planar coordinates and Poincar\'e coordinates are given by respectively
\be\label{eq:sigma in Poincare}
\sigma^\text{dS}= \frac{\eta_1^2+\eta^2_2-|\vec y_{12}|^2}{2\eta_1 \eta_2}~, \qquad \sigma^{\text{EAdS}}= -\frac{z_1^2+z^2_2+|\vec x_{12}|^2}{2z_1 z_2}~.
\ee
This shows that the range of the two-point function invariants are $\sigma^\text{dS}\in \Real$ and $\sigma^\text{EAdS} \in (-\infty,-1)$.
In the main text, we drop the superscript of $\sigma^\text{dS}$ as we are focusing on de Sitter spacetime. The Wick rotation in~(\ref{subsec:wick}) transforms $\sigma^\text{dS} \to \sigma^\text{EAdS}$. 

Similarly to de Sitter we use the index-free notation to represent the traceless symmetric tensors in EAdS by contracting the spin $J$ tensor indicies with $J$ auxiliary vectors: $W^A$. They satisfy tangential and null conditions in EAdS:
\be
W\cdot X = W^2 = 0~.
\ee
In embedding space, the lightcone in $\mathbb{R}^{d+1,1}$ is the boundary of both EAdS and of dS. We thus use the same symbol $P$ to indicate null rays:
\be
P^2 = 0~, \quad P\sim\alpha P
\ee
for $\alpha \in \Real$. In Poincar\'e coordinates, this corresponds to the $z= 0$ plane where the EAdS generators reduce to generators of a $d$-dimensional conformal theory on a Eudclidean plane spanned by $\vec x$.

\subsection{Harmonic functions}
Here we summarize some facts about Euclidean AdS harmonic functions in embedding space, following the notation of \cite{Costa_2014}, e.g. taking $R=1$. Harmonic functions are defined as the regular divergence-free eigenfunctions of the Laplacian operator in EAdS:
\ba
\left(\nabla^2_1+\frac{d^2}{4}+\lambda^2+\ell\right) \, \, \Omega_{\lambda,\ell}(X_1,X_2;W_1,W_2) &= 0\\
\nabla_1\cdot K_1 \, \, \Omega_{\lambda,\ell}(X_1,X_2;W_1,W_2)&=0~.
\label{eq:integralX}
\ea
They are proportional to a difference of two EAdS propagators $\Pi_{\Delta,J}(X_1,X_2;W_1,W_2)$ with an overall factor
\be
 \Omega_{\lambda,\ell}(X_1,X_2;W_1,W_2)=\frac{i\lambda}{2\pi} \left(\Pi_{\hd+i\lambda,\ell}(X_1,X_2;W_1,W_2) - \Pi_{\hd-i\lambda,\ell}(X_1,X_2;W_1,W_2) \right)
 \label{eq:omegafrompis}
\ee
One can explicitly check from the short distance limit of the bulk-to-bulk propagators that the harmonic functions are regular at coincident points. They also satisfy the orthogonality relation 
\be\label{eq:Omega ortho}
\begin{aligned}
\frac{1}{\ell! (\frac{d-1}{2})_\ell} \int_X  \Omega_{\lambda,\ell}(X_1,X;W_1,K) & \Omega_{\lambda',\ell}(X,X_2;W,W_2) \\
&= \half \left[\delta(\lambda-\lambda')+\delta(\lambda+\lambda')\right]\Omega_{\lambda,\ell}(X_1,X_2;W_1,W_2)
\end{aligned}
\ee
where we introduce the short hand notation for integrating over EAdS defined as
\begin{equation}
\int_{X}\equiv\int d^{d+2}X \,\delta(X^2+1)\theta(X^0)\,.
\label{eq:intX}
\end{equation}
where the term $\theta(X^0)$ encodes that we picked the upper hyperboloid ($X^0>0$) in our definition for EAdS. Harmonic functions can equivalently be defined as a product of bulk-to-boundary propagators integrated over the boundary point
\be
\Omega_{\lambda,\ell}(X_1,X_2;W_1,W_2) = \frac{\lambda^2}{\pi \ell! (\hd-1)_\ell} \int_P\,\, \Pi_{\hd+i\lambda,\ell}(X_1,P;W_1,D_Z) \Pi_{\hd-i\lambda,\ell}(P,X_2;Z,W_2)\,.
\label{eq:defsplitrep}
\ee
We refer to this as the split representation. Bulk-to-boundary propagators are defined and normalized through
\be
\begin{aligned}
\Pi_{\D,\ell}(X,P;W,Z) &= \mathfrak{C}_{\D,\ell} \frac{\left((-2P\cdot X) (W\cdot Z)+ 2(W\cdot P)(Z\cdot X)\right)^\ell}{\left(-2P\cdot X\right)^{\D+\ell}}\,,\\
\mathfrak{C}_{\Delta,\ell}&=\frac{(\ell+\Delta-1)\Gamma(\Delta)}{2\pi^{\frac{d}{2}}(\Delta-1)\Gamma(\Delta+1-\frac{d}{2})}\,,
\label{eq:defbulktoboundaryEAdS}
\end{aligned}
\ee
and the action of the differential operator $D_Z$ that contracts the boundary indices is defined in (\ref{eq:DZoperator}). In EAdS, the action of the $K$ operators which contract the bulk indices is similar to the one in dS (\ref{eq:actionofKdS}), up to some signs which are necessary to keep its action internal to the manifold defined by $X^2=-1$,
\begin{equation}
\begin{aligned}
    K_A =& \frac{d-1}{2}\left[\frac{\partial}{\partial W^A} +X_A\left(X\cdot\frac{\partial}{\partial W}\right)\right]+ \left(W\cdot\frac{\partial}{\partial W}\right) \frac{\partial}{\partial W^A}  \\
& +X_A \left(X\cdot\frac{\partial}{\partial W}\right) \left(W\cdot\frac{\partial}{\partial W}\right) - \half \,W_A \left[\frac{\partial^2}{\partial W\cdot\partial W } +\left(X\cdot\frac{\partial}{\partial W}\right) \left(X\cdot\frac{\partial}{\partial W}\right) \right]\,.
\label{eq:actionofKAdS}
\end{aligned}
\end{equation}
The action on $W$ vectors is analogous to (\ref{eq:KWWW}):
\begin{equation}
    \frac{1}{J!(\frac{d-1}{2})_J}K_{A_1}\cdots K_{A_J}W^{B_1}\cdots W^{B_J}=\frac{1}{J!}\sum_\pi G_{A_{\pi_1}}^{\ B_1}\cdots G_{A_{\pi_J}}^{\ B_J}-\text{traces}\,, 
    \label{eq:KWWWAdS}
\end{equation}
with
\begin{equation}
    \qquad G_{AB}=\eta_{AB}+X_AX_B\,.
\end{equation}
The following commutator will be useful for our computations
\begin{equation}
\begin{aligned}
        [K\cdot\nabla,(W\cdot\nabla)^n]    =&\frac{n}{2}(W\cdot\nabla)^{n-1}(d+n+2\mathcal{D}_W-2)\\
        &\times(1-n-(n+\mathcal{D}_W-1)(d+n+\mathcal{D}_W-2)+\nabla^2)\,.
        \label{eq:commuteKW}
\end{aligned}
\end{equation}
\normalsize
where $\CD_W=W\cdot \partial_W$.
Importantly, $\nabla\cdot K=K\cdot\nabla$. 

From the embedding space point of view, the harmonic functions in EAdS and the bulk-to-bulk propagators in dS satisfy the same  Laplacian equation up to the Wick rotation discussed in section \ref{subsec:wick}. They are also divergence-free and regular at $\sigma\to -1$. This is another way to see that they have to be proportional to each other as mentioned in~\ref{eq:G to Omega}.

\subsection{Explicit form of harmonic functions up to $J=2$}\label{sec:omegas}
We have reviewed the definition of  the harmonic functions and how they can be expressed through a split representation. But to carry out numerical checks of the \KL\ decomposition, it is useful to have their explicit expressions. Consider the fact that, for $SO(d+1,1)$ invariance, any two-point function in the index free formalism has to organize itself in terms of polynomials of $W_1\cdot W_2$ and $(W_1\cdot X_2)(W_2\cdot X_1)$ as follows
\begin{equation}
    G_{\mathcal{O}^{(J)}}(X_1,X_2;W_1,W_2)=\sum_{n=0}^J(W_1\cdot W_2)^{J-n}((W_1\cdot X_2)(W_2\cdot X_1))^{n}\mathcal{G}_{\mathcal{O}^{(J)}}^{(n)}(\sigma)\,,
    \label{eq:twopointstructures}
\end{equation}
with $\mathcal{G}_{\mathcal{O}^{(J)}}^{(n)}(\sigma)$ being scalar functions of $\sigma=X_1\cdot X_2\,.$
This is true also for derivatives of harmonic functions
\begin{equation}
    ((W_1\cdot\nabla_1)(W_2\cdot\nabla_2))^{J-\ell}\Omega_{\lambda,\ell}(X_1,X_2;W_1,W_2)=\sum_{n=0}^J(W_1\cdot W_2)^{J-n}((W_1\cdot X_2)(W_2\cdot X_1))^{n}h_{\ell,n}^{(J)}(\sigma)\,,
    \label{eq:harmonicstructures}
\end{equation}
for some scalar functions $h_{\ell,n}^{(J)}(\sigma)$.
If we plug (\ref{eq:twopointstructures}) and (\ref{eq:harmonicstructures}) into the formula for the \KL\ decomposition analytically continued to EAdS (\ref{eq:KLharmonic}), we can match the coefficients of the tensor structures on both sides and obtain some scalar equations
\begin{equation}
    \mathcal{G}^{(n)}_{\mathcal{O}^{(J)}}(\sigma)=\sum_{\ell=0}^{J}\int_{\mathbb{R}}d\lambda\ \Gamma(\pm i\lambda)\rho_{\mathcal{O}^{(J)}}^{\mathcal{P},\ell}(\lambda)h_{\ell,n}^{(J)}(\sigma)\,, 
    \label{eq:KLNumeric}
\end{equation}
where we only wrote the principal series part for simplicity. These are the integrals which we carry out numerically and with which we check the validity of the various spectral densities we have derived. Notice that under the Wick rotation described in \ref{subsec:wick}, $\sigma^{\text{dS}}\to\sigma^{\text{EAdS}}$ so that (\ref{eq:KLNumeric}) are valid both in EAdS ($\sigma\in(-\infty,-1)$) and in de Sitter ($\sigma\in(-\infty,\infty)$) provided one gives a small imaginary part to $\sigma$ when going above the cut at timelike separation.
For example, consider the case of a spin $2$ CFT primary of conformal weight $\mathbf{\Delta}$. As argued in section \ref{subsubsec:cft2}, we have $\mathcal{G}^{(2)}_T(\sigma)=c_T(1-\sigma)^{-\mathbf{\Delta}}$. (\ref{eq:KLNumeric}) then reads
\begin{equation}
    c_T(1-\sigma)^{-\mathbf{\Delta}}=\int_{\mathbb{R}}d\lambda\ \Gamma(\pm i\lambda)\left(\rho_T^{\mathcal{P},0}(\lambda)h^{(2)}_{0,0}(\sigma)+\rho_T^{\mathcal{P},1}(\lambda)h^{(2)}_{1,0}(\sigma)+\rho_T^{\mathcal{P},2}(\lambda)h^{(2)}_{2,0}(\sigma)\right)\,,
\end{equation}
with the spectral densities in (\ref{eq:rhoCFT2}), and we choose $d\geq 2$.
In the rest of this subsection, we report the explicit expression of the $h_{\ell,n}^{(J)}(\sigma)$ 
functions for $J=0,1,2$ and $n,\ell\in[0,J]$, so that all these kinds of integrals can be checked to hold by numerical evaluation. For $J=0$, we have the well known expression of the scalar harmonic function
\begin{equation}
    \Omega_{\lambda,0}(X_1,X_2)=h^{(0)}_{0,0}(\sigma)=\frac{\Gamma(\frac{d}{2}\pm i\lambda)}{2^{d+1}\pi^{\frac{d+1}{2}}\Gamma(\pm i\lambda)}\mathbf{F}\left(\frac{d}{2}-i\lambda,\frac{d}{2}+i\lambda,\frac{d+1}{2},\frac{1+\sigma}{2}\right)
    \label{eq:scalarharmonic}
\end{equation}
Let us explicitly show how to compute $h^{(1)}_{1,n}$ for $n=0$ and $n=1$. The other functions are obtained with the same mechanical steps. We start by considering the relevant harmonic function and its split representation 
\begin{equation}
    \Omega_{\lambda,1}(X_1,X_2;W_1,W_2)=\frac{\lambda^2}{\pi(\frac{d-2}{2})}\int_P\Pi_{\Delta,1}(X_1,P;W_1,D_Z)\Pi_{\bar\Delta,1}(X_2,P;W_2,Z)\,,
    \label{eq:spin1split}
\end{equation}
where the apparent $d=2$ pole is actually canceled by the action of $D_Z$ on $Z$, we are using the notation $\Delta=\frac{d}{2}+i\lambda$ in this appendix and bulk-to-boundary propagators are defined in (\ref{eq:defbulktoboundaryEAdS}). Contracting the boundary indices we obtain
\small
    \begin{equation}
    \begin{aligned}
\Omega_{\lambda,1}(X_1,X_2;W_1,W_2)=&\frac{\lambda^2\mathfrak{C}_{\Delta,1}\mathfrak{C}_{\bar\Delta,1}}{\pi}\int_P\Bigg[\frac{W_1\cdot W_2}{(-2P\cdot X_1)^{\Delta}(-2P\cdot X_2)^{\bar\Delta}}+\frac{(P\cdot W_2)(W_1\cdot X_2)}{(-2P\cdot X_1)^{\Delta}(-2P\cdot X_2)^{\bar\Delta+1}}\\
    &\qquad\quad+\frac{(P\cdot W_1)(W_2\cdot X_1)}{(-2P\cdot X_1)^{\Delta+1}(-2P\cdot X_2)^{\bar\Delta}}+\frac{(P\cdot W_1)(P\cdot W_2)(X_1\cdot X_2)}{(-2P\cdot X_1)^{\Delta+1}(-2P\cdot X_2)^{\bar\Delta+1}}\Bigg]\,.
\end{aligned}
\end{equation}
\normalsize
We can trade factors of $P$ for derivatives with respect to $X_1$ and $X_2$ and obtain
\small
\begin{equation}
    \begin{aligned}
    \Omega_{\lambda,1}(X_1,X_2;W_1,W_2)=&\frac{\mathfrak{C}_{\Delta,1}\mathfrak{C}_{\bar\Delta,1}}{\Delta\bar\Delta}\Bigg(\Delta\bar\Delta W_1\cdot W_2+\frac{\Delta}{2}(W_1\cdot X_2)(W_2\cdot\partial_{X_2})+\frac{\bar\Delta}{2}(W_2\cdot X_1)(W_1\cdot\partial_{X_1})\\
    &\qquad+\frac{1}{4}(X_1\cdot X_2)(W_1\cdot\partial_{X_1})(W_2\cdot\partial_{X_2})\Bigg)\frac{\lambda^2}{\pi}\int_P\frac{1}{(-2P\cdot X_1)^{\Delta}(-2P\cdot X_2)^{\bar\Delta}}\,.
\end{aligned}
\end{equation}
\normalsize
We can now undo the split representation such that this becomes just a sum of derivatives of the scalar harmonic function 
\small
\begin{equation}
    \begin{aligned}
    \Omega_{\lambda,1}(X_1,X_2;W_1,W_2)&=\frac{\mathfrak{C}_{\Delta,1}\mathfrak{C}_{\bar\Delta,1}}{\Delta\bar\Delta\mathfrak{C}_{\Delta,0}\mathfrak{C}_{\bar\Delta,0}}\Bigg(\Delta\bar\Delta W_1\cdot W_2+\frac{\Delta}{2}(W_1\cdot X_2)(W_2\cdot\partial_{X_2})\\
    &+\frac{\bar\Delta}{2}(W_2\cdot X_1)(W_1\cdot\partial_{X_1})+\frac{1}{4}(X_1\cdot X_2)(W_1\cdot\partial_{X_1})(W_2\cdot\partial_{X_2})\Bigg)\Omega_{\lambda,0}(X_1,X_2)\,.
\end{aligned}
\end{equation}
\normalsize
By carrying out these derivatives explicitly, and collecting the coefficients of $W_1\cdot W_2$ and $(X_1\cdot W_2)(X_2\cdot W_1)$ we can read off
\begin{equation}
    \begin{aligned}
        h^{(1)}_{1,0}(\sigma)&=\mathcal{N}_1^{(1)}\left[2\mathbf{F}^{(0)}(\sigma)+\sigma\mathbf{F}^{(1)}(\sigma)\right]\,,\\
        h^{(1)}_{1,1}(\sigma)&=\mathcal{N}_1^{(1)}\Bigg[d\mathbf{F}^{(1)}(\sigma)+\frac{\sigma}{8}((d+2)^2+4\lambda^2)\mathbf{F}^{(2)}(\sigma)\Bigg]\,,
    \end{aligned}
\end{equation}
where we introduced the shorthand notation
\begin{equation}
    \mathbf{F}^{(a)}(\sigma)\equiv\frac{1}{\Gamma(\frac{d+1}{2}+a)}\ F\left(\frac{d}{2}+i\lambda+a,\frac{d}{2}-i\lambda+a,\frac{d+1}{2}+a,\frac{\sigma+1}{2}\right)\,.
    \label{eq:regularized2F1}
\end{equation}
and defined
\be
\mathcal{N}_1^{(1)}=\frac{\Gamma(\frac{d}{2}\pm i\lambda+1)\lambda\sinh(\pi\lambda)}{2^d\pi^{\frac{d+3}{2}}((d-2)^2+4\lambda^2)}
\ee
We report here the other functions for $J=1$, which can be computed with the same procedure
\begin{equation}
    \begin{aligned}
        h^{(1)}_{0,0}(\sigma)&=\mathcal{N}_0^{(1)}\left(\frac{(d-2)^2}{4}+\lambda^2\right)\mathbf{F}^{(1)}(\sigma)\,,\\
        h^{(1)}_{0,1}(\sigma)&=\frac{1}{2}\mathcal{N}_0^{(1)}\left(\frac{(d-2)^2}{4}+\lambda^2\right)\left(\frac{(d+2)^2}{4}+\lambda^2\right)\mathbf{F}^{(2)}(\sigma)\,.
    \end{aligned}
\end{equation}
with $\mathcal{N}_0^{(1)} = \mathcal{N}_1^{(1)} $.
In the rest of this subsection we report the functions for $J=2$. We start from the functions $h^{(2)}_{\ell,n}$ with $\ell=2$.
\small
\begin{equation}
    \begin{aligned}
        \frac{h^{(2)}_{2,0}(\sigma)}{\mathcal{N}^{(2)}_{2}}=&2d\left(\mathbf{F}^{(0)}(\sigma)+\sigma\mathbf{F}^{(1)}(\sigma)\right)+(d\sigma^2-1)\mathbf{F}^{(2)}(\sigma)\\
        \frac{h^{(2)}_{2,1}(\sigma)}{\mathcal{N}^{(2)}_{2}}=&2(d)_2\mathbf{F}^{(1)}(\sigma)+d\sigma\left(5+3d+\frac{d^2}{4}+\lambda^2\right)\mathbf{F}^{(2)}(\sigma)+(d\sigma^2-1)\left(\frac{d}{2}\pm i\lambda+2\right)\mathbf{F}^{(3)}(\sigma)\\
        \frac{h^{(2)}_{2,2}(\sigma)}{\mathcal{N}^{(2)}_{2}}=&\frac{(d)_3}{2}\mathbf{F}^{(2)}(\sigma)+\left(\frac{d}{2}\pm i\lambda+2\right)\Bigg[\frac{d(d+2)\sigma}{2}\mathbf{F}^{(3)}(\sigma)+\frac{(d\sigma^2-1)}{8}\left(\frac{d}{2}\pm i\lambda+3\right)\mathbf{F}^{(4)}(\sigma)\Bigg]
    \end{aligned}
\end{equation}
\normalsize
with
\begin{equation}
    \mathcal{N}^{(2)}_{2}= \frac{(d+2)^2+4\lambda^2}{d(d^2+4\lambda^2)}\mathcal{N}^{(1)}_{1}\,,
\end{equation}
and $(d)_n$ is the Pochhammer symbol. For $J=2$ and $\ell=1$, instead, we have
\begin{equation}
    \begin{aligned}
        \frac{h^{(2)}_{1,0}}{\mathcal{N}^{(2)}_1}=&\mathbf{F}^{(1)}(\sigma)+\sigma\mathbf{F}^{(2)}(\sigma)\,,\\
        \frac{h^{(2)}_{1,1}}{\mathcal{N}^{(2)}_1}=&\frac{1}{8}\Bigg[(d(d+12)+4(\lambda^2+5))\mathbf{F}^{(2)}(\sigma)+2\sigma((d+4)^2+4\lambda^2)\mathbf{F}^{(3)}(\sigma)\Bigg]\,,\\
        \frac{h^{(2)}_{1,2}}{\mathcal{N}^{(2)}_1}=&\frac{(d+4)^2+4\lambda^2}{128}\Bigg[8(d+2)\mathbf{F}^{(3)}(\sigma)+\sigma((d+6)^2+4\lambda^2)\mathbf{F}^{(4)}(\sigma)\Bigg]\,,
    \end{aligned}
\end{equation}
with 
\begin{equation}
    \mathcal{N}^{(2)}_1=\frac{\Gamma(\frac{d}{2}\pm i\lambda+2)\lambda\sinh(\pi\lambda)}{2^{d}\pi^{\frac{d+3}{2}}((d-2)^2+4\lambda^2)}
\end{equation}
and for $J=2$ with $\ell=0$
\begin{equation}
    \begin{aligned}
        h^{(2)}_{0,0}=&\frac{\Gamma(\frac{d}{2}\pm i\lambda+2)\lambda\sinh(\pi\lambda)}{2^{d+2}\pi^{\frac{d+3}{2}}}\mathbf{F}^{(2)}(\sigma)\,, \quad h^{(2)}_{0,1}=\frac{\Gamma(\frac{d}{2}\pm i\lambda+3)\lambda\sinh(\pi\lambda)}{2^{d+2}\pi^{\frac{d+3}{2}}}\mathbf{F}^{(3)}(\sigma)\,,\\
        &\qquad\qquad\qquad\qquad h^{(2)}_{0,2}=\frac{\Gamma(\frac{d}{2}\pm i\lambda+4)\lambda\sinh(\pi\lambda)}{2^{d+5}\pi^{\frac{d+3}{2}}}\mathbf{F}^{(4)}(\sigma)\,.
    \end{aligned}
\end{equation}
\label{subsec:hfunc}

\section{Explicit expressions of the inversion formula}
\label{sec:explicitinversion}
\subsection{Spin 0}
The \Kallen decomposition of scalar two-point functions only has the $\ell=0$ term. The inversion formula~\reef{eq:inversionformulal} then takes the simple form
\be\label{eq:rho of J0}
\rho^{\mathcal{P},0}_{\mathcal{O}^{(0)}}(\lambda) = \frac{2^{d+1}\pi^{\frac{d+1}{2}}\Gamma(\frac{d+1}{2})}{\Gamma(\frac{d}{2}\pm i\lambda)} \int_{X_1}\Omega_{\lambda,0}(X_2,X_1) G_{\mathcal{O}^{(0)}}(X_1,X_2)~.
\ee
The scalar two-point function $G_{\mathcal{O}^{(0)}}(Y_1,Y_2)$ and its Wick rotation to EAdS $G_{\mathcal{O}^{(0)}}(X_1,X_2)$ only depend on the two-point invariants $\sigma^{\text{dS}}\equiv Y_1\cdot Y_2$ and $\sigma^{\text{EAdS}}\equiv X_1\cdot X_2$ which we discuss more in detail in Appendix \ref{sec:harmonicanalysis}.
We can thus use $G_{\mathcal{O}^{(0)}}(\sigma)$ as a short-hand notation for the scalar two-point function of some bulk scalar operator $\mathcal{O}^{(0)}$.
Since $\rho^{\mathcal{P},0}_{\mathcal{O}^{(0)}}(\lambda)$ does not depend on $X_2$, we are free to place it anywhere in EAdS and in particular we can pick the origin, which makes the angular part of integral over $X_1$ trivial. We choose global coordinates in EAdS, given by~\reef{eq:globalAdScoords}, for which we have
\be 
\sigma^\text{EAdS} = -\cosh r~, \qquad \int_{X_1} = \text{vol}(S^d) \int_0^\infty dr\, (\sinh r)^d 
\ee
where we performed the integration over the angular coordinates, which leads to a factor of $\text{vol}(S^d) = \frac{2\pi^{\frac{d+1}{2}}}{\Gamma(\frac{d+1}{2})}$. With the change of variable $r \to \sigma^\text{EAdS}$ and replacing the explicit value of $\Omega_{\lambda,0}$ from eq.~\reef{eq:scalarharmonic}, 
\be\label{eq:rhoAdSApp}
\rho^{\mathcal{P},0}_{\mathcal{O}^{(0)}}(\lambda) = \frac{2\pi^\frac{d+1}{2}}{\Gamma(\pm i\lambda)} \int_{-\infty}^{-1} d\sigma \, (\sigma^2-1)^\frac{d-1}{2} \, \bFF{\hd+i\lambda}{\hd-i\lambda}{\frac{d+1}{2}}{\frac{1+\sigma}{2}} \, G_{\mathcal{O}^{(0)}}(\sigma)~.
\ee
Here, we omit the label EAdS in $\sigma$, not only to avoid clutter but also to emphasise that this formula can be seen as an integration over the point-function in de Sitter as well.
\subsection{Spin 1}
We state the generic form of a spinning two-point function in~\reef{eq:gen GG}. In particular for spin-1 fields we have:
\be
G_{\CO^{(1)}} (Y_1,Y_2;W_1,W_2)=  (W_1\cdot W_2) \mathcal{G}_{\mathcal{O}^{(1)}}^{(0)}(\sigma) +  (W_1\cdot Y_2) (W_2\cdot Y_1) \mathcal{G}_{\mathcal{O}^{(1)}}^{(1)}(\sigma)\,,
\ee
where again we used shorthand notation for the two-point invariant $\sigma = Y_1\cdot Y_2$. After Wick rotating this two-point function as discussed in \ref{subsec:wick}, we plug it into~\reef{eq:inversionformulal} and we carry out all the index contractions through the application of the $K$ operators on the $W$ vectors (\ref{eq:actionofKAdS}). Using the explicit expressions for $\Omega_{\lambda,1}$ and $\Omega_{\lambda,0}$ from appendix~\ref{sec:omegas}, we then place $X_2$ at the origin of EAdS as done in the scalar case and using elementary hypergeometric identities we end up obtaining
\small
\begin{align}
    \rho^{\CP,1}_{\CO^{(1)}} (\lambda)&= \pi^\frac{d-1}{2}\lambda\sinh(\pi\lambda)\int_{-\infty}^{-1} d\sigma\, (\sigma^2-1)^{\frac{d-1}{2}}\ \left(2\mathbf{F}^{(0)}(\sigma)\mathcal{G}_{\mathcal{O}^{(1)}}^{(0)}(\sigma)- (\sigma^2-1)\mathbf{F}^{(1)}(\sigma)\mathcal{G}_{\mathcal{O}^{(1)}}^{(1)}(\sigma)\right)~, \nonumber\\
    \rho^{\CP,0}_{\CO^{(1)}} (\lambda)&=\frac{32\pi^{\frac{d-1}{2}}\lambda\sinh(\pi\lambda)}{\left(d^2+4\lambda^2\right)}\int_{-\infty}^{-1} d\sigma\, (\sigma^2-1)^{\frac{d-1}{2}}\ \mathbf{F}^{(0)}(\sigma)\,f(\sigma,\mathcal{G}_{\mathcal{O}^{(1)}}^{(0)}(\sigma),\mathcal{G}_{\mathcal{O}^{(1)}}^{(1)}(\sigma))~,\label{eq:inversion spin 1 explicit}
\end{align}
\normalsize
where
\small
\begin{align}
f(\sigma,\mathcal{G}_{\mathcal{O}^{(1)}}^{(0)}(\sigma),\mathcal{G}_{\mathcal{O}^{(1)}}^{(1)}(\sigma)) =& \left[\left(d+1\right)^2 \sigma + \left((2d+3)\sigma^2-(d+2)\right)\partial_\sigma +\sigma(\sigma^2-1)\partial^2_\sigma\right] \mathcal{G}_{\mathcal{O}^{(1)}}^{(0)}(\sigma)\\
&+ \left[(d+2)\left((d+2)\sigma^2\!-\!1\right)+(2d+5)\sigma(\sigma^2\!-\!1)\partial_\sigma+(\sigma^2\!-\!1)^2\partial^2_\sigma\right]\mathcal{G}_{\mathcal{O}^{(1)}}^{(1)}(\sigma)~.\nonumber
\end{align}
\normalsize
We stress the fact that these integrals can now be interpreted as being carried out in a physical region of spacelike separation in de Sitter.

\section{Inversion integrals}
\label{sec:masterintegrals}
In this appendix, we show how to carry out the integrals that are encountered when applying the inversion formula to the examples we have considered throughout this work. We will start with some general remarks about these integrals and then show all the specific examples.
\subsection{Free QFTs}
\label{subsec:masterfreeqft}
In the main text, we have considered two-point functions of composite operators made of two fundamental fields with $\Delta_1=\frac{d}{2}+i\lambda_1$ and spin $m$ and $\Delta_2=\frac{d}{2}+i\lambda_2$ and spin $J-m$. In total generality, in a free theory these two-point functions factorize
\small
\begin{equation}
 \begin{aligned}
    \langle\phi_1^{(m)}\phi_2^{(J-m)}(Y_1;W_2)\phi_1^{(m)}&\phi_2^{(J-m)}(Y_2;W_2)\rangle\\
    &=\langle\phi_1^{(m)}(Y_1;W_2)\phi_1^{(m)}(Y_2;W_2)\rangle\langle\phi_2^{(J-m)}(Y_1;W_2)\phi_2^{(J-m)}(Y_2;W_2)\rangle\,.
\end{aligned}
\end{equation}
\normalsize
In every case we studied, after carrying out derivatives and index contractions, we can reduce the inversion formulas for the spectral densities to linear combinations of integrals of the following form
\begin{equation}
\begin{aligned}
    \rho^{\mathcal{P},\ell}_{\phi_1^{(m)}\phi_2^{(J-m)}}(\lambda)&=\sum_{k_1,k_2,k_3}c_{k_1,k_2,k_3}\int_{P_1,P_2,P_3}\frac{\prod_{j=1}^3\Pi_{\bar\Delta_j-k_j,0}(X_2,P_j)}{{(P_{12})^{\Delta_{123,k}}(P_{13})^{\Delta_{132,k}}(P_{23})^{\Delta_{231,k}}}}\\
    &\equiv\sum_{k_1,k_2,k_3}c_{k_1,k_2,k_3}\mathcal{I}^{\text{QFT}}_{k_1,k_2,k_3}(\lambda_1,\lambda_2)\,,
   \label{eq:masterQFTintegraldef}
\end{aligned}
\end{equation}
where
\begin{equation}
\begin{aligned}
    \Delta_{ijl,k}&\equiv\frac{\Delta_i+k_i+\Delta_j+k_j-\Delta_l-k_l}{2}\,,\qquad
    &&P_{ij}\equiv-2P_i\cdot P_j\,, \\
    \Delta_3&\equiv\frac{d}{2}+i\lambda_3\,, \qquad &&\lambda_3\equiv\lambda\,,
    \label{eq:definitionofa}
\end{aligned}
\end{equation}
and $c_{k_1,k_2,k_3}$ are some coefficients which we determined case by case, and which we will show how to find in the following subsections. For now, let us focus on the integral $ \mathcal{I}_{0,0,0}^{\rm QFT}$. Other 
$ \mathcal{I}_{k_1,k_2,k_3}^{\rm QFT}$ can be easily obtained by making the shift $\Delta_i\to\Delta_i+k_i$ in $ \mathcal{I}_{0,0,0}^{\rm QFT}$. The seed integral $ \mathcal{I}_{0,0,0}^{\rm QFT}$
was first solved in \cite{Bros:2009bz} by a brute force computation in local coordinates of the boundary. It was also computed in \cite{Penedones_2011} by using multiple Schwinger parameterizations. We report here a more covariant approach in the more modern language of harmonic analysis, which exploits the underlying representation structure. Since we focus on the $k_i=0$ case, it is convenient to use the simplified notation $\Delta_{ijl}=\Delta_{ijl, 0}=\frac{\Delta_i+\Delta_j-\Delta_l}{2}$. Following the conventions of \cite{Costa_2014,Karateev_2019}, we also define
\begin{align}
\langle\CO_{\Delta_1}(P_1)\CO_{\Delta_2}(P_2)\CO_{\Delta_3}(P_3)\rangle =  \frac{1}{(P_{12})^{\Delta_{123}}(P_{13})^{\Delta_{132}}(P_{23})^{\Delta_{231}}}~,
\end{align}
which represents the standard CFT three-point function of scalar primaries.
Let us first consider the $P_1$ and $P_2$ integrals in $ \mathcal{I}_{0,0,0}^{\rm QFT}$, namely
\begin{align}
    \CJ(P_3)\equiv \int_{P_1, P_2} \langle\CO_{\Delta_1}(P_1)\CO_{\Delta_2}(P_2)\CO_{\Delta_3}(P_3)\rangle\Pi_{\bar\Delta_1}(X_2, P_1)\Pi_{\bar\Delta_2}(X_2, P_2)~.
\end{align}
$\CJ(P_3)$ defined in this way is a scalar function of $P_3$ and $X_2$, and is homogeneous in $P_3$ of degree $-\Delta_3$. So it has to take the form \begin{align}\label{CJreduce}
 \CJ(X_2,P_3)= c (\Delta_1,\Delta_2,\Delta_3)\, \Pi_{\Delta_3}(X_2,P_3)~,
\end{align}
where $c (\Delta_1,\Delta_2,\Delta_3)$ is a constant to be fixed. To extract this constant, we integrate $\CJ(X_2,P_3)$ against another bulk-to-boundary propagator $\Pi_{\bar\Delta_4}(X_2, P_4)$
\begin{align}\label{tCJ}
\widetilde \CJ(P_3, P_4)&\equiv \int_{X_2}\, \CJ (X_2, P_3) \Pi_{\bar\Delta_4}(X_2, P_4),\,\,\,\,\, \Delta_{4}=\frac{d}{2}+i\lambda_4 ~.
\end{align}

\begin{figure}
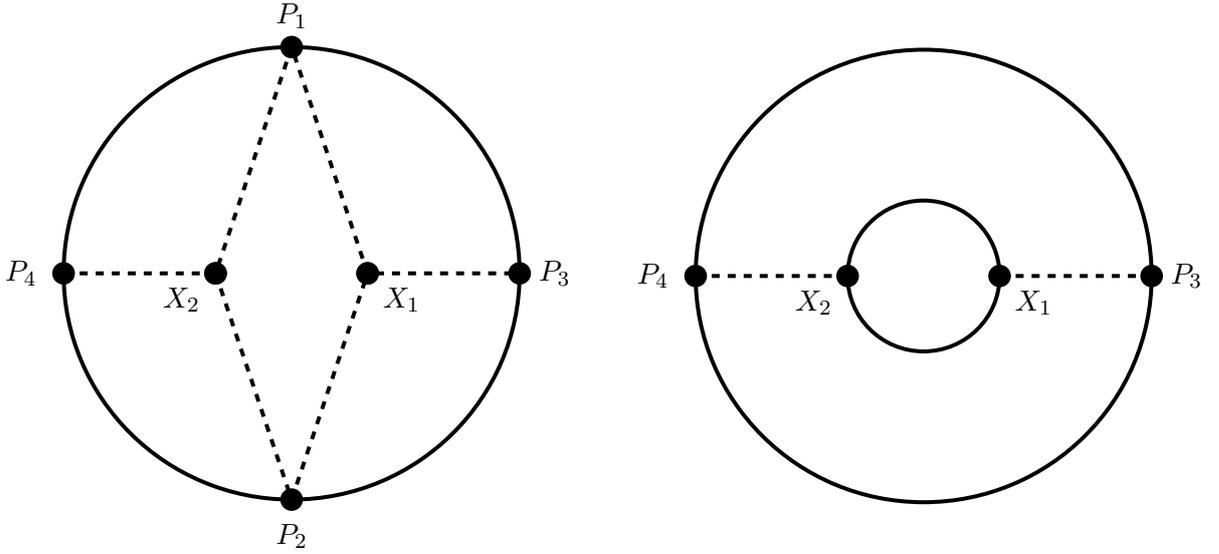

  \centering
  \begin{subfigure}[height=2cm]{0.45\textwidth}
    \centering
    \begin{wittendiagram}
      \draw[dashed,line width=1.5pt] (-3,0) node[vertex]{} coordinate[label= {[shift={(-0.2,0)}]left:$P_4$}]  --
      (-1,0) node[vertex]{} coordinate[label= {[shift={(-0.05,-0.35)}]left:$X_2$}] --
      (0,3) node[vertex]{} coordinate[label= {[shift={(0,0.1)}]above:$P_1$}]  --
      (1,0) node[vertex]{} coordinate[label= {[shift={(0.05,-0.35)}]right:$X_1$}]  --
      (3,0) node[vertex]{} coordinate[label= {[shift={(0.1,0)}]right:$P_3$}] ;
     \draw[dashed,line width=1.5pt]  (-1,0) node[vertex]-- (0,-3) node[vertex]{} coordinate[label= {[shift={(0,-0.8)}]above:$P_2$}]  --(1,0) node[vertex];
    \end{wittendiagram}
    \label{fig:wittendiagram2}
  \end{subfigure}
  \hfill
  \begin{subfigure}[height=2cm]{0.45\textwidth}
    \centering
    \begin{wittendiagram}
    \draw[line width=1.5pt] (0,0) circle (1cm);;
    \draw[dashed,line width=1.5pt] (-3,0) node[vertex]{} coordinate[label= {[shift={(-0.2,0)}]left:$P_4$}]  --
      (-1,0) node[vertex]{} coordinate[label= {[shift={(-0.05,-0.35)}]left:$X_2$}];
     \draw[dashed,line width=1.5pt] (1,0) node[vertex]{} coordinate[label= {[shift={(0.05,-0.35)}]right:$X_1$}]  --
      (3,0) node[vertex]{} coordinate[label= {[shift={(0.1,0)}]right:$P_3$}] ;
    \end{wittendiagram}
    \label{fig:wittendiagram1}
  \end{subfigure}
  \caption{ Witten diagrams representing $\widetilde \CJ(P_3, P_4)$. The dashed lines are bulk-to-boundary propagators $\Pi_{\Delta, 0}$, and the solid internal lines are AdS harmonics $\Omega_{\lambda, 0}$.}
  \label{fig:wittendiagrams}
\end{figure}

\noindent{}If we treat the CFT three-point function $\langle\CO_{\Delta_1}(P_1)\CO_{\Delta_2}(P_2)\CO_{\Delta_3}(P_3)\rangle$ in $\CJ(P_3)$ as a bulk integral of three bulk-to-boundary propagators  $\prod_{j=1}^3 \Pi_{\Delta_j}(X_1, P_j) $, then $\widetilde \CJ(P_3,P_4)$ can be diagrammatically represented by the left panel of fig. \ref{fig:wittendiagrams}, with integrations over $(X_1, X_2, P_1, P_2)$ understood. Performing the  $P_1$ and $P_2$ integrals first would yield a one-loop Witten diagram \footnote{Here  ``Witten diagram'' is an abuse of terminology since the internal lines are harmonic functions instead of AdS Green's functions.} as shown in the right panel of fig. \ref{fig:wittendiagrams}. Such a Witten diagram was considered in \cite{Giombi_2018}. Here, we are effectively considering a different order, namely integrating out the bulk points $X_1$ and $X_2$ first. In  eq.(\ref{tCJ}), the bulk integral over $X_2$ gives another CFT three-point function 
$\langle\CO_{\bar\Delta_1}(P_1)\CO_{\bar\Delta_2}(P_2)\CO_{\bar\Delta_4}(P_4)\rangle$, multiplied by a constant \cite{Costa_2014}
\small
\begin{align}\label{b0}
    b(\bar\Delta_1,\bar\Delta_2,\bar\Delta_4,0)=
    \frac{\pi^{\frac{d}{2}}\Gamma\left(\frac{\bar\Delta_1\!+\!\bar\Delta_2\!+\!\bar\Delta_4\!-\!d}{2}\right)\Gamma\left(\frac{\bar\Delta_1\!+\!\bar\Delta_2\!-\!\bar\Delta_4}{2}\right)\Gamma\left(\frac{\bar\Delta_1\!+\!\bar\Delta_4\!-\!\bar\Delta_2}{2}\right)\Gamma\left(\frac{\bar\Delta_2\!+\!\bar\Delta_4\!-\!\bar\Delta_1}{2}\right)}{2\Gamma(\bar\Delta_1)\Gamma(\bar\Delta_2)\Gamma(\bar\Delta_4)}\mathfrak C_{\bar\Delta_1, 0} \mathfrak C_{\bar\Delta_2, 0} \mathfrak C_{\bar\Delta_4, 0}
\end{align}
\normalsize
and hence  $\widetilde \CJ(P_3, P_4)$ reduces to 
\small
\begin{align}
    \widetilde \CJ(P_3, P_4)=b(\bar\Delta_1,\bar\Delta_2,\bar\Delta_4,0)
    \int_{P_1, P_2}\langle\CO_{\Delta_1}(P_1)\CO_{\Delta_2}(P_2)\CO_{\Delta_3}(P_3)\rangle\langle\CO_{\bar\Delta_1}(P_1)\CO_{\bar\Delta_2}(P_2)\CO_{\bar\Delta_4}(P_4)\rangle~.
    \end{align}
    \normalsize
This integral (and also its higher spin generalization) was originally computed by \cite{Dobrev:1977qv}, and is recently reviewed by \cite{Karateev_2019}. Without loss of generality, we assume $\lambda_3, \lambda_4>0$, and then get 
\begin{align}\label{CJ1}
    \widetilde \CJ(P_3, P_4)=\frac{2\pi^{\frac{d}{2}}}{\Gamma(\frac{d}{2})}\frac{2\pi^{d+1}\Gamma(\pm i\lambda_3)}{\Gamma(\Delta_3)\Gamma(\bar\Delta_3)}b(\bar\Delta_1,\bar\Delta_2,\bar\Delta_3,0)\delta(\lambda_3-\lambda_4)\delta(P_3,P_4)~.
\end{align}
On the other hand, combining eq. (\ref{CJreduce}) and eq. (\ref{tCJ}) yields
\begin{align}\label{CJ2}
    \widetilde \CJ(P_3, P_4)
=c (\Delta_1,\Delta_2,\Delta_3)\, \int_{X_2}\Pi_{\Delta_3}(X_2,P_3)\Pi_{\bar\Delta_4}(X_2,P_4)\end{align}
where the bulk integral over $X_2$ has been carried out in \cite{Costa_2014}:
\begin{align}
    \int_{X_2}\Pi_{\Delta_3}(X_2,P_3)\Pi_{\bar\Delta_4}(X_2,P_4)=\mathfrak C_{\Delta_3} \mathfrak C_{\bar\Delta_4}\frac{2\pi^{d+1}\Gamma(\pm i\lambda_3)}{\Gamma(\Delta_3)\Gamma(\bar\Delta_3)}\delta(\lambda_3-\lambda_4)\delta(P_3,P_4)~.
\end{align}
Comparing eq. (\ref{CJ1}) and eq. (\ref{CJ2}), we obtain 
\begin{align}
    c (\Delta_1,\Delta_2,\Delta_3)=\frac{2\pi^{\frac{d}{2}}}{\Gamma(\frac{d}{2})}\frac{b(\bar\Delta_1,\bar\Delta_2,\bar\Delta_3,0)}{\mathfrak C_{\Delta_3} \mathfrak C_{\bar\Delta_3}} ~.
\end{align}
Once the constant $c (\Delta_1,\Delta_2,\Delta_3)$ is fixed, the remaining integral over $P_3$ in $\mathcal{I}_{0,0,0}^{\text{QFT}}$ is of the form
\begin{align}\label{IY}
\int_P \frac{1}{(-2 P\cdot Y)^d}=\frac{\pi^{\frac{d}{2}}\Gamma(\frac{d}{2})}{\Gamma(d)}\frac{1}{(-Y^2)^{\frac{d}{2}}}~.
\end{align}
Altogether, the final expression of $\mathcal{I}_{0,0,0}^{\text{QFT}}$ is 
\begin{align}\label{000}
    \mathcal{I}_{0,0,0}^{\text{QFT}}
&=\frac{2\pi^{d}}{\Gamma(d)}b(\bar\Delta_1,\bar\Delta_2,\bar\Delta_3,0)=\frac{\Gamma(\frac{d}{2}-\Delta_{123})\Gamma(\frac{d}{2}-\Delta_{132})\Gamma(\frac{d}{2}-\Delta_{231})\Gamma(d-\sum_{j=1}^3\frac{\Delta_j}{2})}{8\Gamma(d)\Gamma(1-i\lambda_1)\Gamma(1-i\lambda_2)\Gamma(1-i\lambda_3)}~.
\end{align}
For arbitrary $ \mathcal{I}_{k_1,k_2,k_3}^{\text{QFT}}$, it suffices to make the substitution $\Delta_i\to\Delta_i+k_i$ in eq. (\ref{000})
\begin{equation}
    \mathcal{I}_{k_1,k_2,k_3}^{\text{QFT}}=\frac{\Gamma(\frac{d}{2}-\Delta_{123,k})\Gamma(\frac{d}{2}-\Delta_{132,k})\Gamma(\frac{d}{2}-\Delta_{231,k})\Gamma(d-\sum_{j=1}^3\frac{\Delta_j+k_j}{2})}{8\Gamma(d)\Gamma(1-i\lambda_1-k_1)\Gamma(1-i\lambda_2-k_2)\Gamma(1-i\lambda_3-k_3)}\,.
    \label{eq:freeQFTintegral}
\end{equation}

\subsubsection{Scalar Free QFT}
The first nontrivial case we have explored in section \ref{sec:applications} is the two-point function of the composite operator $\phi_1\phi_2$ in a free theory. We showed that (eq. (\ref{eq:phi1phi2step}))
\begin{equation}
    \rho^{\mathcal{P}}_{\phi_1\phi_2}(\lambda)=\frac{\Gamma(\pm i\lambda_1)\Gamma(\pm i\lambda_2)}{\mathcal{N}_{0,0}}b(\Delta_1,\Delta_2,\Delta,0)\frac{\lambda_1^2\lambda_2^2\lambda^2}{\pi^3}\mathcal{I}^{\text{QFT}}_{0,0,0}(\lambda_1,\lambda_2)\,,
\end{equation} 
Applying our integral identity to this case is thus immediate. After simplifying, we obtain 
\begin{equation}
    \rho^{\mathcal{P}}_{\phi_1\phi_2}(\lambda)=\frac{\lambda\sinh(\pi\lambda)\Gamma(\frac{d+1}{2})}{2^{6-d}\pi^{\frac{d+7}{2}}\Gamma(d)\Gamma(\frac{d}{2}\pm i\lambda)}\prod_{\pm,\pm,\pm}\Gamma\left(\frac{\frac{d}{2}\pm i\lambda\pm i\lambda_1\pm i\lambda_2}{2}\right)\,.
\end{equation}
\subsubsection{Spin 1 Free QFT}
We studied two spin 1 correlators of composite operators in free theory. For the operator $V\phi$ made of a vector and a scalar, the principal series contributions are given by 
\small
\begin{align}
       \rho_{V\phi}^{\mathcal{P},0}(\lambda)&=\frac{\Gamma(\pm i\lambda_\phi)\Gamma(\pm i\lambda_V)}{\mathcal{N}_{1,0}}\int_{X_1}\Omega_{\lambda,0}(X_2,X_1)(K_1\!\cdot\!\nabla_1)(K_2\!\cdot\!\nabla_2)\Omega_{\lambda_V,1}(X_1,\!X_2;W_1,\!W_2)\Omega_{\lambda_\phi,0}(X_1,X_2)\,,\\
        \rho_{V\phi}^{\mathcal{P},1}(\lambda)&=\frac{\Gamma(\pm i\lambda_\phi)\Gamma(\pm i\lambda_V)}{\mathcal{N}_{1,1}}\int_{X_1}\Omega_{\lambda,1}(X_2,X_1;K_2,K_1)\Omega_{\lambda_V,1}(X_1,X_2;W_1,W_2)\Omega_{\lambda_\phi,0}(X_1,X_2)\,.\label{eq:rhoVphi1harmonics}
    \end{align}
\normalsize
Let us start from $\rho_{V\phi}^{\mathcal{P},0}$. We begin by using the split representation for the harmonic functions and carrying out the integral over $X_1$. This kind of integrals is solved in \cite{Costa_2014}, eq. (126) there
\small
\begin{equation}
\begin{aligned}
    \frac{1}{J!\left(\frac{d-1}{2}\right)_J}\int_X\Pi_{\Delta_2}(X,P_1)&\Pi_{\Delta,J}(X,P_2;K,Z)(W\cdot\nabla)^J\Pi_{\Delta_1}(X,P_3)\\
    &=b(\Delta_1,\Delta_2,\Delta,J)\frac{((Z\cdot P_3)P_{12}-(Z\cdot P_1)P_{23})^J}{P_{13}^{\frac{\Delta_1+\Delta_2-\Delta+J}{2}}P_{23}^{\frac{\Delta_1-\Delta_2+\Delta+J}{2}}P_{12}^{\frac{-\Delta_1+\Delta_2+\Delta+J}{2}}}\,,
\end{aligned}
\end{equation}
\normalsize
with 
\small
\begin{equation}
    b(\Delta_1,\Delta_2,\Delta,J)=\frac{\Gamma\left(\frac{\Delta_1\!+\!\Delta_2\!+\!\Delta\!-\!d+J}{2}\right)\Gamma\left(\frac{\Delta_1\!+\!\Delta_2\!-\!\Delta+J}{2}\right)\Gamma\left(\frac{\Delta_1\!+\!\Delta\!-\!\Delta_2+J}{2}\right)\Gamma\left(\frac{\Delta_2\!+\!\Delta\!-\!\Delta_1+J}{2}\right)}{2^{1-J}\pi^{-\frac{d}{2}}\Gamma(\Delta_1)\Gamma(\Delta_2)\Gamma(\Delta+J)}\mathfrak C_{\Delta_1, 0} \mathfrak C_{\Delta_2, 0} \mathfrak C_{\Delta, J}
\end{equation}
\normalsize
being the generalization of (\ref{b0}). Carrying out all derivatives and index contractions, we can write
\small
\begin{equation}
\begin{aligned}
    \rho_{V\phi}^{\mathcal{P},0}(\lambda)=\widetilde{\mathcal{N}}^{V\phi}_{0}\int_{P_1,P_2,P_3}\frac{P_{13}(-2P_2\cdot X)+P_{12}(-2P_3\cdot X)-P_{23}(-2P_1\cdot X)}{(-2P_1\cdot X)^{\bar\Delta_\phi}(-2P_2\cdot X)^{\bar\Delta_V+1}(-2P_3\cdot X)^{\bar\Delta+1}(P_{12})^{\alpha}(P_{13})^\beta(P_{23})^\gamma}\,,
    \label{eq:rho0Vphistep}
\end{aligned}
\end{equation}
\normalsize
with
\small
\begin{equation}
\begin{aligned}
    \alpha=\frac{\Delta_V+\Delta_\phi-\Delta+1}{2}\,, \qquad
    \beta=\frac{\Delta+\Delta_\phi-\Delta_V+1}{2}\,, \qquad
    \gamma=\frac{\Delta+\Delta_V-\Delta_\phi-1}{2}\,.
\end{aligned}
\end{equation}
\normalsize
and
\begin{equation}
    \widetilde{\mathcal{N}}^{V\phi}_{0}=\frac{\lambda^2\lambda_\phi^2\lambda_V^2\Gamma(\pm i\lambda_V)\Gamma(\pm i\lambda_\phi)}{4\pi^3\mathcal{N}_{1,0}}(d-1)^2\bar\Delta_\lambda\mathfrak{C}_{\bar\Delta_\lambda,0}\mathfrak{C}_{\bar\Delta_V,1}\mathfrak{C}_{\bar\Delta_\phi,0}b(\Delta,\Delta_\phi,\Delta_V,1)
\end{equation}
The three terms in the sum in (\ref{eq:rho0Vphistep}) are exactly of the form (\ref{eq:masterQFTintegraldef}), so that we can write
\begin{equation}
\begin{aligned}
    \rho_{V\phi}^{\mathcal{P},0}(\lambda)&=\widetilde{\mathcal{N}}^{V\phi}_{0}\left(\mathcal{I}^{\text{QFT}}_{-1,0,0}(\lambda_V,\lambda_\phi)+\mathcal{I}^{\text{QFT}}_{0,0,-1}(\lambda_V,\lambda_\phi)-\mathcal{I}^{\text{QFT}}_{-1,1,-1}(\lambda_V,\lambda_\phi)\right)\\
    &=\frac{\pi^{-3-\frac{d}{2}}\lambda\sinh(\pi\lambda)}{2(\Delta_V-1)(\bar\Delta_V-1)(d^2+4\lambda^2)\Gamma(\frac{d}{2})\Gamma(\frac{d}{2}\pm i\lambda+1)}\prod_{\pm,\pm,\pm}\Gamma\left(\frac{\frac{d}{2}+1\pm i\lambda\pm i\lambda_V\pm i\lambda_\phi}{2}\right)\,.
\end{aligned}
\end{equation}
Then, we continue with $\rho_{V\phi}^{\mathcal{P},1}$. After using the split representation on the harmonic functions in (\ref{eq:rhoVphi1harmonics}), let us focus on the resulting $X_1$ integral
\small
\begin{equation}
    \begin{aligned}
        &\int_{X_1}\Pi_{\Delta_\phi}(X_1,P_1)\Pi_{\Delta,1}(X_1,P_2;K_1,Z_2)\Pi_{\Delta_V,1}(X_1,P_3;W_1,Z_3)\\
    &\propto\int_{X_1}\frac{\left((-2P_2\cdot X_1)(Z_2\cdot K_1)+2(X_1\cdot Z_2)(P_2\cdot K_1)\right)\left((-2P_3\cdot X_1)(W_1\cdot Z_3)+2(X_1\cdot Z_3)(P_3\cdot W_1)\right)}{(-2X_1\cdot P_1)^{\Delta_\phi}(-2X_1\cdot P_2)^{\Delta+1}(-2X_1\cdot P_3)^{\Delta_V+1}}\,.
\end{aligned}
\end{equation}
\normalsize
We can trade all factors of $X_1$ in the numerator for derivatives with respect to boundary points. In this way, the $X_1$ integral becomes an integral over a product of three scalar bulk-to-boundary propagators, again leading to a CFT three point function
\begin{equation}
    \begin{aligned}
    &\int_{X_1}\Pi_{\Delta_\phi}(X_1,P_1)\Pi_{\Delta,1}(X_1,P_2;K_1,Z_2)\Pi_{\Delta_V,1}(X_1,P_3;W_1,Z_3)\\
    &=\frac{1}{\Delta\Delta_V}\left((P_2\cdot P_3)Z_2\cdot\partial_{P_2}Z_3\cdot\partial_{P_3}+\Delta_VZ_3\cdot P_2Z_2\cdot\partial_{P_2}+\Delta Z_2\cdot P_3Z_3\cdot \partial_{P_3}+\Delta\Delta_V Z_2\cdot Z_3\right)\\
    &\qquad\qquad\times\frac{b(\Delta_\phi,\Delta,\Delta_V,0)}{(P_{12})^{\Delta_{123}}(P_{13})^{\Delta_{132}}(P_{23})^{\Delta_{231}}}\,.
    \label{eq:rho1tradingders}
    \end{aligned}
\end{equation}
Carrying out the derivatives and substituting this back into (\ref{eq:rhoVphi1harmonics}), we can write the result as a linear combination of $\mathcal{I}^{\text{freeQFT}}$:
\begin{equation}
\begin{aligned}
    \rho_{V\phi}^{\mathcal{P},1}(\lambda)=\widetilde{\mathcal{N}}^{V\phi}_{1}\Big[&((\Delta-\Delta_V)^2-\Delta_\phi^2)\Big(2\mathcal{I}_{0,1,-1}+2\mathcal{I}_{-1,1,0}-\mathcal{I}_{1,0,-1}-\mathcal{I}_{-1,0,1}-\mathcal{I}_{-1,2,-1}\Big)\\
    &+2(\Delta_\phi+\Delta(2\Delta_V-1)-\Delta_V)\Big((d-2)\mathcal{I}_{0,0,0}+2\mathcal{I}_{-1,0,-1}\Big)\Big]
\end{aligned}
\end{equation}
where we kept the notation abbreviated and the subscripts of $\mathcal{I}_{k_1,k_2,k_3}$ indicate, in order, the integers $k_j$ to add to $\Delta_V$, $\Delta_\phi$ and $\Delta\equiv\frac{d}{2}+i\lambda\,.$ Moreover,
\begin{equation}
    \widetilde{\mathcal{N}}^{V\phi}_{1}=\frac{\Gamma(\pm i\lambda_\phi)\Gamma(\pm i\lambda_V)}{\mathcal{N}_{1,1}}\frac{\lambda^2\lambda_V^2\lambda_\phi^2}{16\pi^3\Delta\Delta_V}(d-1)^2b(\Delta_\phi,\Delta,\Delta_V,0)\frac{\mathfrak{C}_{\Delta,1}\mathfrak{C}_{\Delta_V,1}}{\mathfrak{C}_{\Delta,0}\mathfrak{C}_{\Delta_V,0}}\,.
\end{equation}
Assembling all the pieces together and simplifying, we obtain
\begin{equation}
    \begin{aligned}
     \rho^{\mathcal{P},1}_{V\phi}(\lambda)=\frac{2^{-12}\pi^{-3-\frac{d}{2}}\lambda\sinh(\pi\lambda)f_{\lambda,\lambda_V,\lambda_\phi}}{\Gamma(\frac{d+2}{2})(\Delta_V-1)(\bar\Delta_V-1)\Gamma(\frac{d}{2}\pm i\lambda+1)}\prod_{\pm,\pm,\pm}\Gamma\left(\frac{\frac{d}{2}\pm i\lambda\pm i\lambda_\phi\pm i\lambda_V}{2}\right)\,,
    \end{aligned}
\end{equation}
with 
\begin{align}
f_{\lambda,\lambda_V,\lambda_\phi}=&16 \left(\lambda_\phi^2-(\lambda^2+\lambda_V^2)\right)^2+64 (d-1) \lambda ^2 \lambda_V^2\nonumber\\
&+8 d (3 d-4) \lambda_\phi ^2+8 d \left(2 d^2-5 d+4\right) \left(\lambda ^2+\lambda_V^2\right)+d^3 \left(4 d^2-11 d+8\right)\,.
\end{align}
For the operator $\phi_1\nabla\phi_2$, the steps are analogous, so we only report the linear combination in terms of the standard master integral. 
We have
\begin{equation}
\begin{aligned}
    \rho_{\phi_1\nabla\phi_2}^{\mathcal{P},0}(\lambda)&=\widetilde{\mathcal{N}}^{\phi_1\nabla\phi_2}_{0}\left(2\bar\Delta_1\mathcal{I}^{\text{QFT}}_{-1,-1,0}+(\Delta_1+\Delta_2-d)\mathcal{I}^{\text{QFT}}_{0,0,0}\right)\\
    &=\frac{(d^2+4(\lambda^2-\lambda_1^2+\lambda_2^2))^2\Gamma(\frac{d+1}{2})}{2^{8-d}\pi^{\frac{d+7}{2}}(d^2+4\lambda^2)^2\Gamma(d)\Gamma(\frac{d}{2}\pm i\lambda)}\lambda\sinh(\pi\lambda)\prod_{\pm,\pm,\pm}\Gamma\left(\frac{\frac{d}{2}\pm i\lambda\pm i\lambda_1\pm i\lambda_2}{2}\right)\,,
\end{aligned}
\end{equation}
with
\small
\begin{equation}
    \widetilde{\mathcal{N}}^{\phi_1\nabla\phi_2}_{0}=\frac{\Gamma(\pm i\lambda_1)\Gamma(\pm i\lambda_2)}{2\pi^3\mathcal{N}_{1,0}}(d-1)^2\lambda^2\lambda_1^2\lambda_2^2\Delta_2\bar\Delta_2((\bar\Delta_1-\Delta_2)b(\Delta,\Delta_1,\Delta_2,0)+2\Delta_1b(\Delta,\Delta_1+1,\Delta_2+1,0))
\end{equation}
\normalsize
and
\begin{equation}
\begin{aligned}
    \rho_{\phi_1\nabla\phi_2}^{\mathcal{P},1}(\lambda)&=\widetilde{\mathcal{N}}^{\phi_1\nabla\phi_2}_{1}\left(\mathcal{I}^{\text{QFT}}_{0,0,-1}-\mathcal{I}^{\text{QFT}}_{1,-1,-1}+\mathcal{I}^{\text{QFT}}_{0,-1,0}\right)\\
    &=\frac{\Gamma(-\frac{d}{2}\pm i\lambda)(\cosh(2\pi\lambda)-(-1)^d)}{2^5\pi^{5+\frac{d}{2}}\Gamma(\frac{d+2}{2})}\lambda\sinh(\pi\lambda)\prod_{\pm,\pm,\pm}\Gamma\left(\frac{\frac{d}{2}+1\pm i\lambda\pm i\lambda_1\pm i\lambda_2}{2}\right)\,.
\end{aligned}
\end{equation}
with
\begin{equation}
    \widetilde{\mathcal{N}}^{\phi_1\nabla\phi_2}_{1}=\frac{\Gamma(\pm i\lambda_1)\Gamma(\pm i\lambda_2)}{4\pi^3\Delta\mathcal{N}_{1,1}}(d-1)^2\Delta_2\bar\Delta_2(\Delta+\Delta_1-\Delta_2-1)\lambda^2\lambda_1^2\lambda_2^2\frac{\mathfrak{C}_{\Delta,1}}{\mathfrak{C}_{\Delta,0}}b(\Delta,\Delta_1,\Delta_2+1,0)\,.
\end{equation}
\subsection{CFTs}
\label{subsec:mastercft}
Another class of two-point functions we considered in this work are two-point functions of spin $J$ primary bulk CFT operators with conformal dimension $\mathbf{\Delta}$ in de Sitter, which as argued in section \ref{subsec:cfts}, are of the form
\begin{equation}
\begin{aligned}
    \langle \CO^{(J)}(Y_1,W_1)\CO^{(J)}(Y_2,W_2)\rangle&=c_{\CO}\frac{[(W_1\cdot W_2)(1-Y_1\cdot Y_2)+(Y_1\cdot W_2)(Y_2\cdot W_1)]^J}{2^{\mathbf{\Delta}}(1-Y_1\cdot Y_2)^{\mathbf{\Delta}+J}}\\
    &=\frac{c_{\CO}}{2^{\mathbf{\Delta}}}\sum_{m=0}^J\binom{J}{m}\frac{(W_1\cdot W_2)^m[(Y_1\cdot W_2)(Y_2\cdot W_1)]^{J-m}}{(1-Y_1\cdot Y_2)^{\mathbf{\Delta}+J-m}}
\end{aligned}
\end{equation}
Applying the inversion formula (\ref{eq:inversionformulal}) we can retrieve the principal series contribution
\small
\begin{equation}
\begin{aligned}
    \rho^{\mathcal{P},\ell}_{\CO^{(J)}}(\lambda)=\frac{1}{\mathcal{N}_{J,\ell}}\int_{X_1}\Omega_{\lambda,\ell}(X_2,X_1;K_2,K_1)[(K_1\cdot\nabla_1)(K_2\cdot\nabla_2)]^{J-\ell}\langle \CO^{(J)}(X_1,W_1)\CO^{(J)}(X_2,W_2)\rangle\,.
\end{aligned}
\end{equation}
\normalsize
After carrying out all the index contractions and the derivatives, in all the examples we explored the result can always be written as
\begin{equation}
\begin{aligned}
     \rho^{\mathcal{P},\ell}_{O^{(J)}}(\lambda)&=\sum_{n=-\ell}^\ell\sum_{k=0}^{J-|n|}c_{n,k}(\mathbf{\Delta},\lambda)\int_{X_1}\Omega_{\lambda+in,0}(X_2,X_1)(1-X_1\cdot X_2)^{-\mathbf{\Delta}-J+k}\\
     &\equiv\sum_{n=-\ell}^\ell\sum_{k=0}^{J-|n|}c_{n,k}(\mathbf{\Delta},\lambda)\mathcal{I}^{(J)}_{\text{CFT},n,k}(\mathbf{\Delta},\lambda)\,,
     \label{eq:CFTdensities}
\end{aligned}
\end{equation}
with some coefficients $c_{n,k}(\mathbf{\Delta},\lambda)$ which we determined case by case and which we will show in the following subsections of this Appendix. They appear to satisfy a symmetry $c_{n,k}(\mathbf{\Delta},\lambda)=c_{-n,k}(\mathbf{\Delta},\lambda)\,.$ Let us focus on the integral $\mathcal{I}^{(J)}_{\text{CFT},n,k}(\mathbf{\Delta},\lambda)$. By conformal invariance, we can fix $X_2$ in the origin of EAdS, which in global coordinates is given by $r_2=0$. In the same coordinates, we have $\sigma=X_1\cdot X_2=-\cosh r_1$ and 
\begin{equation}
\int_{X}=\int_0^\infty\mathrm{d}r\sinh^d r\int \mathrm{d}\Omega_d=S^d\int_{-\infty}^{-1}\mathrm{d}\sigma(\sigma^2-1)^{\frac{d-1}{2}}\,.
\end{equation}
The integral is thus
\small
\begin{equation}
    \mathcal{I}^{(J)}_{\text{CFT},n,k}(\mathbf{\Delta},\lambda)=C_\Omega S^d\int_{-\infty}^{-1}\mathrm{d}\sigma(\sigma^2-1)^{\frac{d-1}{2}}\ _2F_1\left(\Delta-n,\bar\Delta+n,\frac{d+1}{2},\frac{1+\sigma}{2}\right)(1-\sigma)^{-\mathbf{\Delta}-J+k}\,,
\end{equation}
\normalsize
where $\Delta\equiv\frac{d}{2}+i\lambda$.
To solve this, we resort to the Mellin Barnes representation of the hypergeometric function (which is the inverse Mellin transformation of (\ref{MF}))
\begin{equation}
    _2 F_1(a,b,c,z)=\frac{\Gamma(c)}{\Gamma(a)\Gamma(b)}\int_{-i\infty}^{i\infty}\frac{ds}{2\pi i}\frac{\Gamma(a+s)\Gamma(b+s)\Gamma(-s)}{\Gamma(c+s)}(-z)^s\,,
\end{equation}
and we change variables to $u=\frac{\sigma+1}{2}$:
\small
\begin{equation}
    \mathcal{I}^{(J)}_{\text{CFT},n,k}(\mathbf{\Delta},\lambda)=\tilde c\int_{-i\infty}^{i\infty}\frac{ds}{2\pi i}\frac{\Gamma(\Delta-n+s)\Gamma(\bar\Delta+n+s)\Gamma(-s)}{\Gamma(\frac{d+1}{2}+s)}\int_{0}^{\infty}\mathrm{d}u(1+u)^{\frac{d-1}{2}+k-J-\mathbf{\Delta}}u^{\frac{d-1}{2}+s}\,,
\end{equation}
\normalsize
with
\begin{equation}
    \tilde c=\frac{2^{k-J-\mathbf{\Delta}}(i\lambda-n)\sin(\pi(n-i\lambda))}{\pi\Gamma(\frac{d+1}{2})}\,.
\end{equation}
The integral over $u$ gives some gamma functions, one of which crucially cancels the denominator in the Mellin integral, giving
\begin{equation}
    \mathcal{I}^{(J)}_{\text{CFT},n,k}(\mathbf{\Delta},\lambda)=\tilde c\int\frac{ds}{2\pi i}\frac{\Gamma(-s)\Gamma(\bar\Delta+n+s)\Gamma(s+\Delta-n)\Gamma(-d-k-s+J+\mathbf{\Delta})}{\Gamma(\frac{1-d}{2}-k+J+\mathbf{\Delta})}\,.
    \label{eq:ICFTstep}
\end{equation}
The resulting Mellin integral can be carried out with Barnes' first lemma, which states
\begin{equation}
    \int_{-i\infty}^{i\infty}\frac{ds}{2\pi i}\Gamma(a+s)\Gamma(b+s)\Gamma(c-s)\Gamma(d-s)=\frac{\Gamma(a+c)\Gamma(a+d)\Gamma(b+c)\Gamma(b+d)}{\Gamma(a+b+c+d)}\,.
\end{equation}
Applying this to (\ref{eq:ICFTstep}), we obtain
\begin{equation}
    \mathcal{I}^{(J)}_{\text{CFT},n,k}(\mathbf{\Delta},\lambda)=\tilde c\frac{\Gamma(\bar\Delta+n)\Gamma(\Delta-n)\Gamma(-k+l+n+\mathbf{\Delta}-\Delta)\Gamma(-k+l-n+\mathbf{\Delta}+\bar\Delta)}{\Gamma(-k+l+\mathbf{\Delta})\Gamma(\frac{1}{2}-\frac{d}{2}-k+l+\mathbf{\Delta})}\,.
    \label{eq:CFTMasterIntegral}
\end{equation}
Substituting this into (\ref{eq:CFTdensities}) we can find the spectral densities for any spin $J$ CFT two-point function. To avoid clutter in the next subsections, we introduce the convenient shorthand notation
\begin{equation}
    \bar{\mathcal{I}}_{\text{CFT},n,k}^{(J)}\equiv\frac{\mathcal{I}_{\text{CFT},n,k}^{(J)}(\mathbf{\Delta},\lambda)}{\mathfrak{C}_{\frac{d}{2}+i\lambda-n,0}\mathfrak{C}_{\frac{d}{2}-i\lambda+n,0}(\lambda+in)^2}
\end{equation}

\subsubsection{Scalar CFT}
Let us start from the scalar case
\begin{equation}
    \langle \CO(Y_1)\CO(Y_2)\rangle=\frac{c_{\CO}}{2^{\mathbf{\Delta}}(1-Y_1\cdot Y_2)^{\mathbf{\Delta}}}\,.
\end{equation}
The inversion formula for the principal series contribution reads
\begin{equation}
    \rho^{\mathcal{P}}_\CO(\lambda)=\frac{c_{\CO}}{2^{\mathbf{\Delta}}\mathcal{N}_{0,0}}\int_{X_1}\Omega_{\lambda,0}(X_1,X_2)(1-X_1\cdot X_2)^{-\mathbf{\Delta}}\,.
\end{equation}
This is already of the form in (\ref{eq:CFTdensities}), with $n=0$ and $k=0$ and the coefficient being $c_{0,0}(\mathbf{\Delta},\lambda)=\frac{c_{\CO}}{2^{\mathbf{\Delta}}\mathcal{N}_{0,0}}$. We can thus simply apply (\ref{eq:CFTMasterIntegral}) and obtain the spectral density
\begin{equation}
    \rho^{\mathcal{P}}_\CO(\lambda)=\frac{c_{\CO}}{2^{\mathbf{\Delta}}\mathcal{N}_{0,0}}\mathcal{I}_{\text{CFT},0,0}^{(0)}(\mathbf{\Delta},\lambda)= c_{\CO}\frac{2^{1+d-2 \mathbf{\Delta}} \pi ^{\frac{d-1}{2}} \Gamma \left(-\frac{d}{2}+\mathbf{\Delta}\pm i \lambda \right) }{\Gamma (\mathbf{\Delta} ) \Gamma \left(\frac{1-d}{2}+\mathbf{\Delta}\right)}\lambda\sinh (\pi  \lambda )
\end{equation}
\subsubsection{Spin 1 CFT}
Moving on to the spin $1$ case, we have
\begin{equation}
    \langle J(Y_1;W_1)J(Y_2;W_2)\rangle=\frac{c_J}{2^{\mathbf{\Delta}}}\left[\frac{W_1\cdot W_2}{(1-Y_1\cdot Y_2)^{\mathbf{\Delta}}}+\frac{(Y_1\cdot W_2)(Y_2\cdot W_1)}{(1-Y_1\cdot Y_2)^{\mathbf{\Delta}+1}}\right]\,.
\end{equation}
We start by inverting $\rho_{J}^{\mathcal{P},1}(\lambda)$
\begin{equation}
    \rho_{J}^{\mathcal{P},1}(\lambda)=\frac{1}{\mathcal{N}_{1,1}}\int_{X_1}\Omega_{\lambda,1}(X_2,X_1;K_2,K_1)\langle J(X_1;W_1)J(X_2;W_2)\rangle\,.
    \label{eq:rho1CFT1part1}
\end{equation}
First, we use the split representation (\ref{eq:defsplitrep}) on $\Omega_{\lambda,1}$ and carry out the contraction of the boundary indices following the action of $D_Z$ on $Z$ (\ref{eq:DZoperator})
\small
\begin{equation}
\begin{aligned}
    &\Omega_{\lambda,1}(X_2,X_1;K_2,K_1)=\frac{\lambda^2}{\pi(\frac{d-2}{2})}\int_P\Pi_{\Delta}(X_1,P;K_1,D_Z)\Pi_{\bar\Delta}(X_2,P;K_2,Z)\\
    &=\frac{\mathfrak{C}_{\Delta,1}\mathfrak{C}_{\bar\Delta,1}\lambda^2}{\pi(\frac{d-2}{2})}\int_P\frac{((K_1\cdot P)(X_1\cdot D_Z)-(P\cdot X_1)(K_1\cdot D_Z))((K_2\cdot P)(X_2\cdot Z)-(P\cdot X_2)(K_2\cdot Z))}{(-2P\cdot X_1)^{\Delta+1}(-2P\cdot X_2)^{\bar\Delta+1}}\\
    &=\frac{\mathfrak{C}_{\Delta,1}\mathfrak{C}_{\bar\Delta,1}}{\pi\lambda^{-2}}\int_P\frac{P\cdot X_1(K_1\cdot K_2P\cdot X_2-P\cdot K_2X_2\cdot K_1)+P\cdot K_1(K_2\cdot PX_1\cdot X_2-P\cdot X_2X_1\cdot K_2)}{(-2P\cdot X_1)^{\Delta+1}(-2P\cdot X_2)^{\bar\Delta+1}}\,.
\end{aligned}
\end{equation}
\normalsize
Plugging this into (\ref{eq:rho1CFT1part1}) and computing the action of the $K$ operators over the $W$ vectors (\ref{eq:actionofKAdS}) we obtain 
\begin{equation}
\begin{aligned}
    \rho_{J}^{\mathcal{P},1}(\lambda)&=\widetilde{\mathcal{N}}_{1,1}^{\text{CFT}}\int_{X_1,P}\frac{((P_1\cdot X_1)^2+(P_1\cdot X_2)^2+P_1\cdot X_1P_1\cdot X_2(d-(d-2)X_1\cdot X_2))}{(1-X_1\cdot X_2)^{\mathbf{\Delta}+1}(-2P\cdot X_1)^{\Delta+1}(-2P\cdot X_2)^{\bar\Delta+1}}\\
    &=\frac{1}{2}\widetilde{\mathcal{N}}_{1,1}^{\text{CFT}}\left(\bar{\mathcal{I}}_{\text{CFT},1,0}^{(1)}+\bar{\mathcal{I}}_{\text{CFT},-1,0}^{(1)}+2\bar{\mathcal{I}}_{\text{CFT},0,0}^{(1)}+(d-2)\bar{\mathcal{I}}_{\text{CFT},0,1}^{(1)}\right)
    \label{eq:rho1Jstep}
\end{aligned}
\end{equation}
where 
\begin{equation}
    \widetilde{\mathcal{N}}_{1,1}^{\text{CFT}}\equiv\frac{c_J(d-1)^2\mathfrak{C}_{\Delta,1}\mathfrak{C}_{\bar\Delta,1}\lambda^2}{2^{\mathbf{\Delta}}\pi\mathcal{N}_{1,1}}\,.
\end{equation}
To the second line of (\ref{eq:rho1Jstep}) we have carried out the $P$ integral and retrieved scalar harmonic functions, and then organized the sum as a polynomial in $(1-X_1\cdot X_2)\,.$ That put the expression in a form where (\ref{eq:CFTMasterIntegral}) is applicable to each term in the sum. By substituting the expression for $\mathcal{I}_{\text{CFT}}$ we obtain
\begin{equation}
    \rho_{J}^{\mathcal{P},1}(\lambda)=c_J\frac{2^{1+d-2\mathbf{\Delta}}\pi^{\frac{d-1}{2}}(\mathbf{\Delta}-1)\Gamma(-\frac{d}{2}+\mathbf{\Delta}\pm i\lambda)}{\Gamma(\mathbf{\Delta}+1)\Gamma(\frac{1-d}{2}+\mathbf{\Delta})}\lambda\sinh(\pi\lambda)\,.
\end{equation}
We do the analogous steps for $\rho_{J}^{\mathcal{P},0}(\lambda)$, starting from the inversion formula
\begin{equation}
    \rho_{J}^{\mathcal{P},0}(\lambda)=\frac{1}{\mathcal{N}_{1,0}}\int_{X_1}\Omega_{\lambda,0}(X_2,X_1)(\nabla_1\cdot K_1)(\nabla_2\cdot K_2)\langle J(X_1;W_1)J(X_2;W_2)\rangle\,.
\end{equation}
We carry out the derivatives and the contractions with the $K$ operators and obtain
\begin{equation}
\begin{aligned}
    \rho_{J}^{\mathcal{P},0}(\lambda)&=\frac{c_J(d-1)^2(\mathbf{\Delta}-d)}{2^{2+\mathbf{\Delta}}\mathcal{N}_{1,0}}\int_{X_1}\frac{\Omega_{\lambda,0}(X_2,X_1)}{(1-X_1\cdot X_2)^{\mathbf{\Delta}+1}}(1+\mathbf{\Delta}+(\mathbf{\Delta}-d)X_1\cdot X_2)\\
    &=\frac{c_J(d-1)^2(\mathbf{\Delta}-d)}{2^{2+\mathbf{\Delta}}\mathcal{N}_{1,0}}\left((1-d+2\mathbf{\Delta})\mathcal{I}_{\text{CFT},0,0}^{(1)}+(d-\mathbf{\Delta})\mathcal{I}_{\text{CFT},0,1}^{(1)}\right)
\end{aligned}
\end{equation}
Substituting the expression (\ref{eq:CFTMasterIntegral}) and simplifying, we obtain what we presented in the main text
\begin{equation}
    \rho_{J}^{\mathcal{P},0}(\lambda)=c_J\frac{2^{3+d-2\mathbf{\Delta}}\pi^{\frac{d-1}{2}}(\mathbf{\Delta}-d)\Gamma\left(-\frac{d}{2}+\mathbf{\Delta}\pm i\lambda\right)}{(d^2+4\lambda^2)\Gamma(\mathbf{\Delta}+1)\Gamma(\frac{1-d}{2}+\mathbf{\Delta})}\lambda\sinh(\pi\lambda)\,.
\end{equation}
\subsubsection{Spin 2 CFT}
To treat the spin 2 case, the logic is the same.
\small
\begin{equation}
    \langle T(Y_1;W_1)T(Y_2;W_2)\rangle=\frac{c_T}{2^{\mathbf{\Delta}}}\Big[\frac{(W_1\cdot W_2)^2}{(1-Y_1\cdot Y_2)^{\mathbf{\Delta}}}+2\frac{(W_1\cdot W_2)(Y_1\cdot W_2)(Y_2\cdot W_1)}{(1-Y_1\cdot Y_2)^{\mathbf{\Delta}+1}}
    \!+\!\frac{[(Y_1\cdot W_2)(Y_2\cdot W_1)]^2}{(1-Y_1\cdot Y_2)^{\mathbf{\Delta}+2}}\Big]\,.
\end{equation}
\normalsize
We start from $\rho_{T}^{\mathcal{P},2}(\lambda)$ 
\begin{equation}
    \rho_{T}^{\mathcal{P},2}(\lambda)=\frac{1}{\mathcal{N}_{2,2}}\int_{X_1}\Omega_{\lambda,2}(X_2,X_1;K_2,K_1)\langle T(X_1;W_1)T(X_2;W_2)\rangle\,.
\end{equation}
We follow the same identical steps as in the spin 1 example: we use the split representation on $\Omega_{\lambda,2}$, carry out the contractions between $D_Z$ and $Z$ and between $K$ and $W$. We land on a linear combination of scalar harmonic functions which we can express in terms of $\mathcal{I}_{\text{CFT}}$
\begin{equation}
\begin{aligned}
    \rho_{T}^{\mathcal{P},2}(\lambda)=&\frac{\lambda^2(d+1)^2(d-1)^3\mathfrak{C}_{\Delta,2}\mathfrak{C}_{\bar\Delta,2}}{2^{3+\mathbf{\Delta}}d\ \mathcal{N}_{2,2}}\sum_{\pm}\Big(2\bar{\mathcal{I}}_{\text{CFT},\pm2,0}^{(2)}+8\bar{\mathcal{I}}_{\text{CFT},\pm1,0}^{(2)}+2(d-2)\bar{\mathcal{I}}_{\text{CFT},\pm1,1}^{(2)}\\
    &\qquad\qquad\qquad\qquad\qquad\qquad+12\bar{\mathcal{I}}_{\text{CFT},0,0}^{(2)}+4(d-2)\bar{\mathcal{I}}_{\text{CFT},0,1}^{(2)}
    +d(d-2)\bar{\mathcal{I}}_{\text{CFT},0,2}^{(2)}\Big)\,,
\end{aligned}
\end{equation}
which gives
\begin{equation}
    \rho^{\mathcal{P},2}_{T}(\lambda)=c_T\frac{2^{1+d-2\mathbf{\Delta}}\pi^{\frac{d-1}{2}}(\mathbf{\Delta}-1)\mathbf{\Delta}\Gamma(-\frac{d}{2}+\mathbf{\Delta}\pm i\lambda)}{\Gamma(\mathbf{\Delta}+2)\Gamma(\frac{1-d}{2}+\mathbf{\Delta})}\lambda\sinh(\pi\lambda)\,.
\end{equation}
For $\rho_{T}^{\mathcal{P},1}(\lambda)$ instead, we have
\begin{equation}
    \rho_{T}^{\mathcal{P},1}(\lambda)=\frac{1}{\mathcal{N}_{2,1}}\int_{X_1}\Omega_{\lambda,1}(X_2,X_1;K_2,K_1)(K_1\cdot\nabla_1)(K_2\cdot\nabla_2)\langle T(X_1;W_1)T(X_2;W_2)\rangle\,.
\end{equation}
After applying the split representation and carrying out derivatives and index contractions, we obtain
\begin{equation}
\begin{aligned}
    \rho_{T}^{\mathcal{P},1}(\lambda)=\widetilde{\mathcal{N}}^{\text{CFT}}_{2,1}\sum_{\pm}\Big(&(4(\mathbf{\Delta}-1)+d(6\mathbf{\Delta}-4+d(d-2\mathbf{\Delta}-5))\bar{\mathcal{I}}^{(2)}_{\text{CFT},0,1}\\
    &+(d+1+2d^2-\mathbf{\Delta}-3d\mathbf{\Delta})\left(2\bar{\mathcal{I}}_{\text{CFT},0,0}^{(2)}+\bar{\mathcal{I}}_{\text{CFT},\pm1,0}^{(2)}\right)\\
    &-(d+1)(d+1-\mathbf{\Delta})\left((d-2)\bar{\mathcal{I}}^{(2)}_{\text{CFT},0,2}-\bar{\mathcal{I}}^{(2)}_{\text{CFT},\pm1,1}\right)\,.
\end{aligned}
\end{equation}
with 
\begin{equation}
\widetilde{\mathcal{N}}^{\text{CFT}}_{2,1}\equiv\frac{\lambda^2(d+1-\mathbf{\Delta})(d+1)(d-1)^2\mathfrak{C}_{\Delta,1}\mathfrak{C}_{\bar\Delta,1}}{2^{3+\mathbf{\Delta}}\mathcal{N}_{2,1}}\,.
\end{equation}
Explicitly,
\begin{equation}
\rho_{T}^{\mathcal{P},1}(\lambda)=c_T\frac{2^{4+d-2\mathbf{\Delta}}\pi^{\frac{d-1}{2}}(1-\mathbf{\Delta})(d+1-\mathbf{\Delta})\Gamma(-\frac{d}{2}+\mathbf{\Delta}\pm i\lambda)}{((d+2)^2+4\lambda^2)\Gamma(\mathbf{\Delta}+2)\Gamma(\frac{1-d}{2}+\mathbf{\Delta})}\lambda\sinh(\pi\lambda)\,,
\end{equation}
Finally, we have
\begin{equation}
\begin{aligned}
    \rho_{T}^{\mathcal{P},0}(\lambda)&=\frac{1}{\mathcal{N}_{2,0}}\int_{X_1}\Omega_{\lambda,0}(X_2,X_1)(K_1\cdot\nabla_1)^2(K_2\cdot\nabla_2)^2\langle T(X_1;W_1)T(X_2;W_2)\rangle\\
    &=\widetilde{\mathcal{N}}^{\text{CFT}}_{2,0}\Big((d^2+3+8\mathbf{\Delta}+4\mathbf{\Delta}^2-4d(\mathbf{\Delta}+1))\mathcal{I}_{\text{CFT},0,0}^{(2)}\\
    &\qquad\qquad-(d-\mathbf{\Delta})\left(2(d-1-2\mathbf{\Delta})\mathcal{I}_{\text{CFT},0,1}^{(2)}+(\mathbf{\Delta}-d-1)\mathcal{I}_{\text{CFT},0,2}\right)\Big)\\
    &=c_T\frac{2^{5+d-2\mathbf{\Delta}}(d+1)\pi^{\frac{d-1}{2}}(d-\mathbf{\Delta})(d+1-\mathbf{\Delta})\Gamma(-\frac{d}{2}+\mathbf{\Delta}\pm i\lambda)}{d(d^2+4\lambda^2)((d+2)^2+4\lambda^2)\Gamma(\mathbf{\Delta}+2)\Gamma(\frac{1-d}{2}+\mathbf{\Delta})}\lambda\sinh(\pi\lambda)\,,\\
\end{aligned}
\end{equation}
where
\begin{equation}
    \widetilde{\mathcal{N}}^{\text{CFT}}_{2,0}\equiv\frac{(d-1)^2d(d+1)(d+d^2-2d\mathbf{\Delta}+\mathbf{\Delta}(\mathbf{\Delta}-1))}{2^{2+\mathbf{\Delta}}\mathcal{N}_{2,0}}~.
\end{equation}

\section{Diagrammatics of de Sitter}
\label{sec:ininformalism}
In this section we review the in-in formalism and we show the details of the computation in section \ref{subsec:qftint}. To perform computations in the in-in formalism, we find it convenient to analytically continue to EAdS, as done in~\cite{Sleight:2021plv,Sleight:2019mgd,DiPietro:2021sjt}, such that we can exploit the large amount of mathematical results that are already known for Witten diagrams. In this subsection, we will only be interested in scalar fields, and as such we will omit spin labels. $G_{\lambda}(Y_1,Y_2)$ will indicate a spin 0 free propagator which we otherwise refer to as $G_{\lambda,0}(Y_1,Y_2)\,.$
\subsection{In-in formalism}\label{sec:In-in formalism}
\begin{figure}[t]
\centering
\begin{tikzpicture}[scale=1.2]
\coordinate (origintwo) at (9,0);
\draw[line width=0.3mm,black,->] (origintwo) -- ++(2,0);
\draw[line width=0.3mm,black,->] (origintwo) -- ++(0,3);
\draw[line width=0.3mm,red] (origintwo) -- ++(-4,0) node (zminus) {};
\draw[line width=0.3mm,black,-] (origintwo) -- ++(0,-1) node (xminus) {};

\draw [line width=0.3mm,blue!70] (9,0.32) to[out=0,in=90] (9.35,0) ;
\draw [line width=0.3mm,blue!70] (9,-0.32) to[out=0,in=-90] (9.35,0);
\draw [line width=0.3mm,blue!70] (9,0.32) -- (5,0.32) node [black!50 ,above] {} --(7,0.32) ;
\draw [line width=0.3mm,blue!70] (9,-0.32) -- (5,-0.32) ;
\draw[blue!70, thick,->] (8,0.32) --(8.05,0.32)node[red!80, above] {\tiny{~right}};
\draw (8,-0.32) node[cross,black]{};
\draw (6.5,0.32) node[cross,black]{};
\draw[blue!70, thick,<-] (6,-0.32)--node[red!80, below]{\tiny{left}}(6.05,-0.32);
\draw[black!50, thick,->] (5,0.32)--(5,0)node[black!50, above left] {$\epsilon$};
\draw[black!50, thick,<-] (5,0.32)--(5,0);

\draw[line width=0.2mm,black,-] (10.5,2.5)--(10.85,2.5);
\draw[line width=0.2mm,black,-] (10.5,2.85)--(10.5,2.5)node [black, above right] {\large{$\eta$}};
\small
\draw (8,-0.32) node[black,below]{$\mathcal{O}(\eta_1,\vec y_1)$};
\draw (6.5,0.32) node[black,above]{$\mathcal{O}(\eta_2,\vec y_2)$};
\normalsize
\end{tikzpicture}
\caption{The contour integral of the in-in formalism, when computing the Wightman function $G^{lr}(Y_1,Y_2)$. The ordering in time of $\eta_1$ and $\eta_2$ does not matter.}
\label{fig:In-in}
\end{figure}
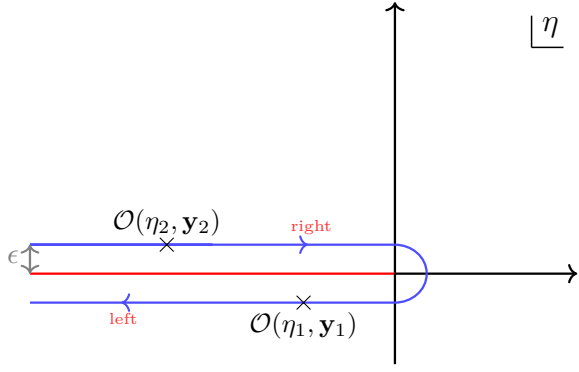

The in-in (or Schwinger-Keldysh) formalism \cite{LINDBLAD1976393} has been used to compute physical observables in QFT in de Sitter since the seminal works \cite{Weinberg:2005vy,Maldacena:2002vr}. We are interested in using it to compute bulk two-point functions on the interacting Bunch-Davies vacuum
\begin{equation}\
\langle\Omega|\mathcal{O}(\eta_1,\vec y_1)\mathcal{O}(\eta_2,\vec y_2)|\Omega\rangle\,.
\end{equation}
More explicitly, in the interaction picture, we are computing
\small
\begin{equation}\label{eq:In-in correlator}
\langle\Omega|\mathcal{O}(\eta_1,\vec y_1)\mathcal{O}(\eta_2,\vec y_2)|\Omega\rangle = \frac{\langle 0|U^\dag_I(\eta_1,-\infty)\mathcal{O}_I(\eta_1,\vec y_1) U^\dag_I(0,\eta_1)U_I(0,\eta_2)\mathcal{O}_I(\eta_2,\vec y_2)U_I(\eta_2,-\infty)|0\rangle}{\langle 0|U^\dag_I(0,-\infty)U_I(0,-\infty)|0\rangle}
\end{equation} 
\normalsize
where 
\begin{equation}
\begin{aligned}
    U_I(\eta_1,\eta_2)&=T\left[\text{exp}\left(-i\int_{\eta_2(1-i\epsilon)}^{\eta_1(1-i\epsilon)}d\eta\ H_I(\eta)\right)\right]\,,
\end{aligned}
\end{equation}
is the time evolution operator with the interacting part of the Hamiltonian $H_I(\eta)$ (which in de Sitter explicitly depends on time), $\mathcal{O}_I(\eta,\vec{y})$ is the operator $\mathcal{O}(\eta,\vec y)$ in the interaction picture 
and $|0\rangle$ is the free Bunch-Davies vacuum. Concretely, we will be interested in the case where $\mathcal{O}_I=\phi^2$ and $\phi$ is an elementary field in the following theory
\begin{equation}
    \mathcal{L}=-\frac{1}{2}g^{\mu\nu}\partial_\mu\phi\partial_\nu\phi-\frac{1}{2}m^2\phi^2-\frac{g}{4!}\phi^4\,.
\end{equation}
The full contour time integral in (\ref{eq:In-in correlator}) is represented pictorially in fig. \ref{fig:In-in}.

Expanding the exponentials in ~\reef{eq:In-in correlator} for weak couplings and carrying out all the possible Wick contractions results in a set of diagrammatic rules, which are for instance explained in the appendix of~\cite{Weinberg:2005vy}. We review them here for completeness, and to introduce the notation that we use in \ref{subsec:qftint} 
\begin{itemize}
\item There are ``right'' $(r)$ and ``left'' $(l)$ vertices coming from operators in the interacting Hamiltonian, depending on whether they are, respectively, in the time ordered or the anti-time ordered part of the contour. The right vertices are multiplied by $-i$,  while the left vertices are multiplied by $+i$. One then sums over each vertex being $l$ or $r$.
\item Wick contractions between operators on the time ordered part of the contour lead to time ordered propagators $G^{rr}_{\lambda_\phi}(Y_1,Y_2)=\langle T\phi(Y_1)\phi(Y_2)\rangle$, with $\phi$ being some free field with $\Delta=\frac{d}{2}+i\lambda_\phi$ and where we are now using embedding space notation for the coordinates in de Sitter.
\item Wick contractions between operators on the anti-time ordered part of the contour lead to anti-time ordered propagators $G^{ll}_{\lambda_\phi}(Y_1,Y_2)=\langle \bar T\phi(Y_1) \phi(Y_2)\rangle$. 
\item Wick contractions between operators inserted on two different branches of the time contour lead to Wightman functions
\begin{equation}
G^{lr}_{\lambda_\phi}(Y_1,Y_2)=\langle \phi(Y_1) \phi(Y_2)\rangle \,, \qquad G^{rl}_{\lambda_\phi}(Y_1,Y_2)=\langle \phi(Y_2) \phi(Y_1)\rangle.
\end{equation}
\end{itemize}
These definitions imply the following relations
\begin{equation}
\begin{aligned}
    G^{ll}_{\lambda_\phi}(Y_1,Y_2)=\theta(\eta_1-\eta_2)G^{rl}_{\lambda_\phi}(Y_1,Y_2)+\theta(\eta_2-\eta_1)G^{lr}_{\lambda_\phi}(Y_1,Y_2)\,,\\
    G^{rr}_{\lambda_\phi}(Y_1,Y_2)=\theta(\eta_1-\eta_2)G^{lr}_{\lambda_\phi}(Y_1,Y_2)+\theta(\eta_2-\eta_1)G^{rl}_{\lambda_\phi}(Y_1,Y_2)\,.
    \label{eq:timeorderings}
\end{aligned}
\end{equation}

\subsection{EAdS-dS dictionary}
In this section, we review the dictionary between in-in de Sitter diagrams and Witten diagrams in EAdS. We translate the rules discussed in \cite{Sleight:2021plv,DiPietro:2021sjt} in our own embedding space notation. The Wick rotation chosen to open up the in-in contour and analytically continue to EAdS is the following, in planar coordinates (see section \ref{subsec:coordinates} for a discussion on coordinate systems in dS and EAdS)
\begin{equation}
    \eta^l\rightarrow e^{i\frac{\pi}{2}}\eta^l\,, \qquad \eta^r\rightarrow e^{-i\frac{\pi}{2}}\eta^r\,,
\end{equation}
and then identifying the absolute value of $\eta$ with the radial coordinate $z$ in EAdS. The authors of \cite{Sleight:2021plv,DiPietro:2021sjt} have shown that, under this continuation,
\begin{equation}
\begin{aligned}
    G^{ll}_{\lambda}(Y_1,Y_2)&\rightarrow\frac{i\lambda}{2\pi}\Gamma(\pm i\lambda)\left(e^{i\pi\Delta_\lambda}\Pi_{\Delta_\lambda}(X_1,X_2)-e^{i\pi\bar\Delta_\lambda}\Pi_{\bar\Delta_\lambda}(X_1,X_2)\right)\,,\\
    G^{rr}_{\lambda}(Y_1,Y_2)&\rightarrow\frac{i\lambda}{2\pi}\Gamma(\pm i\lambda)\left(e^{-i\pi\Delta_\lambda}\Pi_{\Delta_\lambda}(X_1,X_2)-e^{-i\pi\bar\Delta_\lambda}\Pi_{\bar\Delta_\lambda}(X_1,X_2)\right)\,,\\
    G^{lr}_{\lambda}(Y_1,Y_2)&\rightarrow\Gamma(\pm i\lambda)\Omega_{\lambda}(X_1,X_2)\,,\\
    G^{rl}_{\lambda}(Y_1,Y_2)&\rightarrow\Gamma(\pm i\lambda)\Omega_{\lambda}(X_1,X_2)\,,
    \label{eq:dStoAdSrules}
\end{aligned}
\end{equation}
where $\Pi_{\Delta}(X_1,X_2)$ is an EAdS bulk-to-bulk propagator and $\Delta_\lambda\equiv\frac{d}{2}+i\lambda$. 
Under this continuation, the integrals appearing in perturbative computations also rotate accordingly
\begin{equation}
    i\int_{Y_l}(\cdots)\rightarrow e^{-i\frac{\pi}{2}(d-1)}\int_X(\cdots)\,, \qquad -i\int_{Y_r}(\cdots)\rightarrow e^{i\frac{\pi}{2}(d-1)}\int_X(\cdots)\,,
\end{equation}
where 
\begin{equation}
    \int_{Y_\alpha}(\cdots)\equiv\int\frac{d\eta^\alpha d^dy}{(-\eta^\alpha)^{d+1}}(\cdots)
\end{equation}
with $\alpha=l$ or $\alpha=r$ and $\int_X$ is defined in (\ref{eq:integralX}). 
\subsection{Details of the anomalous dimensions computation}
\label{subsec:anomalousappendix}
We report here the details of the computation of the anomalous dimensions presented in \ref{subsubsec:anomalous}. Let us start from the in-in formalism sum for the order $g^2$ contribution to the Wightman two-point function of $\phi^2$, where we selected $Y_1\in l$ and $Y_2\in r$. We have only one vertex, so we obtain two terms, since we have to consider the case in which this vertex comes from the interaction Hamiltonian in the time ordered ($r$) and in the anti-time ordered ($l$) part of (\ref{eq:In-in correlator}):
\begin{equation}
\langle\phi^2(Y_1)\phi^2(Y_2)\rangle^{lr}_{(g)}=ig\left[\int_{Y^l}(G^{ll}_{\lambda_\phi}(Y_1, Y))^2(G^{lr}_{\lambda_\phi}(Y,Y_2))^2-\int_{Y^r}(G^{lr}_{\lambda_\phi}(Y_1,Y))^2(G^{rr}_{\lambda_\phi}(Y,Y_2))^2\right]\,,
\label{eq:diagramininstep1}
\end{equation}
see Figure \ref{fig:Diagram} for the associated diagram.
Analytically continuing to EAdS, we apply the rules (\ref{eq:dStoAdSrules}) and obtain
\small
\begin{align}
    \langle\phi^2(X_1)\phi^2(X_2)\rangle_{(g)}^{lr}=&\mathcal{N}_{(g)}\Big[e^{-i\frac{\pi}{2}(d-1)}\int_X\Big(2e^{i\pi d}\Pi_{\Delta_\phi}(X_1,X)\Pi_{\bar\Delta_\phi}(X_1,X)-e^{2i\pi\Delta_\phi}\Pi^2_{\Delta_\phi}(X_1,X)\nonumber\\
    &\qquad\qquad\qquad\qquad\qquad\qquad-e^{2i\pi\bar\Delta_\phi}\Pi_{\bar\Delta_\phi}^2(X_1,X)\Big)\Omega_{\lambda_\phi}^2(X,X_2)\label{eq:ininSTEP1}\\
    &\qquad+e^{i\frac{\pi}{2}(d-1)}\int_X\Big(2e^{-i\pi d}\Pi_{\Delta_\phi}(X,X_2)\Pi_{\bar\Delta_\phi}(X,X_2)-e^{-2i\pi\Delta_\phi}\Pi^2_{\Delta_\phi}(X,X_2)\nonumber\\
    &\qquad\qquad\qquad\qquad\qquad\qquad-e^{-2i\pi\bar\Delta_\phi}\Pi_{\bar\Delta_\phi}^2(X,X_2)\Big)\Omega_{\lambda_\phi}^2(X_1,X)\Big]\,.\nonumber
\end{align}
\normalsize
where the normalization factor is
\begin{equation}
    \mathcal{N}_{(g)}\equiv\frac{g\lambda^2_\phi}{(2\pi)^2}\Gamma(\pm i\lambda_\phi)^4\,.
\end{equation}
Therefore, we need to evaluate a bulk integral that involves two bulk-to-bulk propagators and two harmonic functions.
To make progress, we express $\Omega^2_{\lambda_\phi}$ appearing in (\ref{eq:ininSTEP1}) as an integral over one single harmonic function. It  has been effectively done in section \ref{sec:applications}, and takes the following form
\begin{equation}
    (\Omega_{\lambda_\phi}(X_1,X_2))^2=\int_\mathbb{R}d\lambda\  \rho^{\mathcal{P}}_{\Omega}(\lambda)\Omega_{\lambda}(X_1,X_2)+\sum_{n=0}^N\rho_{\Omega}^{\mathcal{C}}(n)\Omega_{2\lambda_\phi+i(\frac{d}{2}+2n)}(X_1,X_2)\,,
    \label{eq:omega2}
\end{equation}
where the sum appears only if $\lambda_\phi$ is imaginary and $\frac{d}{4}+N<i\lambda_\phi<\frac{d}{4}+N+1$, and  
\small
\begin{equation}
\begin{aligned}
    \rho_{\Omega}^{\mathcal{P}}(\lambda)&=\frac{\Gamma(\pm i\lambda)} {2\Gamma(\pm i\lambda_\phi)^2}\rho_{\phi^2,\text{free}}^{\mathcal{P}, 0}(\lambda)=\frac{\lambda_\phi^2\sinh^2(\pi\lambda_\phi)}{32\pi^{4+\frac{d}{2}}\Gamma(\frac{d}{2})\Gamma(\frac{d}{2}\pm i\lambda)}\Gamma\left(\frac{\frac{d}{2}\pm i\lambda}{2}\right)^2\prod_{\pm,\pm}\Gamma\left(\frac{\frac{d}{2}\pm i\lambda\pm 2i\lambda_\phi}{2}\right)\,,\\
    \rho_{\Omega}^{\mathcal{C}}(n)&=\frac{\lambda_\phi^2(\frac{d}{2})_n\Gamma(\frac{d}{2}+n-i\lambda_\phi)^2\Gamma(-n+i\lambda_\phi)^2\Gamma(\frac{d}{2}+n-2i\lambda_\phi)\Gamma(-n+2i\lambda_\phi)\sinh^2(\pi\lambda_\phi)}{4n!(-1)^n\Gamma(d+2n-2i\lambda_\phi)\Gamma(-2n+2i\lambda_\phi)}~.
\label{eq:rhoOmega2}
\end{aligned}
\end{equation}
\normalsize
The density $\rho_{\phi^2,\text{free}}^{\mathcal{P}, 0}(\lambda)$ for free theory is given by eq. (\ref{rhofree}).

Then, the remaining integral to be evaluated is of the type 
\begin{align}\label{PIPIO}
    \int_{X}\, \Pi_{\Delta_1}(X_1, X)\Pi_{\Delta_2}(X_1, X) \Omega_\lambda (X,X_2)~.
\end{align}
where $\Delta_1$ and $\Delta_2$ are equal to either $\Delta_\phi$ or $\bar\Delta_\phi$. This integral is convergent for $d<3$, and has a UV divergence when $d\ge 3$, which can be easily seen from the coincident limit $\Pi_\Delta(X_1, X_2) \sim |1+X_1\cdot X_2|^{\frac{1-d}{2}}$. We will regularize this UV divergence in dS$_4$ by using dimensional regularization, namely taking $d=3-\epsilon$. With the regularization scheme specified, we proceed to compute the integral in (\ref{PIPIO}) with the help of  \KL\, decomposition in AdS \cite{Fitzpatrick_2012}
\begin{align}\label{afunction}
   &\Pi_{\Delta_1}(X_1, X)\Pi_{\Delta_2}(X_1, X) = \sum_{n\ge 0} a_{\Delta_1, \Delta_2} (n) \Pi_{\Delta_1+\Delta_2+2n}(X_1, X)~,\nonumber\\
       & a_{\Delta_1,\Delta_2}(n)\equiv\frac{(\frac{d}{2})_n(\Delta_1+\Delta_2+n+1-d)_n (\Delta_1+\Delta_2+2n)_{\frac{2-d}{2}}}{2\pi^{\frac{d}{2}}n!(\Delta_1+n)_{\frac{2-d}{2}}(\Delta_2+n)_{\frac{2-d}{2}}(\Delta_1+\Delta_2+n-\frac{d}{2})_n}~.
\end{align}
The resulting integral involves only one bulk-to-bulk propagator and one harmonic function, i.e. $\int_X\Pi_\Delta(X_1, X)\Omega_\lambda(X, X_2)$. Such an integral is equivalent to the harmonic decomposition of $\Pi_\Delta$ \cite{Costa_2014}
\begin{align}\label{Pi=O}
    \Pi_\Delta (X_1, X)=\int_{\mathbb R}\,d\lambda\, \frac{\Omega_\lambda(X_1, X)}{\lambda^2+\left(\Delta-\frac{d}{2}\right)^2}~,
\end{align}
where the real part of $\Delta$ should be larger than $\frac{d}{2}$. 
Applying the orthogonality relation (\ref{eq:Omega ortho}) of the harmonics functions to (\ref{Pi=O}) yields 
\begin{align}
    \int_X\Pi_\Delta(X_1, X)\Omega_\lambda(X, X_2) = \frac{\Omega_\lambda(X_1, X_2)}{\lambda^2+\left(\Delta-\frac{d}{2}\right)^2}, \,\,\,\,\, \Re \Delta>\frac{d}{2}~.
\end{align}
Putting all the ingredients together, we obtain 
\begin{align}\label{Pi2Omega2}
    \int_{X}\Pi_{\Delta_1}(X_1, X) \Pi_{\Delta_2}(X_2, X) \Omega_{\lambda_\phi}^2 (X, X_2) =\int_{\mathbb R}d\lambda\, \rho^\CP_\Omega(\lambda)\, B_{\Delta_1, \Delta_2} (\lambda)\, \Omega_{\lambda}(X_1, X_2)~,
\end{align}
where 
\begin{align}\label{eq:sumB}
    B_{\Delta_1, \Delta_2}(\lambda)\equiv \sum_{n= 0}^\infty\frac{a_{\Delta_1,\Delta_2}(n)}{\lambda^2+\left(\Delta_1+\Delta_2+2n-\frac{d}{2}\right)^2}
\end{align}
is known as the bubble function \cite{DiPietro:2021sjt,Sleight:2021plv}. The infinite sum defining $B_{\Delta_1, \Delta_2}$ is divergent when $d\ge 3$, because the leading large $n$ behavior of its summand is $2^{-d-1}\pi^{-\frac{d}{2}} n^{d-4}$. This is a UV divergence. In the convergence region, re-summation can be performed, obtaining a $_7F_6$ hypergeometric function \cite{Sleight:2021plv}, and when $d=2$, the result can be further simplified in terms of $\psi$ functions \cite{Carmi:2018qzm}. However, these re-summed expressions are not directly useful for our purposes, since we are mainly interested in the residues of $B_{\Delta_1, \Delta_2}$. For dS$_4$, the dimensional regularization $d=3-\epsilon$ is implicitly implemented. In dimensional regularization, $ B_{\Delta_1, \Delta_2}(\lambda)$ has a $\lambda$-independent $\frac{1}{\epsilon}$ divergence, and the analytical properties of its finite part is insensitive to the renormalization scheme. Therefore, we will  still use eq. (\ref{eq:sumB}) formally in dS$_4$, without specifying any renormalization.

\noindent{}Altogether, combining eq. (\ref{eq:ininSTEP1}) and eq. (\ref{Pi2Omega2}) and rotating back to de Sitter, we get the leading order correction to the \KL\, decomposition of $\phi^2$:
\begin{equation}
    \langle\phi^2(Y_1)\phi^2(Y_2)\rangle_{(g)}^{lr}=\int_{\mathbb{R}}d\lambda\ \rho^{\mathcal{P}}_{\phi^2,g}(\lambda)G^{lr}_{\lambda}(Y_1,Y_2)\,,
\end{equation}
where
\begin{equation}
\begin{aligned}
    \rho^{\mathcal{P}}_{\phi^2,g}(\lambda)=&g\frac{\rho^{\CP,0}_{\phi^2,\text{free}}(\lambda)}{4\sinh^2(\pi\lambda_\phi)}\Bigg[\sin\left(\pi\left(\frac{d}{2}+2i\lambda_\phi\right)\right)B_{\Delta_\phi,\Delta_\phi}(\lambda)
    \\
    &+\sin\left(\pi\left(\frac{d}{2}-2i\lambda_\phi\right)\right)B_{\bar\Delta_\phi,\bar\Delta_\phi}(\lambda)-2\sin\left(\frac{d\pi}{2}\right)B_{\Delta_\phi,\bar\Delta_\phi}(\lambda)\Bigg]\,.
    \label{eq:rhoPint}
\end{aligned}
\end{equation}
Now to compute the anomalous dimensions of $[\mathcal{O}\mathcal{O}]_n$ and $[\widetilde{\mathcal{O}}\widetilde{\mathcal{O}}]_n$, we need to extract the coefficient of the double poles at $\Delta=2\Delta_\phi+2n$ and $\Delta=2\bar\Delta_\phi+2n$ in (\ref{eq:rhoPint}), where $\Delta\equiv\frac{d}{2}+i\lambda$, as explained in section \ref{subsec:qftint}. From eq. (\ref{eq:sumB}), we know that the bubble function $B_{\Delta_\phi, \Delta_\phi}$ ($B_{\bar\Delta_\phi, \bar\Delta_\phi}$) has a single pole at $2\Delta_\phi+2n$ ($2\bar\Delta_\phi+2n$). In addition, $\rho^{\CP,0}_{\phi^2,\text{free}}(\lambda)$ also has  single poles at these points. Therefore, in this case, the coefficient $c_2$ defined by eq. (\ref{c2def}) should be 
\begin{equation}\label{c2rho}
\begin{aligned}
    c_2^{[\mathcal{O}\mathcal{O}]_n}&=\frac{\sin\left(\pi\left(\frac{d}{2}+2i\lambda_\phi\right)\right)}{4\sinh^2(\pi\lambda_\phi)}\frac{a_{\Delta_\phi,\Delta_\phi}(n)}{d-4n-4\Delta_\phi} c_0^{[\mathcal{O}\mathcal{O}]_n} \,,\\
c_2^{[\widetilde{\mathcal{O}}\widetilde{\mathcal{O}}]_n}&=\frac{\sin\left(\pi\left(\frac{d}{2}-2i\lambda_\phi\right)\right)}{4\sinh^2(\pi\lambda_\phi)}\frac{a_{\bar\Delta_\phi,\bar\Delta_\phi}(n)}{d-4n-4\bar\Delta_\phi}c_0^{[\widetilde{\mathcal{O}}\widetilde{\mathcal{O}}]_n}\,,
\end{aligned}
\end{equation}
 where
\begin{align}\label{c0rho}
     c_0^{[\mathcal{O}\mathcal{O}]_n}=\underset{\Delta=2\Delta_\phi+2n}{\text{Res}}\rho^{\CP,0}_{\phi^2,\text{free}}(\lambda), \,\,\,\,\,   c_0^{[\widetilde{\mathcal{O}}\widetilde{\mathcal{O}}]_n}=\underset{\Delta=2\bar\Delta_\phi+2n}{\text{Res}}\rho^{\CP,0}_{\phi^2,\text{free}}(\lambda)~.
\end{align}
Plugging eq. (\ref{c2rho}) into eq. (\ref{eq:anomalousdimsgen}), we obtain the anomalous dimensions of $[\mathcal{O}\mathcal{O}]_n$ and $[\widetilde{\mathcal{O}}\widetilde{\mathcal{O}}]_n$ respectively
\begin{align}\label{gammaina}
&\gamma^{[\mathcal{O}\mathcal{O}]_n}=g\frac{c_2^{[\mathcal{O}\mathcal{O}]_n}}{c_0^{[\mathcal{O}\mathcal{O}]_n}}=g\,\frac{\sin\left(\pi\left(\frac{d}{2}+2i\lambda_\phi\right)\right)}{4\sinh^2(\pi\lambda_\phi)}\frac{a_{\Delta_\phi,\Delta_\phi}(n)}{d-4n-4\Delta_\phi}
\nonumber\\
&\gamma^{[\widetilde{\mathcal{O}}\widetilde{\mathcal{O}}]_n}=g\,\frac{c_2^{[\widetilde{\mathcal{O}}\widetilde{\mathcal{O}}]_n}}{c_0^{[\widetilde{\mathcal{O}}\widetilde{\mathcal{O}}]_n}}=g\,\frac{\sin\left(\pi\left(\frac{d}{2}-2i\lambda_\phi\right)\right)}{4\sinh^2(\pi\lambda_\phi)}\frac{a_{\bar\Delta_\phi,\bar\Delta_\phi}(n)}{d-4n-4\bar\Delta_\phi}\,,
\end{align}
where $a_{\Delta_1,\Delta_2}(n)$ is given by eq. (\ref{afunction}).

\bibliographystyle{JHEP}
\bibliography{bibliography}
\end{document}